\documentclass{aa}  
\usepackage{txfonts}
\usepackage{graphicx,natbib}
\bibpunct{(}{)}{;}{a}{}{,} 
\newcommand{\ltsima} {$\; \buildrel < \over \sim \;$}  
\newcommand{\gtsima} {$\; \buildrel > \over \sim \;$}  
\newcommand{\lta} {\lower.5ex\hbox{\ltsima}}  
\newcommand{\gta} {\lower.5ex\hbox{\gtsima}}  
\newcommand{\ha} {H$\alpha$}

\newcommand{\kms}{$\rm{\,km \,s}^{-1}$}

\newcommand{\forb}[2]{\mbox{$[{\rm #1\, #2}]$}}
\newcommand{\oiii}{\forb{O}{III}}

\begin{document}

\title{The MURALES survey. VI.} \subtitle{Properties and origin of the
  extended line emission structures in radio galaxies}
\author{Barbara Balmaverde\inst{1} 
                \and Alessandro Capetti\inst{1}
                \and R.D. Baldi\inst{2}
                \and S. Baum\inst{3}
                \and M. Chiaberge\inst{4,5}
                \and R. Gilli\inst{6}
                \and Ana Jimenez-Gallardo\inst{7,1,8,9}
                \and Alessandro Marconi\inst{10,11}
                \and Francesco Massaro\inst {7,1,9}
                \and E. Meyer\inst{12}
                \and C. O$'$Dea\inst{3}
                \and G. Speranza\inst{13,14}
                \and E. Torresi\inst{6} 
                \and Giacomo Venturi\inst{15,11}                
} \institute {INAF - Osservatorio Astrofisico di Torino, Via
  Osservatorio 20, I-10025 Pino Torinese, Italy \and INAF- Istituto di
  Radioastronomia, Via Gobetti 101, I-40129 Bologna, Italy \and
  Department of Physics and Astronomy, University of Manitoba,
  Winnipeg, MB R3T 2N2, Canada \and Space Telescope Science Institute,
  3700 San Martin Dr., Baltimore, MD 21210, USA \and Johns Hopkins
  University, 3400 N. Charles Street, Baltimore, MD 21218, USA \and
  INAF - Osservatorio di Astrofisica e Scienza dello spazio di
  Bologna, Via Gobetti 93/3, I-40129 Bologna, Italy \and Dipartimento
  di Fisica, Universit\`a degli Studi di Torino, via Pietro Giuria 1,
  I-10125 Torino, Italy \and European Southern Observatory, Alonso de
  C\`ordova 3107, Vitacura, Regi\`on Metropolitana, Chile \and
  Istituto Nazionale di Fisica Nucleare, Sezione di Torino, I-10125
  Torino, Italy \and Dipartimento di Fisica e Astronomia, Universit\`a
  di Firenze, via G. Sansone 1, 50019 Sesto Fiorentino (Firenze),
  Italy \and INAF - Osservatorio Astrofisico di Arcetri, Largo Enrico
  Fermi 5, I-50125 Firenze,Italy \and University of Maryland Baltimore
  County, 1000 Hilltop Circle, Baltimore, MD 21250, USA \and Instituto
  de Astrof\`isica de Canarias, Calle V\`ia L\`actea, s/n, 38205, La
  Laguna, Tenerife, Spain \and Departamento de Astrof\`isica,
  Universidad de La Laguna, 38206, La Laguna, Tenerife, Spain \and
  Instituto de Astrof\`isica, Facultad de F\`isica, Pontificia
  Universidad Cat\`olica de Chile, Casilla 306, Santiago 22, Chile }
\offprints{barbara.balmaverde@inaf.it} \date{} 

\abstract{This is the sixth paper presenting the results of the MUse
  RAdio Loud Emission line Snapshot (MURALES) survey. We observed 37
  radio sources from the 3C sample with $z<0.3$ and a declination
  $<20^\circ$ with the Multi Unit Spectroscopic Explorer (MUSE)
  optical integral field spectrograph at the Very Large
  Telescope. Here, we focus on the properties of the extended emission
  line regions (EELRs) that can be studied with unprecedented detail
  thanks to the depth of these observations. Line emission in the ten
  FR~Is is, in most cases, confined to within $\lesssim 4$kpc, while
  large-scale ($\gtrsim 4$ kpc) ionized gas is seen in all but two of
  the 26 FR~IIs. It usually takes the form of elongated or filamentary
  structures, typically extending between 10 and 30 kpc, but also
  reaching distances of $\sim$80 kpc. We find that the large-scale
  ionized gas structures show a tendency to be oriented at large
  angles from the radio axis, and that the gas on a scale of a few
  kiloparsecs from the nucleus often shows ordered rotation with a
  kinematical axis forming a median angle of 65$^\circ$ with the radio
  axis. We also discuss the velocity field and ionization properties
  of the EELRs. The observed emission line structures appear to be
  associated with gaseous ``superdisks'' that formed after a gas-rich
  merger. The different properties of the EELR can be explained with a
  combination of the source evolutionary state and the orientation of
  the superdisk with respect to the radio axis. The general alignment
  between the superdisks and radio axis might be produced by stable
  and coherent accretion maintained over long timescales.}
  
  \keywords{Galaxies: active -- Galaxies: ISM --
  Galaxies: nuclei -- galaxies: jets}

\titlerunning{The MURALES survey. VI.} 
\authorrunning{B. Balmaverde et al.}
 \maketitle

\section{Introduction}
\label{intro}

Radio galaxies (RGs) are among the most energetic manifestations of
active galactic nuclei (AGNs) and harbor the most supermassive black
holes (SMBHs) in the Universe, typically hosted in the brightest
galaxies at the center of clusters or groups. They are therefore
extraordinarily relevant when addressing important unknowns relating the
interaction between SMBHs to their environment. It is in fact becoming
increasingly clear that feedback from AGNs, that is 
the exchange of energy and matter between the active nucleus and the
surrounding medium, is a fundamental ingredient in the formation and
evolution of large-scale astrophysical structures and for the
luminosity function of galaxies (see, e.g., \citealt{fabian12}).

The evidence of {\rm kinetic} AGN feedback is often witnessed in local
radio galaxies, for example from the presence of cavities inflated by
the radio emitting gas seen in the X-ray images (e.g.,
\citealt{birzan04,mcnamara12}). However, we still lack a comprehensive
view of the effects that highly collimated jets and nuclear emission
have on the host and its immediate environment. In particular,
questions remain about the coupling between radio outflows and ionized
gas, and whether the AGN outflows enhance or quench star formation
(positive or negative feedback, e.g., \citealt{carniani16}) and under
which conditions.

In jetted AGNs, the study of the properties of ionized gas
represents a powerful tool to explore AGN feedback. Although 
warm gas is a relatively minor constituent of the interstellar medium
(ISM), the optical emission lines that it produces provide us with
several key diagnostics, as it enables us to study its kinematics,
measure physical quantities (such as density and mass), and to identify
its ionization mechanism.

Large-scale structures of optical line-emitting gas have been observed
around quasistellar objects (QSOs) since the 1970s
\citep{wampler75,stockton76,richstone77}. Follow-up surveys of
low-redshift QSOs (z$<$0.45) showed that optical extended emission is
mostly found around steep-spectrum radio-loud quasars
\citep{stockton83,stockton87}. In some cases, the ionized gas is
distributed in complex filaments, which is apparently unrelated to the stars
in the host galaxy or the radio structure
\citep{stockton02,stockton06}.  \citet{stockton87} noted that the
presence of close companion galaxies or a disturbed morphology in
optical images suggests that the extended ionized gas could be
associated with tidal debris resulting from interactions or
mergers. The origin of the gas and the physical mechanism that
produces the distribution and the ionization properties of the gas in
the extended emission line regions (EELRs) are still uncertain.

\citet{chambers87} and \citet{mccarthy87} first demonstrated that the
stellar continuum and the emission lines are both aligned along the
radio axis in sources at a redshift $\gtrsim$0.6. A tendency for a
spatial relationship between the gas in the EELRs and the radio
source, in both radio galaxies and radio-loud, steep-spectrum quasars
in lower redshift objects (z$<0.6$) was also found by \citet{baum89a}.
In these sources, the extent of the radio source is similar to or
greater than the extent of the emission-line gas. These authors
pointed out that large \citet{fanaroff74} type II (FR~II) radio
sources ($>$ 150 kpc) show the brightest emission line features along
an axis that is skewed relative to the radio source axis and they
detected emission-line filaments perpendicular to the radio structure
in several sources. \citet{mccarthy89} demonstrated that the degree of
alignment increases with redshift.

The primary mechanisms for the emission-line gas alignment are thought
to be shocks induced by the radio jets and photoionization from the
central AGN (e.g., \citealt{baum89,baum89b,baum92,mccarthy93,best00}).
\citet{tadhunter98} suggested that the density of emitting line clouds
is enhanced at higher redshift because of the compression effect of
shocks driven through the ISM by the radio jets becomes
increasingly important as the redshift and radio power increase.
Therefore, the alignment depends on both the presence of extended gas
in the environment as well as the ability of the radio source to
influence its environment (see also \citealt{inskip02,inskip05}).
There are a few examples of emission-line gas at low z in which bright
compact emitting line regions are cospatial to radio jets or radio
knots  (e.g., \citealt{stockton87} and \citealt{tadhunter00}).

\begin{figure*}  
\centering{ 
\includegraphics[width=2\columnwidth]{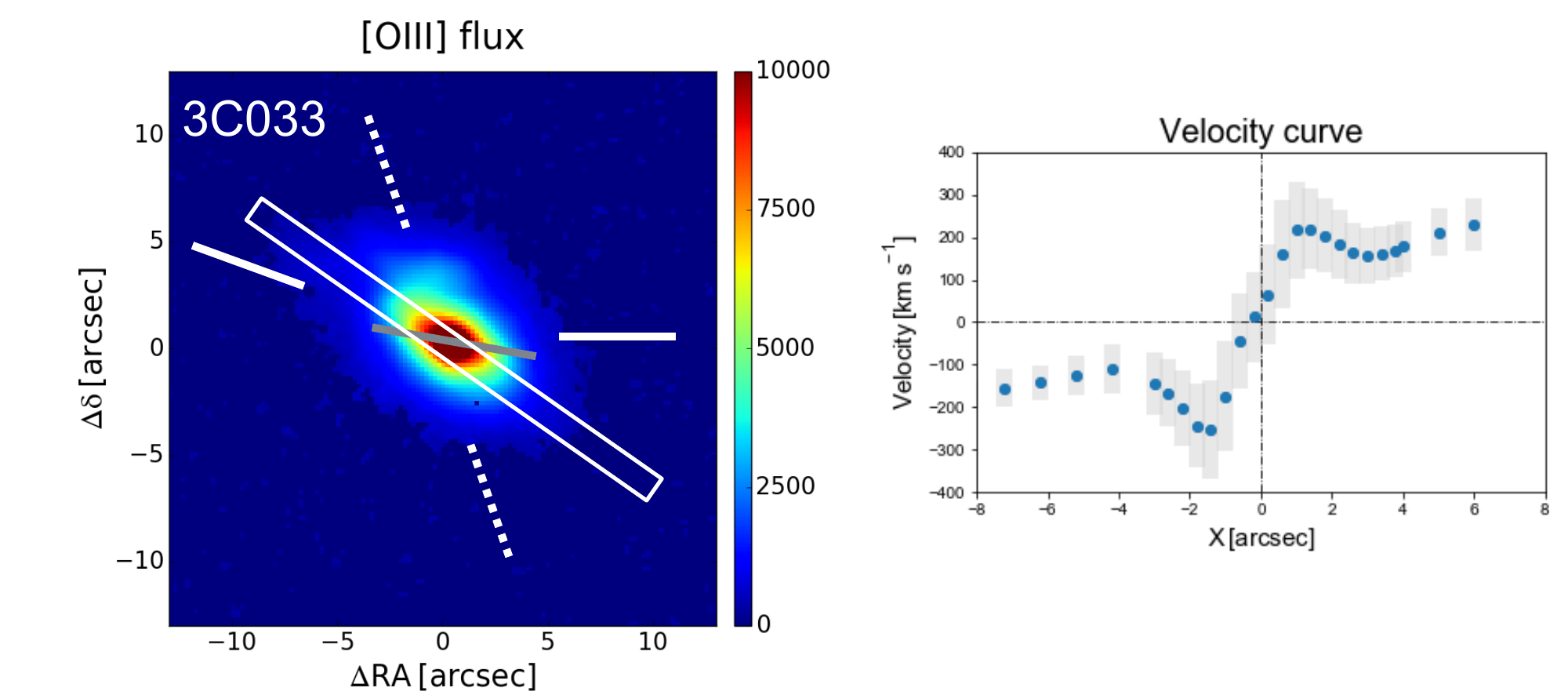}
\caption{Example of the emission line properties. Left: [O~III]
  emission line image of 3C~33. The white dotted lines mark the radio
  position angle, the solid white lines show the orientation of the
  emission line structures on both sides of the nucleus, and the gray
  line marks the orientation of the kinematical axis in the inner
  regions measured with {kinemetry} (see Sect. \ref{kin}). The
  rectangle represents the region from which we extracted the velocity
  curve shown in the right panel. Data were extracted on each 0\farcs2
  pixel in the innermost regions and then at distances of
  1\arcsec. The length of the gray bars represents the line width at
  each location.}}
\label{ppt}
\end{figure*}

\begin{figure}  
\centering{\includegraphics[width=9.0cm]{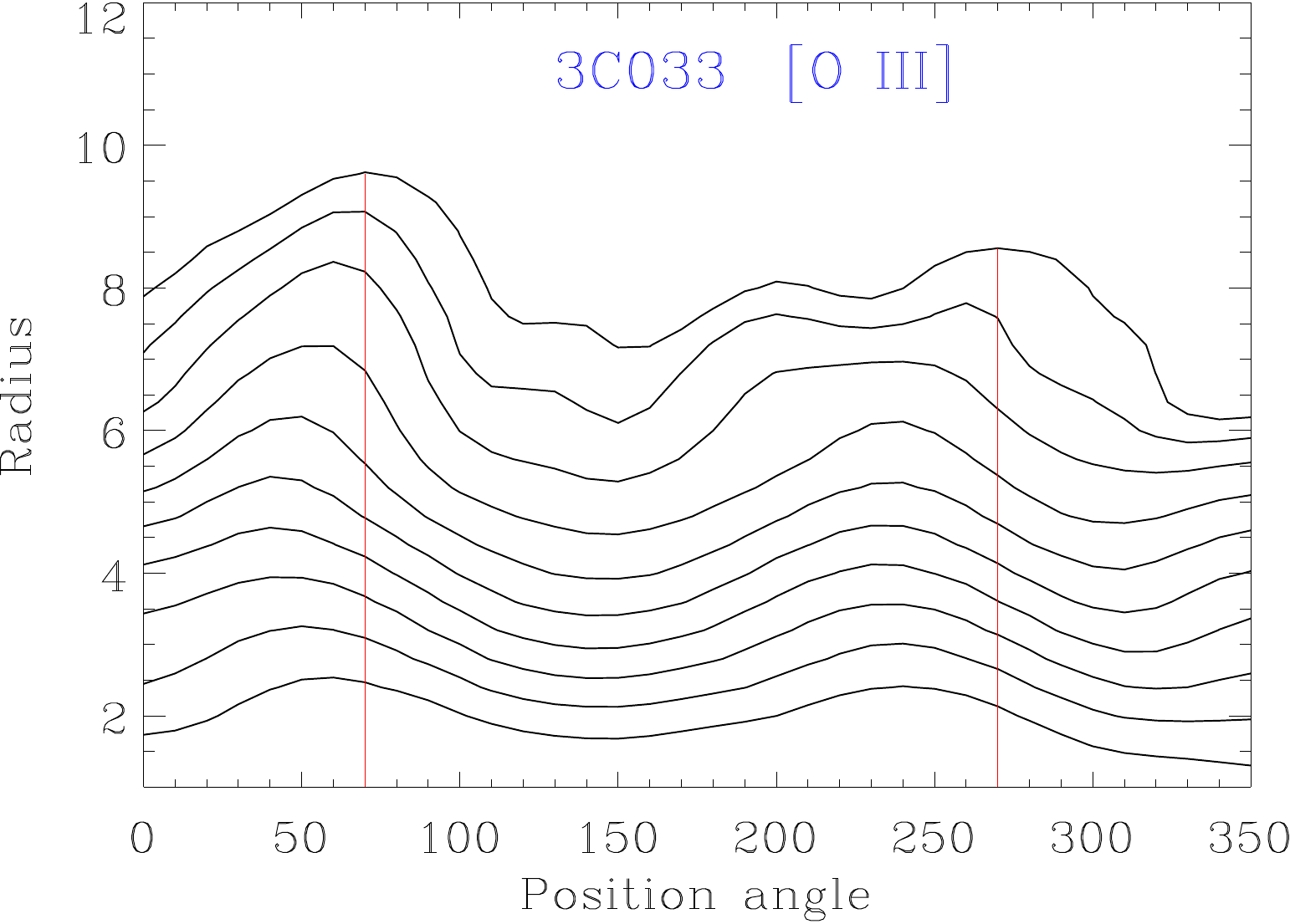}}
\caption{Polar diagram of the [O~III] emission line in 3C~33. Contour
  levels are drawn starting from three times the root mean square of the
  images and increase in geometric progression with a common ratio of
  two. The red solid lines mark the ionized gas position angle on each
  side of the nucleus.}
\label{polar}
\end{figure}

\begin{figure}  
\centering{ 
\includegraphics[width=9cm]{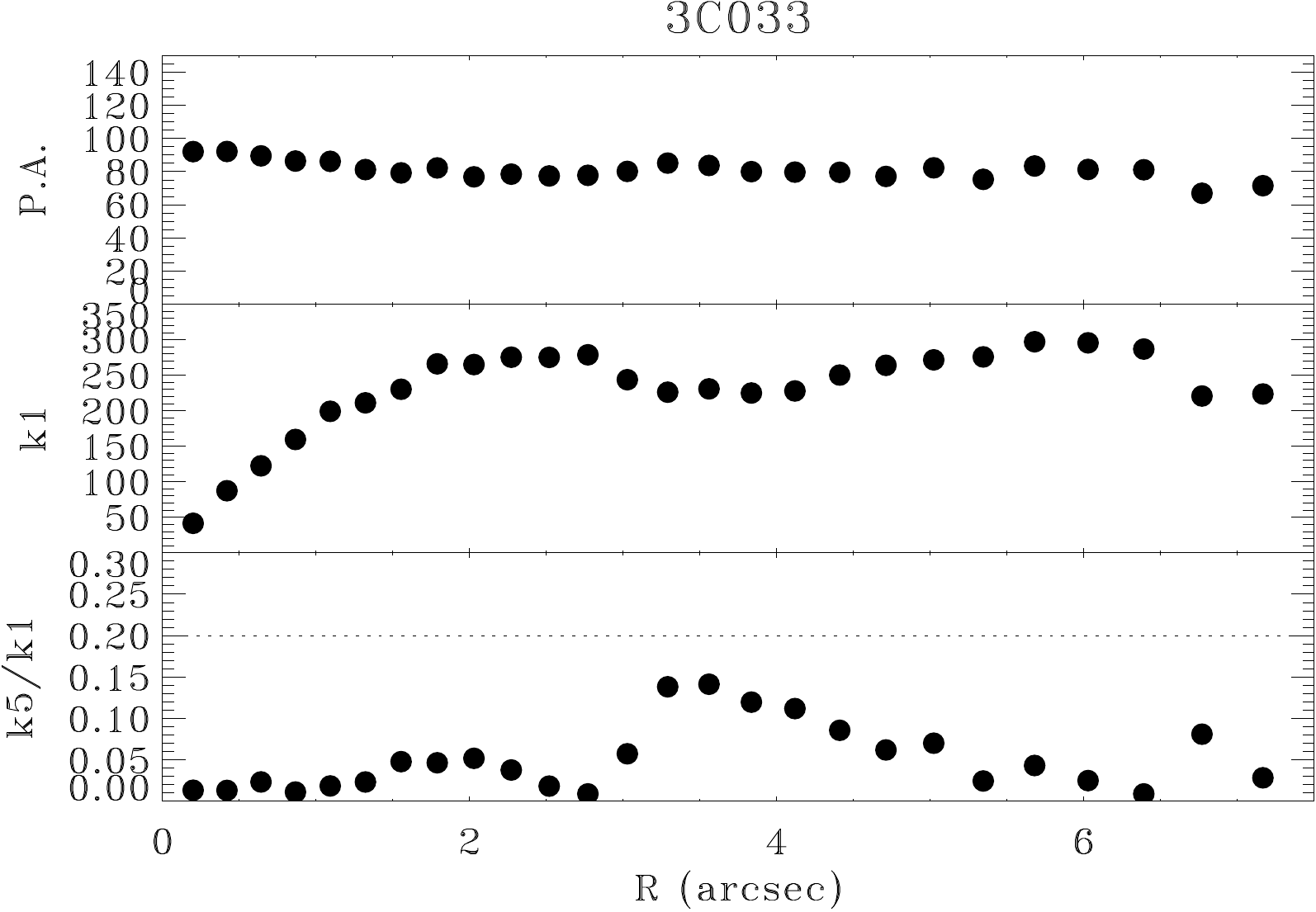}
\caption{Results obtained for 3C~33 with the {kinemetry} software
  \citep{krajnovic06}. From top to bottom: Kinematical position
  angle $P.A.$ (in degrees), the {\sl k1} parameter, corresponding to
  the amplitude of the rotation curve (in \kms), and the ratio between
  the fifth and first coefficient $k5/k1$, which quantifies the
  deviations from ordered rotation.}
\label{kinemetry403}}
\end{figure}   

\begin{figure*}  
\includegraphics[width=9.35cm]{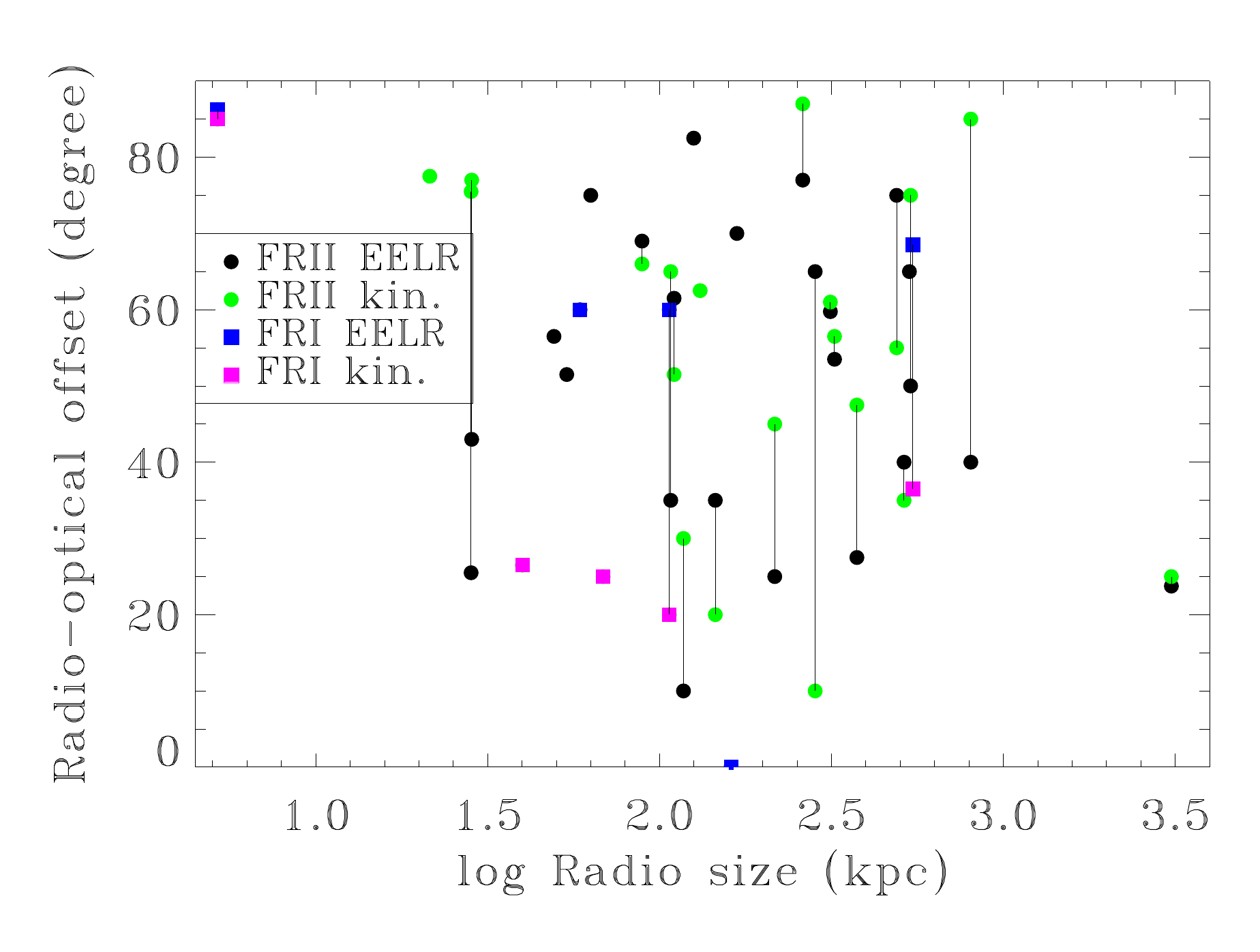}
  \includegraphics[width=9.22cm]{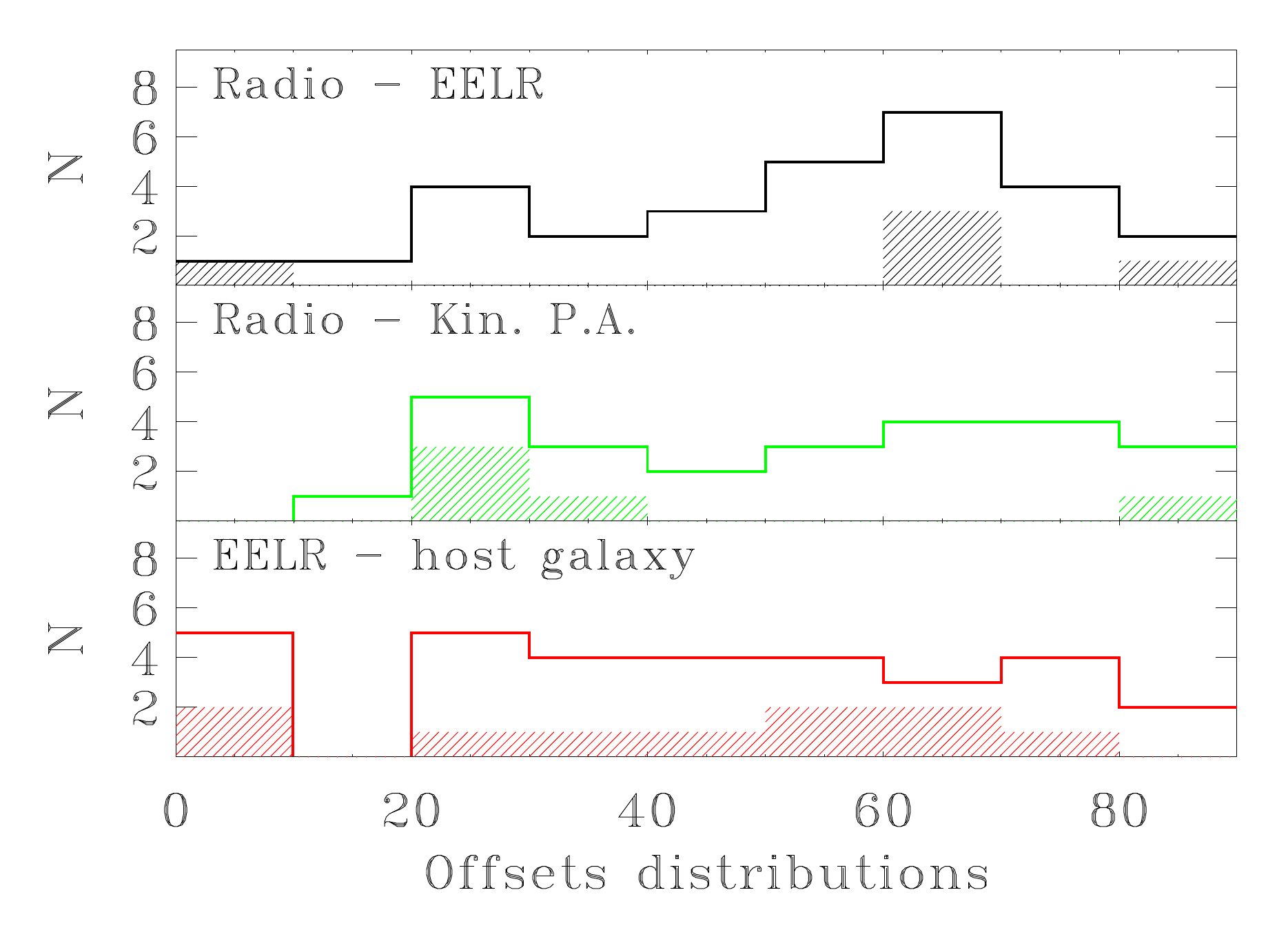}
\caption{Comparison of the radio and optical axis. Left: Offsets
  between the position angle of the most extended line structure and
  the radio axis against the L.A.S. of the radio sources. FR~Is are
  marked with blue squares. The green circles are the offsets between
  the radio and the kinematical axis derived from {kinemetry} for the
  FR~IIs (magenta squares are for the FR~Is). The vertical lines
  connect the two offsets measurements for the same source when both
  are available. Right: Distributions of the offsets of the radio axis
  from (top) the emission line structures, averaged between the two
  sides, (center) the kinematical axis, and (bottom) the host's major
  axis. The shaded area of the histograms is the contribution of the
  FR~Is.}
\label{offset}
\end{figure*}

\begin{table*}
\caption{Sample properties.}
\begin{tabular}{ l c c l | r r | l r r r | r | r c}
\hline
Name & redshift &  FR & Exc. & \multicolumn{2}{c}{Radio} &
\multicolumn{4}{|c|}{Line emission} & Kinemetry & Host \\
     &  &   & class &        P.A.     & L.A.S. &      & P.A. (1) & P.A. (2) & r (kpc) & P.A. & P.A. \\
\hline            
3C~015   & 0.073 & I  & LEG  & 152 &  59 & [N~II]    & 120 &   0  &    6.7 & --- &  20  \\ 
3C~017   & 0.220 & II & BLO  & 156 &  49 & [N~II]    & 280 &      &   11.6 & --- &   0  \\  
3C~018   & 0.188 & II & BLO  & 163 & 167 & [O~III]   &  50 & 270  &   20.1 & --- & 165  \\ 
3C~029   & 0.045 & I  & LEG  & 160 & 115 & [O~III]   &     &      &        & --- & 100  \\ 
3C~033   & 0.060 & II & HEG  &  19 & 313 & [O~III]   &  70 & 270  &   10.7 &  80 & 165  \\ 
3C~040   & 0.018 & I  & LEG  &  13 & 450 & [N~II]    &     &      &        & --- & 100  \\ 
3C~063   & 0.175 & II & HEG  &  33 &  53 & H$\alpha$ & 260 &  90  &   31.5 & --- &  75  \\ 
3C~076.1 & 0.032 &  I & --   & 130 & 106 & [N~II]    &  50 & 270  &    4.2 & 110 & 130  \\   
3C~078   & 0.028 &  I & LEG  &  53 &  39 & [N~II]    &     &      &        &  80 & 155  \\  
3C~079   & 0.256 & II & HEG  & 105 & 537 & [O~III]   & 160 & 330  &   35.7 &   0 &  15  \\  
3C~088   & 0.030 & II & LEG  &  55 & 107 & [N~II]    &  90 &      &    3.7 & 120 & 150  \\  
3C~089   & 0.138 &  I & --   & 120 & 161 & [N~II]    & 300 &      &    8.8 & --- &  40  \\  
3C~098   & 0.030 & II & HEG  &  25 & 145 & [O~III]   & 350 & 170  &   16.3 &   5 &  55  \\  
3C~105   & 0.089 & II & HEG  & 127 & 131 & [O~III]   &     &      &        &  10 &  50  \\  
3C~135   & 0.125 & II & HEG  &  75 & 216 & [O~III]   & 230 &  50  &   43.7 &  30 & 130  \\  
3C~180   & 0.220 & II & HEG  &  10 &3080 & [O~III]   &  40 & 210  &   34.9 &  35 &  30  \\  
3C~196.1 & 0.198 & II & LEG  &  45 &  63 & [O~III]   &  60 &      &   14.5 & --- &  50  \\  
3C~198   & 0.081 & II & SF   &  30 & 117 & H$\alpha$ & 220 &      &   35.2 &   0 &  65  \\  
3C~227   & 0.086 & II & BLO  &  85 & 514 & [O~III]   &  50 & 220  &   49.6 & 120 &   0  \\  
3C~264   & 0.021 &  I & LEG  & 120 &  68 & [N~II]    &     &      &        & 145 &  15  \\  
3C~272.1 & 0.003 &  I & LEG  & 175 &   5 & [N~II]    &  80 & 260  &    0.5 &  80 & 130  \\  
3C~287.1 & 0.216 & II & BLO  &  90 & 283 & [O~III]   & 170 &  40  &   27.0 & 100 &  30  \\  
3C~296   & 0.024 &  I & LEG  &  35 & 130 & [N~II]    &     &      &        & --- & 145  \\  
3C~300   & 0.270 & II & HEG  & 132 & 374 & [O~III]   & 130 & 260  &   36.3 &  85 & 125  \\
3C~318.1 & 0.045 & -- &  --  &     &     & [N~II]    & 190 &      &   18.1 & --- & 105  \\  
3C~327   & 0.105 & II & HEG  & 100 & 489 & [N~II]    & 170 &  20  &   16.3 &  45 & 135  \\  
3C~348   & 0.155 & I  & ELEG &  36 & 545 & H$\alpha$ & 280 & 110  &   28.2 &   0 & 145  \\  
3C~353   & 0.030 & II & LEG  &  82 & 125 & H$\alpha$ &   0 & 180  &   15.4 & --- &  75  \\ 
3C~386   & 0.017 & II &  --  &  16 &  88 & [N~II]    &  80 & 270  &   10.1 & 130 & 105  \\  
3C~403   & 0.059 & II & HEG  &  86 & 110 & [O~III]   & 200 &  30  &    8.7 &  35 &  40  \\   
3C~403.1 & 0.055 & II & LEG  & 127 & 260 & H$\alpha$ & 230 &  50  &    5.3 &  20 & 150  \\  
3C~424   & 0.127 & II & LEG  & 155 &  28 & [N~II]    & 130 &      &   29.0 &  80 &  80  \\  
3C~442   & 0.026 & II & LEG  &  63 & 322 & [N~II]    & 190 &      &    4.1 & 120 & 130  \\  
3C~445   & 0.056 & II & BLO  & 173 & 532 & [O~III]   & 235 & 105  &   15.7 & --- & ---  \\  
3C~456   & 0.233 & II & HEG  &  17 &  21 & [O~III]   &     &      &        & 120 & ---  \\   
3C~458   & 0.289 & II & HEG  &  75 & 804 & [O~III]   &  30 & 220  &   81.2 &  20 &  55  \\   
3C~459   & 0.220 & II & BLO  &  97 &  28 & [O~III]   & 320 &      &   59.6 & 120 & ---  \\
  \hline
\end{tabular}
\label{tab1}

\smallskip
\small{Column description: (1) source name; (2) redshift; (3 and 4)
  radio morphological classification and excitation class from
  \citet{buttiglione10} (LEG = low excitation galaxy, HEG = high
  excitation galaxy, BLO = broad line object, SF = star-forming
  galaxy, ELEG = extremely low excitation galaxy); (5 and 6) position
  angle and largest angular size of the radio source with uncertainties
  usually between 5$^\circ$ and 10$^\circ$.; (7 through 9) emission
  line considered and position angles of the ionized gas on each side of
  the nucleus, (10) largest distance of emission line detection, (11)
  perpendicular to the line of nodes derived with the kinemetry
  software ({kinematical axis}), (12) host position angle.}
\end{table*}

Significant progress in understanding the fueling and evolution of the
activity of radio-loud AGNs, the triggering process of the radio
emission, and its impact on the environment has been made studying the
third Cambridge catalog of RGs \citep{laing83,spinrad85}. The 3C
catalog is the premiere statistically complete sample of powerful RGs
selected at 178 MHz with a flux density limit of 9 Jy; it covers a
large range of radio power, it contains all types of RGs from the
point of view of their optical spectrum, radio morphology, and
environment. In the last two decades, a superb suite of ground- and
space-based observations at all accessible wavelengths has been built
with all major observing facilities from Hubble Space Telescope (with
images from the ultraviolet to the near-infrared and also including
emission line imaging), to Chandra, Spitzer, Herschel, Very Large
Array, as well as ground-based optical spectra.

\citet{privon08} and \citet{baldi19} presented HST narrow-band
emission-line images of a large sample of 3CR radio
sources. \citeauthor{privon08} confirmed a weak alignment at low
redshift (z $<$ 0.6) between the radio and optical emission line
structures and suggested that both mechanical and radiant energy are
responsible for this alignment. The dominant mechanism responsible for
the emission line is photoionization by nuclear radiation,
but the relation between the line emission, radio, and X-ray
luminosities indicates that the kinetic energy deposited in the EELRs
by the jets has an important contribution. 

\begin{figure*} [h]
\centering{ 
  \includegraphics[width=0.66\columnwidth]{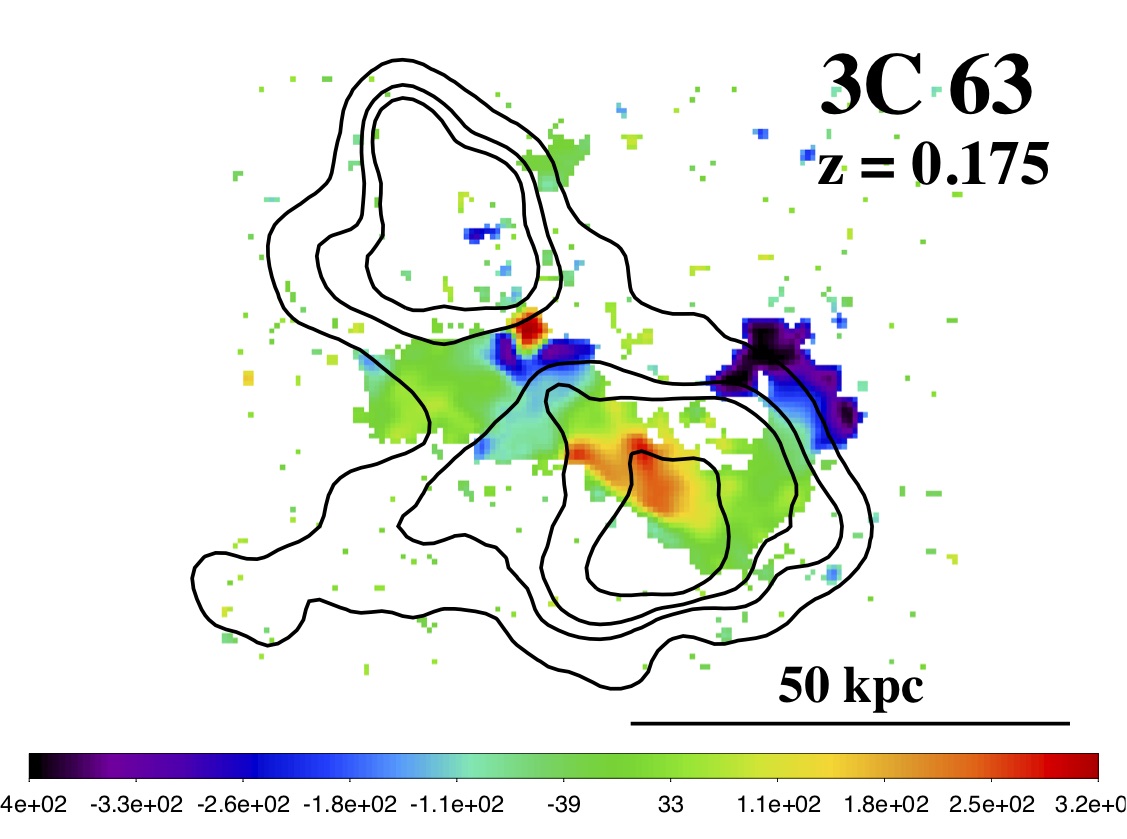}
  \includegraphics[width=0.62\columnwidth]{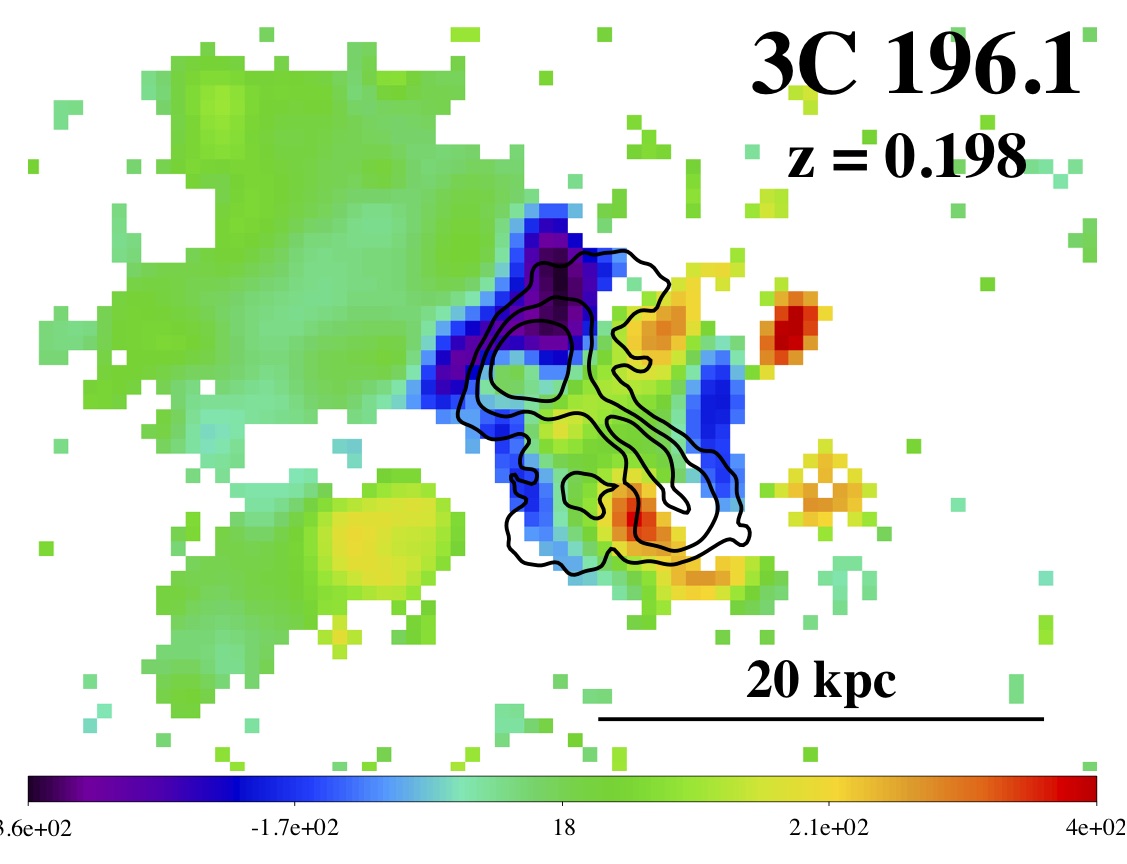}
  \includegraphics[width=0.70\columnwidth]{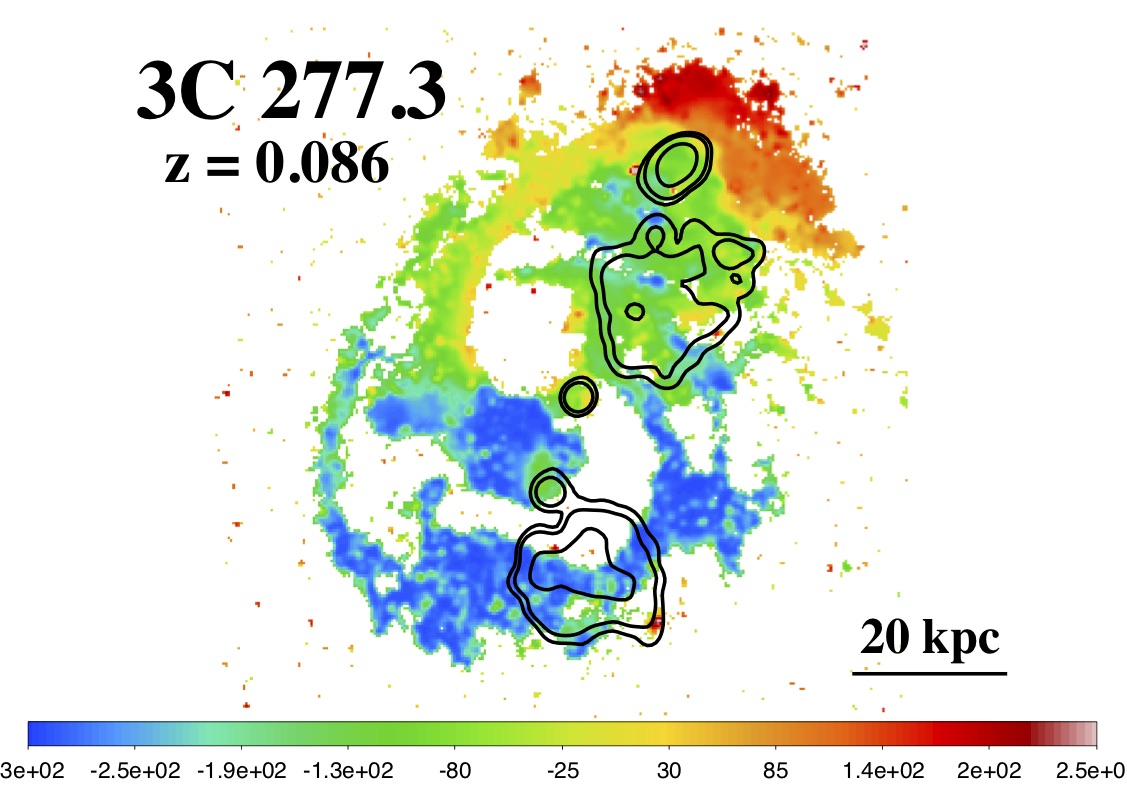}
  \includegraphics[width=0.66\columnwidth]{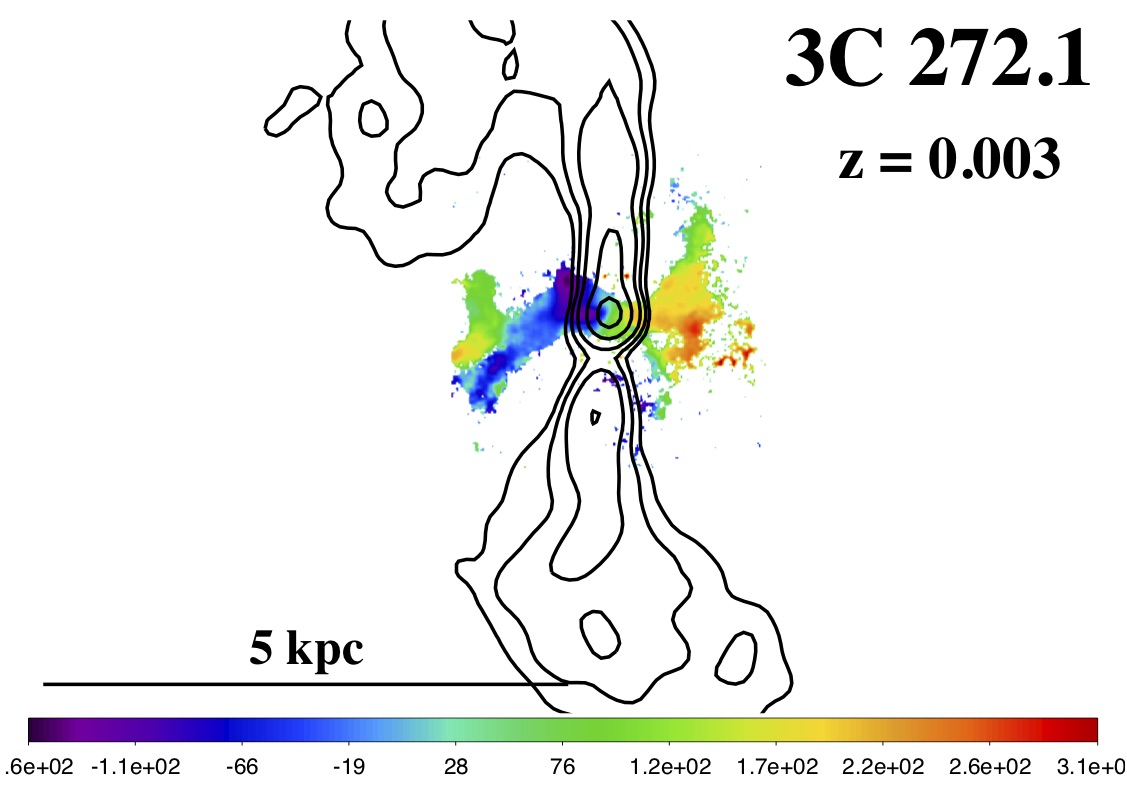}
  \includegraphics[width=0.66\columnwidth]{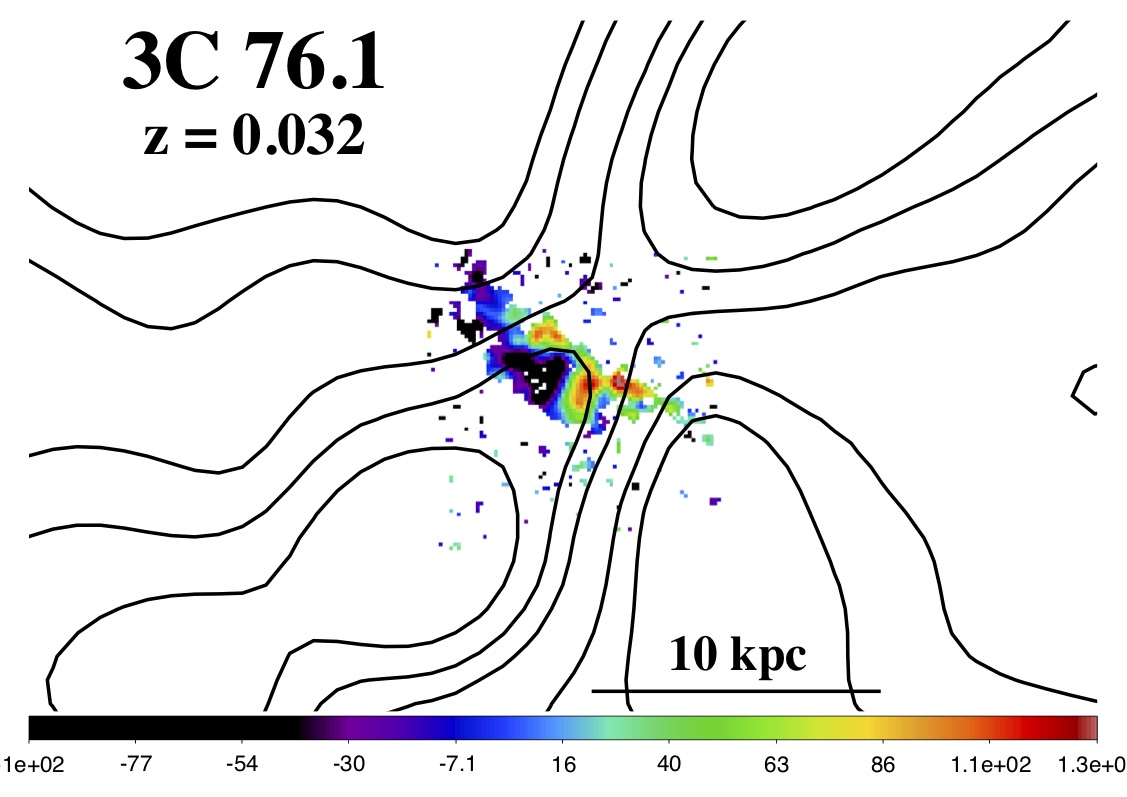}
  \includegraphics[width=0.66\columnwidth]{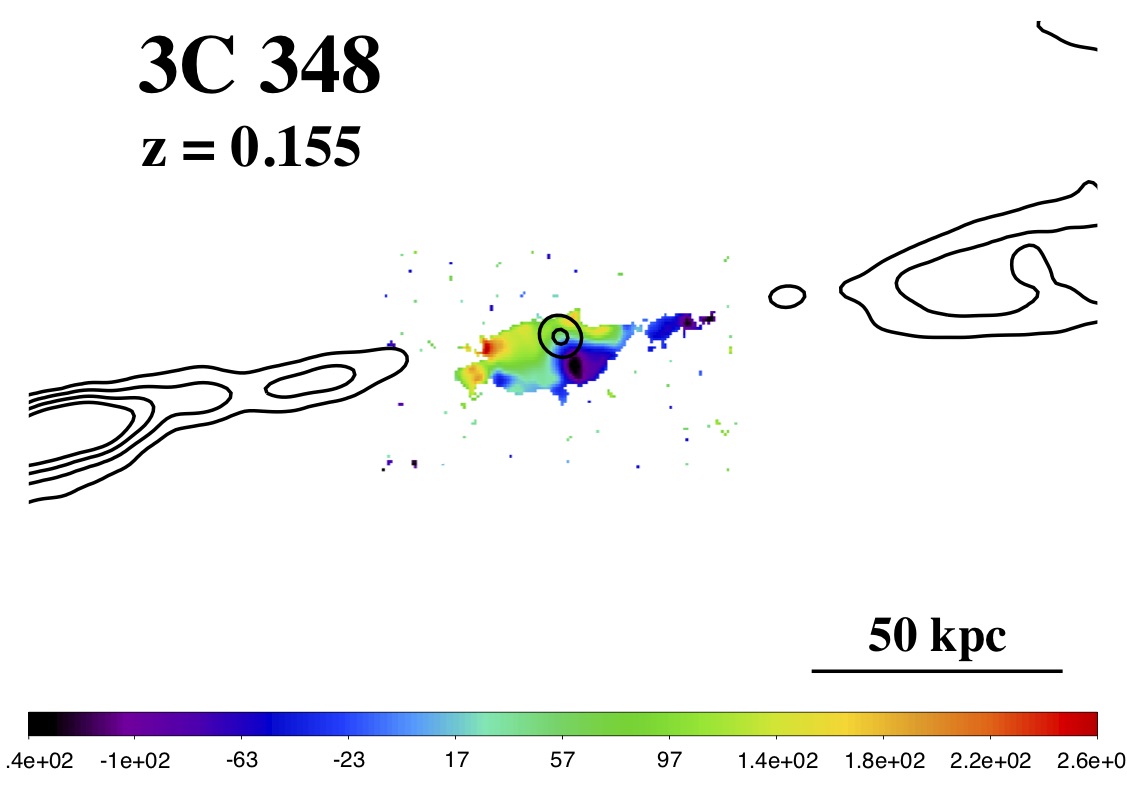}
  \includegraphics[width=0.66\columnwidth]{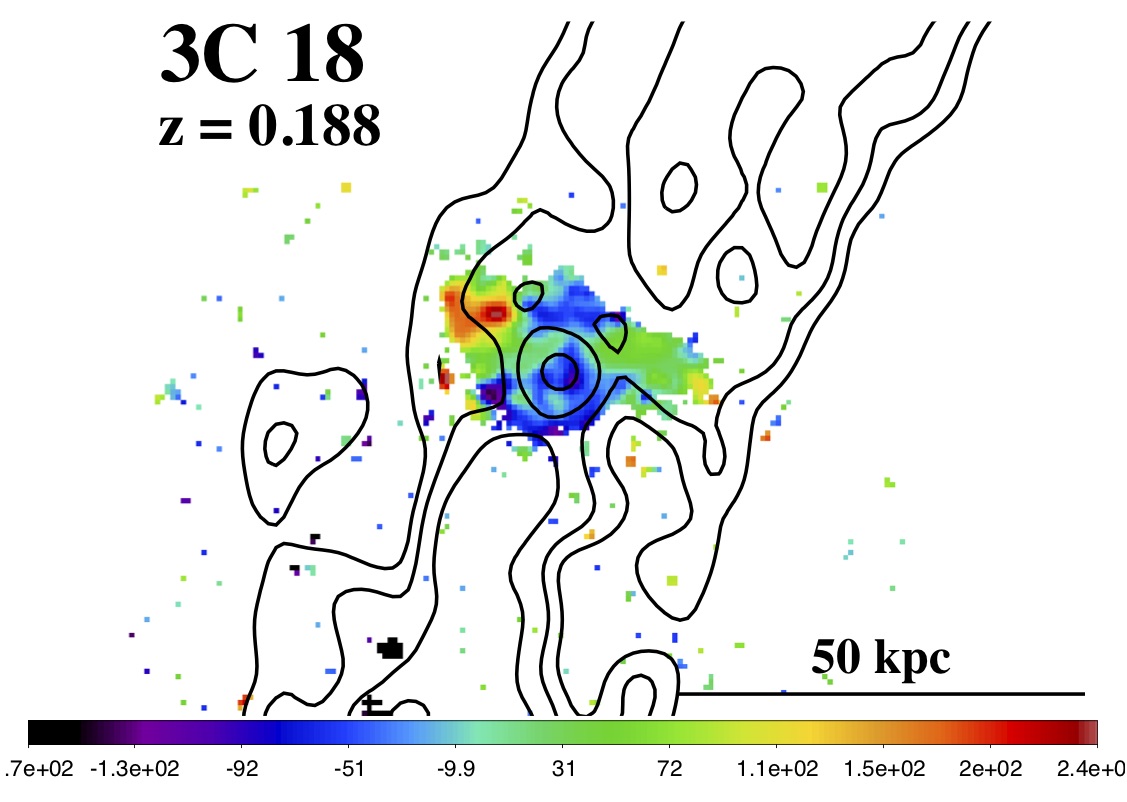}
  \includegraphics[width=0.66\columnwidth]{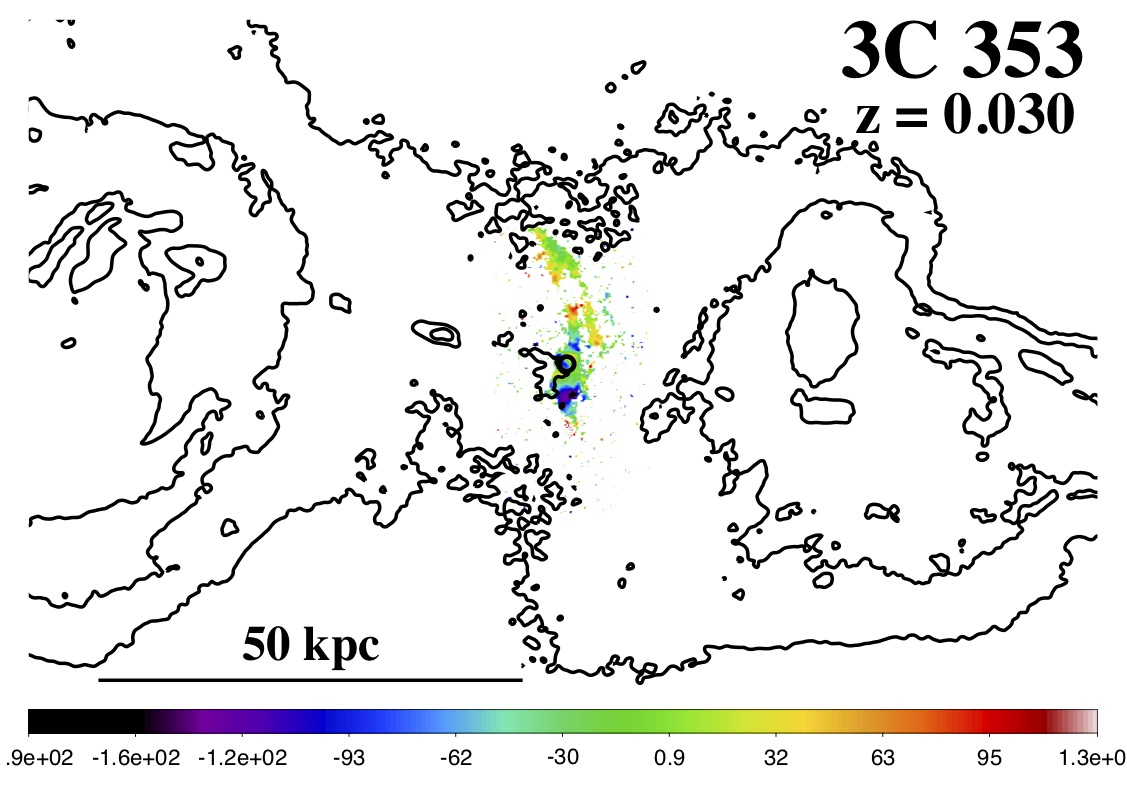}
 \includegraphics[width=0.66\columnwidth]{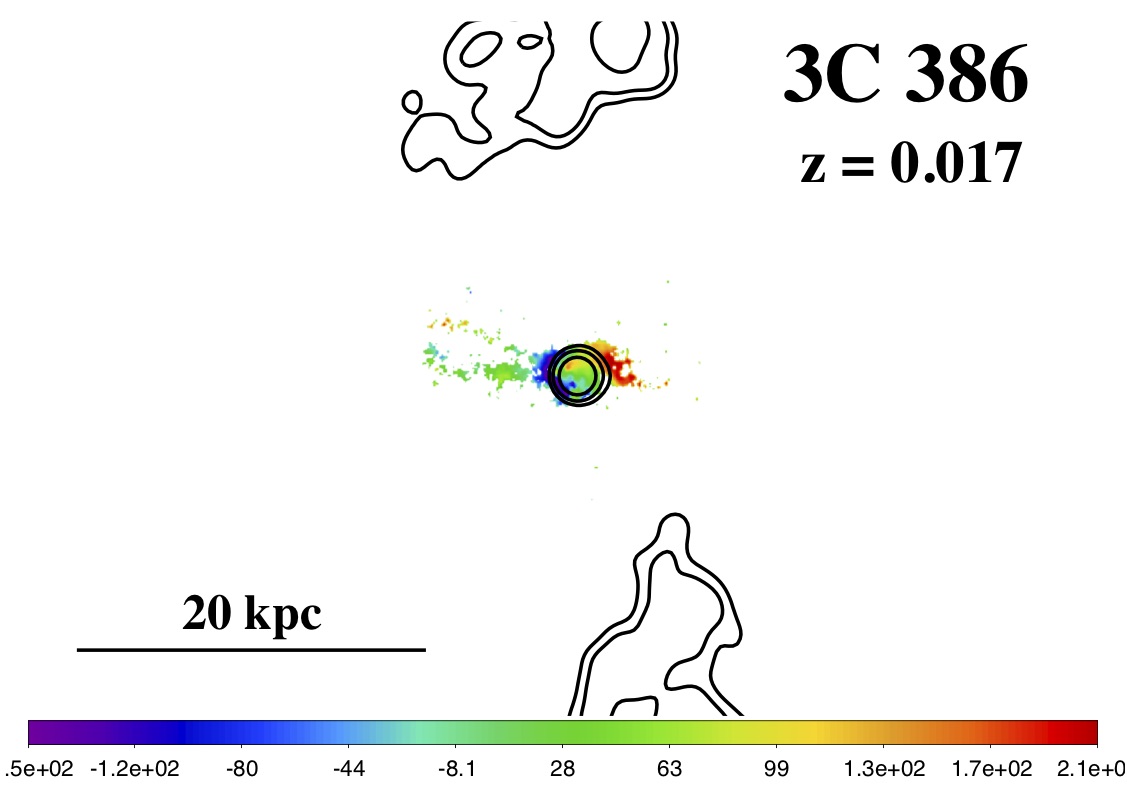}
\includegraphics[width=0.66\columnwidth]{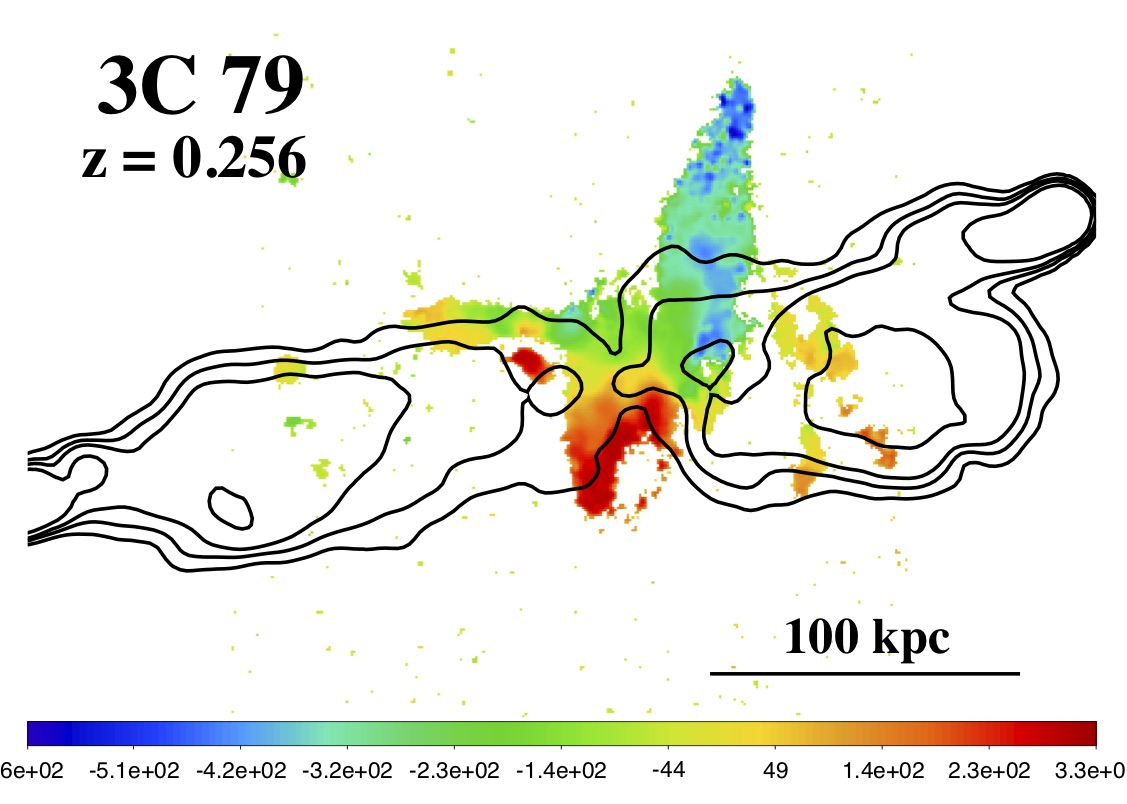}
 \includegraphics[width=0.66\columnwidth]{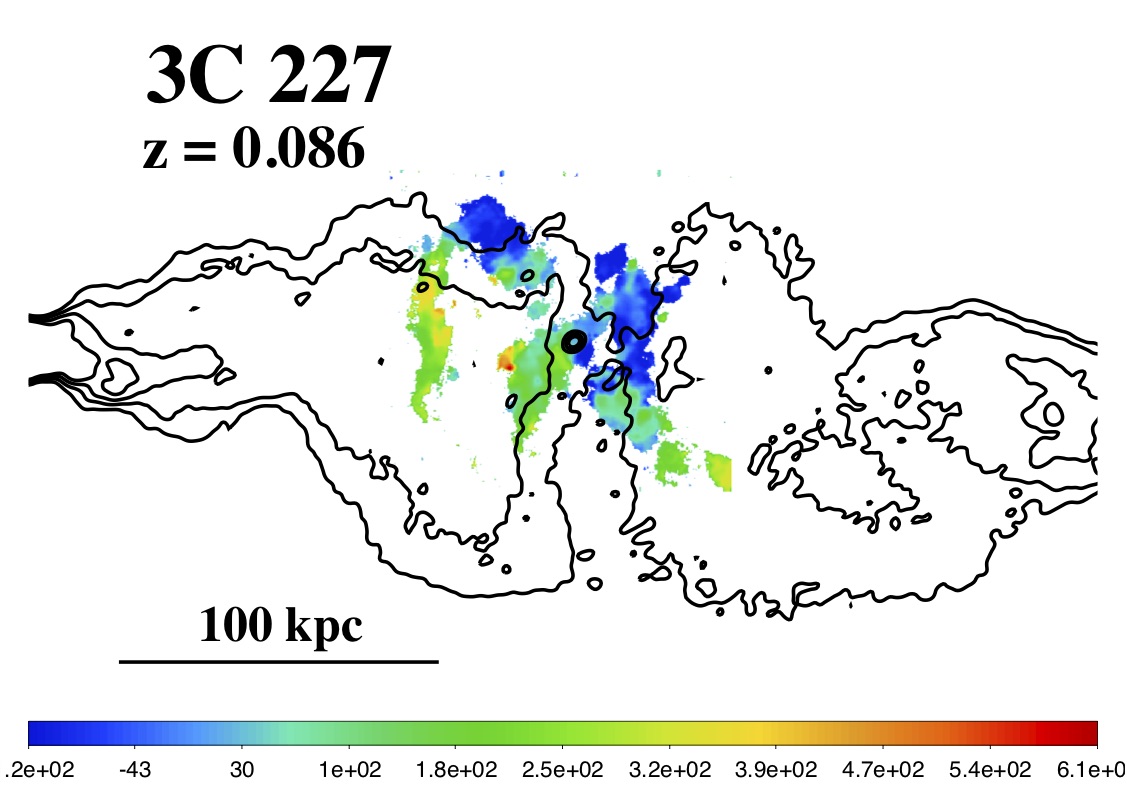}
 \includegraphics[width=0.66\columnwidth]{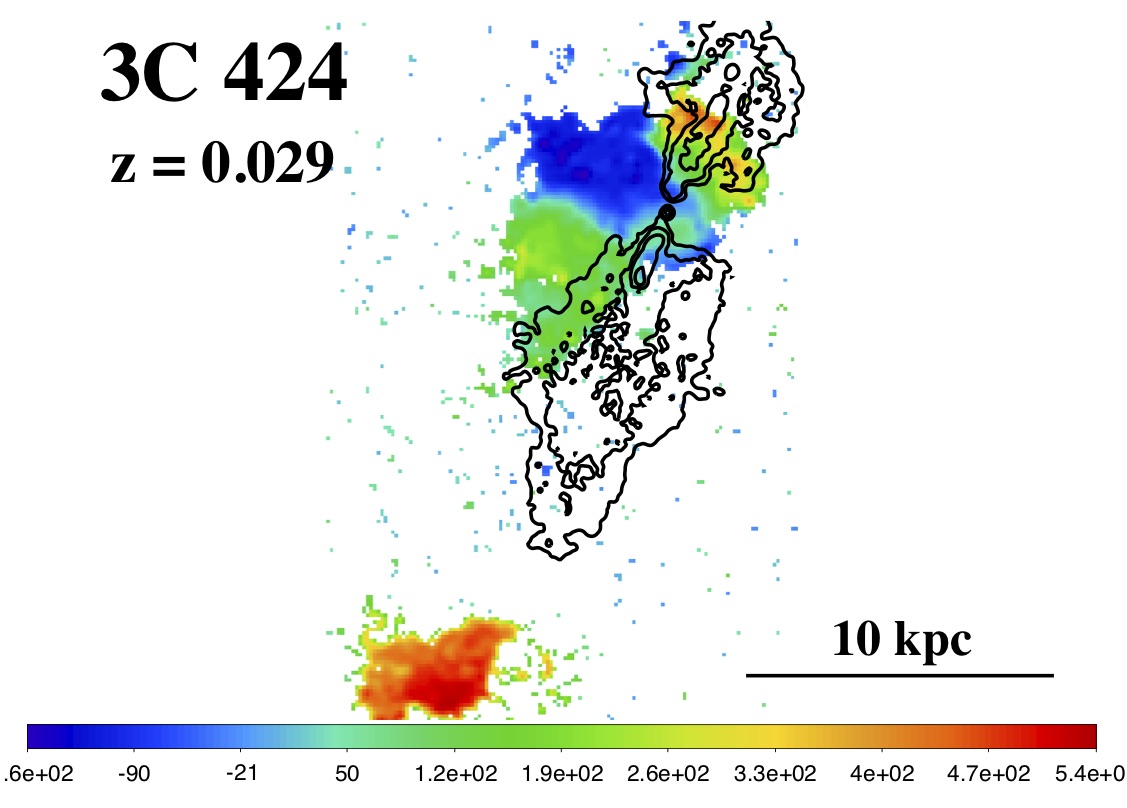}
\includegraphics[width=0.66\columnwidth]{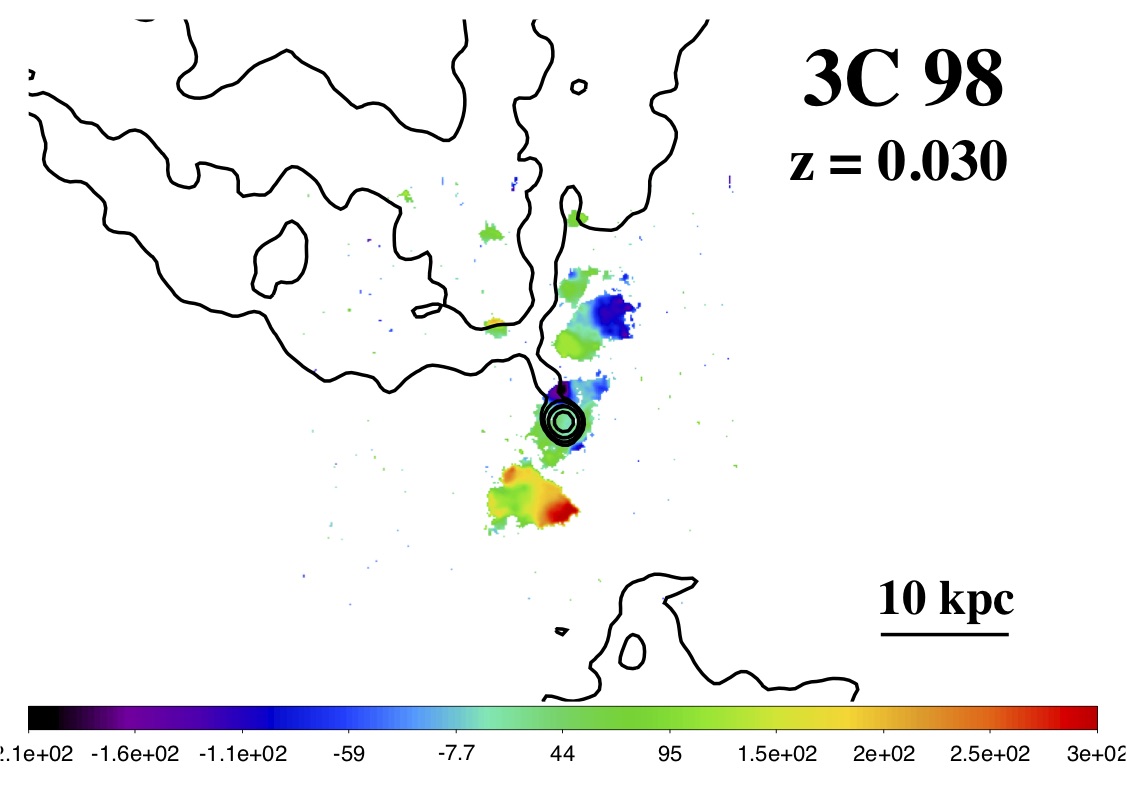}  
\includegraphics[width=0.66\columnwidth]{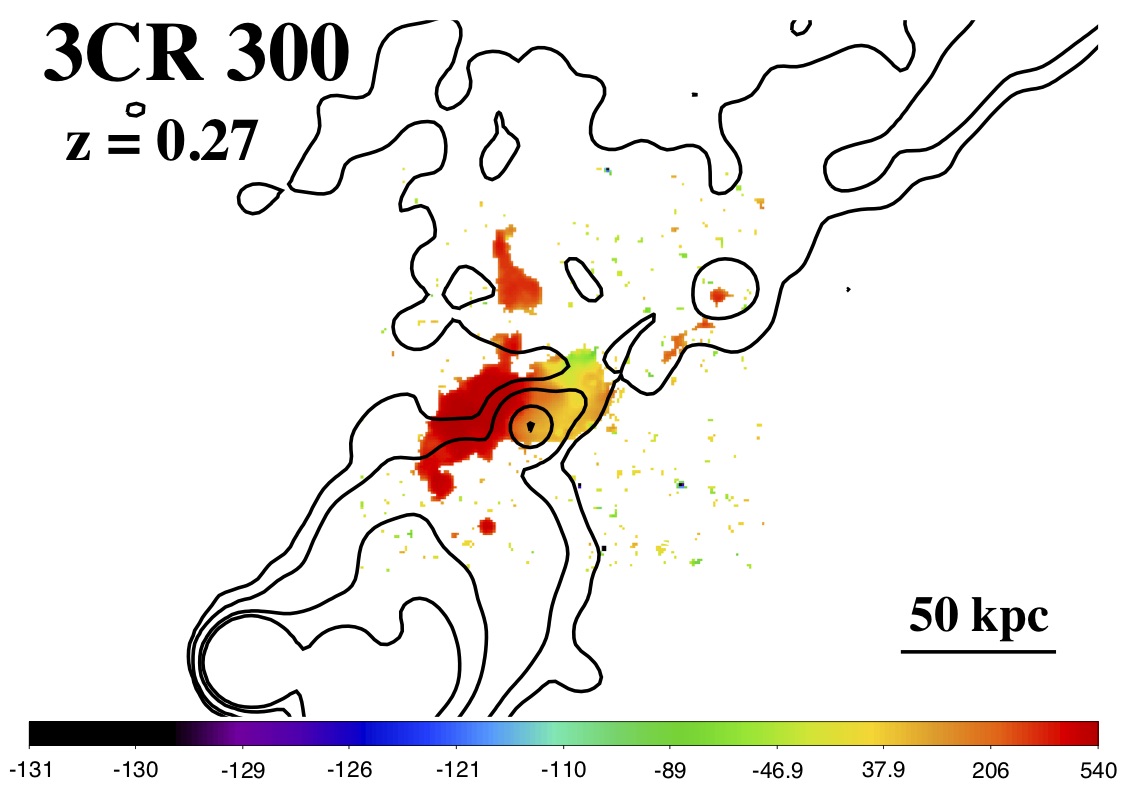}  
 \includegraphics[width=0.66\columnwidth]{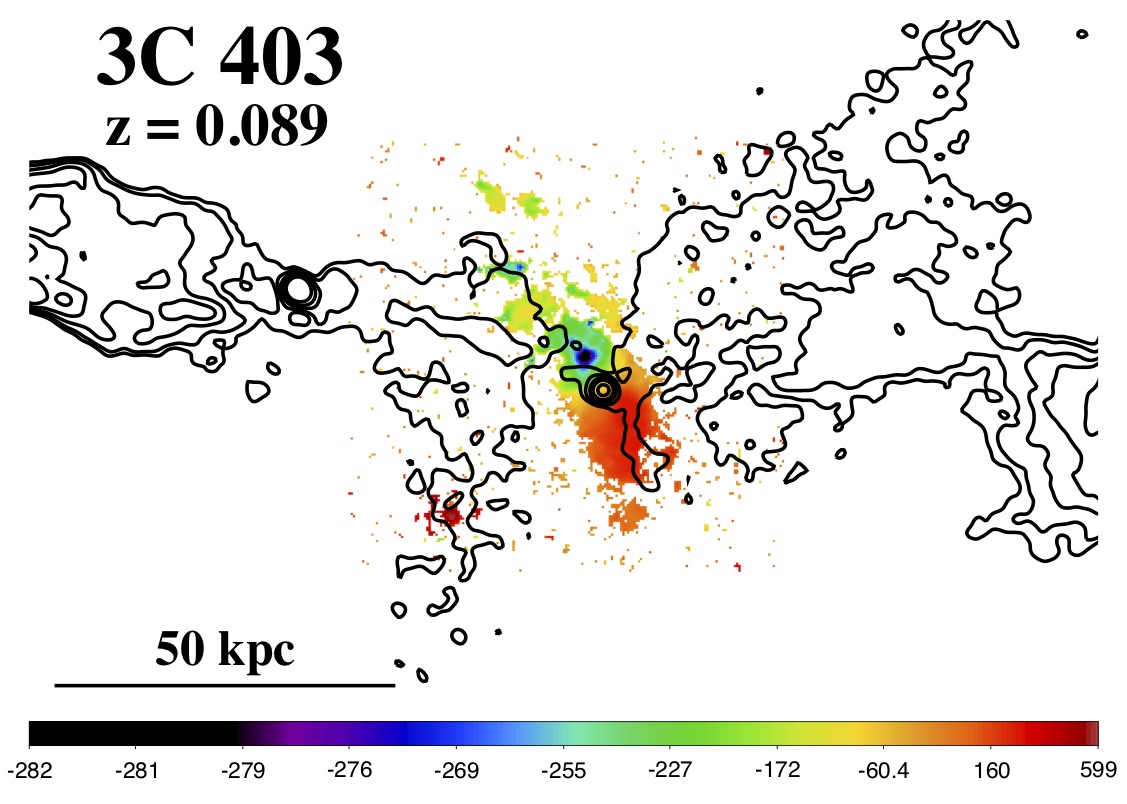}  
 }
\caption{Superposition of the radio contours onto the ionized gas
  velocity maps.}
\label{best}
\end{figure*}  

\begin{table*}
\caption{Synthetic aperture for the off-nuclear spectra and diagnostic line ratios}
\begin{tabular}{ l c r r  r r | r  r c c | r  r  r r r}
\hline
Name (type) &   Size & \multicolumn{3}{c}{Distance ($\arcsec$)}  (kpc)  & Offset  &
{\small [O~III]/H$\beta$} &{\small  [N~II]/H$\alpha$} & {\small [S~II]/H$\alpha$}& {\small [O~I]/H$\alpha$}& Type\\
\hline
3C~017   (BLO)  & 1.2 &  2.4  W   &  0.0 N & 10.5 &  66 &   4.74 & 0.74 & 0.64 & 0.31 & HEG  \\
3C~018   (BLO)  & 2.0 &  3.4  W   &  0.6 N & 13.0 &  63 &  16.72 & 1.11 & 0.60 & 0.13 & HEG  \\ 
3C~033   (HEG)  & 2.0 &  4.0 ~E   &  3.6 N &  6.3 &  29 &  14.54 & 0.34 & 0.26 & 0.12 & HEG  \\ 
3C~063   (HEG)  & 4.0 &  8.6  W   &  1.4 S & 30.1 &  48 &   1.18 & 0.30 & 0.52 & 0.22 & Peculiar \\ 
{\bf 3C~079}   (HEG) \,A& 2.0 &2.6 ~E   &  7.4 S & 56.6 &  56 &  12.43 & 0.38 & 0.25 &      & HEG  \\ 
\,\,\,\,\,\,\,\,\,\,\,\,\,\,\,\,\,\,\,\,\,\,\,\,\,\,\,\,\,\,\,\,\,\,\,\,\,               J& 2.0 & 10.0 ~E   &  1.0 N & 56.0 &  21 &   6.63 & 0.59 & 0.41 & 0.30 & HEG  \\ 
{\bf 3C~098}   (HEG)  & 2.0 &  2.8 ~E   &  9.6 S &  6.0 &  41 &  11.90 & 0.18 & 0.17 &      & HEG  \\
{\bf 3C~135}   (HEG)  & 1.6 & 14.0  W   &  0.0 S & 31.6 &  15 &  13.54 & 0.88 & 0.59 & 0.16 & HEG  \\
{\bf 3C~180}   (HEG)  & 2.0 &  8.8  W   & 14.2 S & 59.8 &  22 &   8.22 & 0.48 & 0.49 & 0.10 & HEG  \\
{\bf 3C~196.1} (LEG)  & 1.2 &  2.0 ~E   &  2.0 N &  9.3 &   0 &   2.33 & 1.23 & 1.00 & 0.42 & LEG  \\
{\bf 3C~198}   (SF)   & 1.2 & 12.6  W   & 17.0 S & 32.6 &   7 &   1.46 & 0.25 & 0.37 & 0.09 & SF   \\
{\bf 3C~227}   (BLO) \,J& 2.0 &28.4 W   &  2.2 N & 46.1 &   9 &   6.94 & 0.40 & 0.43 & 0.12 & HEG  \\ 
\,\,\,\,\,\,\,\,\,\,\,\,\,\,\,\,\,\,\,\,\,\,\,\,\,\,\,\,\,\,\,\,\,\,\, 
                 A& 2.0 &  9.6  W & 14.8 N & 28.6 &  62 &   7.30 & 0.47 & 0.32 & 0.08 & HEG  \\ 
{\bf 3C~300}   (HEG)  & 2.0 &  3.6 ~E   &  2.0 S & 17.2 &  13 &   9.90 & 0.23 & 0.30 & 0.10 & HEG  \\
3C~318.1 (---)  & 1.6 &  0.4 ~E   & 13.0 S & 11.7 & --- &   0.23 & 1.24 & 0.43 & 0.16 & Peculiar \\  
3C~327   (HEG)  & 2.0 &  0.4 ~E   &  3.8 N &  8.1 &  86 &  11.64 & 1.33 & 0.75 &      & HEG  \\  
3C~353   (LEG)  & 4.0 &  0.8 ~E   & 19.8 N & 12.7 &  80 &$<$1.90 & 0.66 & 0.69 &      & LEG  \\ 
3C~386   (---)  & 4.0 & 11.4 ~E   &  0.0 N &  4.2 &  74 &   0.66 & 1.01 & 0.28 &      & Peculiar \\  
3C~403   (HEG)  & 4.0 &  6.8 ~E   &  6.6 N & 11.3 &  40 &$>$9.50 & 1.15 & 0.72 &      & HEG  \\   
3C~424   (LEG)  & 4.0 &  5.8 ~E   &  6.8 S &  2.4 &  15 &   0.77 & 0.68 & 0.58 & 0.18 & Peculiar \\  
3C~442   (LEG)  & 1.2 &  0.4  W   &  4.6 S &  3.0 &  58 &$>$0.52 & 3.00 & 1.74 & 0.40 & LEG  \\  
3C~445   (BLO)  & 4.0 & 11.8  W   &  9.8 S & 17.0 &  57 &   8.20 & 0.15 & 0.35 & 0.08 & HEG  \\  
3C~458   (HEG)  & 2.0 &  7.8 ~E   & 11.6 N & 80.2 &  41 &   5.60 & 0.58 & 0.99 & 0.53 & LEG  \\   
3C~459   (BLO)  & 1.2 & 12.4  W   &  9.8 N & 69.0 &  31 &   5.50 & 0.18 & 0.68 &      & HEG  \\   
  \hline
\end{tabular}
\label{tab2}

\smallskip
\small{Column description: (1) source name (we report the
  sources for which we obtained new measurements in boldface) and spectroscopic
  classification. For 3C~079 and 3C~227 we considered two regions,
  across (A) and along (J) the jet; (2) region size (arcseconds); (3)
  distance of the region considered from the nucleus in arcseconds and
  (4) kpc, (5) angular offset between the radio axis and the fitted
  region (6, 7, 8, and 9) diagnostic line ratios, (10) spectroscopic
  classification of the extended region.}
\end{table*}

\begin{figure*}  
\centering{ 
\includegraphics[width=18.5cm]{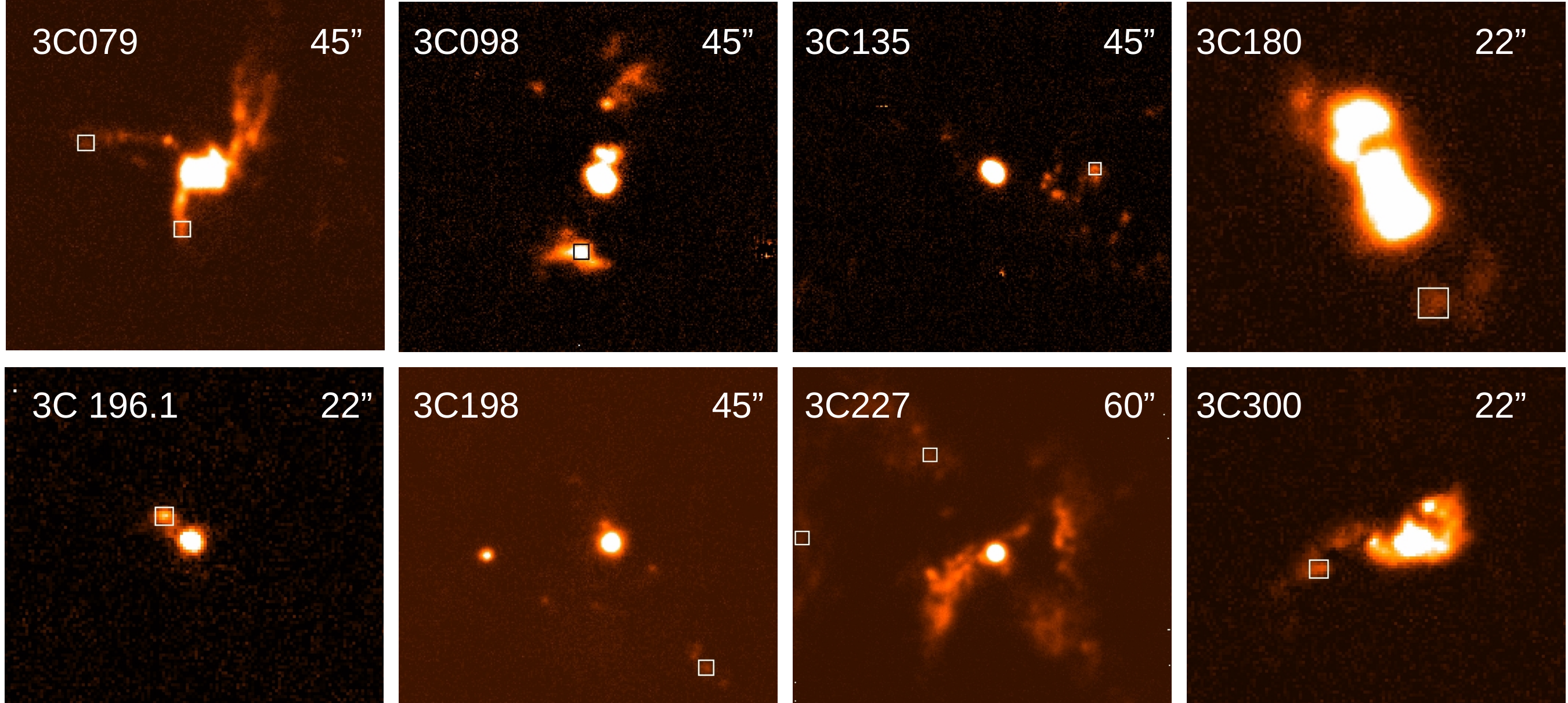}}
\caption{Emission line images (in logarithmic scale) of the eight radio
  galaxies observed presented by \citet{balmaverde21} in which we were
  able to extract off nuclear spectra. Images of the remaining 14
  sources are shown in \citet{balmaverde19}. The fields of view are
  indicated in the upper left corner of each image. The white boxes
  mark the synthetic aperture from which we extracted the spectra.}
\label{franco}
\end{figure*}   

\begin{figure*}  
\centering{ 
\includegraphics[width=18.5cm]{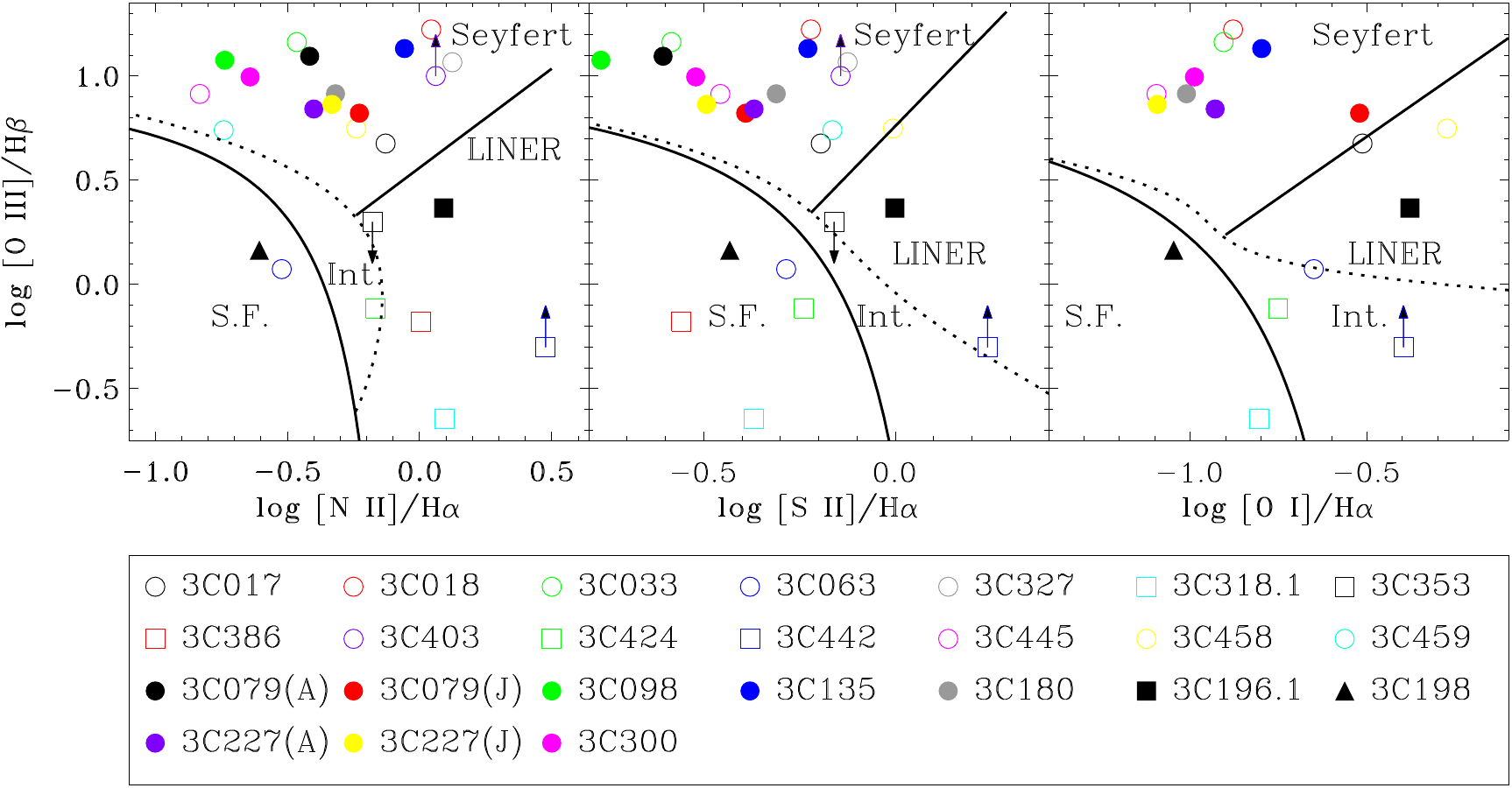}}
\caption{Location of the 22 FR~II radio galaxies with extended line
  emission in the spectroscopic diagnostic diagrams. Circles (squares)
  correspond to sources classified as HEGs (LEGs or unclassified) based
  on their nuclear emission line ratios \citep{buttiglione10}.  The
  solid curves separate star-forming (S.F.) from intermediate galaxies
  (Int.), the dotted curve separates intermediate galaxies from
  AGNs. The black lines separate low and high excitation galaxies, LEGs
  and HEGS, respectively \citep{law21}.}
\label{diagfig}
\end{figure*}

The advent of the Multi Unit Spectroscopic Explorer (MUSE), the wide
field integral field unit mounted at the Very Large Telescope,
enables us to explore the emission line properties with unprecedented
detail.  An example of the capabilities of MUSE to study the
properties of the EELRs and AGN feedback comes from the data obtained
in a pilot study for 3C~317 \citep{balmaverde18a}. In this source, a
network of line filaments enshrouds the northern cavity indicating
that the radio jets strongly affect the gas clouds in the surrounding
medium: they displace a large amount of hot gas and filaments of
line-emitting gas. The gas kinematics show the hallmarks of a shell
expansion, with both blue- and red-shifted regions, with a velocity of
$\sim$ 250 km/s.

As part of the MUse RAdio Loud Emission line
Snapshot (MURALES) survey, we observed a sample of 37 RGs with MUSE. The sample
includes the 3C radio sources limited to $z<0.3$ and
$\delta<20^\circ$. The main aim of this survey is to obtain deep
emission line images and explore the gas properties in the EELRs and their relationship with the
relativistic outflows. This enables us to explore the effects of
feedback in radio-loud AGNs. In \citet{balmaverde19,balmaverde21}, we
presented the main results of these observations. Thanks to their
unprecedented depth (the median 3$\sigma$ surface brightness limit in
the emission line maps is 6$\times$10$^{-18}$ erg s$^{-1}$ cm$^{-2}$
arcsec$^{-2}$), these observations reveal emission line structures
extending to several tens of kiloparsec in most objects.

In \citet{speranza21}, we looked for outflow from the sources by
analyzing the emission line profiles of the 37 MURALES sources. We
found evidence of nuclear outflows in 21 sources, with velocities
between $\sim$400--1000 \kms, while extended outflows were found in 13
sources, with sizes between 0.4 and 20 kpc. Comparing the jet power,
the AGN nuclear luminosity, and the outflow kinetic energy rate, we
concluded that outflows of high excitation galaxies (HEGs) are likely
radiatively powered, while jets likely play a role only in low
excitation galaxies (LEGs). Given the gas masses, velocities, and
energetics involved, the observed ionized outflows have a limited
impact on the gas content or the star formation in the host.

Here we present the analysis of the properties of the EELRs and of
their relationship with the radio outflows. The paper is organized as
follows: in Sect. 2 we study the geometry and kinematics of the
ionized gas, while in Sect. 3 and 4 we explore the connection between
radio and the EELR. In Sect. 5 we discuss the ionization properties of
the EELR. The results are discussed in Sect. 6 and summarized in
Sect. 7.

We adopt the following set of cosmological parameters: $H_{\rm
  o}=69.7$ \kms\ Mpc$^{-1}$ and $\Omega_m$=0.286 \citep{bennett14}. We
use cgs units throughout the whole paper unless stated otherwise.

\section{Properties of the ionized gas}
\label{data}

In Table \ref{tab1} we list the main properties of the sample of the
sources observed with MUSE, derived from the literature and from our
spectroscopic data. More specifically, we provide the source redshift,
its radio morphological classification and its excitation class from
\citet{buttiglione10}, the position angle ($P.A$., measured
anticlockwise starting from the north) and largest angular size (L.A.S.)
of the radio source,\footnote{The radio images were retrieved from the
  NRAO VLA Archive Survey, (c) AUI/NRAO, available at
  http://archive.nrao.edu/nvas/. For each source, we considered the map
  that best shows its radio structure. The radio $P.A.$ was measured
  as the line joining the brightest emission regions on each side of
  the nucleus, while the L.A.S. was measured at three times the noise
  level of each image.} the $P.A.$ of the ionized gas on each side of
the nucleus and the largest distance of emission line detection, the
perpendicular to the line of nodes derived with the {kinemetry}
software (in the following {kinematical axis}, see
Sect. \ref{kin}), and the host position angle.

In Fig. \ref{ppt} we present the main results obtained for one of our
targets, 3C~33, as an example, and all other sources are presented
in the Appendix \ref{appA}. In the left panel, we show the image
obtained for the brightest emission line, as listed in
Tab. \ref{tab1}; to this image, we superpose the orientation of the
emission line structures on both sides of the nucleus (solid white
lines), the orientation of the kinematical major axis in the inner
regions (gray line), and the radio $P.A.$ (dotted white lines). The white
rectangle represents the region from which we extracted the velocity
curve that is shown in the right panel.

\subsection{Emission lines' orientation and size}

In order to estimate the extent and the orientation of the ionized gas
structures, we produced a polar diagram for each source, centered on
the continuum peak. In Fig. \ref{polar} we present the
case of the [O~III] emission line in 3C~33 as an example. The contour levels are
drawn starting from three times the root mean square of the images.  In
each source we measured the distance and position angle of the faintest
emission line structure on both sides of the nucleus. The $P.A.$ and
radius measurements are given in Tab. \ref{tab1}. The uncertainties
are typically of $\sim 10^\circ$. The polar diagrams for all sources
are presented in Appendix \ref{appB}.

In the sample there are ten FR~I sources: all the FR~Is for which it
is possible to derive a spectroscopic classification are LEGs. In five
of them, the line emission is unresolved or barely resolved and it
extends only to a radius of $\lesssim$ 1 kpc, while in four sources
the EELR reaches distances between 0.5 and 8.8 kpc. The only FR~I with
an EELR extending beyond the size of the host galaxy is 3C~348 (also
known as Hercules A), which shows an elongated morphology reaching a
radius of $\sim 28$ kpc. A detailed morphological study reveals that
this source has a hybrid FR~I/FR~II (e.g., \citealt{gizani05}).  All
but one (3C~318.1, see \citealt{giacintucci07} and
\citealt{jimenez21} for detailed studies of this source) of the
remaining 27 objects are FR~II radio galaxies.

The spectroscopic diagnostic diagrams are commonly used to
  address the gas ionization mechanism by comparing the strength of
  various emission lines
  (e.g., \citealt{heckman80,baldwin81,veilleux87,kewley06,law21}) and to
  classify AGNs into different classes. Here, we adopt the spectral
  classification derived for the sources of the sample derived by
  \citet{buttiglione10,buttiglione11}. From the point of view of their
  optical spectroscopic properties, six of them are classified as
  broad line objects (BLOs), 12 are HEGs, six
  are LEGs, one (3C~198) has optical
  diagnostic line ratios typical of star-forming galaxies, while the
  emission lines in 3C~386 are too weak for such a classification.

All the FR~II show extended line emission, with the only exceptions being
3C~105 and 3C~456. The sizes of the EELRs range from $\sim$4 to
$\sim$80 kpc, with a median of $\sim$16 kpc. As we already discussed
in \citet{balmaverde21}, there are no apparent connections between the
EELRs' sizes and luminosities with the spectral classification.

\subsection{Central gas kinematics}
\label{kin}

We fit the 2D gas velocity field in the innermost regions with the
{kinemetry} software \citep{krajnovic06}, a generalization of
surface photometry ellipse fitting which reproduces the moments of the
line-of-sight velocity distributions.  The parameters of interest
returned by this software are as follows: 1) the kinematical $P.A.$; 2) the
coefficient of the harmonic expansion $k1$ from which one derives the
rotation curve; and 3) the ratio between the fifth and first
coefficient $k5/k1$, which quantifies the deviations from purely
ordered rotation and which is sensitive to the existence of multiple
kinematic components.  As an example, in Fig. \ref{kinemetry403} we
show the results obtained for 3C~33, while the diagrams for all other
sources are in Appendix \ref{appC}.

We considered as reliable only the measurements of the kinematical
parameters in the sources where these are defined, at least in one
point, at a radius $r >$2\arcsec\ (i.e., three times the typical seeing
of the observations) and where
$(k5/k1$+$\sigma_{k5/k1})<$0.2. According to this, we detected central
rotating gas in five out of ten FR~Is and in 20 out of 26 FR~IIs. We
measured the kinematical $P.A.$ as the median of the values fulfilling
the above criteria, a region that extends in most cases to
3\arcsec-5\arcsec. The angles of the kinematical major axis are listed
in Tab. \ref{tab1}. The uncertainties are usually between 5$^\circ$
and 10$^\circ$.

\subsection{Kinematics of the large-scale gas}

The kinematics of the large-scale gas is often complex and usually cannot be described by a simple rotation model. To investigate its
kinematical properties, we extracted a position-velocity diagram and
obtained the velocity curve from a synthetic slit aligned along the
most extended line structure. While the gas on smaller scale (up to a few
kiloparsecs) is very often in ordered rotation, this behavior is rarely seen
on a scale of tens of kiloparsecs. This is the case for 3C~33, 3C~79, 3C~300,
3C~327, and 3C~403, where the whole EELR (extending for between 10 and
35 kpc) follows a smooth rotation curve. In other sources, the central
portion of the emission line region shows ordered rotation, while the
velocity field at larger distances is disturbed (e.g., in 3C~76.1,
3C~272.1, and 3C~348). In other galaxies, there is a general symmetry
in the velocity field, with one side of the EELR being redshifted and
the opposite side blueshifted (e.g., in 3C~98, 3C~180, and 3C~227,
3C~386), but with large changes in the velocity on small scales.

\subsection{Measurements of the host galaxy major axis}

We explore the connection between the radio and host galaxy morphology
by analyzing their optical and infrared continuum images. Most of the
images used were obtained from the near-infrared HST snapshot survey
of the 3C sample \citep{madrid06}, taken with the F160W broadband
filter and less affected by the presence of dust absorption. For some
objects, these are not available and we thus used the images taken
with WFPC2/HST with the broad red filter F702W instead. We fit
elliptical isophotes to each image and derived the $P.A.$ of the
host's major axis reported in Table \ref{tab1}. The uncertainties were
derived from the $P.A.$ variations at different radii and are usually
between 5$^\circ$ and 10$^\circ$.

In two objects (3C~445 and 3C~456), the optical emission is dominated
by their bright nuclei, while in the case of 3C~459 the host
morphology is highly disturbed. For these galaxies no meaningful
estimate of the stellar $P.A.$ is possible.

\section{The geometric connection between radio and line emission}

We analyze the geometric connection between the ionized gas with the
radio emission, also considering the information derived from the 2D
velocity field measured on different scales and the information
derived on the host major axis.  In the left panel of Fig. \ref{offset}, we
show the offset between the radio axis and the gas position angle
as well as the kinematical axis estimated with {kinemetry} listed in
Tab.\ref{tab1}. When measurements are possible for a given source on
both sides of the nucleus, the two offsets values are connected with a
vertical line.

The right panel reproduces the various offsets' distributions. The
Kolmogorov-Smirnov test indicates that, by adopting a 5\%
  threshold, none of them are significantly different from a uniform
distribution.\footnote{The Kolmogorov-Smirnov test returns a
  probability of P=0.20 when comparing the radio-EELR offsets with a
  random distribution, a value that decreases to P=0.16 for the
  FR~IIs. The test applied to the kinematical axis gives P=0.37 for
  the whole sample and P=0.14 for the FR~IIs and P=0.98 for the offset
  with the host major axis (P=0.79 for the FR~IIs).} Nonetheless, the
offset between the radio and EELRs' axis is larger than 45$^\circ$ in 18
out of the 29 sources for which we are able to obtain this
measurement. The probability of having this fraction, or a
  higher fraction, of misaligned sources, estimated via the binomial
  distribution is P=0.13. There are instead 14 galaxies in which the
offset between the radio and kinematic axis is $> 45^\circ$ and ten where
this is $< 45^\circ$(corresponding to P=0.27). By only considering the FR~II sources (13 against six), we obtained a probability of
  P=0.08. Conversely, there is an equal number of galaxies in which
the difference between the $P.A.$ of the EELRs and of the host galaxy
is larger or smaller than $45^\circ$.

In summary, we find a weak preference for the EELRs to be oriented
at a large angle from the radio axis.\ An effect is also found for the
rotation axis of the innermost gas structures in the FR~IIs,
while no connection is found with the host axis.

\section{The 2D connection between radio and line emission.}

The statistical analysis presented in the previous sections does not
completely grasp the full complexity of the connection between the
host galaxy, the radio ejecta, and the properties of the ionized
gas. In order to explore these issues in more detail, we superposed
radio maps onto the 2D velocity maps obtained from the MUSE data. In
Fig. \ref{best} we show a subset of representative sources, ordered to
present the various behaviors. The remaining images with extended
emission line regions are shown in Appendix \ref{appD} and ordered according to
their right ascension.

The first row shows the three sources in which the emission line
region is located at the edges of the radio lobes. In this
category, we include 3C~63, 3C~196.1, as well as 3C~277.3, a source not included in
the MURALES sample (since its declination is larger than the limit
of our survey), but for which we obtained MUSE observations
(\citealt{capetti21}, but see also \citealt{miley81},
\citealt{tadhunter00} and \citealt{solorzano03} for previous emission
line images of this source). A similar behavior is also seen in 3C~317
\citep{balmaverde18a}. Both the morphology and the kinematics of the
emission-line gas strongly suggest that in these sources we are
seeing the effects of the expanding radio lobes: the radio ejecta
compress the external gas, increasing its emissivity and forming
shells of line emission with a bubble-like morphology surrounding the
radio structure. They are all FR~II sources, with a L.A.S. of $\sim 50-60$ kpc, which is three times smaller than the median
size of the FR~IIs in the sample.

In the second row, we show three FR~I sources. In the first two
objects (3C~76.1 and 3C~272.1), we detect a small-scale structure
($\lesssim 4$ kpc in size) of ionized gas perpendicular to the radio
jets, showing an ordered rotation with a kinematic axis aligned with
the jets. The emission line images of two further FR~Is (3C~78 and
3C~264) are barely resolved, but similarly show an ordered velocity
field at the center, with an axis displaced by $\sim 25^\circ$ from
the radio axis. In 3C~348, the gas is also in ordered rotation, but it
is oriented along the jets.

A very similar behavior is seen in the three FR~II sources shown in
the third row, with quasi-linear structures, close to perpendicular to
the radio jets. While the velocity field in 3C~18 is very complex, the
remaining sources show a general symmetry. An analogous morphology is
seen in 3C~327 that is also characterized by a symmetric rotation
curve. The main difference with respect to the FR~Is presented above
is the larger size of these features, with radii between 15 and 40
kpc.

Similar structures are also seen in the three FR~IIs presented in the
fourth row. In these cases, however, additional emission line
structures are also present. They are mostly located at the edges of
the inner portion of the radio lobes. For example, in the case of
3C~79, the emission line region is dominated by a large-scale ($\sim$
70 kpc in size) rotating gaseous structure, but there is also a
filament of ionized gas wrapping around the northern edge of the
eastern radio lobe, reaching a distance of $\sim 80$ kpc.  The effect
of the lateral expansion of the radio lobes far from the radio jet
axis has been observed in other radio galaxies (e.g., in 3C~171 and
3C~265, \citealt{solorzano03}).  A somewhat similar structure is also seen in 3C~227 and 3C~424. In these cases, the emission line region
surrounding the radio lobe is the most extended structure: the
measurements of the EELR position angle derived in the previous
section refer to these components that are aligned within
25$^\circ$-40$^\circ$ to the radio axis. These sources can be
interpreted as a morphological combination of the objects presenting
shells surrounding the radio lobes and those dominated by an ionized
gas perpendicular to the radio jets.

Finally, in the fifth row, we show three sources (3C~98, 3C~300, and
3C~403) in which the brightest structures of ionized gas are oriented
at a small angle ($\Delta P.A. \lesssim 40^\circ$) from the radio
jets. A similar morphology is seen in 3C~33 and 3C~180. Three of them
have a clumpy morphology and the velocity field is complex, while
3C~33 and 3C~403 have a smooth disk-like morphology and show well-ordered rotation (for 3C~33, see also
\citealt{couto17}). Interestingly, 3C~403 is an X-shaped radio source
and the emission line is perpendicular to the secondary radio wings.
A statistical tendency for the very extended emission-line gas to
align with the radio axis has been observed preferentially in small
radio sources (\citealt{baum89}).

\section{Ionization properties of the extended emission line regions}

Thanks to integral field spectroscopic data, we can explore the gas
ionization properties in different portions of the EELRs by comparing
the strength of the various emission lines (e.g.,
\citealt{singh13}). Since this approach requires one to detect several
emission lines in each spaxel, we extracted spectra from synthetic
apertures as far as possible from the nuclei, but still in regions with
 a sufficient signal to derive useful measurements. In
\citet{balmaverde19}, we already presented this analysis for a
subsample of 20 3C sources: we discarded the sources in which the
emission line region is only marginally extended to obtain off-nuclear
measurements and we were left with useful data for 14 sources. By applying
the same method to the new list of 17 objects, we can study the
ionization properties of the off-nuclear gas in eight further sources. In
3C~079 and 3C~227, the EELR extends in both the parallel and
perpendicular direction from the radio axis and we then extracted
spectra in two off-nuclear regions. The spectra of these regions are
shown in Appendix \ref{appE}. The selected areas, which are located at a
median distance of $\sim$15 kpc from the nucleus, but reaching
distances of $\sim$80 kpc, are marked in Fig. \ref{franco}. In
Tab. \ref{tab2} we list the line ratios for each region and derive a
spectroscopic classification based on the location in the diagnostic
diagrams defined by \citet{kewley06} (Fig. \ref{diagfig}).

Most of the sources (15 out of 22, all of them FR~IIs) considered are
HEGs; this also includes the BLOs whose narrow emission line ratios
are characteristic of high excitation gas in this class. In many of
them, the off-nuclear region we considered for this analysis is
located at a large angle with respect to the radio axis. In 3C~327,
for example, it is almost perpendicular to the radio axis, while in
four other sources the offset is $\sim 60^{\circ}$. In 13 of them, the
large-scale gas also has a HEG spectrum. Similarly, in three of the
four LEGs, the extended gas has a low excitation spectrum; also, in
the only star-forming RG of the sample (3C~198), the large-scale
emission has line ratios consistent with photoionization by young
stars.

However, there is one HEG, 3C~458, in which the emission line ratios
measured on the extended emission correspond to a different
spectroscopic type, for which we derived a LEG classification. In this
case, the extended emission line region is located at a very large
distance from the nucleus (80 kpc). The transition from high to low
ionization appears to occur gradually based on the decrease of the
\oiii/\ha\ ratio with distance.

Finally, there are four sources with a peculiar behavior, namely 3C~63
(a HEG), 3C~318.1 and 3C~386 (both of uncertain spectral type), and
3C~424 (a LEG): they are located in different regions of the
diagnostic diagrams depending on the panel considered, an indication
that we might be observing regions in which photoionization from an
active nucleus or from young stars is not the dominant process. This
is reminiscent of the results found with MUSE observations by
\citet{balmaverde18a} for 3C~317: we argued that ionization in that
source is due to slow shocks \citep{dopita95} or collisional heating
from cosmic rays \citep{ferland08,ferland09,fabian11}. The general
small velocity dispersion of the emission lines in these regions,
$\sim 50 - 100$ \kms, argues against the importance of shocks, thus
favoring ionization from energetic particles.

In the two RGs in which the EELRs have an X-shaped morphology (namely,
3C~079 and 3C~227), we were able to extract off-nuclear spectra along
different angles, almost parallel and perpendicular to the radio jet,
respectively. In both cases, the emission line ratios in the two
regions are extremely similar.  In particular, in 3C~227, the spectrum
of the region located at a projected angle of 9$^{\circ}$ from the
radio axis share essentially the same location in the diagnostic
diagrams of the region 62$^{\circ}$ away from it.

\section{Discussion}
\label{discussion}

The kinematics and the distribution of ionized gas and its connection
with the radio emission provides us with a powerful tool to explore
its origin and the effects of the active nucleus onto the external
medium, that is, the process of feedback in radio-loud AGNs.

\subsection{Gas geometry and kinematics}

From the point of view of gas geometry, we confirm the previous
results that, in low redshift RGs, the radio and line alignment is rather
weak, contrarily to what is observed in more distant and luminous
sources. More specifically, we find a preference of the more distant
emission line structure to be located at an angle larger than
45$^\circ$ relative to the radio axis; this occurs in 18 out of the 29
sources in which we were able to perform this comparison. In most
cases of these ``misaligned'' sources, the EELR takes the form of a
disk-like (elongated or filamentary) structure, usually extending for
several tens of kiloparsecs. In some of the ``aligned'' objects, bubbles of gas
enshroud the radio lobes, clearly indicative of a direct impact of the
jets on the external medium; in these galaxies, the radio source is
three times smaller ($\sim 50-60$ kpc) than the average size of the
sample, suggesting that this process occurs in the early phase of
their evolution. The smaller size of these sources can also be caused
by the denser environment in which the jets propagate. In addition,
there is a group of sources in which, similarly, filamentary
structures follow the edges of the radio source leading to a general
alignment, but disk-like structures are present on a smaller
scale. These structures can be interpreted as ``remnant'' bubbles,
produced in earlier phases of the source expansion, and left behind
now that the radio jets reached sizes of several hundreds of kiloparsecs.

From the point of view of gas kinematics, the analysis performed
with the {kinemetry} code revealed gas in ordered rotation,
suggestive of the presence of a disk of a few kiloparsecs, in 77\% of the
FR~IIs (20 out of 26) and in 50\% of the FR~Is (five out of ten, but in
several FR~Is the emission is unresolved and no kinematic study is
possible). The misalignment between the radio and rotation axis is
stronger than was found for the morphology, in particular for the FR~IIs
where 13/19 sources show an offset larger than $\sim 45^\circ$. This,
again, is in line with previous results \citep{baum89b,solorzano03}.

\subsection{The gas ionization mechanism}

A significant point is the identification of the ionization mechanism
of the warm extended gas. We measured the off-nuclear emission (with a
median distance of 15 kpc) line ratios in 22 sources of the sample
where the line emission is sufficiently well resolved. In most cases
(15), these are HEGs from the point of view of their nuclear properties
and we also found emission line ratios typical of HEGs in the EELR of
13 sources. This suggests that on both scales, we are observing gas
photoionized by the central source. However, the angle that formed by the
radio axis (most likely, coincident with the axis of the nuclear
ionizing radiation field) and that leading to the off-nuclear region
is large, between 30$^\circ$ and 85$^\circ$, with a median of
60$^\circ$.  Along these lines of sight, the nuclear ionizing
radiation should be blocked by the circumnuclear torus
\citep{antonucci93}. The presence of this absorbing material is
confirmed by the X-ray nuclear spectra of several of these sources
showing large columns of cold gas \citep{massaro12b,massaro15}. The
study of \citet{baldi13}, based on the relative number of 3C sources
with $z<0.3$ with and without broad lines in their optical spectra,
indicates a covering factor of the circumnuclear matter of 65\% that
corresponds, adopting a torus geometry, to an opening angle of
50$\pm5^\circ$. The regions outside this bicone should not be
illuminated by the nuclear radiation field, contrary to what is
observed. However, projection effects might play a significant role,
particularly considering that the ionized gas is most likely only a
small component of a much more massive gaseous structure. It is also
possible that we are witnessing the effects of the expanding radio
lobes that are compressing and ionizing the intergalactic medium:
sufficiently fast shocks (with velocities of several hundreds of km
s$^{-1}$) produce line ratios compatible with those seen in the EELR
of the RGs \citep{allen08}. However, at these large angles from the
path of the radio jets, such high velocities are unlikely to be
reached. Overall, we must therefore conclude that the ionization
mechanism of the EELR remains unclear in general.

\subsection{Evidence for superdisks in radio galaxies.}

A novel result of our analysis is the widespread presence, with only a
few exceptions, of large structures of ionized gas with sizes up to
$\sim$80 kpc, usually with a highly elongated and filamentary
morphology.  \citet{gopal00} argued that various pieces of evidence
point to the presence of a large gaseous disk associated with
radio-loud AGNs: the dust absorption features seen in several nearby
RGs, the sharp edges of the radio lobes on the side facing the host,
the asymmetry of Ly$\alpha$ nebulae in high redshift radio
galaxies. Given the estimated sizes of such structures ($\sim 75
$kpc$\times 25$ kpc), they defined them as superdisks. They suggested
that these structures are tidally stretched remnants of gas-rich
mergers.  Large disks of neutral hydrogen are commonly seen in
early-type galaxies, often stretching to many tens of kiloparsecs
\citep{serra12}. The results of the MUSE observations provide direct
evidence that such superdisks are indeed present in the majority of
the nearby RGs. Both the morphology (with filamentary and irregular
structures) and the kinematics (in about half of the sources the gas
kinematics is complex, with large velocity changes on small scales) of
the ionized gas indicates that the gas is not in a relaxed state. This
is probably due, in part, to the long timescale for the gravitational
settling of these large-scale structures, but also to the impact of
the radio emitting outflows.

\subsection{Jets and gas interactions and evolution of radio galaxies}

\begin{figure*}[h] 
\centering{ 
\includegraphics[width=18.5cm]{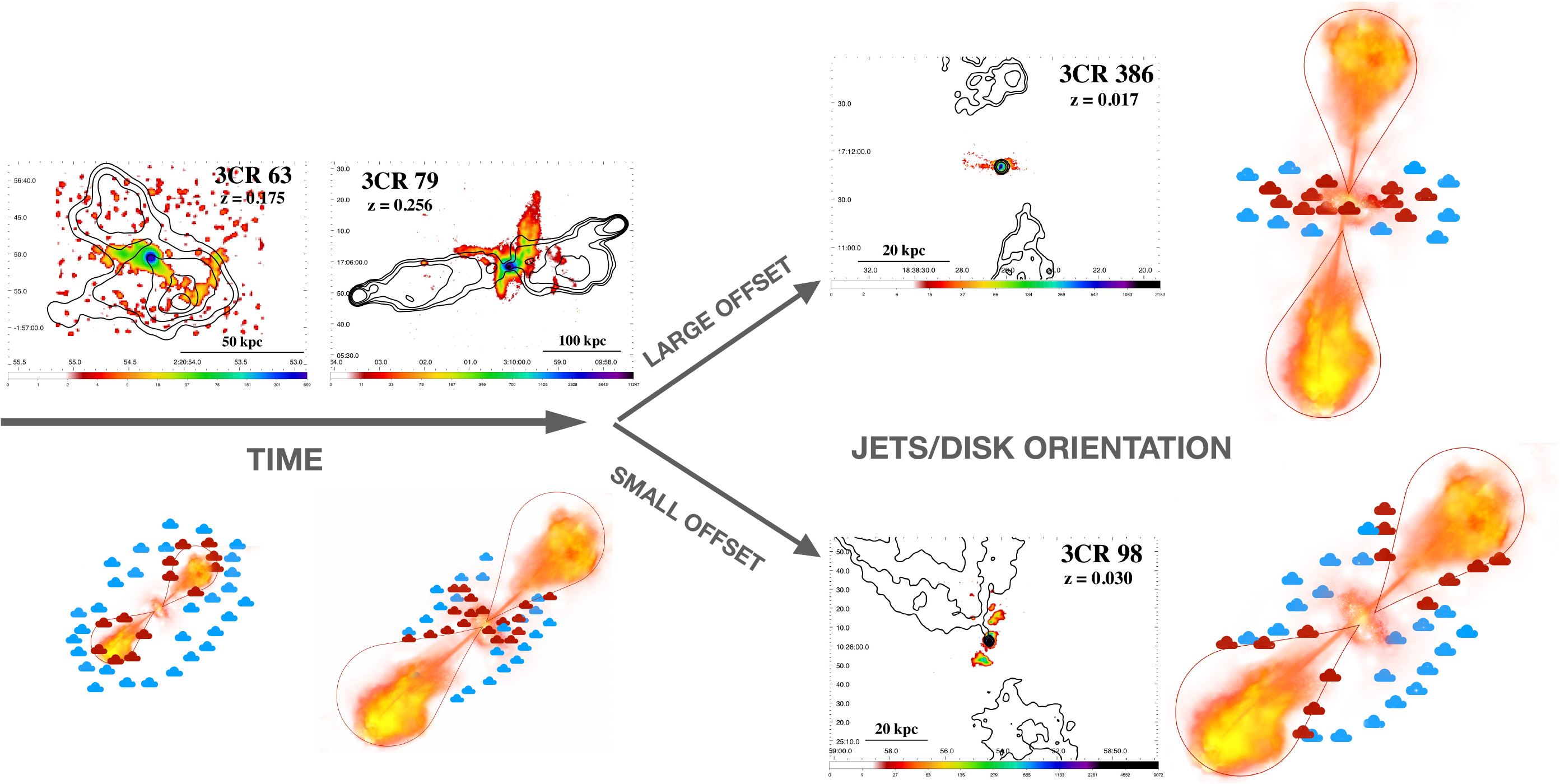}}
\caption{Schematic view of the evolution of the ionized gas
  structure. In the initial stages, the expanding radio lobes form a
  bubble of ionized gas (red clouds). When the radio source has
  escaped the superdisk, the remnants of the bubble are still visible
  as filaments around the base of the lobes, but the main structure is
  a rotating disk. The final state depends on the orientation of the
  superdisk with respect to the radio axis: when they are
  perpendicular to each other, a linear rotating structure is present;
  whereas, when the jets form a small angle with the superdisk, clumpy
  line emission with a complex velocity field is seen, indicative of
  ongoing strong jet-cloud interactions.}
\label{cartoon}
\end{figure*}

Overall, the different properties of the EELR can be explained with a
combination of the source evolutionary state and the orientation of
the superdisk with respect to the radio axis (see Fig. \ref{cartoon}
for a schematic view).  As long as the radio source is smaller than
the vertical height of the superdisk, it propagates into a dense
environment and it forms bubbles of gas, visible as expanding shells
in the emission line images (for example in 3C~63, 3C~196.1, 3C~277.3,
and 3C~317). This situation is reminiscent of the process at the
origin of the radio and line alignment commonly observed on smaller
scales in the young, compact steep spectrum sources (CSS,
\citealt{devries97,devries99,axon00}). In some cases, the bubbles that
formed in the initial stages are still visible in larger radio sources
as filaments surrounding the lobes (e.g., in 3C~79). The subsequent
evolution depends on the orientation of the radio jets with respect to
the superdisk axis. At high inclination, when the jets have escaped
from the denser regions, the EELR is dominated by a disk structure; at
least on a scale of a few kiloparsecs, it shows ordered rotation at a
large angle from the radio axis.  In sources where the jets propagate
at a small angle with respect to the plane of the disk, strong
jet-cloud interactions occur and the EELR has an
arc-like or clumpy morphology. This is clearly seen in 3C~180 in both
the MUSE and the HST emission line images \citep{baldi19}. A clumpy
EELR, such as observed in 3C~98 and 3C~300, might result when the jet
grazes the superdisk.

The radio and optical alignment observed in high redshift RGs can also be
understood in this scenario. High-z RGs are known to be smaller than
those at low z (e.g., \citealt{saxena17} and references therein) and
they propagate in a denser environment. The putative superdisks
associated with these sources are also likely to be farther from a
settled state. The reduced sizes of high-z RGs and the different
properties of the external medium favor the possibility of enhanced
jet-cloud interactions and might lead to a higher fraction of
radio and optical aligned sources.

The interaction between the relativistic outflow and an external gas
disk presented above is in line with the results obtained from the
numerical simulations performed by \citet{mukherjee18} in which a
relativistic jet interacts with a dense turbulent gaseous disk. The
main difference between the data and the simulations relies on the
scale heights of the superdisks, which are an order of magnitude larger than
those adopted in their model. This implies a correspondingly longer
timescale ($\sim 1$Myr for the simulations) over which the jet-cloud
interactions are effective. Nonetheless, the temporal evolution and the
relevance of the relative jet-gas inclination they derived is fully
consistent with our conclusions. It is unclear whether this picture
also applies to FR~Is. Gaseous and dusty disks are seen in these
sources (e.g., \citealt{sparks00}), but with relatively small sizes with
respect to FR~IIs. This difference can be due to the paucity of gas in
FR~Is, and also to the lack of photons required to ionize
large-scale structures.

Considering the orientation of the ionized gas with respect to the
radio axis, we found a preference for the most extended gas to be at
an offset larger than 45$^\circ$, a situation occurring in about two-thirds of the sources analyzed. This fraction actually underestimates
the number of sources in which there are linear gas structures at
large angles from the jets due to the presence of objects such as 3C~227
and 3C~424 in which we see an elongated gas feature, perpendicular to
the radio axis, but this is not the most extended line structure. In
addition, in the sources characterized by a shell-like EELR, the
observed alignment of the EELR with the radio axis is caused by the
jet-cloud interactions and not by the intrinsic gas distribution.
  
The large angle generally formed between the axis of the radio and the
EELR might be produced (or enhanced) by the pressure exerted on the
superdisks by the jet material back-flowing from the hot spots toward
the nucleus. However, the geometrical link between the structure of
ionized gas and the radio jets can also be the result of an intrinsic
alignment between the axis of the AGN and the gaseous disk
\citep{bardeen75,rees78,scheuer96,natarajan98}. This is what is
expected when a preferred plane of accretion, defined by the orbital
plane of a merging galaxy, is maintained over a large time interval. In
this situation, the SMBH is fueled by material with a coherent and
stable angular momentum (also including any secondary SMBH) and this
tends to align its rotation axis with that of the accreting
matter. The resulting geometry of the system also depends on the
SMBH spin before the merger: slowly rotating black holes are more
easily reoriented, while SMBH of a higher spin maintain their
axis. We speculate that the latter ones are associated with the RGs in
which we observed the smaller offset between the radio and gas axis.

\subsection{AGN feedback in radio galaxies}

A general picture of the interaction between the dense ionized gas and
the radio jets emerges from the analysis of the emission line images
of the nearby RGs of the 3CR sample. In the early phases of the
sources' lifetime, the relativistic jets transfer energy and momentum
from the AGN to the external medium that is compressed, accelerated,
and displaced. As shown by \citet{speranza21}, both nuclear and
kiloparsec scale outflows of ionized gas are commonly detected in
these sources, but they have a limited impact on the gas content or
the star formation in the host. Based on the observed emission line
ratios, typical of AGN ionization over the whole emission line nebula
(with the exception of 3C~277.3 presented in \citealt{capetti21} and
3C~135), this interaction does not produce detectable star-forming
regions. This phase lasts as long as the source is confined within the
central tens of kiloparsecs corresponding to the vertical height of
the superdisks. In the later evolution, the radio source escapes from
the dense central regions and the jet-gas interactions are strongly
reduced. The impact of the radio source on the superdisks' gas is then
limited to the compression of the back-flowing material, but, as long
as the radio source keeps expanding, it will continue to drive shocks
and transfer energy to the external gas.

\section{Summary and conclusions}
\label{summary}

In this sixth paper presenting the MURALES survey, we analyze the results
obtained from the MUSE observations of 37 3C radio sources with
$z<0.3$. In the FR~Is, the ionized gas is confined to within $\lesssim
4$kpc, while elongated or filamentary gas structures are seen in all
but two of the 26 FR~IIs, extending for 10 - 30 kpc, and showing a
broad distribution of offsets from the radio axis.

The gas on a scale of a few kiloparsecs generally shows ordered rotation
around an axis forming a median angle of 65$^\circ$ with the radio
axis. The kinematics of the large-scale gas is more complex: in half
of the sources, the EELR shows a regular velocity field or a general
velocity symmetry; however, in the other galaxies, it is more complex, with
large velocity changes on small spatial scales.

The large-scale gas generally shows ionization properties similar to
those derived for the nucleus. In many sources, these extended
structures are located at a large angle from the radio axis: due to
the presence of circumnuclear obscuring material, they should not be
illuminated by the nucleus. However, shock ionization is unlikely
because this gas is located far from the path of the relativistic
jets. The ionization mechanism of the large-scale gas remains unclear.

The radio and the gas emission presents a variety of geometrical
connections: shells of gas surrounding the radio lobes; linear
or filamentary structures, usually forming a large angle with the
radio axis; and arc-like or clumpy EELRs forming a small angle with the
radio axis. The observed emission line structures in FR~IIs can be
interpreted as due to the presence of gaseous superdisks, which are remnants
of gas-rich mergers. These superdisks, with sizes of several tens of
kiloparsecs, have been previously postulated based on various properties of
RGs. The variety of morphologies of the EELRs can be accounted for with
a combination of the source age and of the orientation of the
superdisk with respect to the radio axis. The bubble-like
morphology arises when the radio source is in its early phases and
confined within the superdisk. At later stages, the EELR properties
depend on the relative orientation of jets and superdisks: the sources
with a large radio-line offset escape more rapidly from the regions
where the ISM is denser; whereas, in those located at smaller angles, the
jet-cloud interactions are active for a longer time.

We speculate that the alignment often observed between the axis of the
AGN and the gaseous disk might be due to the accretion of material
over a large time interval confined to a preferred plane, that is, the
orbital plane of the gas-rich galaxy responsible for the formation of
the superdisks. This effect tends to align the SMBH spin with that of
the accreting matter.

Concerning the AGN feedback, in these nearby radio-loud AGNs, energy is
mainly transferred to the ISM during the early phases of their
lifetime when this is compressed and heated, but it does not generally
lead to star formation. The outflows of ionized gas, commonly detected
in these sources, also have a limited impact on the external
medium. Nonetheless, the presence of extended gaseous structures
possibly associated with superdisks extends the timescale of the
impact of the AGNs onto the ISM phase with respect to the previous
scenario in which this was confined to the size of the host galaxy,
that is, in the CSS phase. However, the warm line-emitting gas represents
only a small component of the ISM. A global view of radio
feedback also requires one to include  the cold ISM phase in the analysis.

Further analysis is also needed concerning the numerical simulations
by considering more extended gas distribution with respect to those
that have been included so far. It is also of great interest to assess
whether superdisks are only present in the powerful radio-loud
AGNs, a result that suggests a strong connection between these sources
with gas-rich mergers, or whether these are also found in radio-quiet
AGNs. Finally, superdisks should also be present in non-active
sources if they are observed after the merger event, but before the
onset of the nuclear activity.

\begin{acknowledgements}
A.M. acknowledges financial support from PRIN-MIUR contract
2017PH3WAT. S.B. and C.O. acknowledge support from the Natural
Sciences and Engineering Research Council (NSERC) of Canada.  Based on
observations made with ESO Telescopes at the La Silla Paranal
Observatory under programme ID 097.B-0766(A). This research has made
use of data obtained from the Chandra Data Archive. The National Radio
Astronomy Observatory is a facility of the National Science Foundation
operated under cooperative agreement by Associated Universities,
Inc. B.B. acknowledge financial contribution from the agreement
ASI-INAF I/037/12/0. Some of the data presented here are based on
observations made with the NASA/ESA Hubble Space Telescope, obtained
from the data archive at the Space Telescope Science Institute. STScI
is operated by the Association of Universities for Research in
Astronomy, Inc. under NASA contract NAS 5-26555.

\end{acknowledgements}

\begin{appendix}

\section{Emission line images and velocity curves.}
\label{appA}  
\begin{figure*}  
\centering{ 
\includegraphics[width=17.cm]{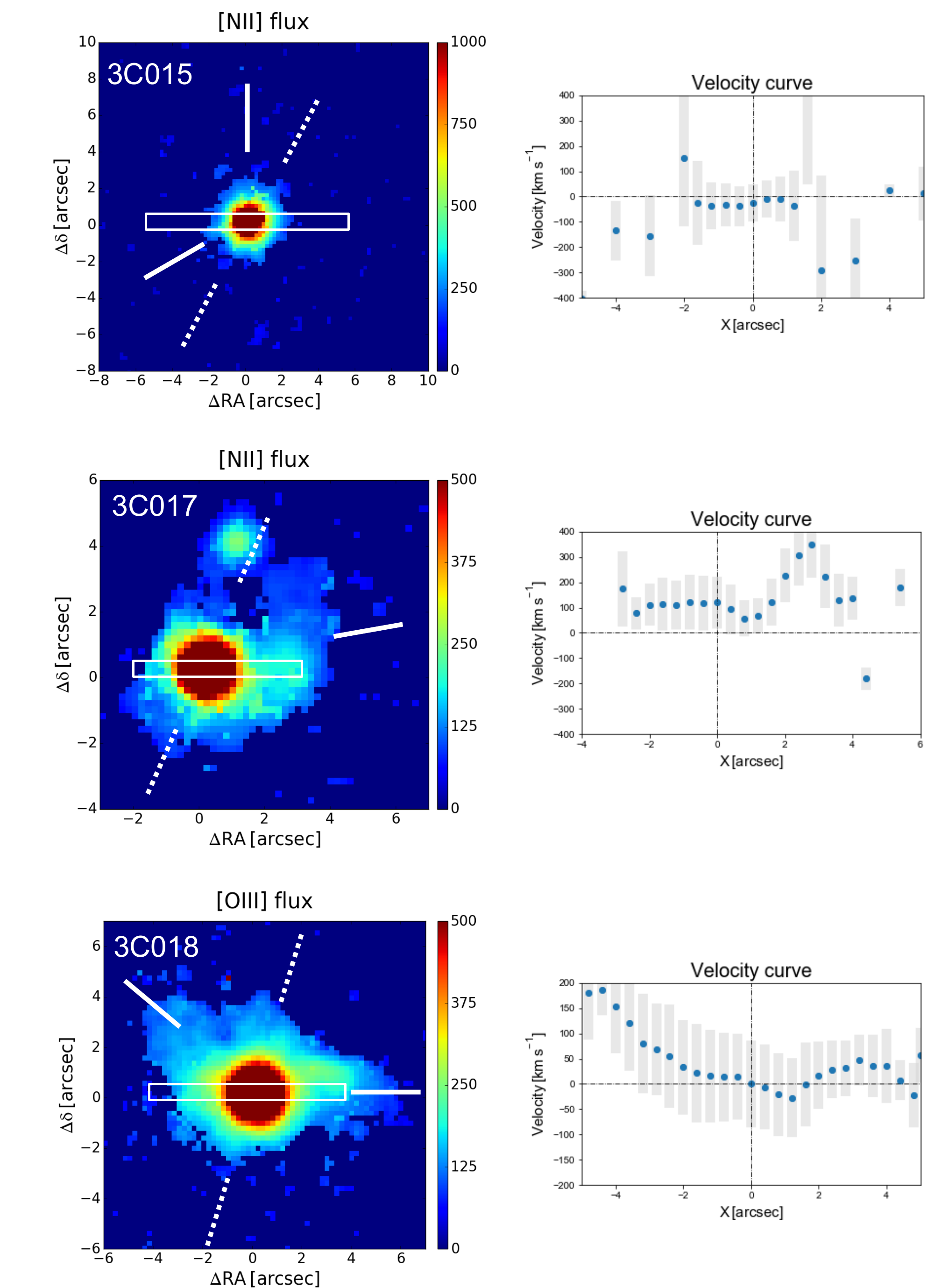}}
\caption{Images of the brightest emission line for the 37
  radiogalaxies observed with MUSE. The white dotted lines mark the
  radio position angle, and the solid white lines show the orientation of the
  emission line structures, when possible, on both sides of the
  nucleus. The gray line is the orientation of the kinematical major
  axis in the inner regions measured with {kinemetry} for all
  sources with data quality sufficient for a robust measurement (see
  text for details). The rectangle represents the regions from which
  we extracted the velocity curve shown in the right panel. The length
  of the gray bars represents the line width at each location.}
\label{ppt2}
\end{figure*}

\begin{figure*}  
\addtocounter{figure}{-1}
\centering{ 
\includegraphics[width=17.5cm]{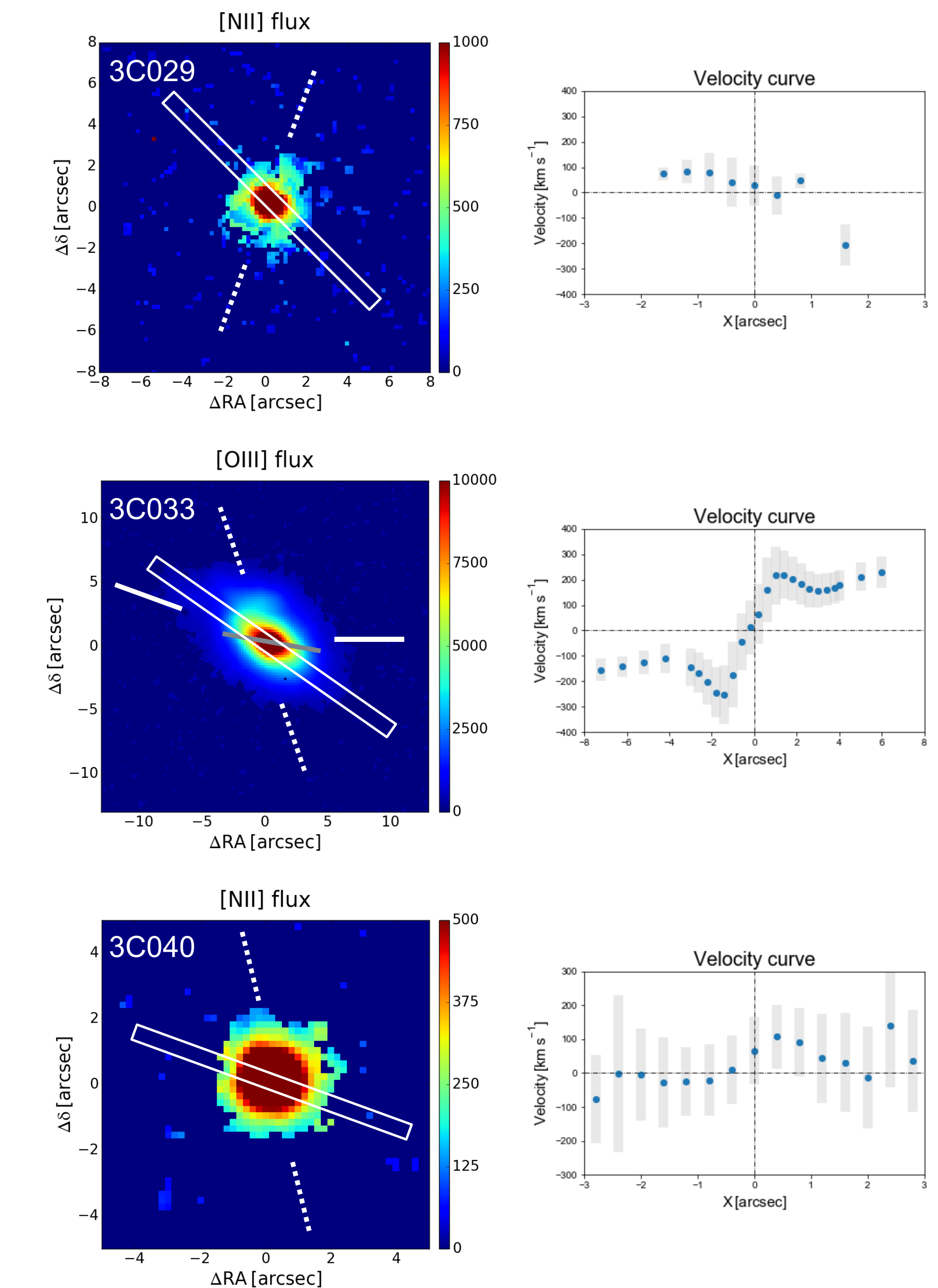}}
\caption{- continued.}
\end{figure*}   

\begin{figure*}  
\addtocounter{figure}{-1}
\centering{ 
\includegraphics[width=17.5cm]{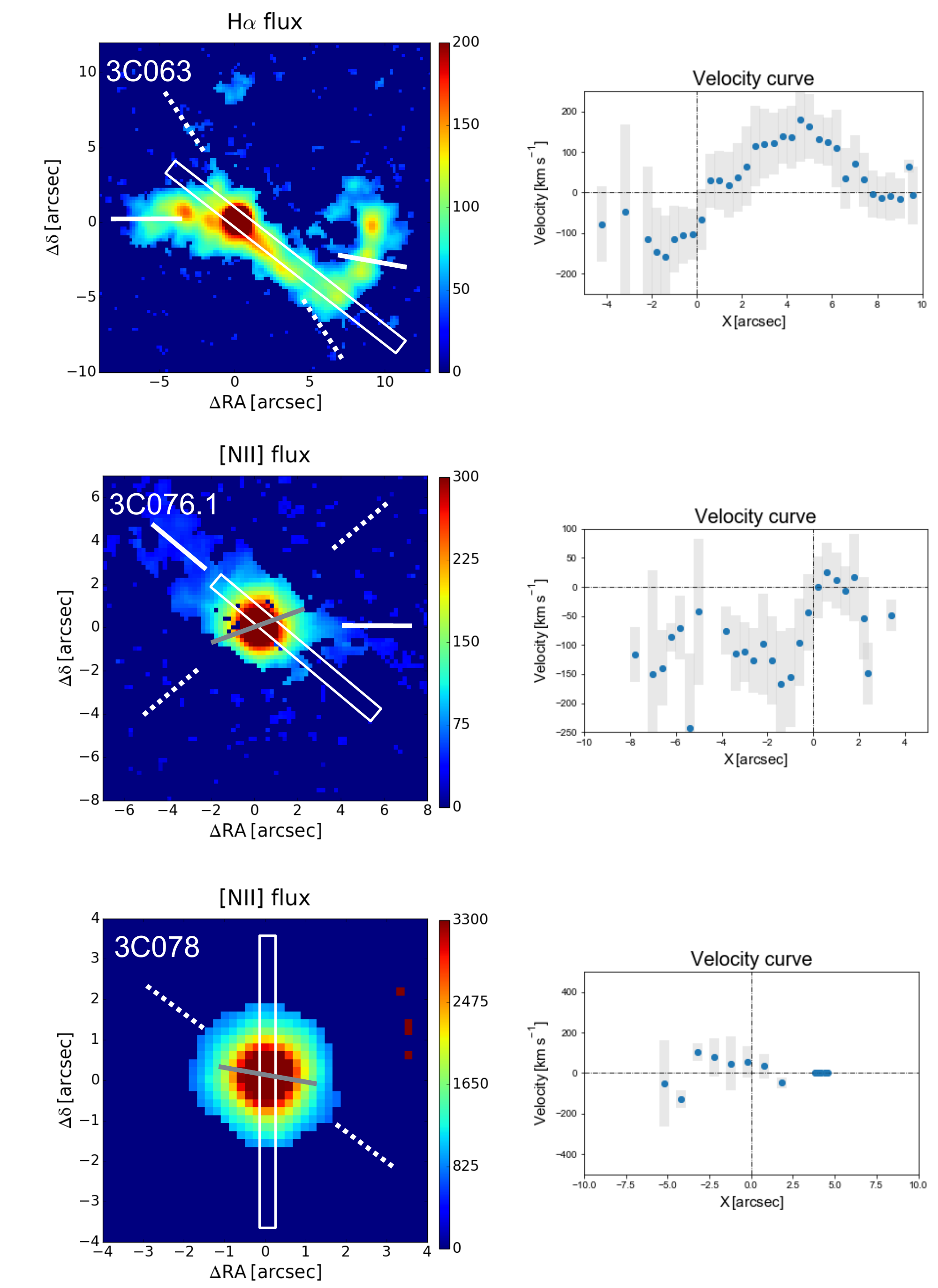}}
\caption{- continued.}
\end{figure*}   

\begin{figure*}  
\addtocounter{figure}{-1}
\centering{ 
\includegraphics[width=17.5cm]{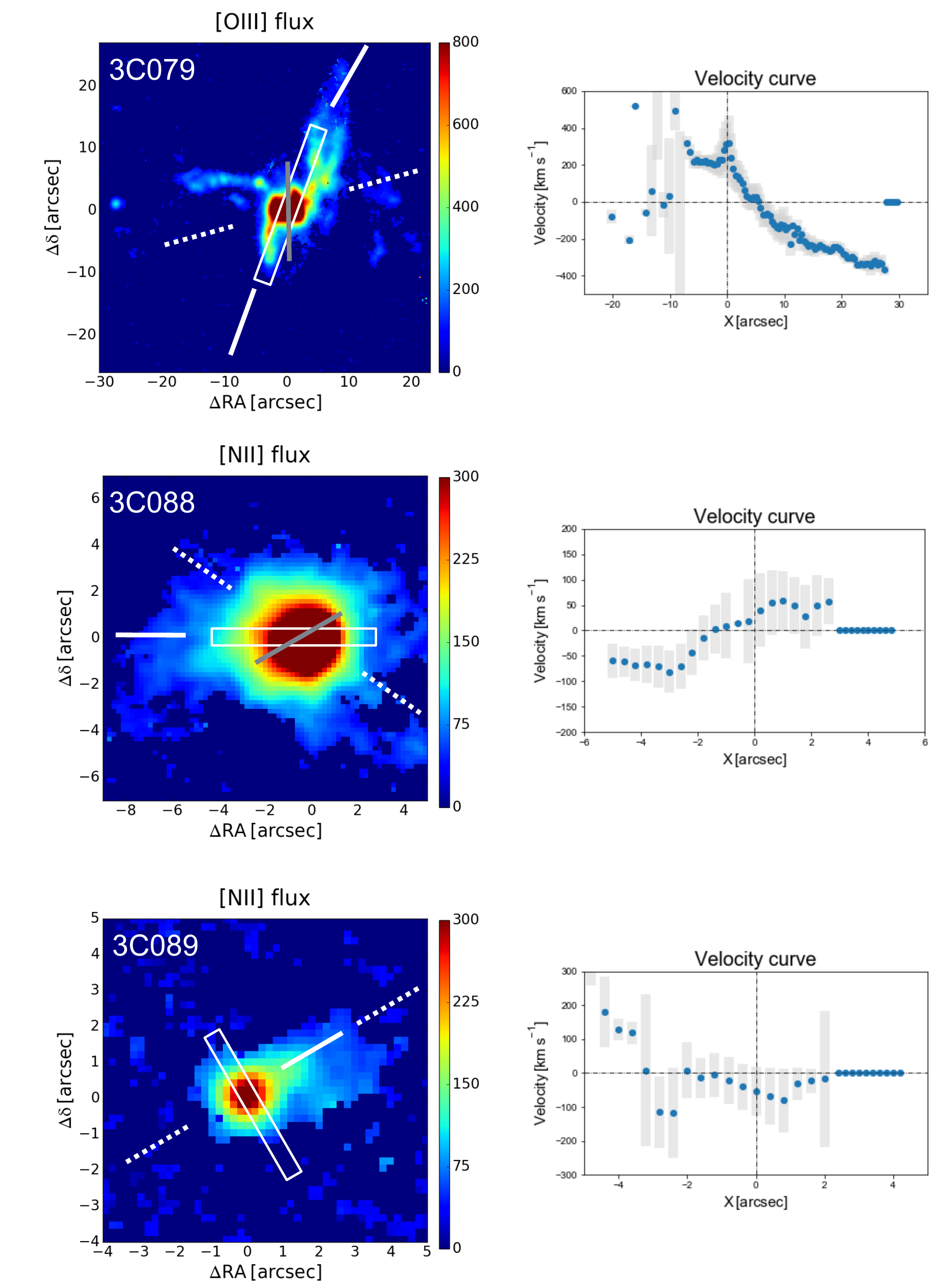}}
\caption{- continued.}
\end{figure*}   

\begin{figure*}  
\addtocounter{figure}{-1}
\centering{ 
\includegraphics[width=17.5cm]{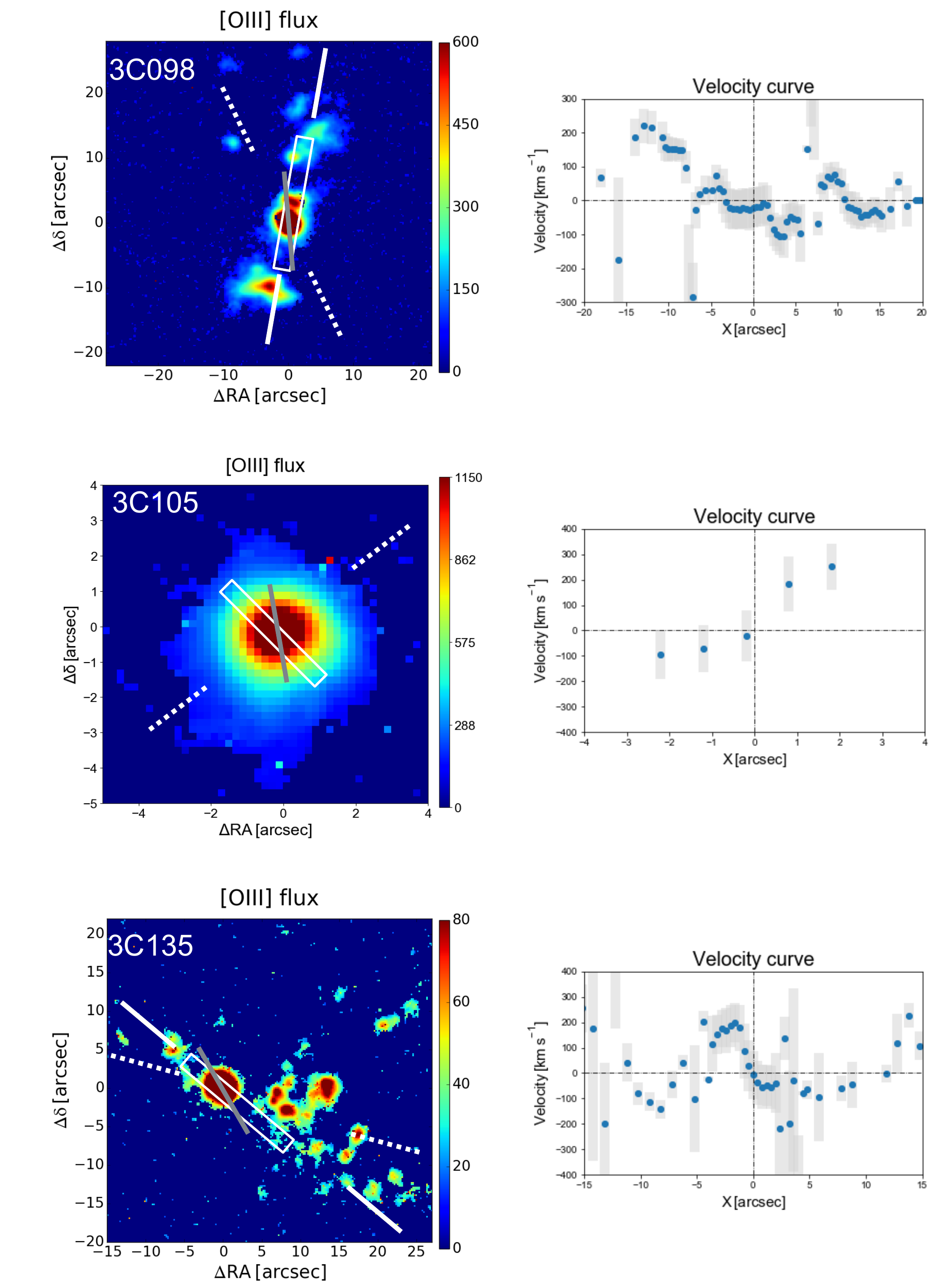}}
\caption{- continued.}
\end{figure*}   

\begin{figure*}  
\addtocounter{figure}{-1}
\centering{ 
\includegraphics[width=17.5cm]{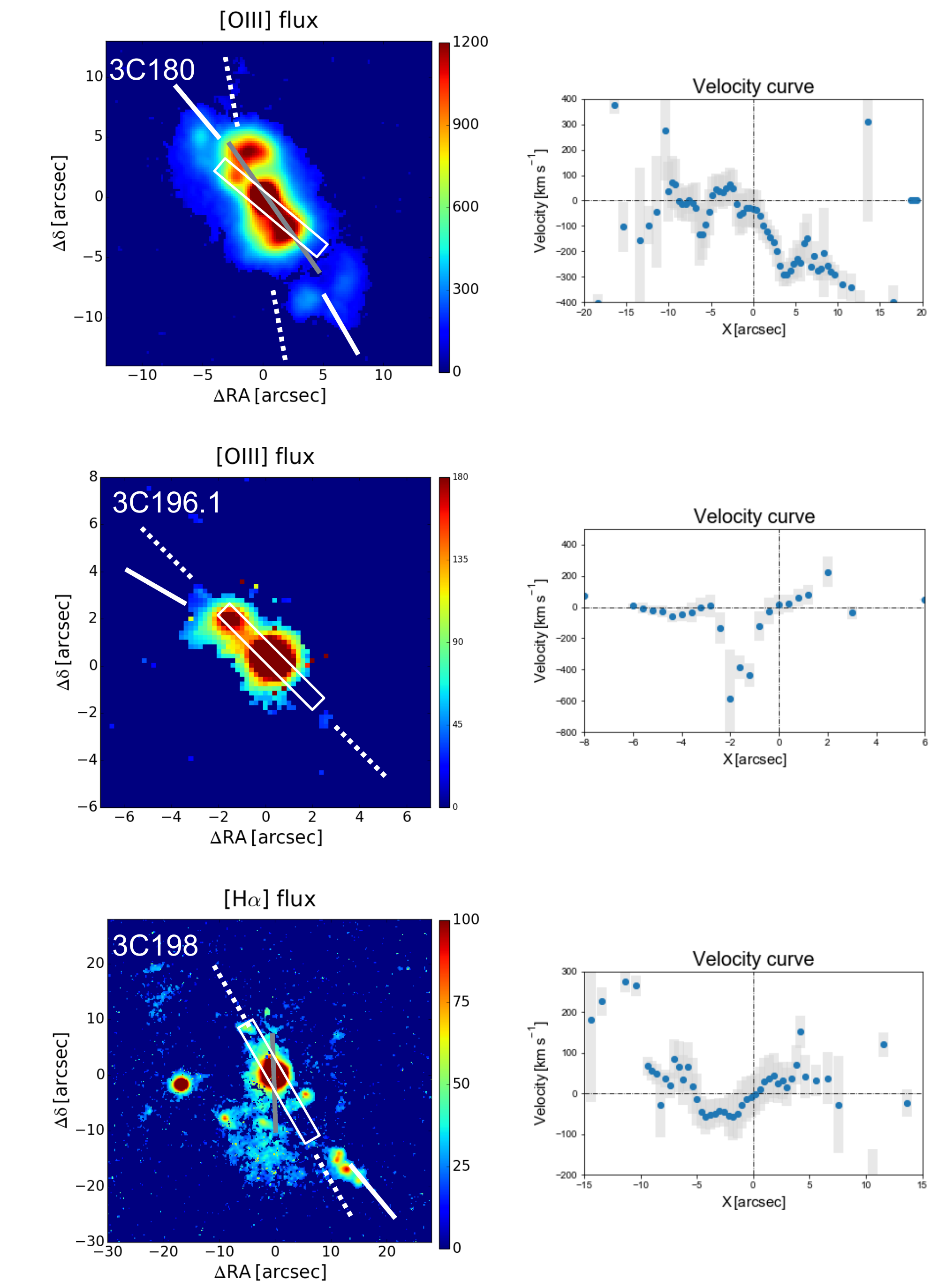}}
\caption{- continued.}
\end{figure*}   

\begin{figure*}  
\addtocounter{figure}{-1}
\centering{ 
\includegraphics[width=17.5cm]{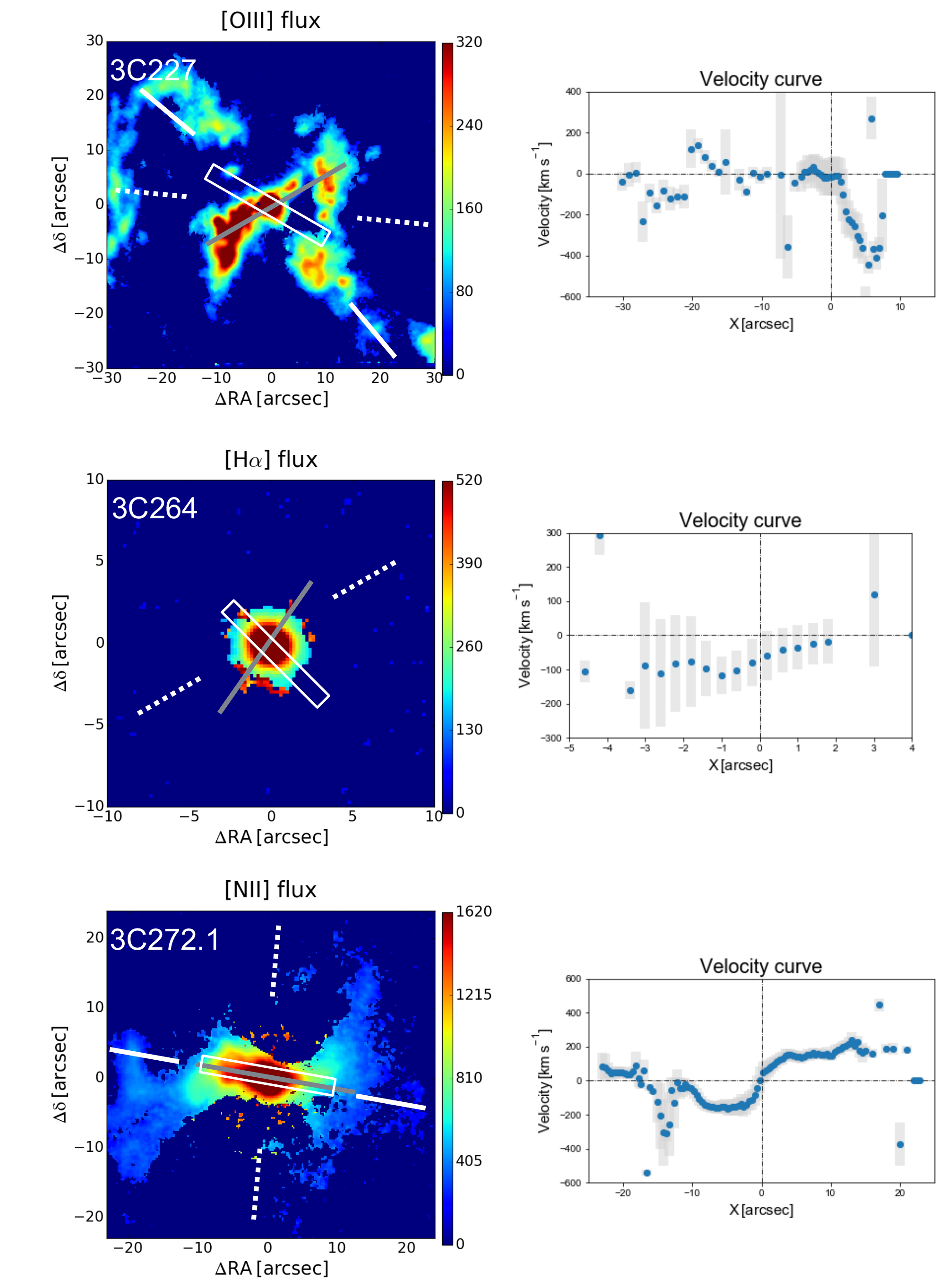}}
\caption{- continued.}
\end{figure*}   

\begin{figure*}  
\addtocounter{figure}{-1}
\centering{ 
\includegraphics[width=17.5cm]{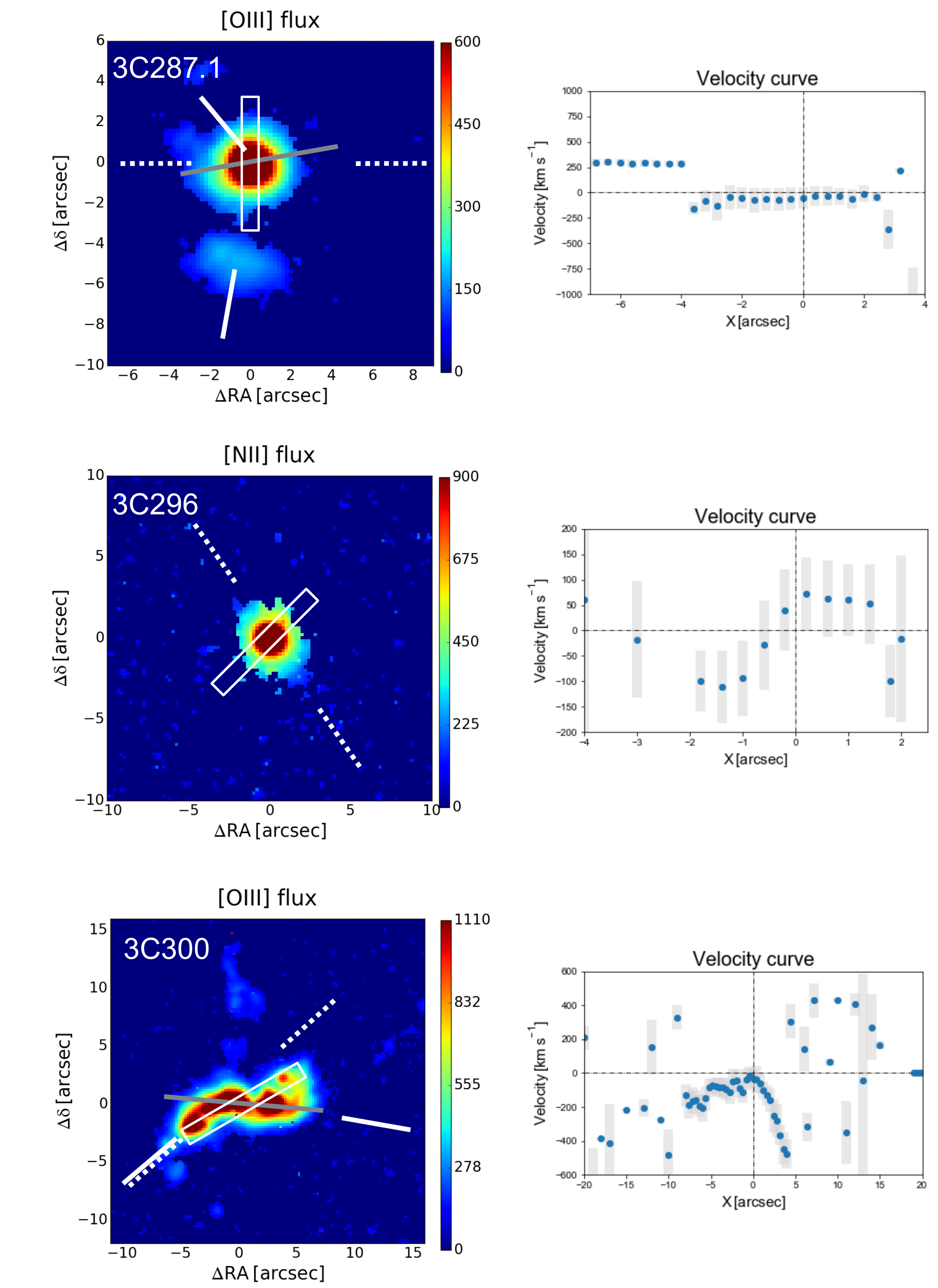}}
\caption{- continued.}
\end{figure*}   

\begin{figure*}  
\addtocounter{figure}{-1}
\centering{ 
\includegraphics[width=17.5cm]{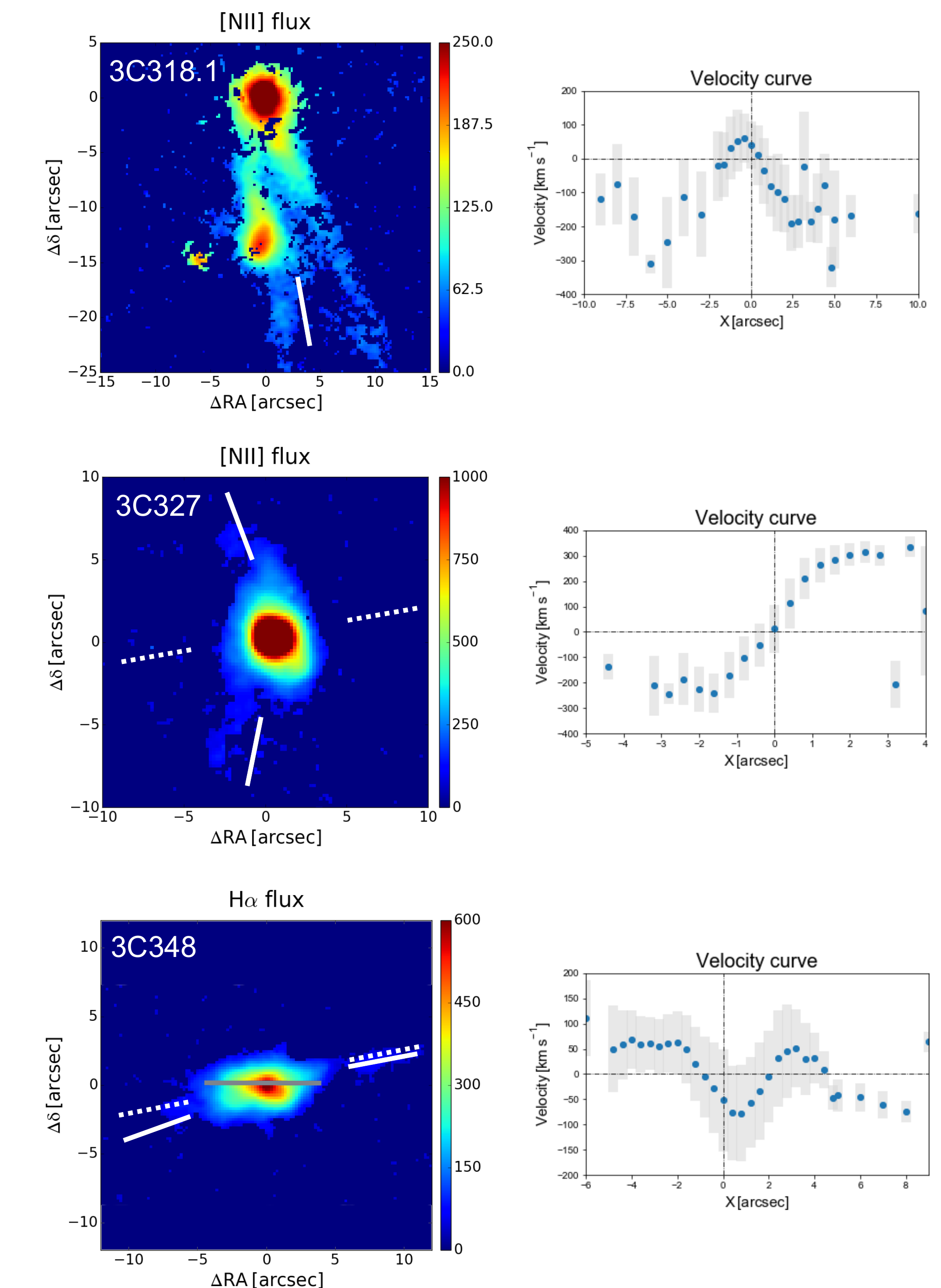}}
\caption{- continued.}
\end{figure*}   

\begin{figure*}  
\addtocounter{figure}{-1}
\centering{ 
\includegraphics[width=17.5cm]{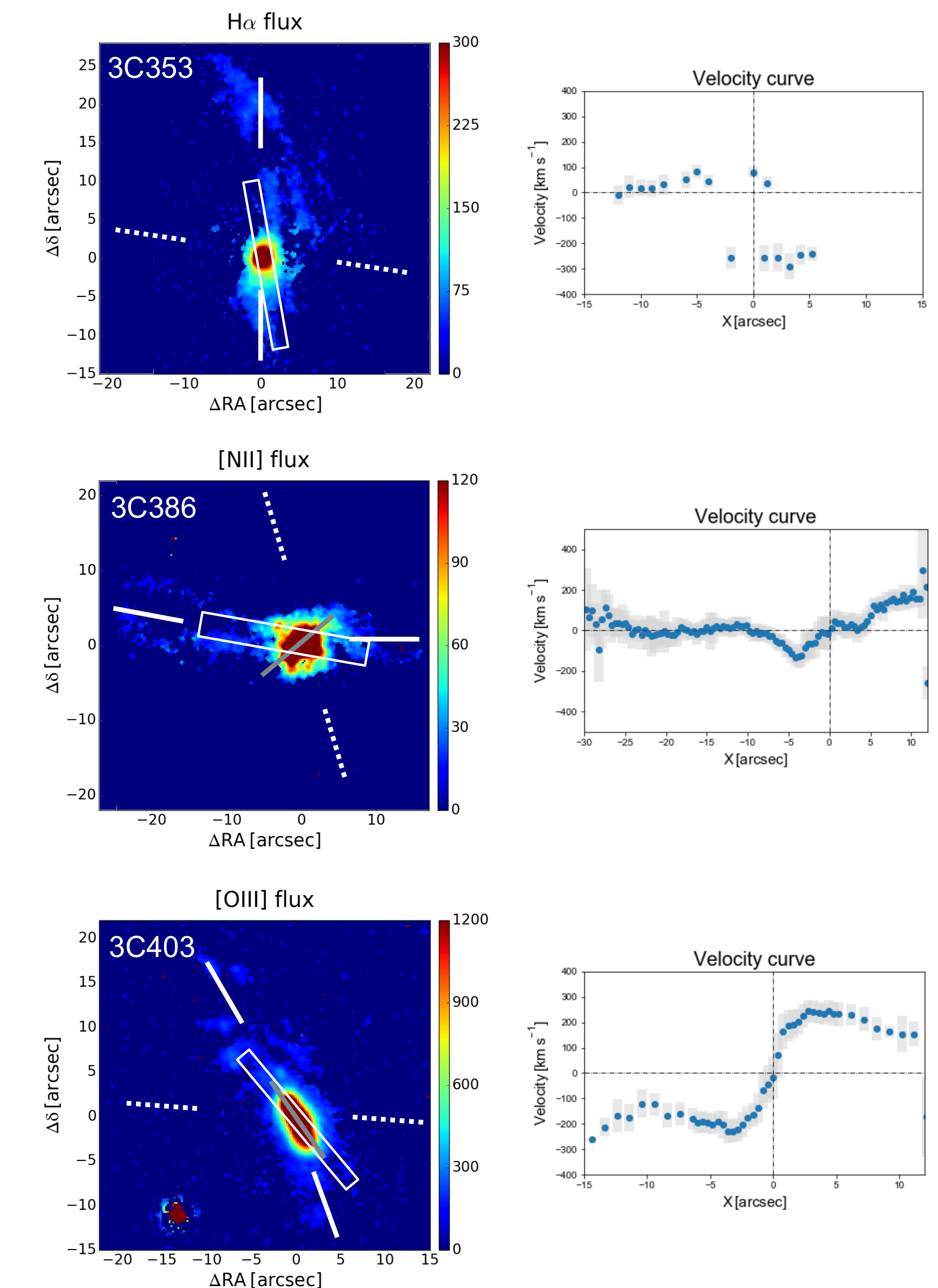}}
\caption{- continued.}
\end{figure*}   

\begin{figure*}  
\addtocounter{figure}{-1}
\centering{ 
\includegraphics[width=17.5cm]{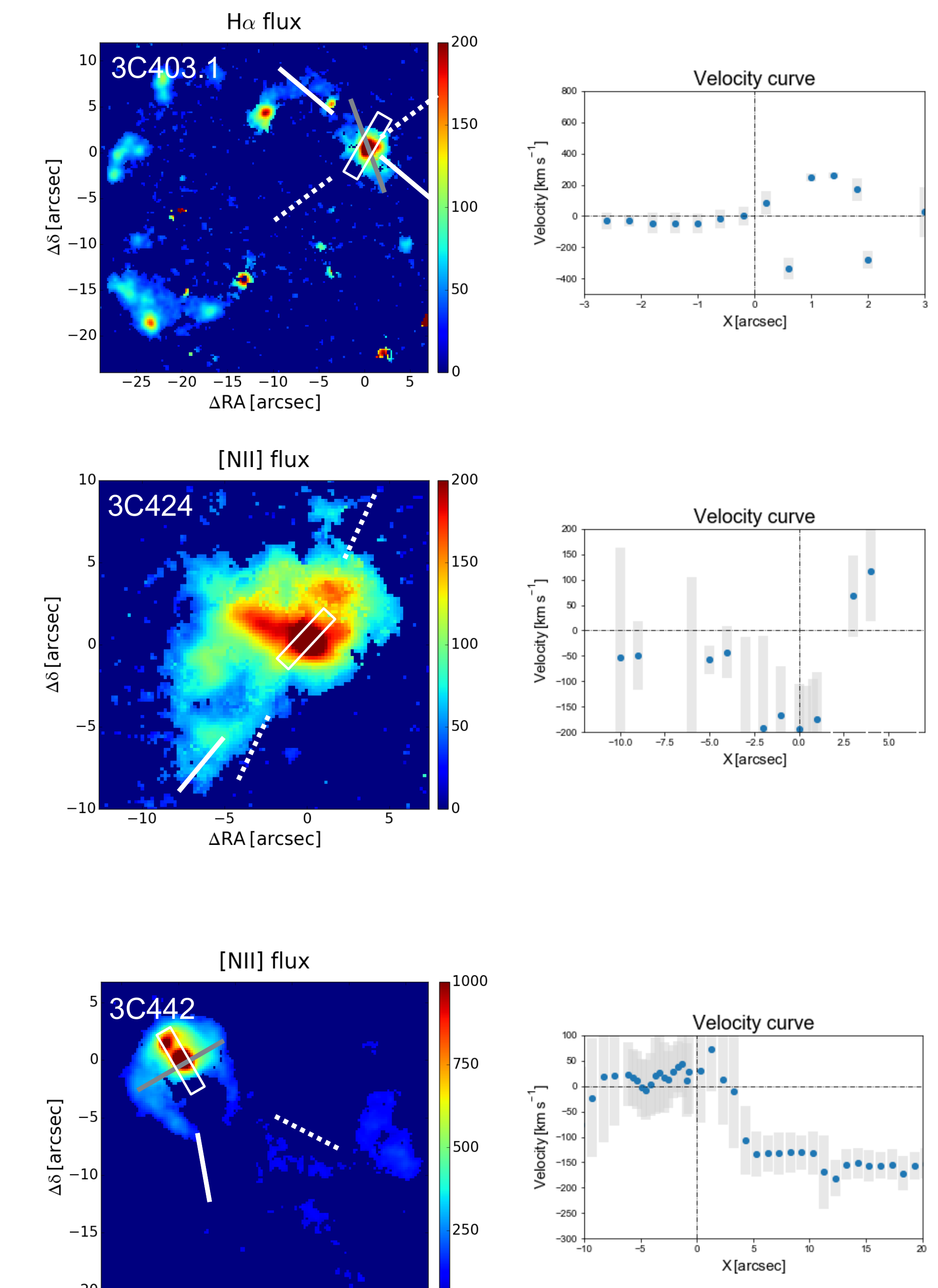}}
\caption{- continued.}
\end{figure*}   

\begin{figure*}  
\addtocounter{figure}{-1}
\centering{ 
\includegraphics[width=17.5cm]{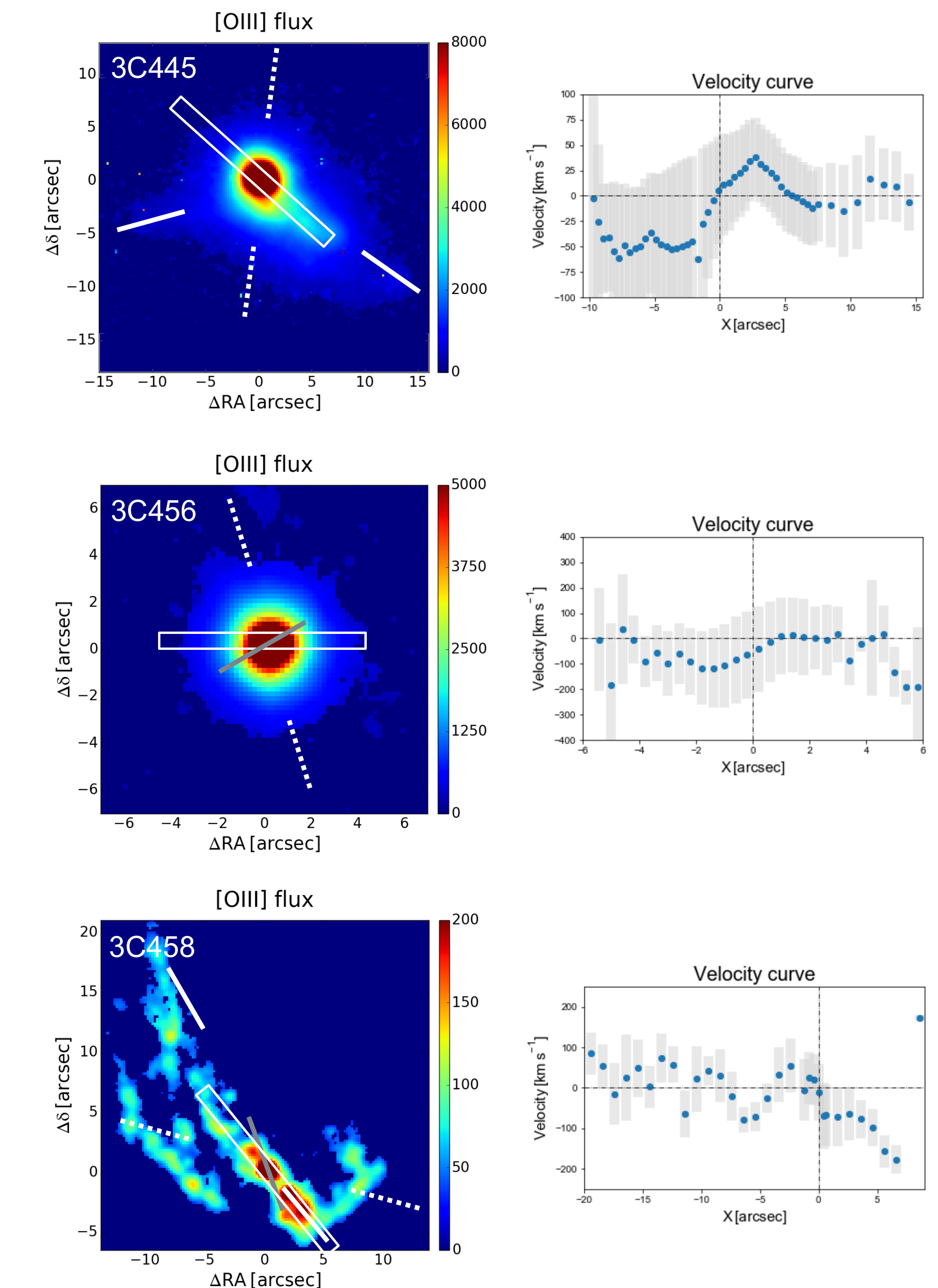}}
\caption{- continued.}
\end{figure*}   

\begin{figure*}  
\addtocounter{figure}{-1}
\centering{ 
\includegraphics[width=17.5cm]{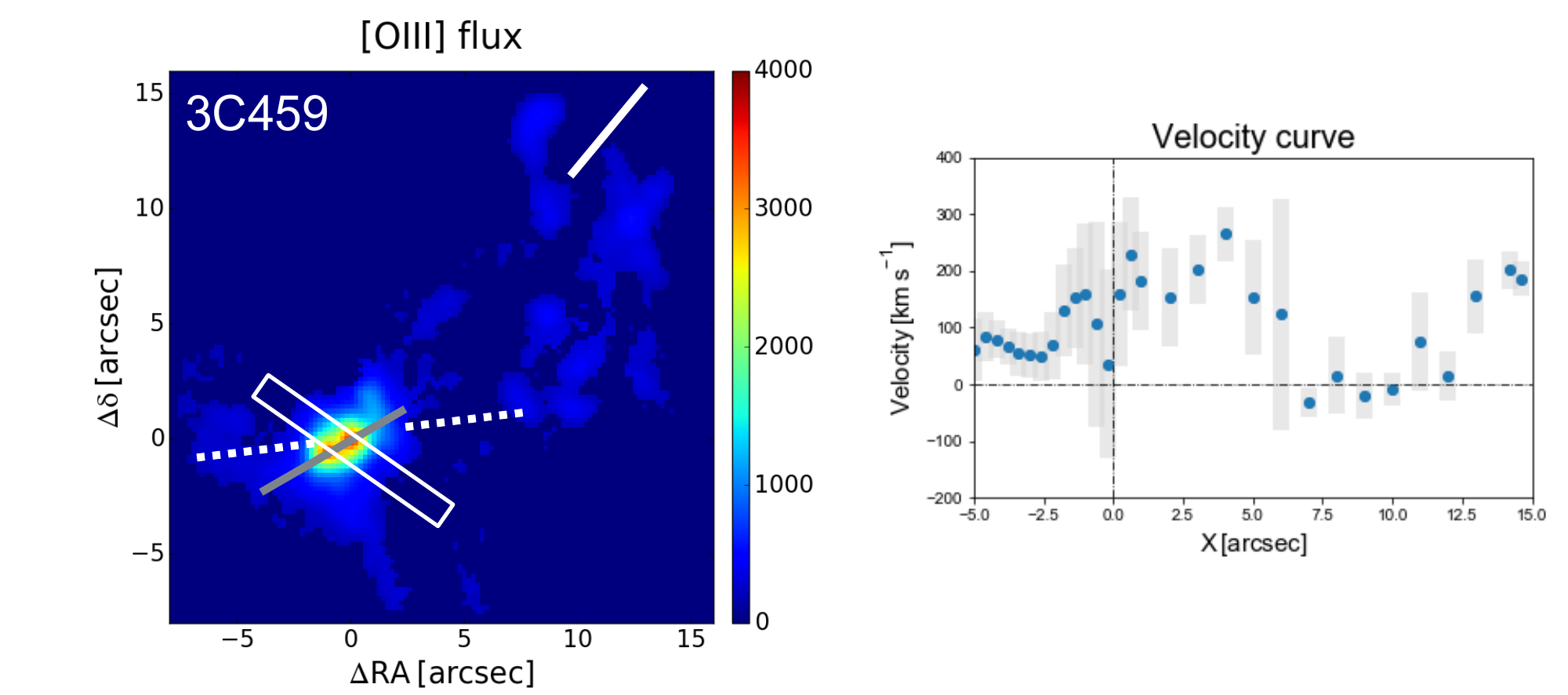}}
\caption{- continued.}
\end{figure*}   

\newpage
\clearpage
\section{Polar diagrams of the emission lines' distribution.}
\label{appB}  
\begin{figure*}
\centering{ 
\includegraphics[width=6.0cm]{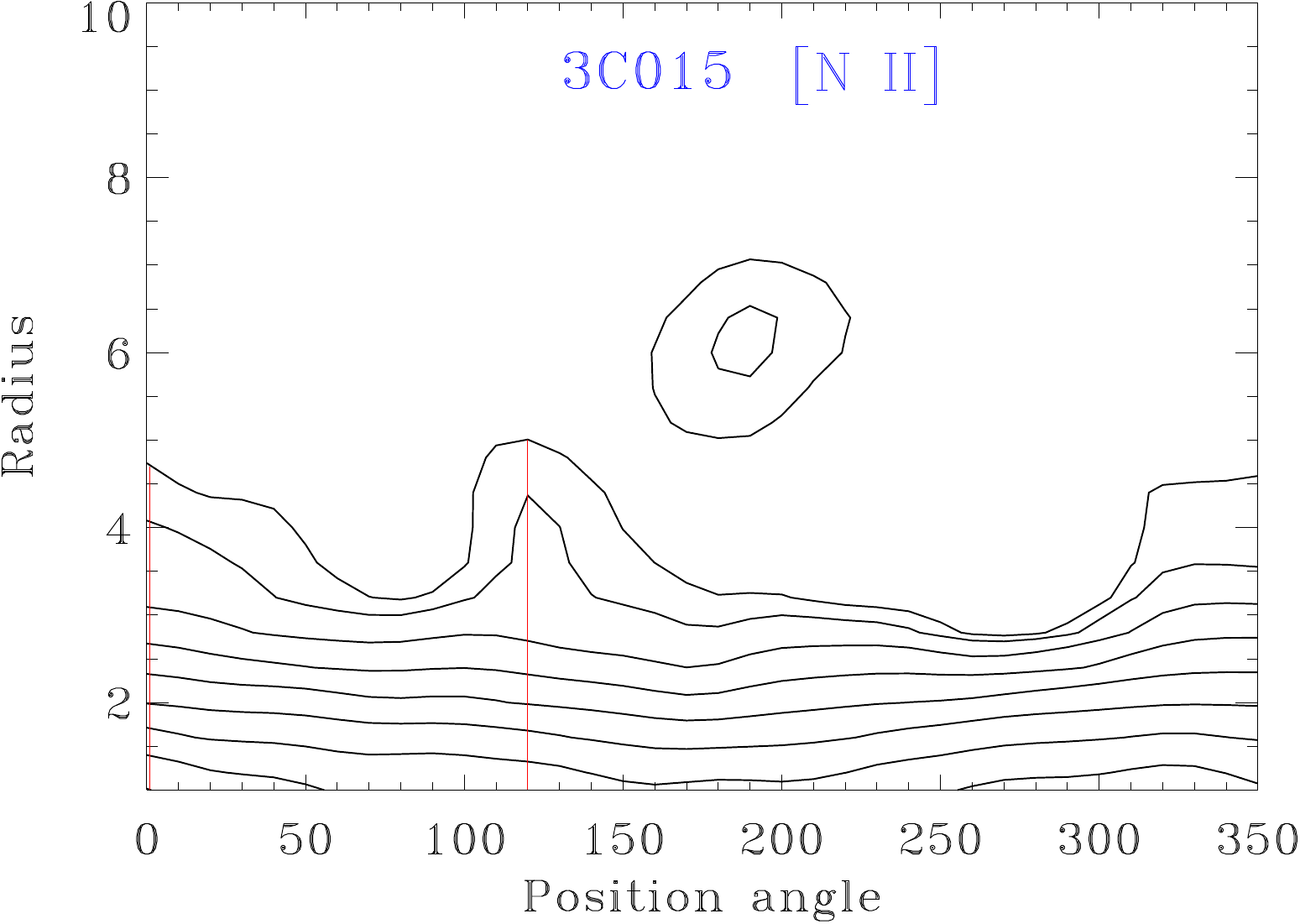}
\includegraphics[width=6.0cm]{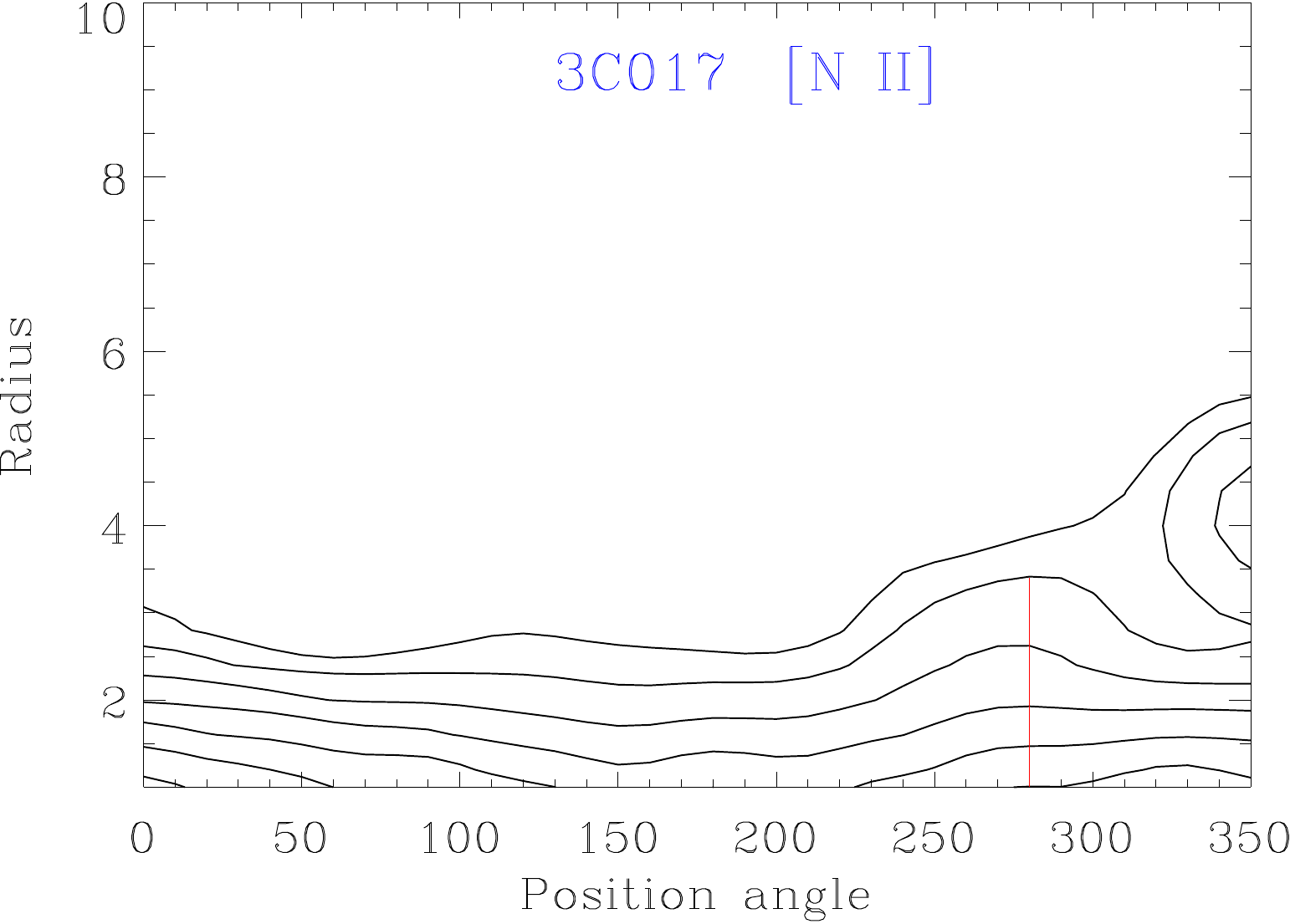}
\includegraphics[width=6.0cm]{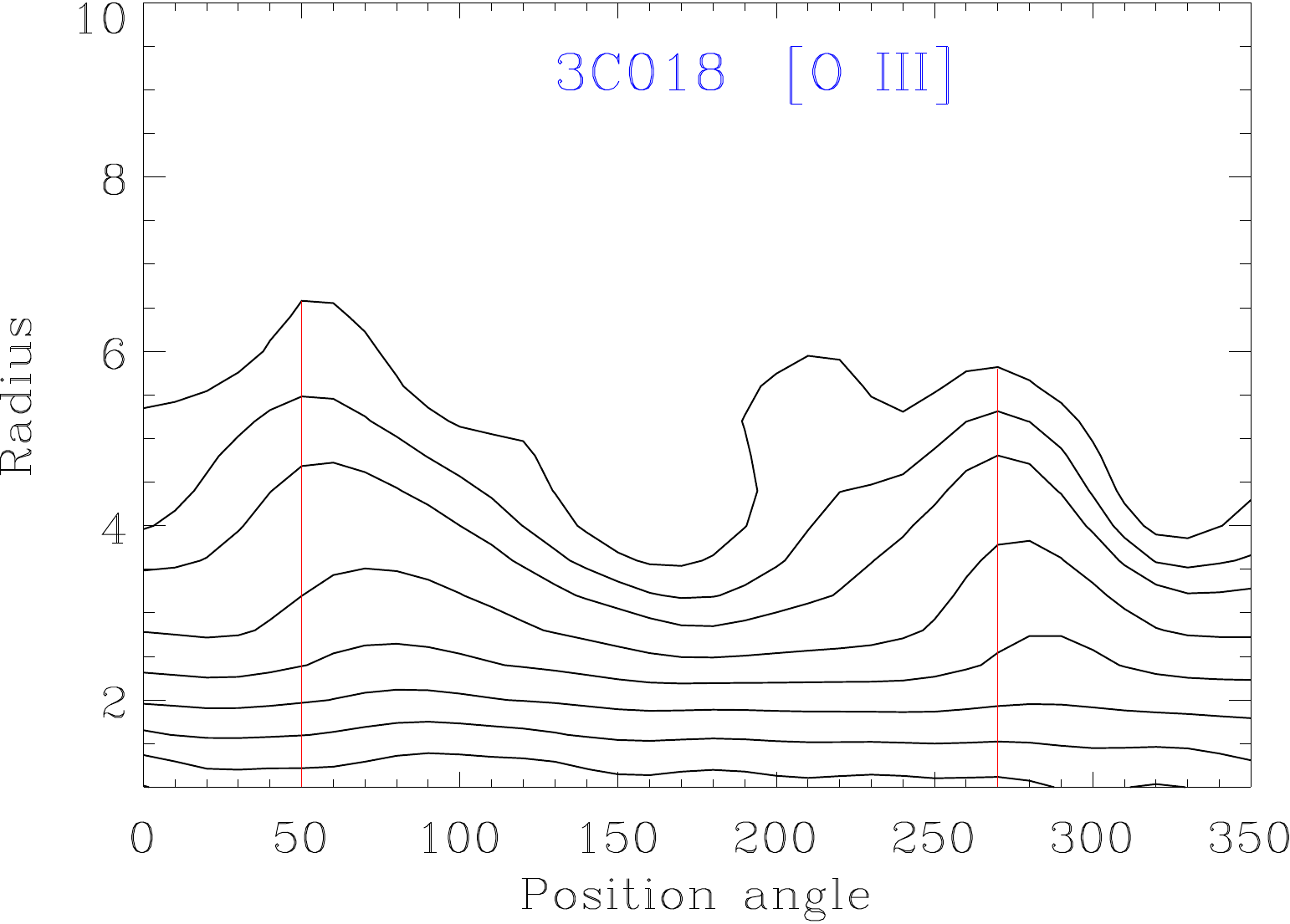}
\includegraphics[width=6.0cm]{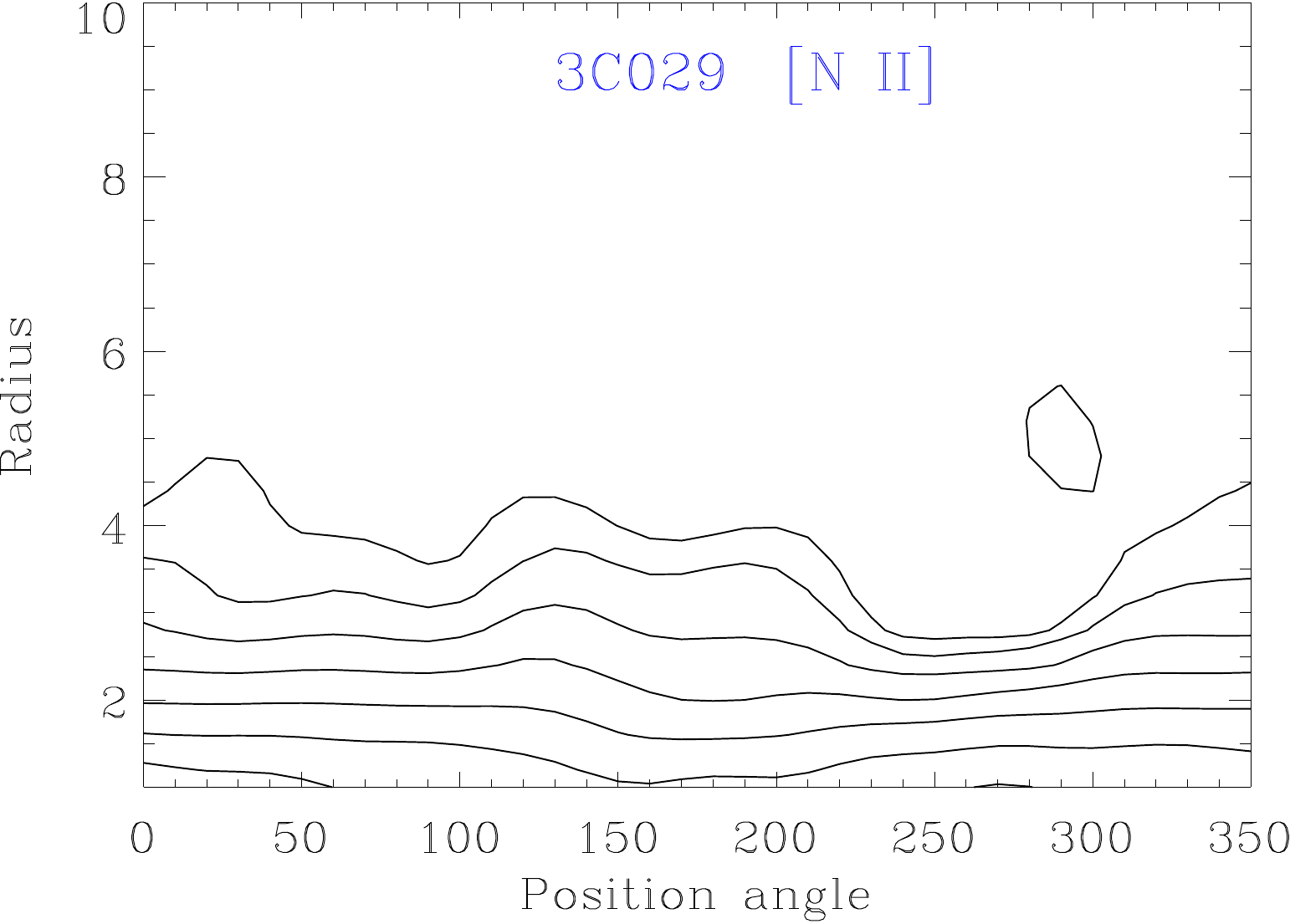}
\includegraphics[width=6.0cm]{3C033o3c.pdf}
\includegraphics[width=6.0cm]{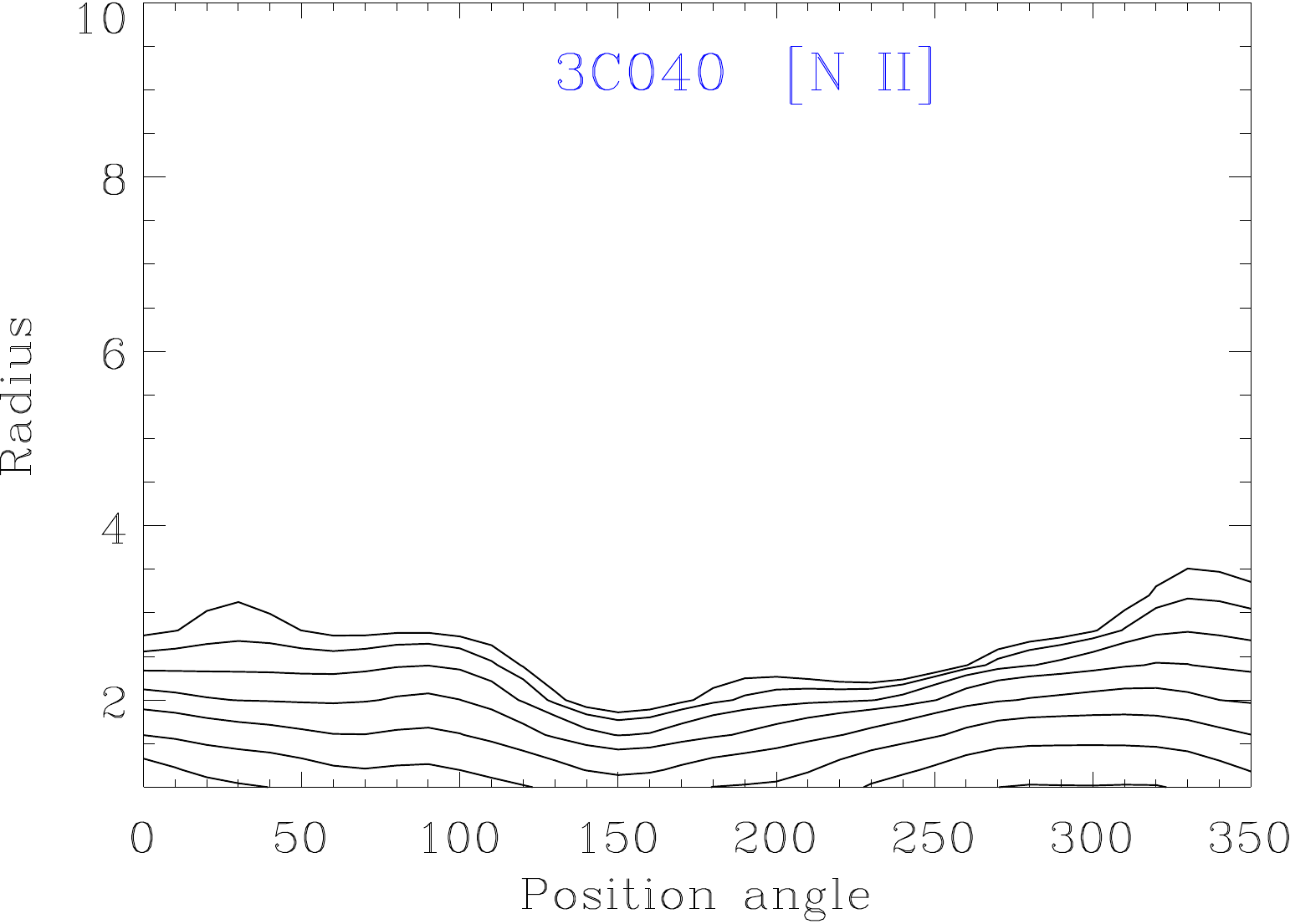}
\includegraphics[width=6.0cm]{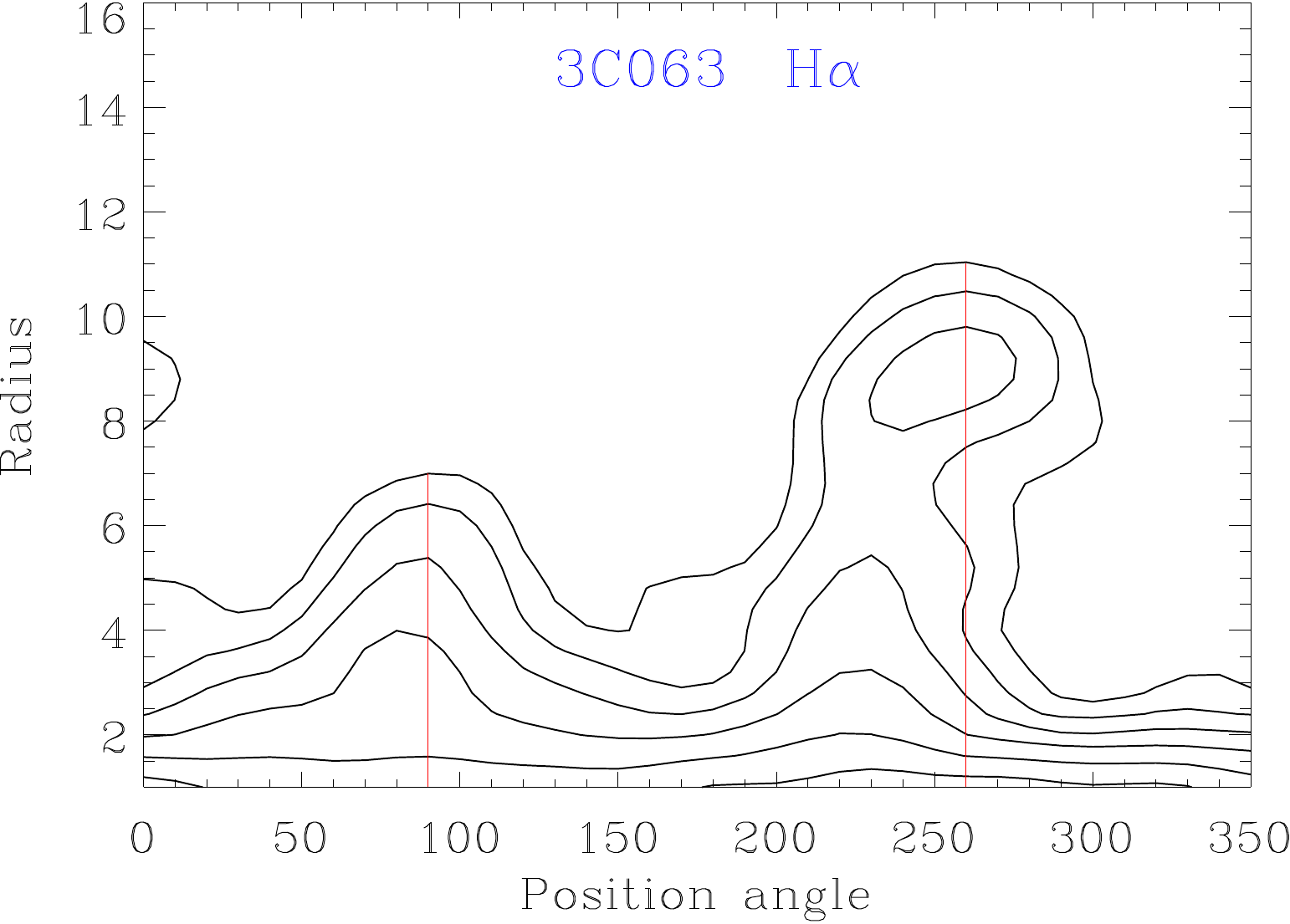}
\includegraphics[width=6.0cm]{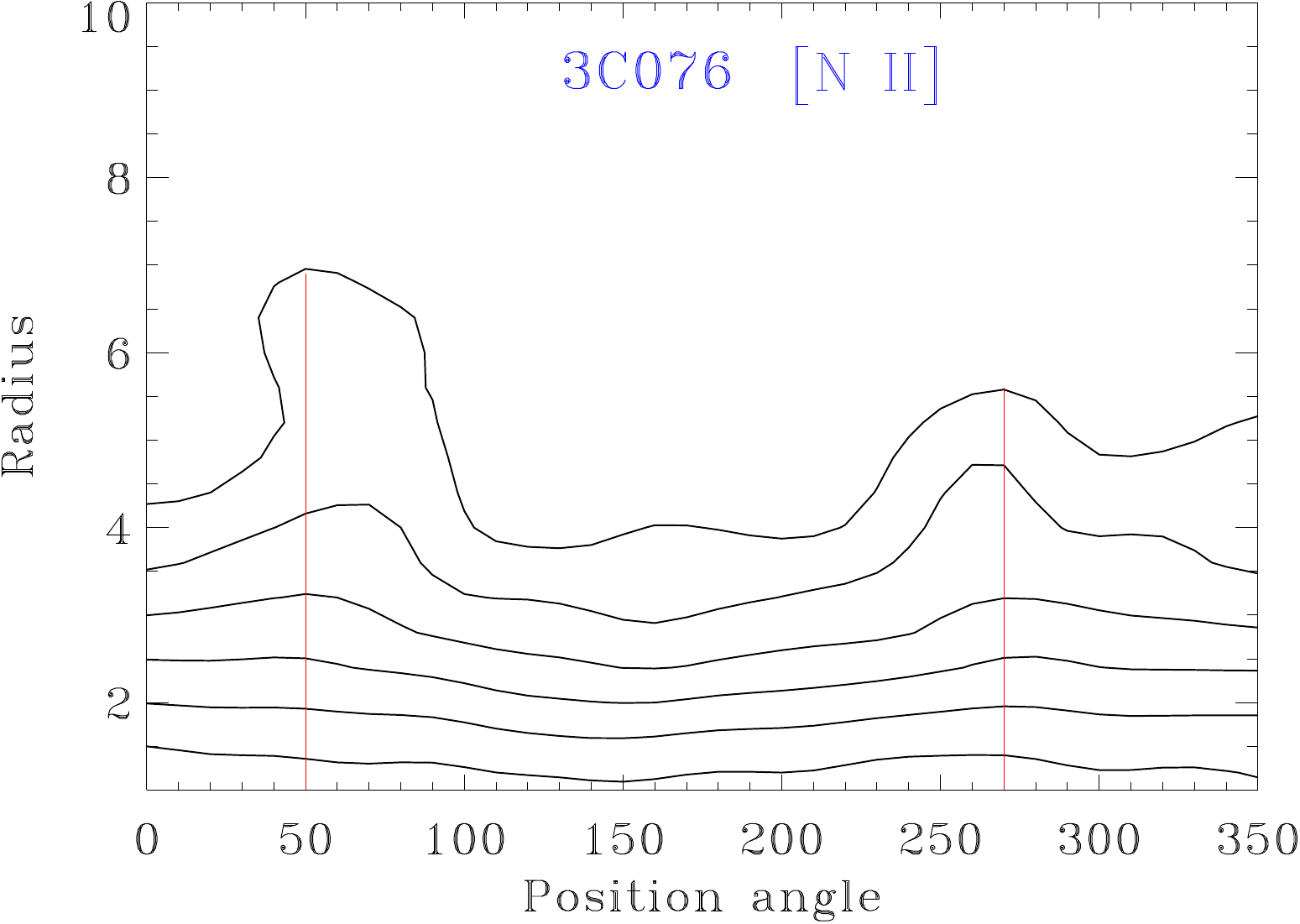}
\includegraphics[width=6.0cm]{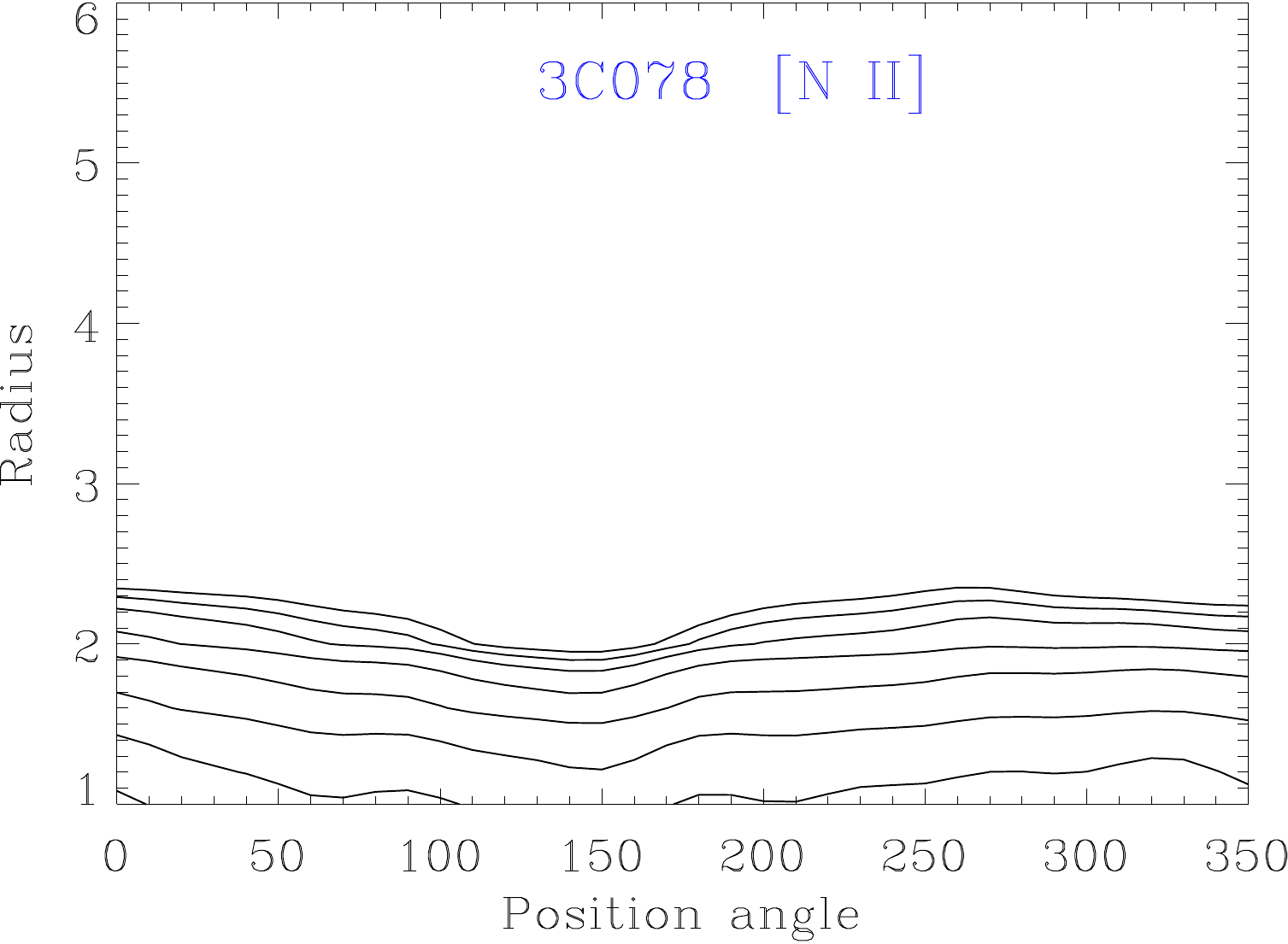}
\includegraphics[width=6.0cm]{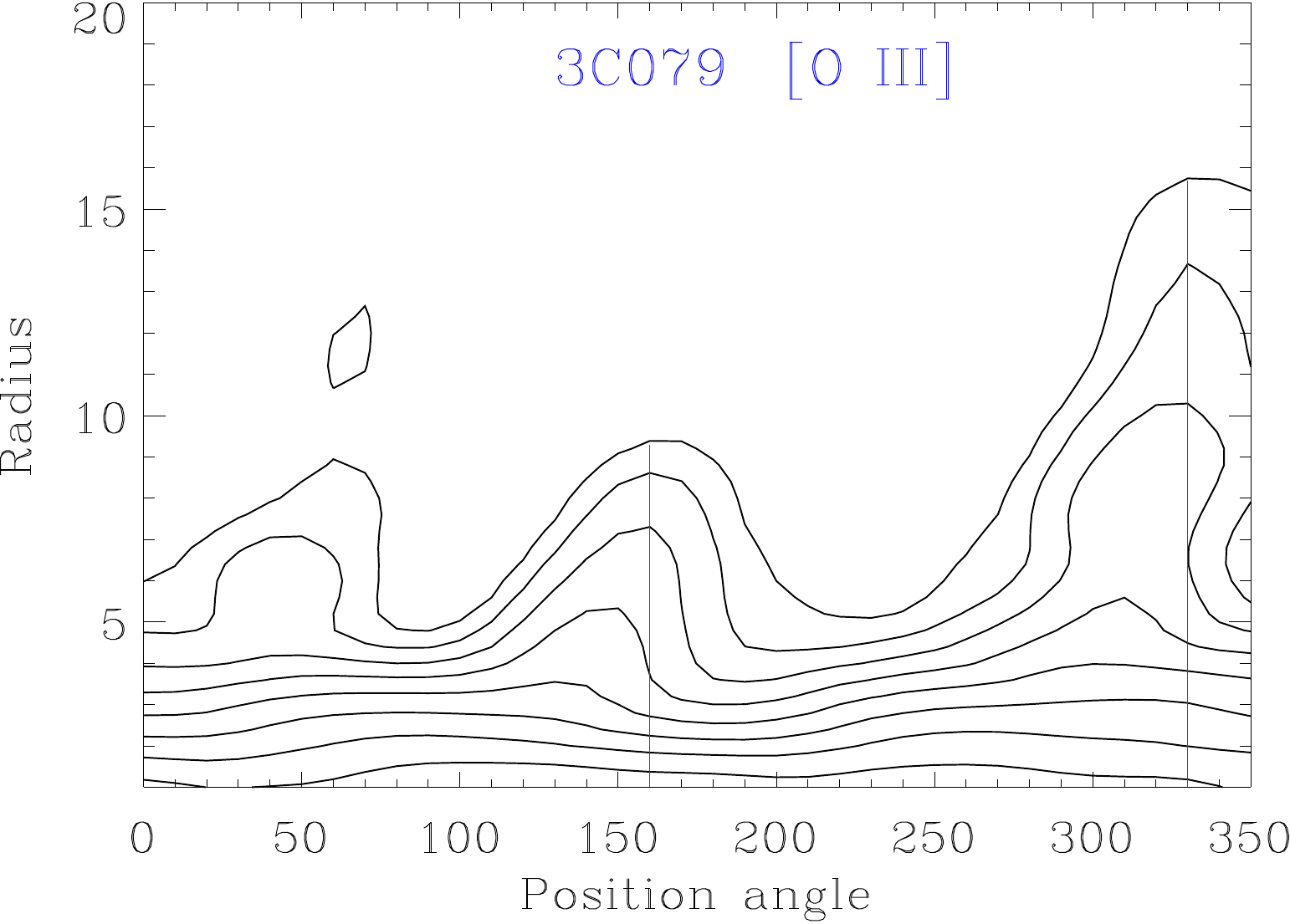}
\includegraphics[width=6.0cm]{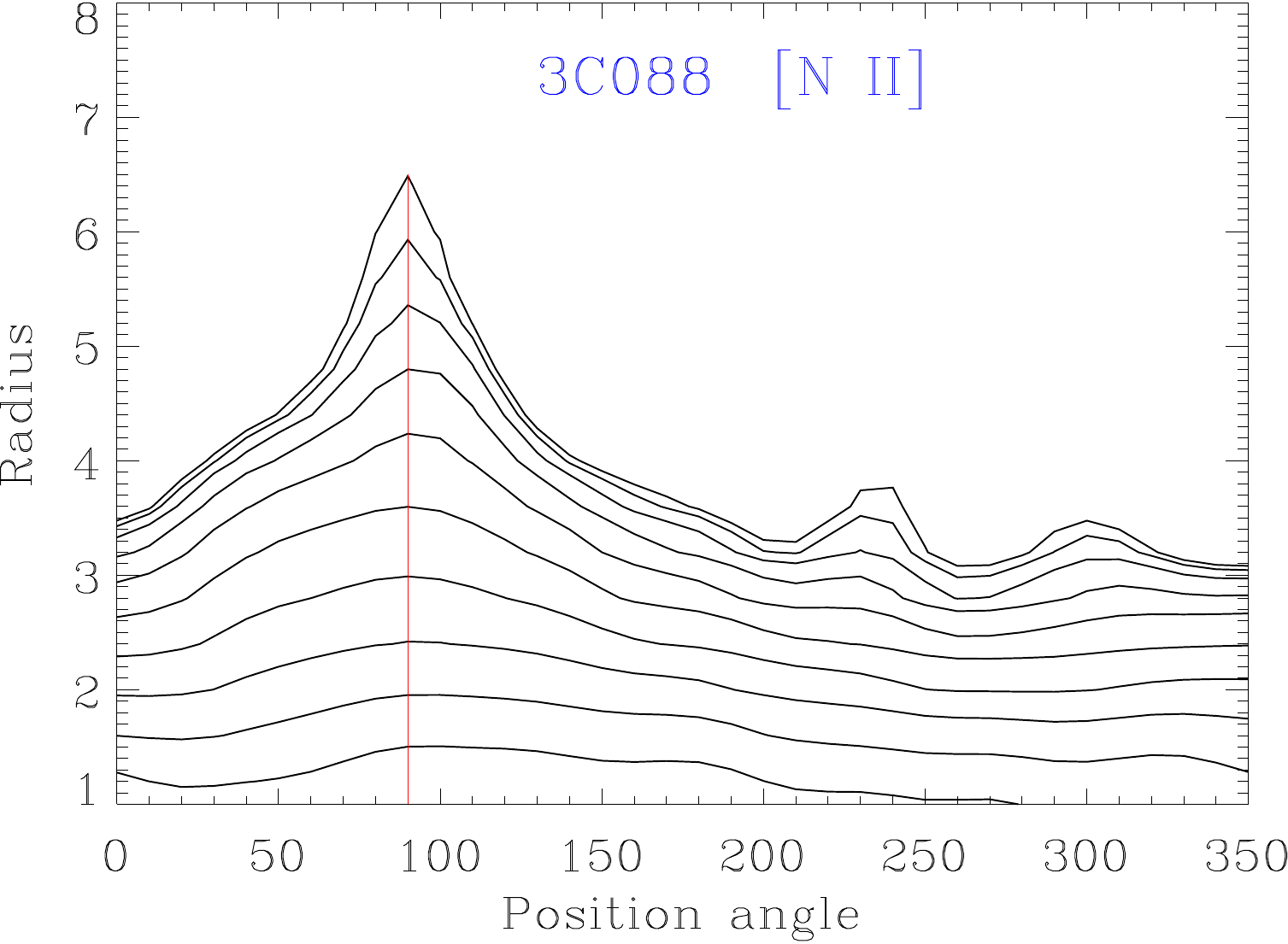}
\includegraphics[width=6.0cm]{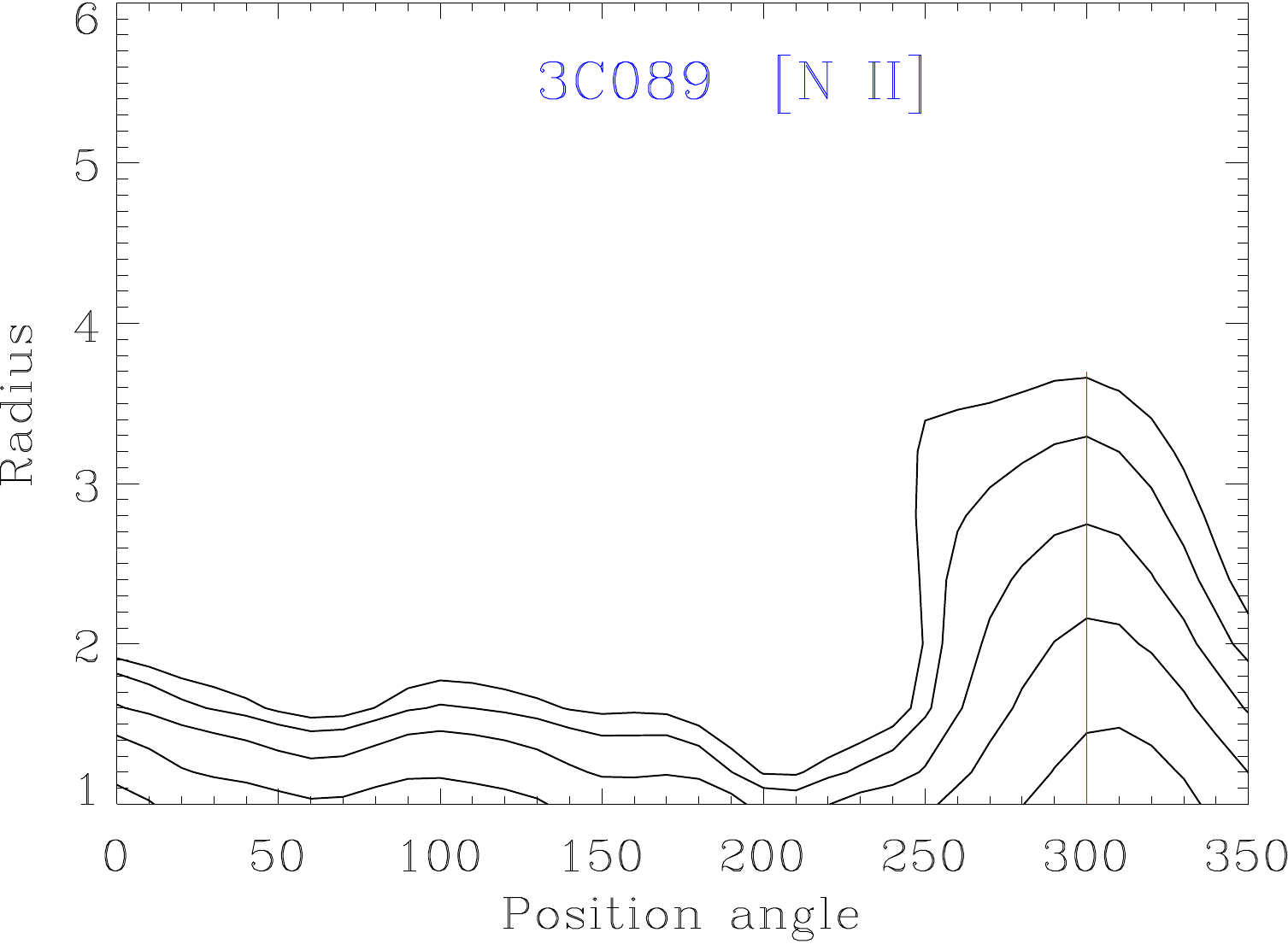}
\includegraphics[width=6.0cm]{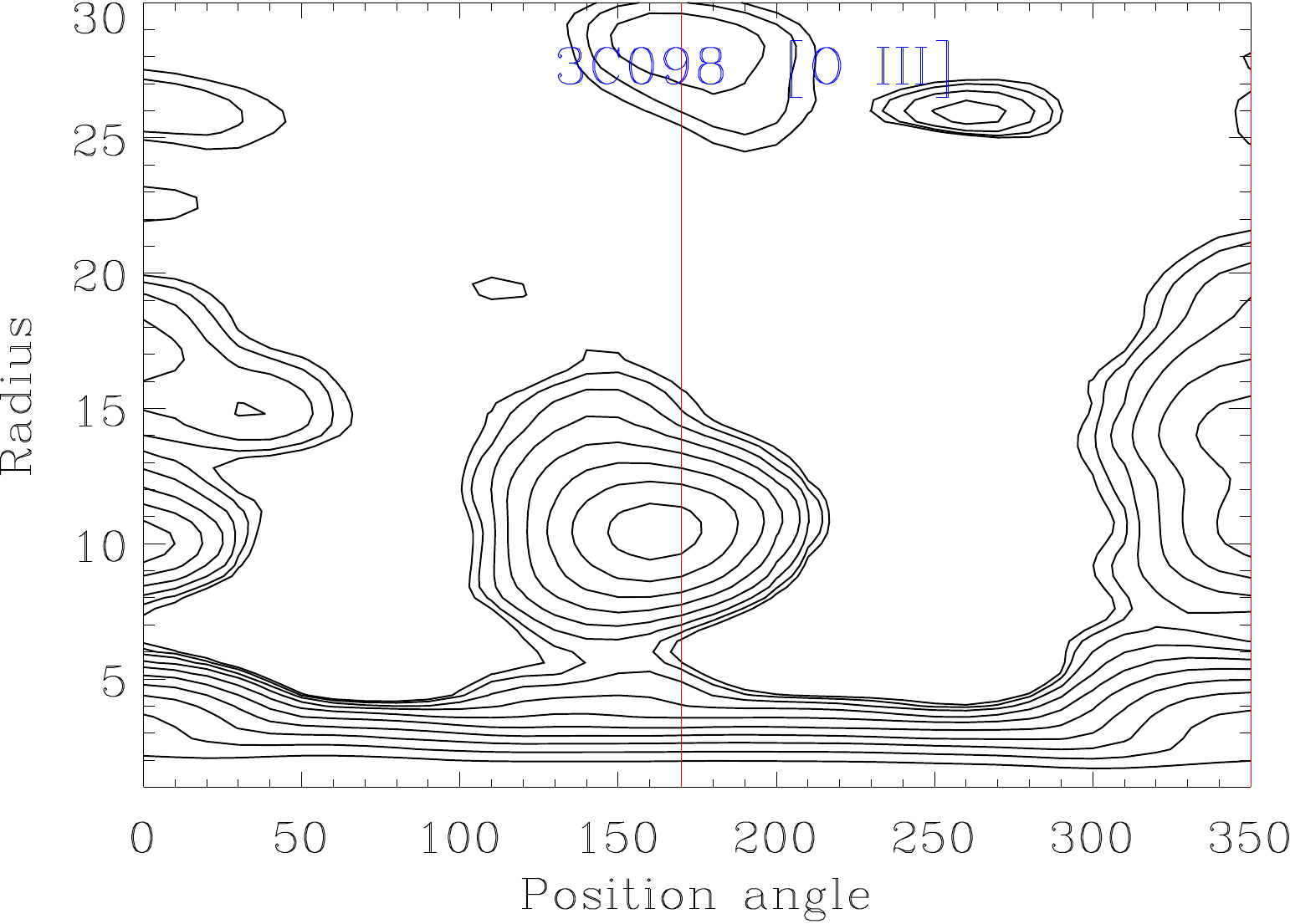}
\includegraphics[width=6.0cm]{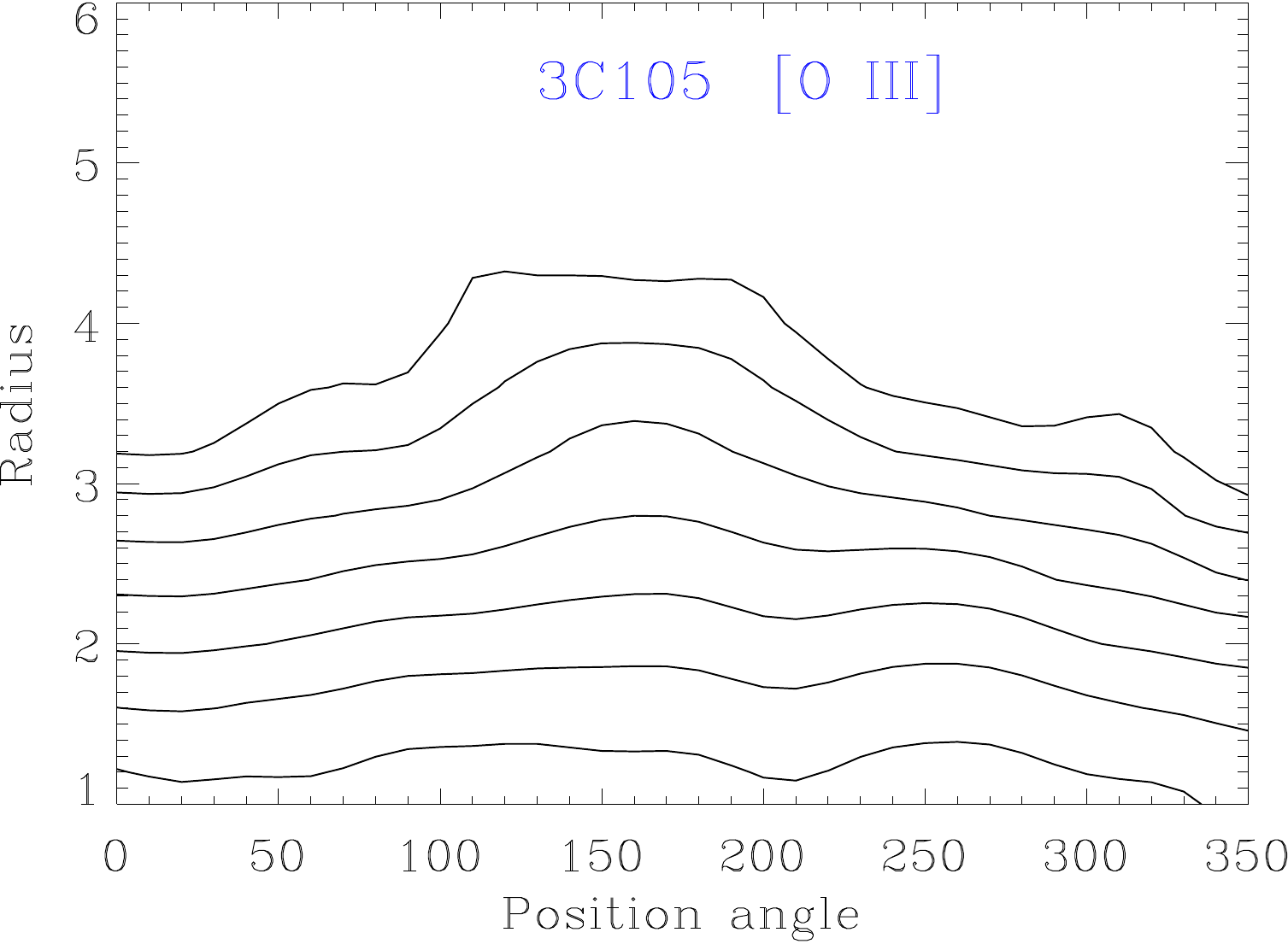}
\includegraphics[width=6.0cm]{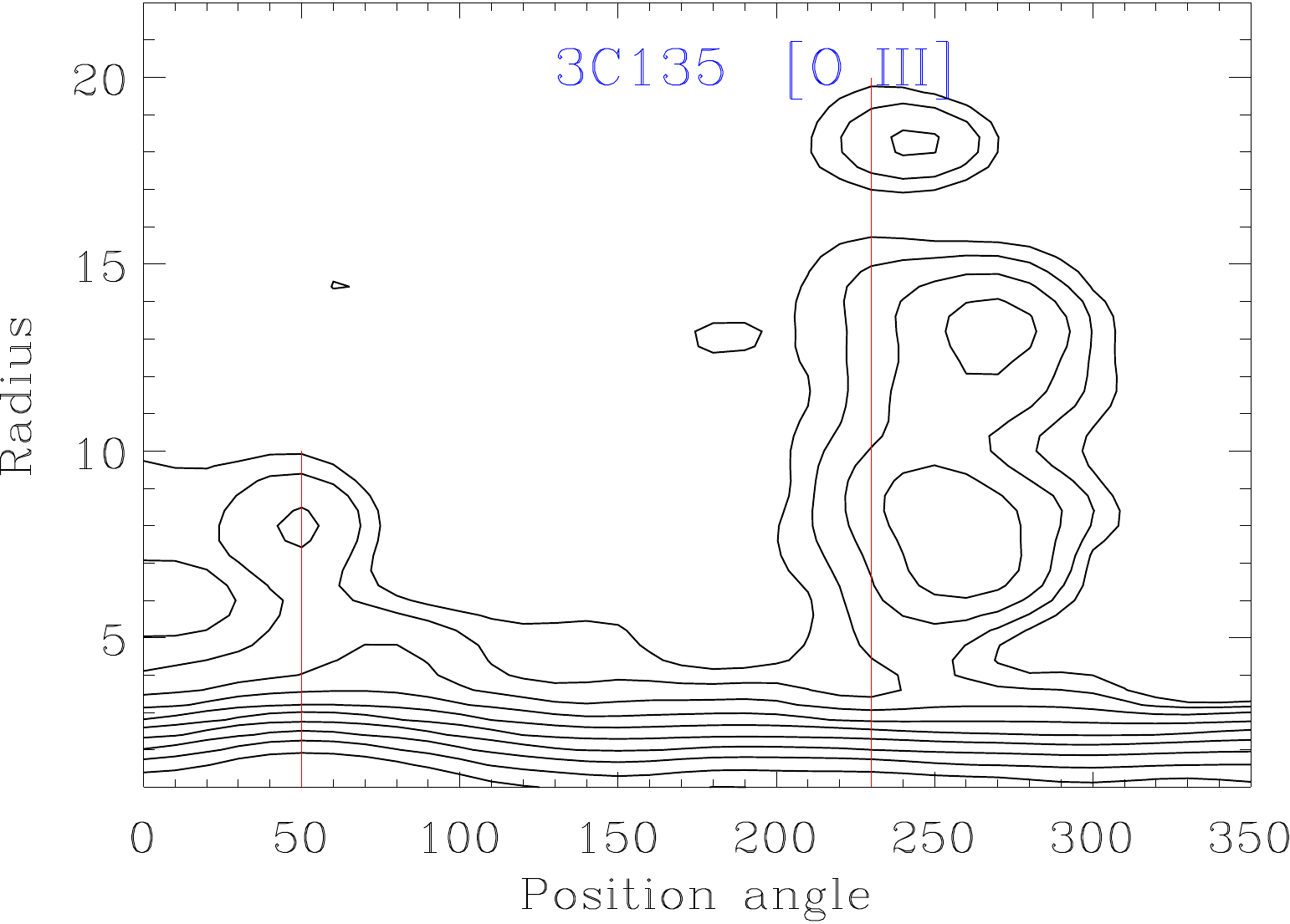}
\caption{Polar diagrams of the emission lines reported in Tab. \ref{tab1} for
  all the 3C radio galaxies. The
solid  red lines mark the ionized gas position angle on each side of the
  nucleus.}}
\end{figure*}

\addtocounter{figure}{-1}
\begin{figure*}
\centering{ 
\includegraphics[width=6.0cm]{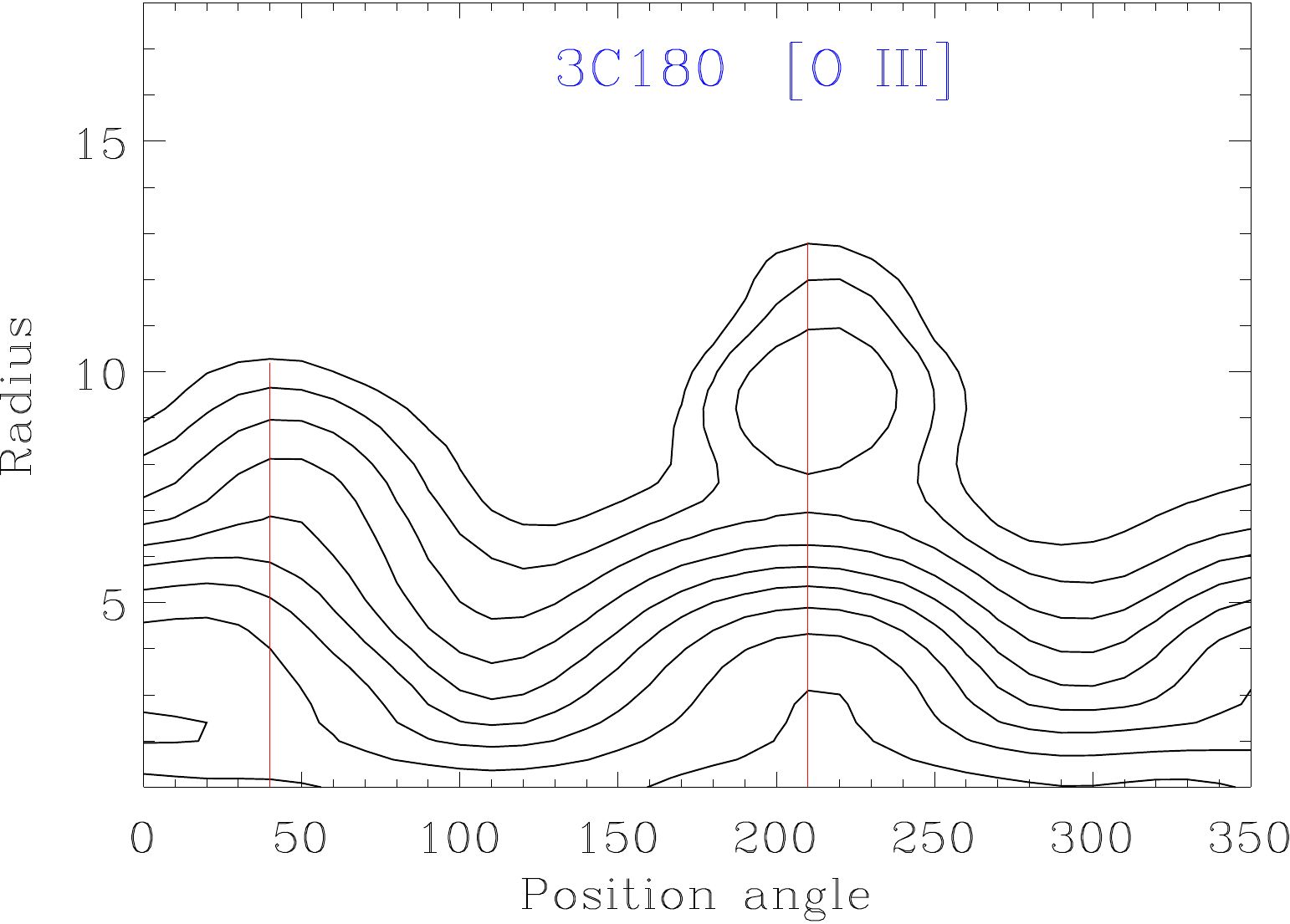}
\includegraphics[width=6.0cm]{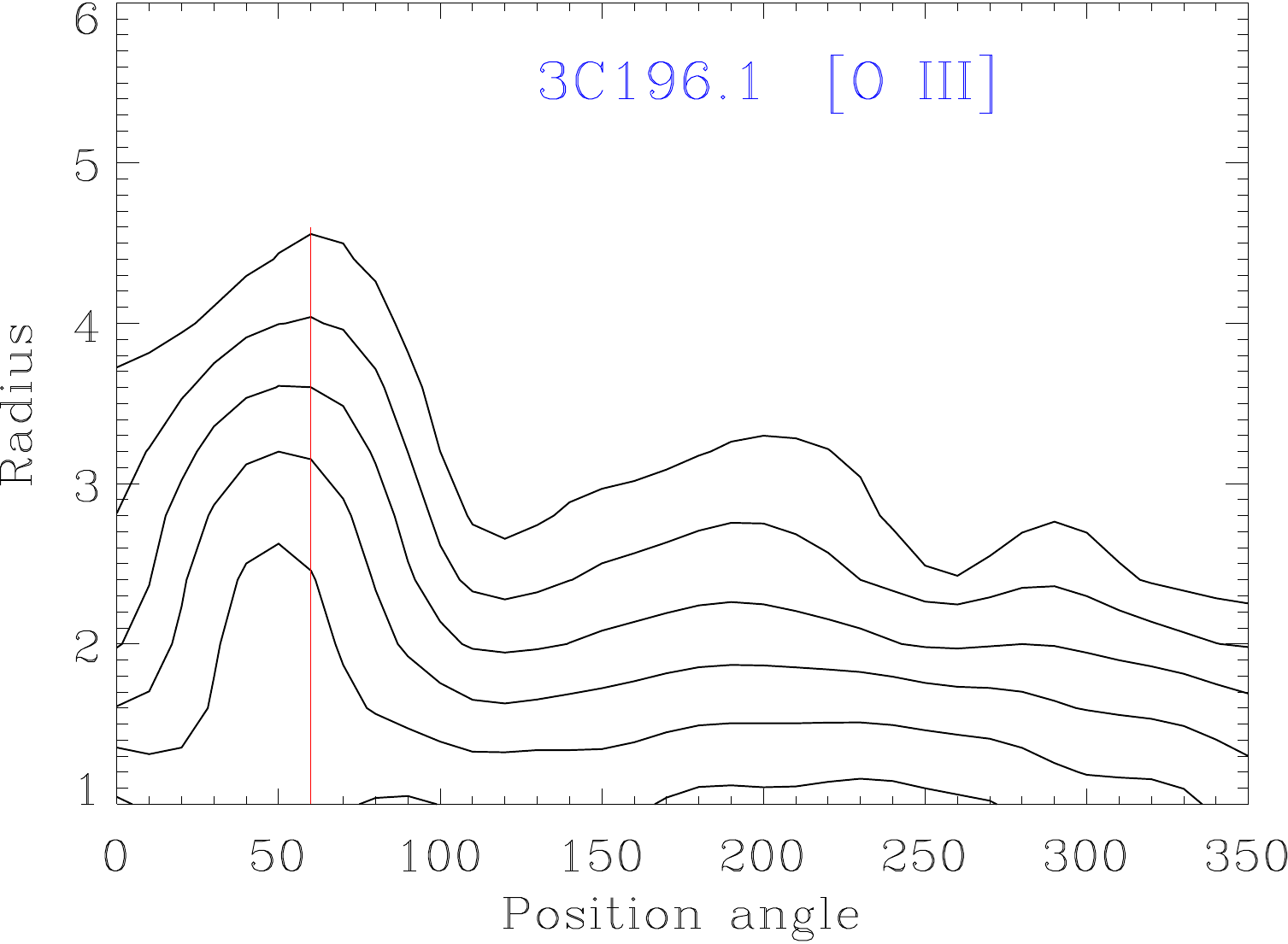}
\includegraphics[width=6.0cm]{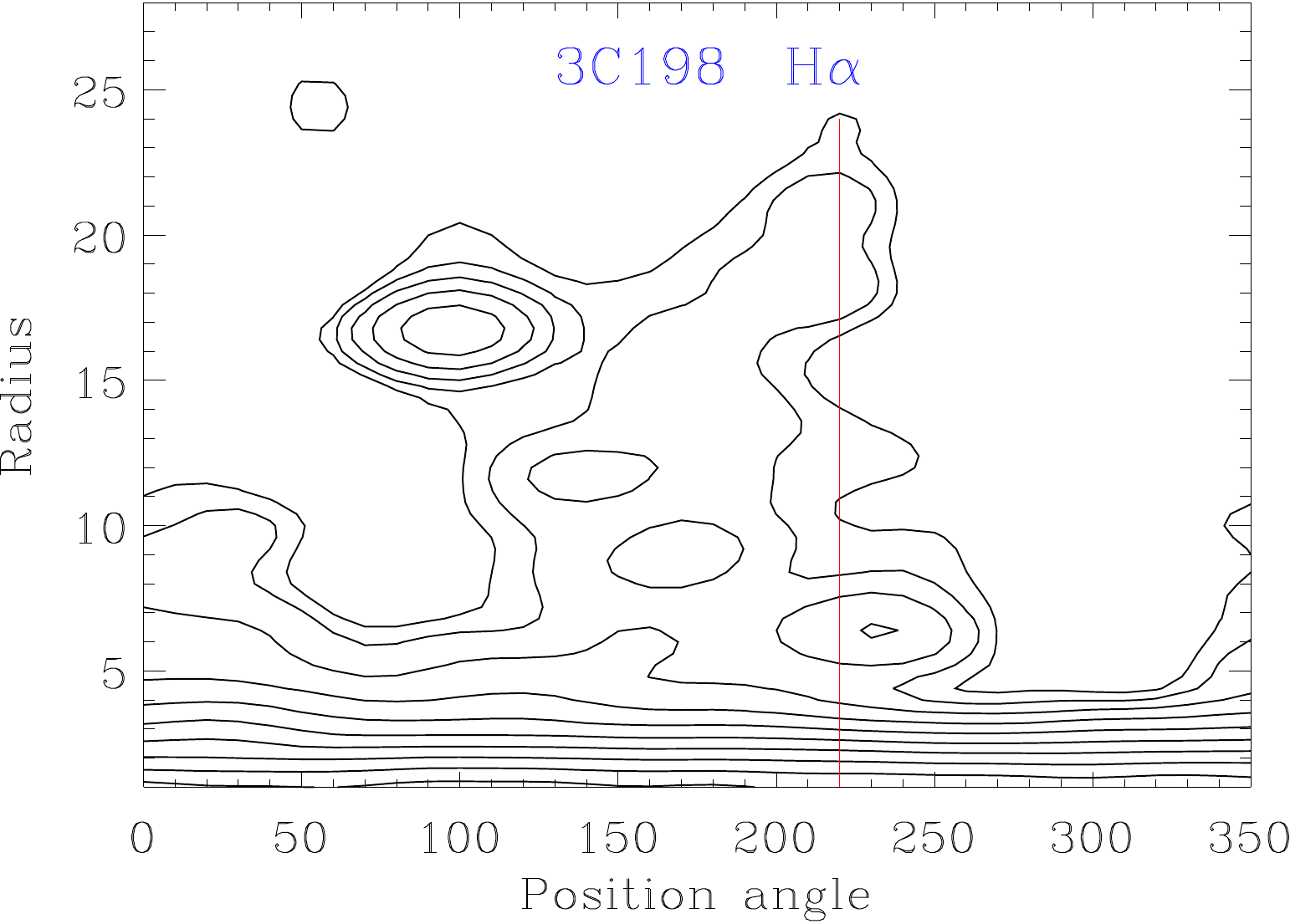}
\includegraphics[width=6.0cm]{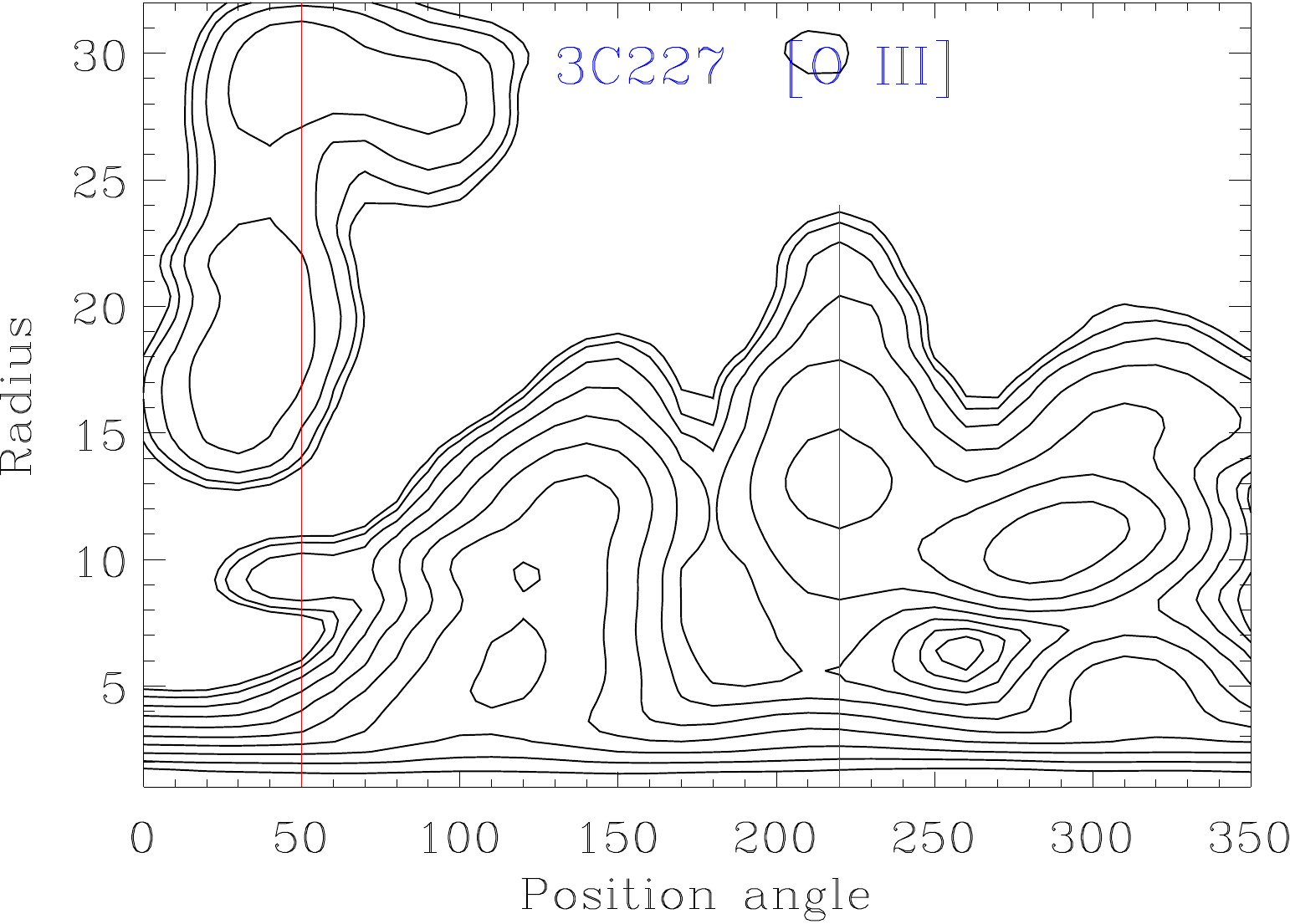}
\includegraphics[width=6.0cm]{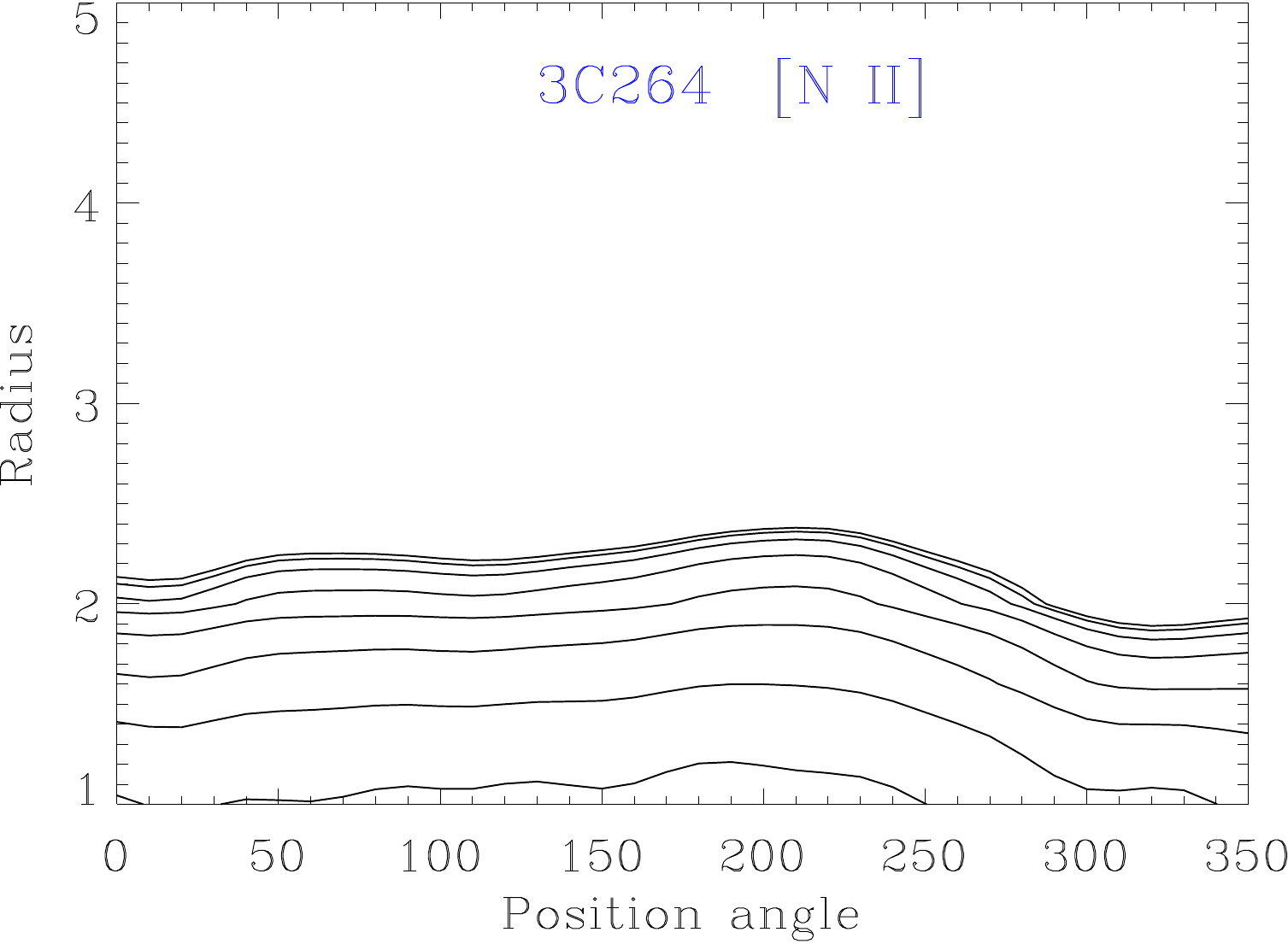}
\includegraphics[width=6.0cm]{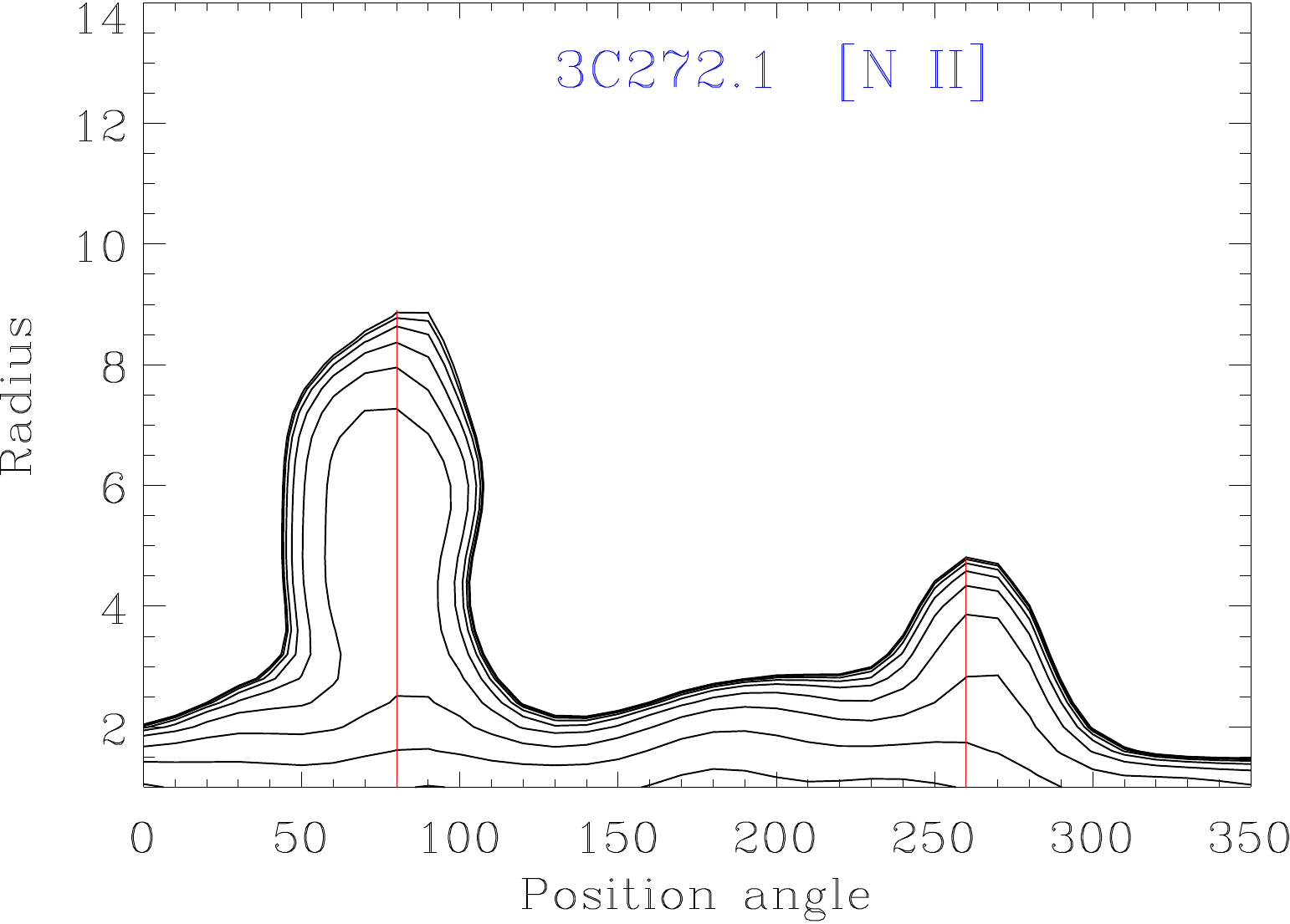}
\includegraphics[width=6.0cm]{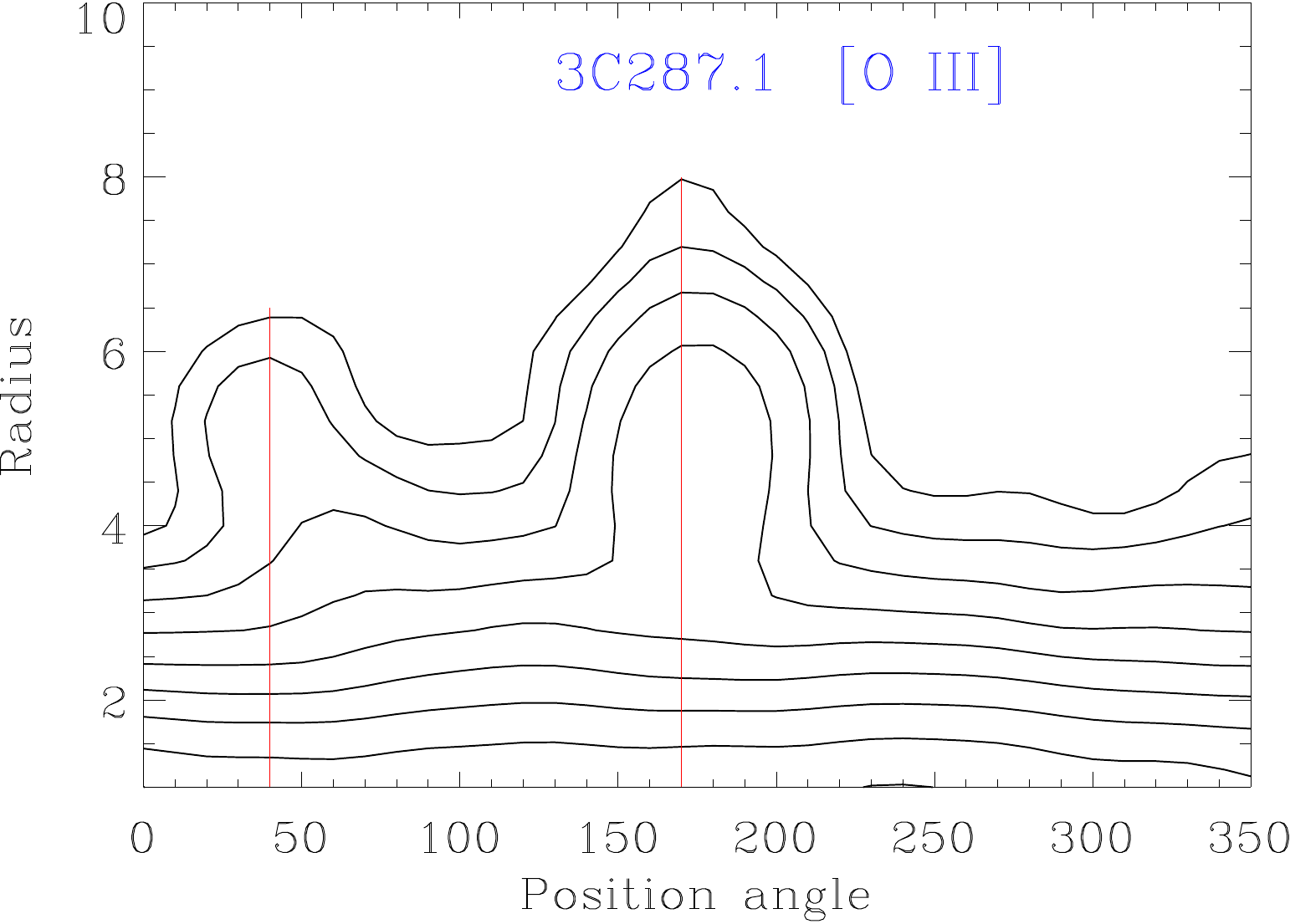}
\includegraphics[width=6.0cm]{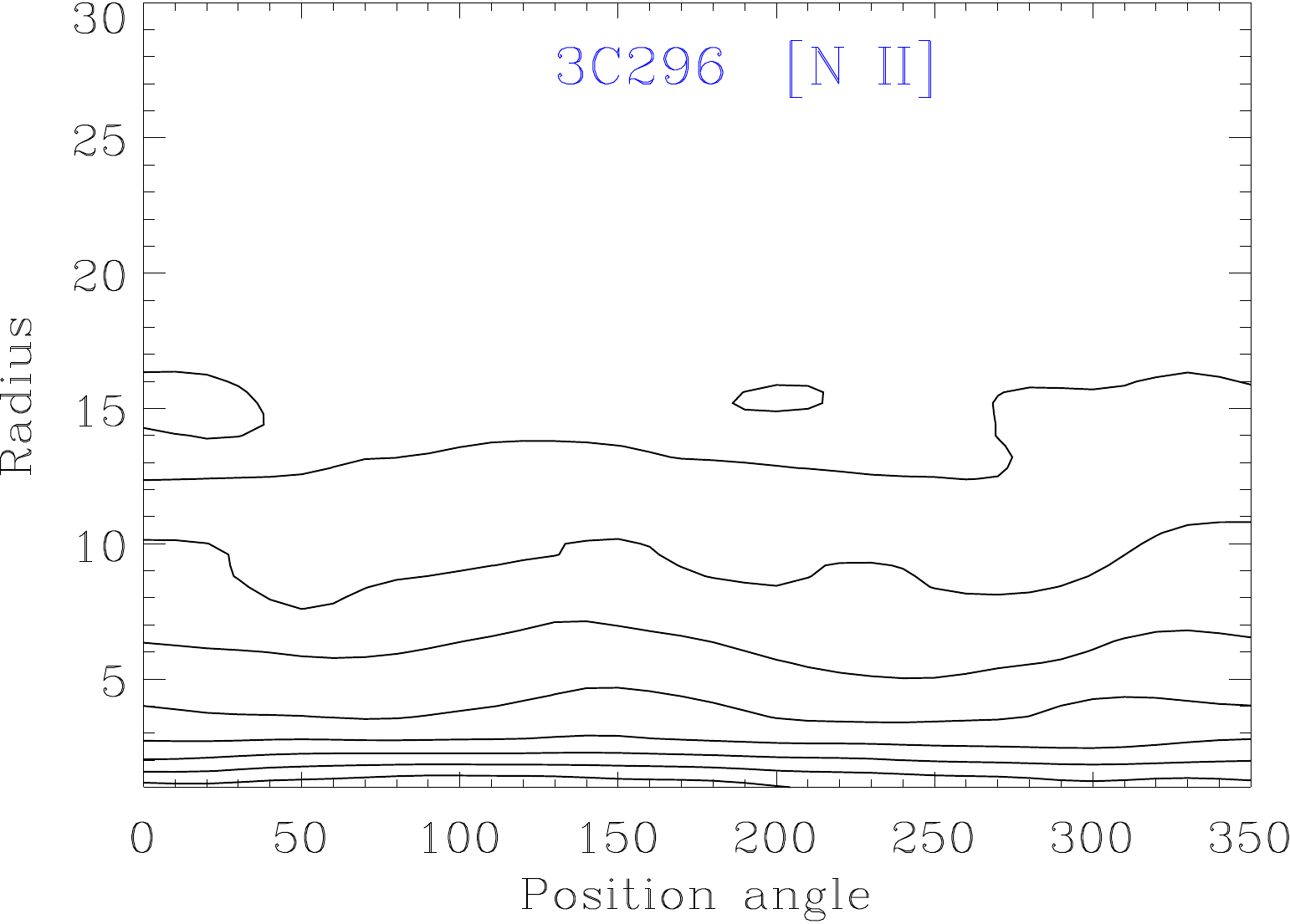}
\includegraphics[width=6.0cm]{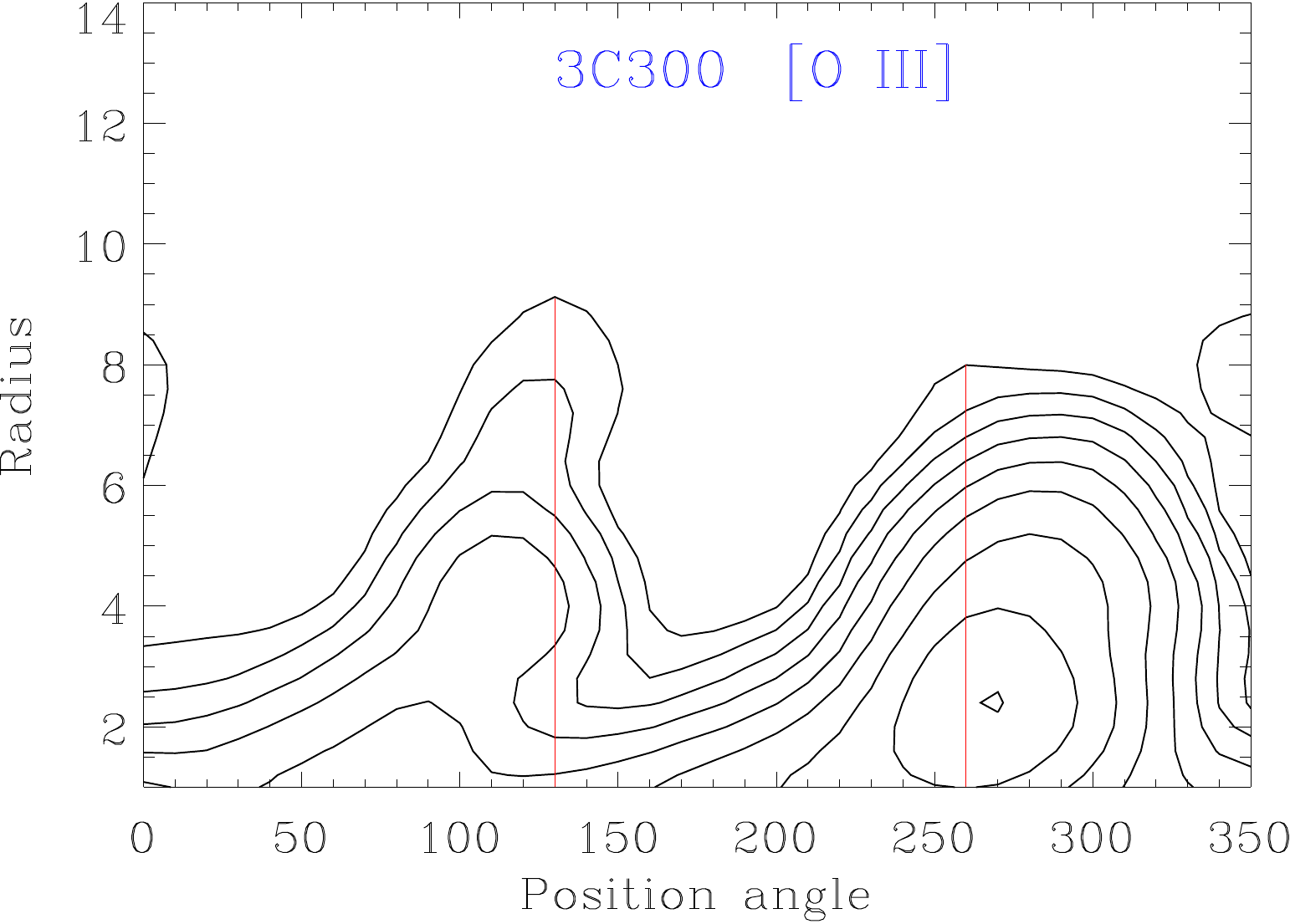}
\includegraphics[width=6.0cm]{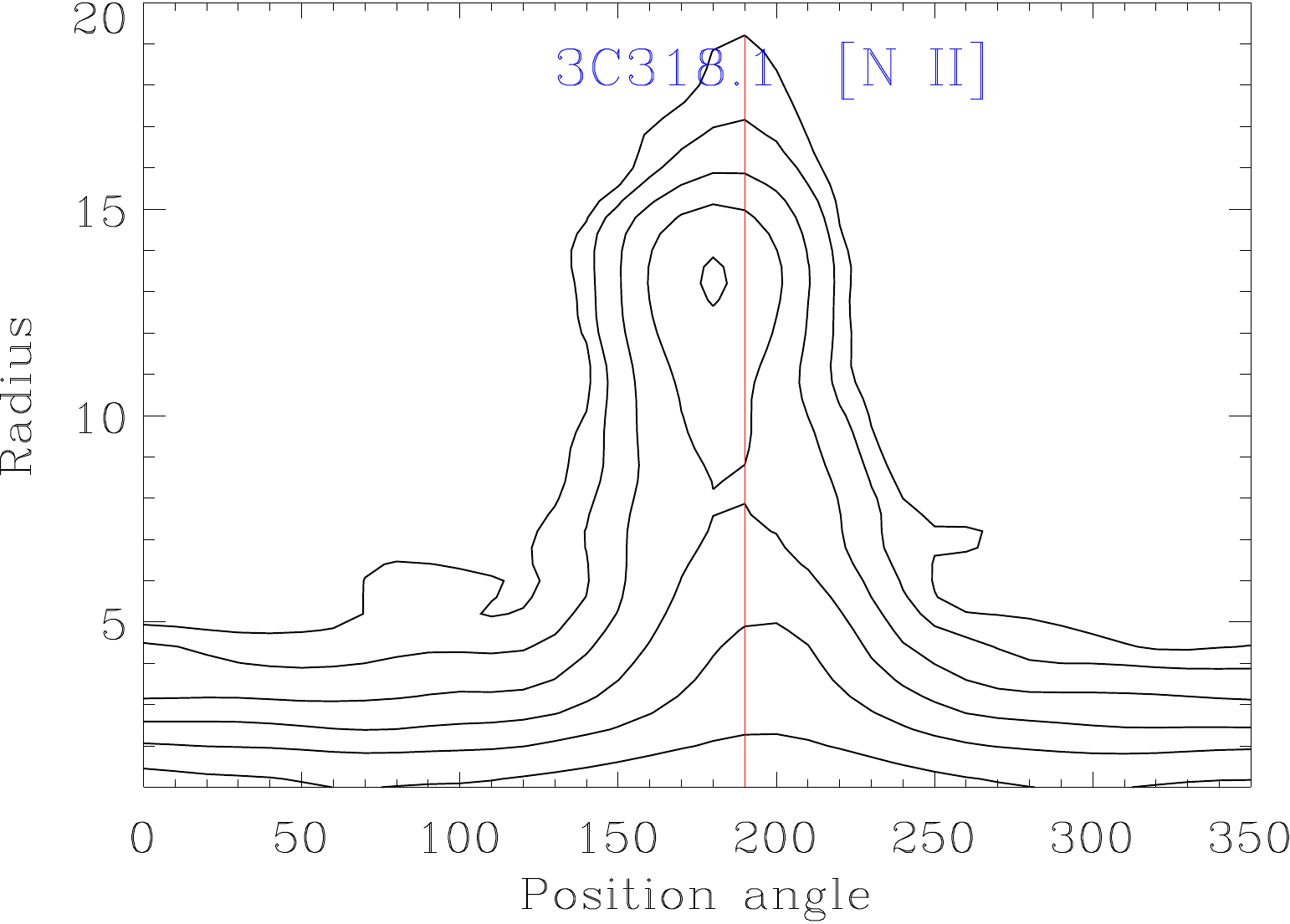}
\includegraphics[width=6.0cm]{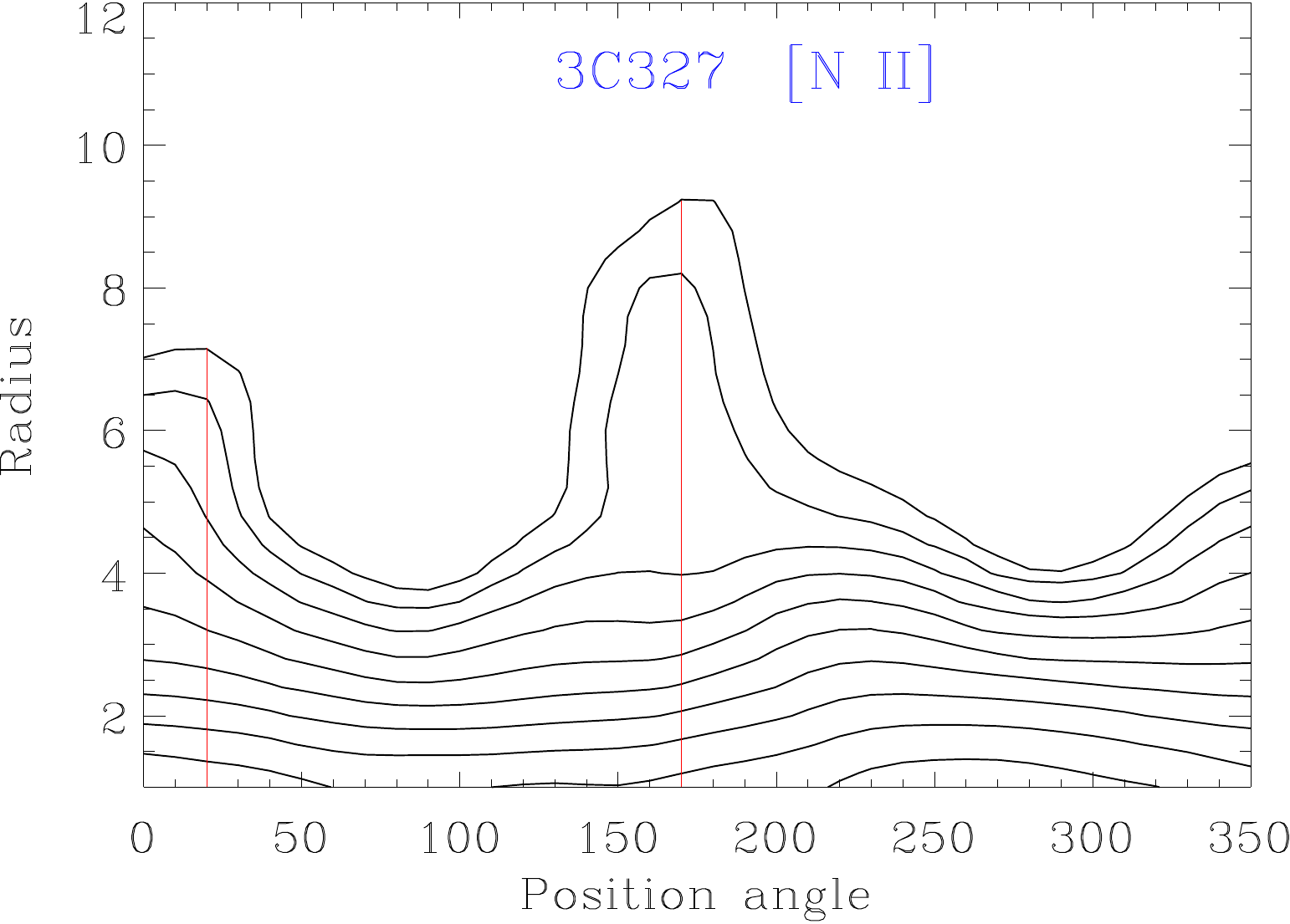}
\includegraphics[width=6.0cm]{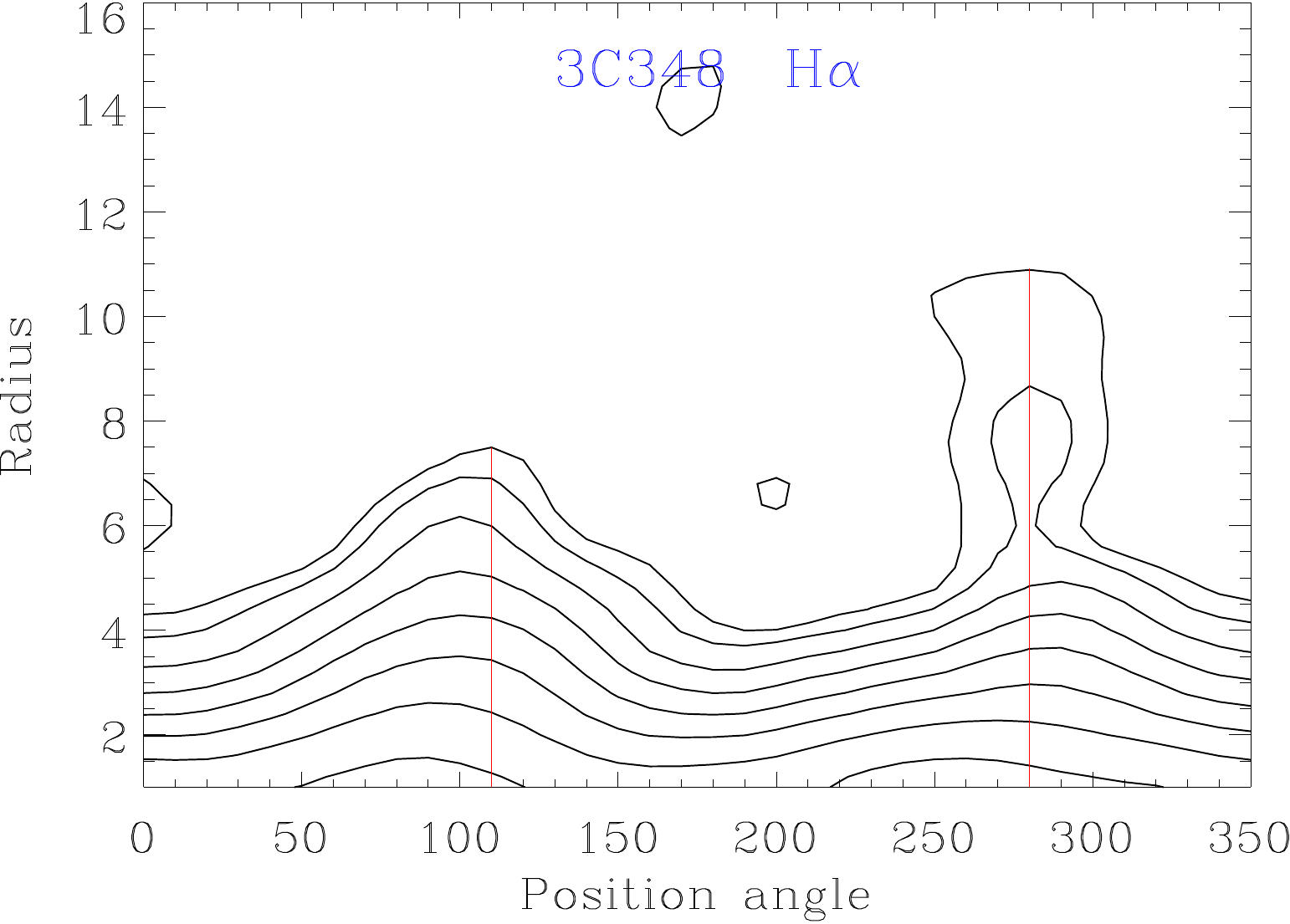}
\includegraphics[width=6.0cm]{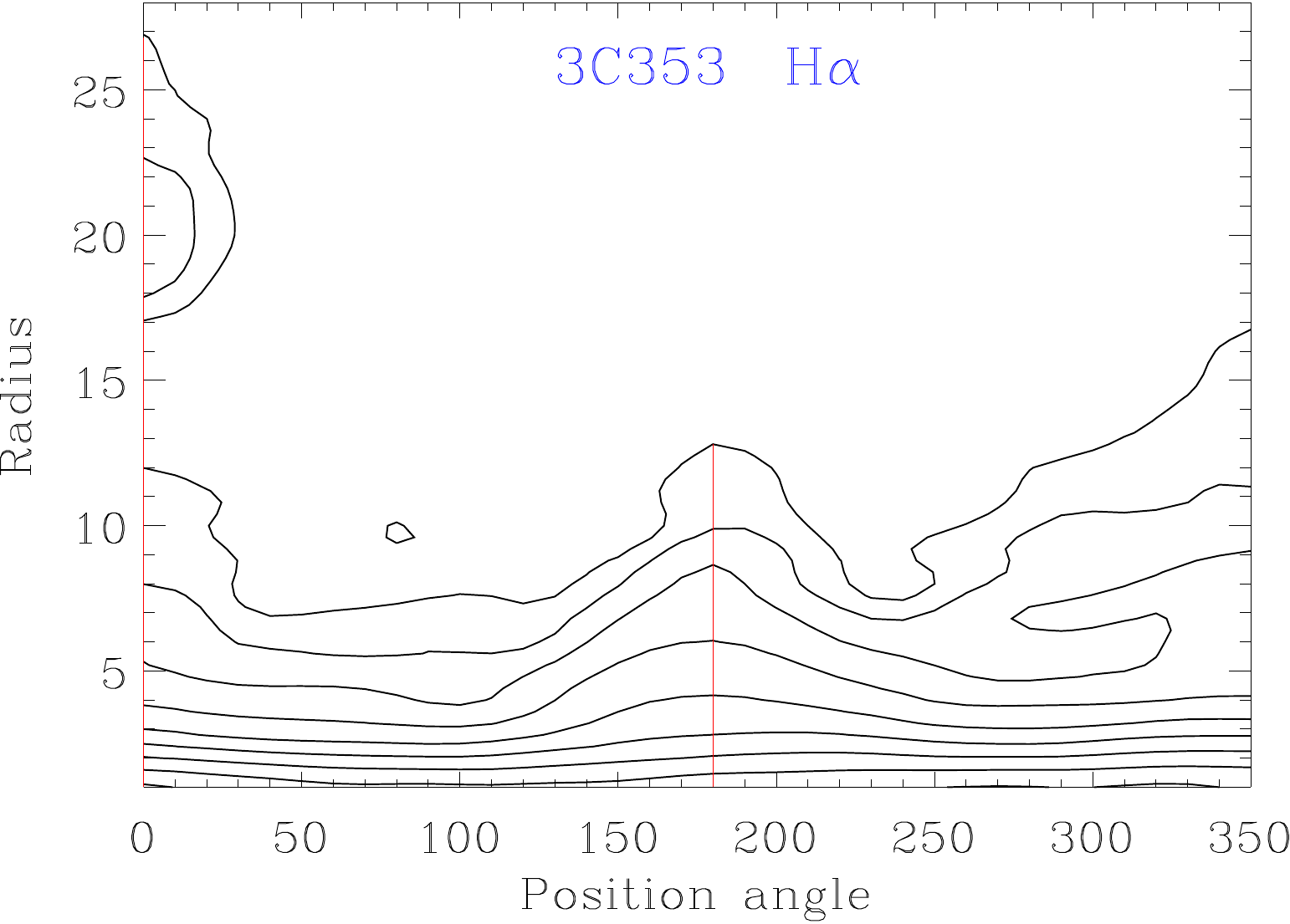}
\includegraphics[width=6.0cm]{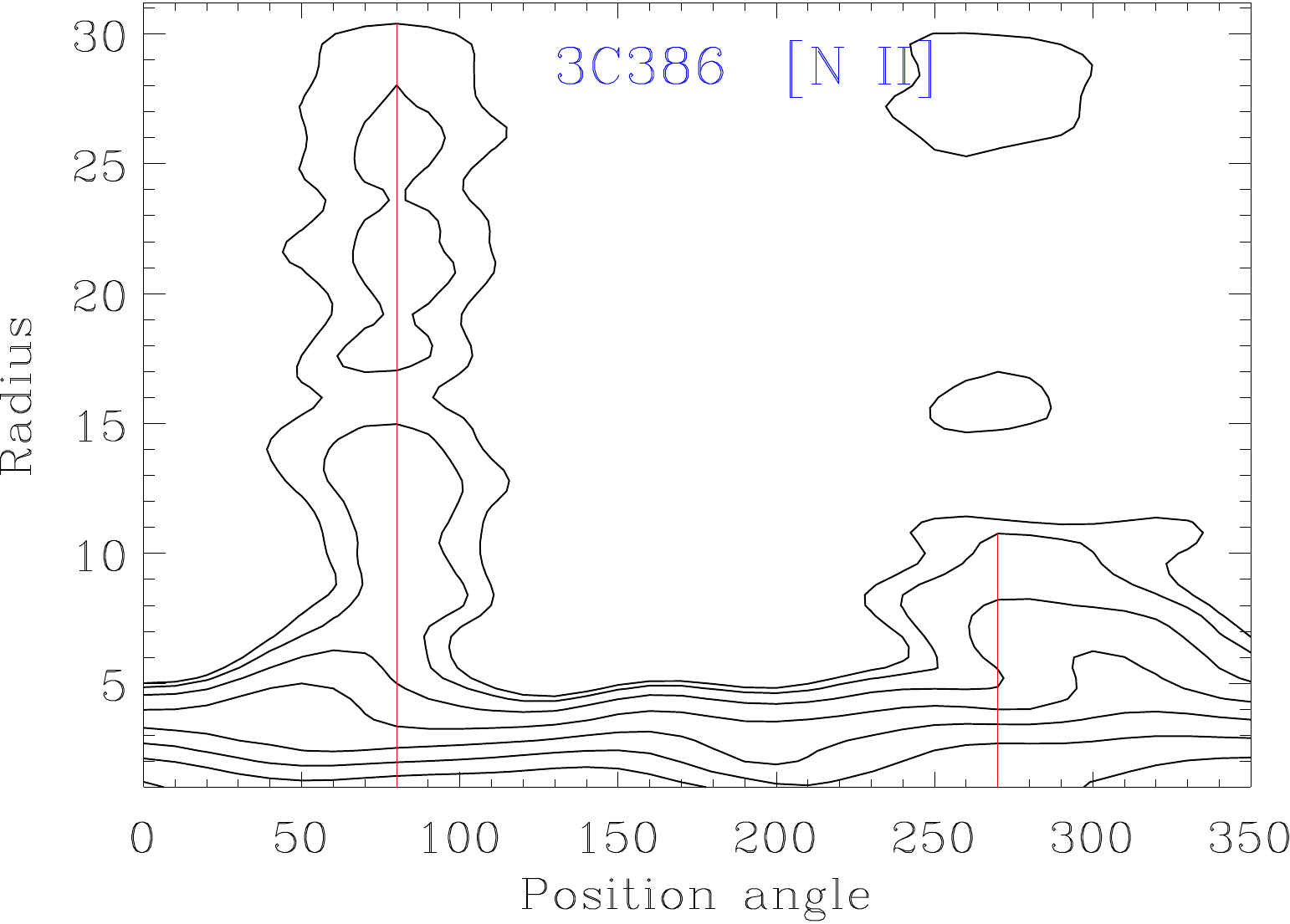}
\includegraphics[width=6.0cm]{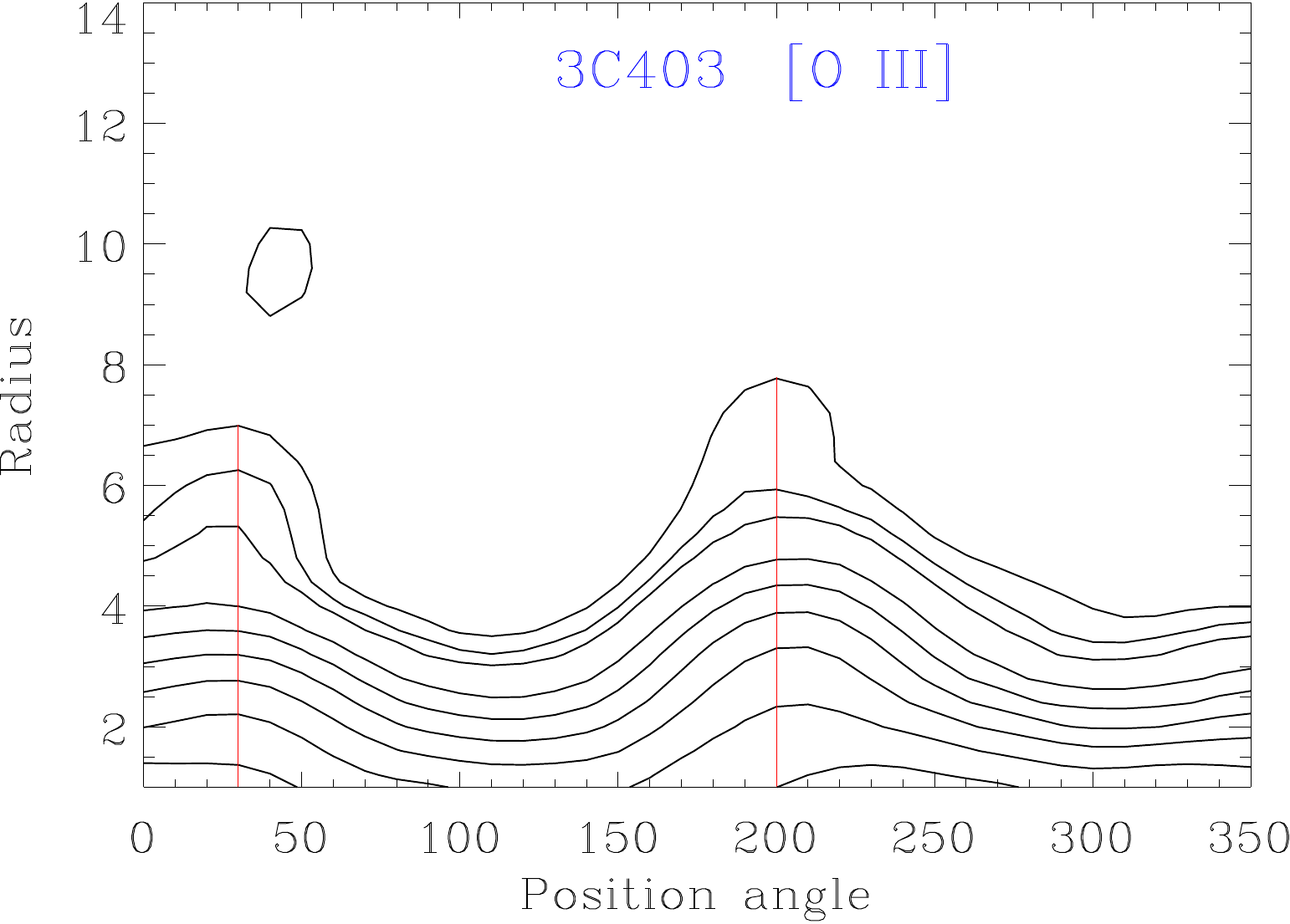}
\caption{- continued.}}
\end{figure*}

\addtocounter{figure}{-1}
\begin{figure*}
\centering{ 
\includegraphics[width=6.0cm]{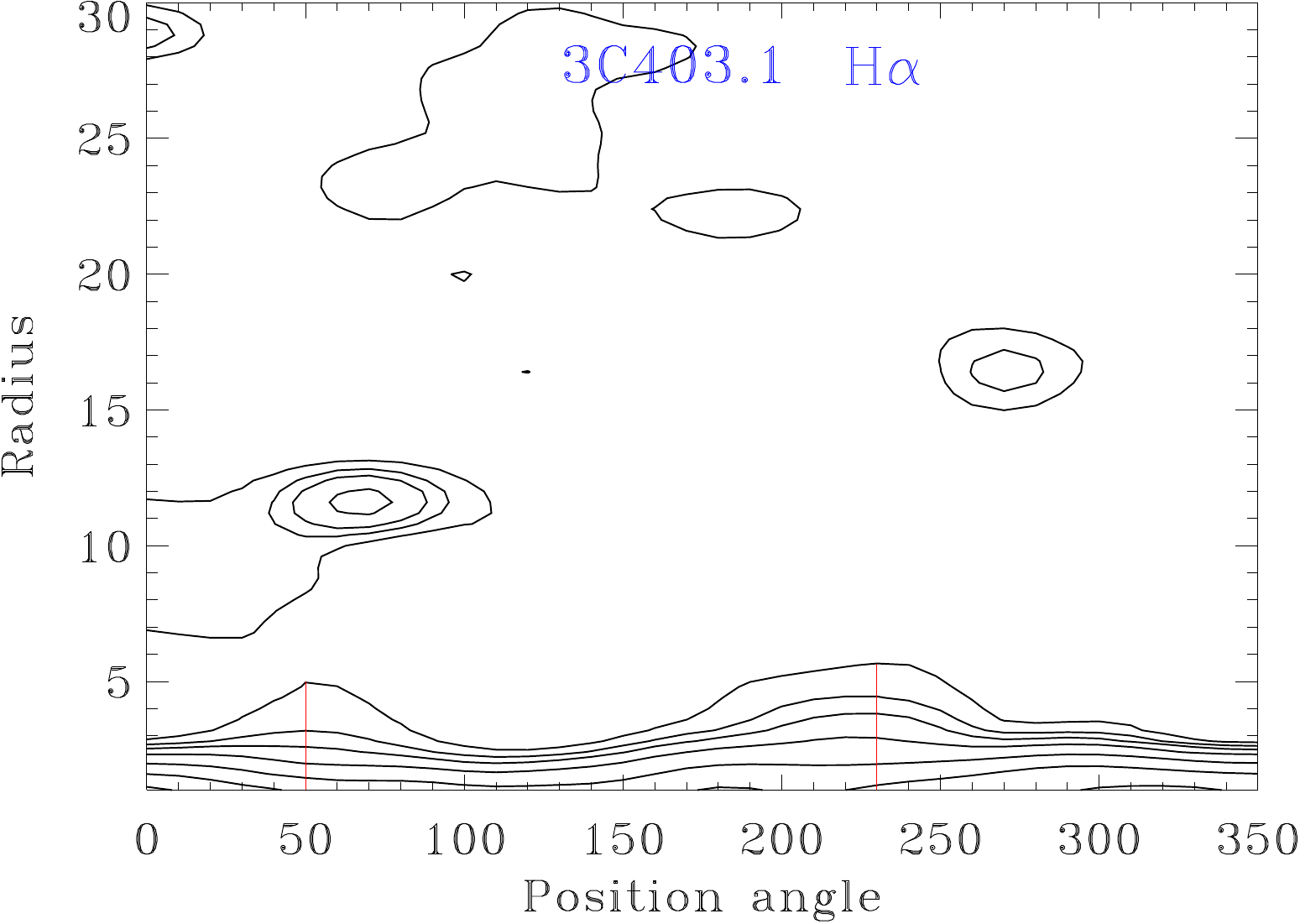}
\includegraphics[width=6.0cm]{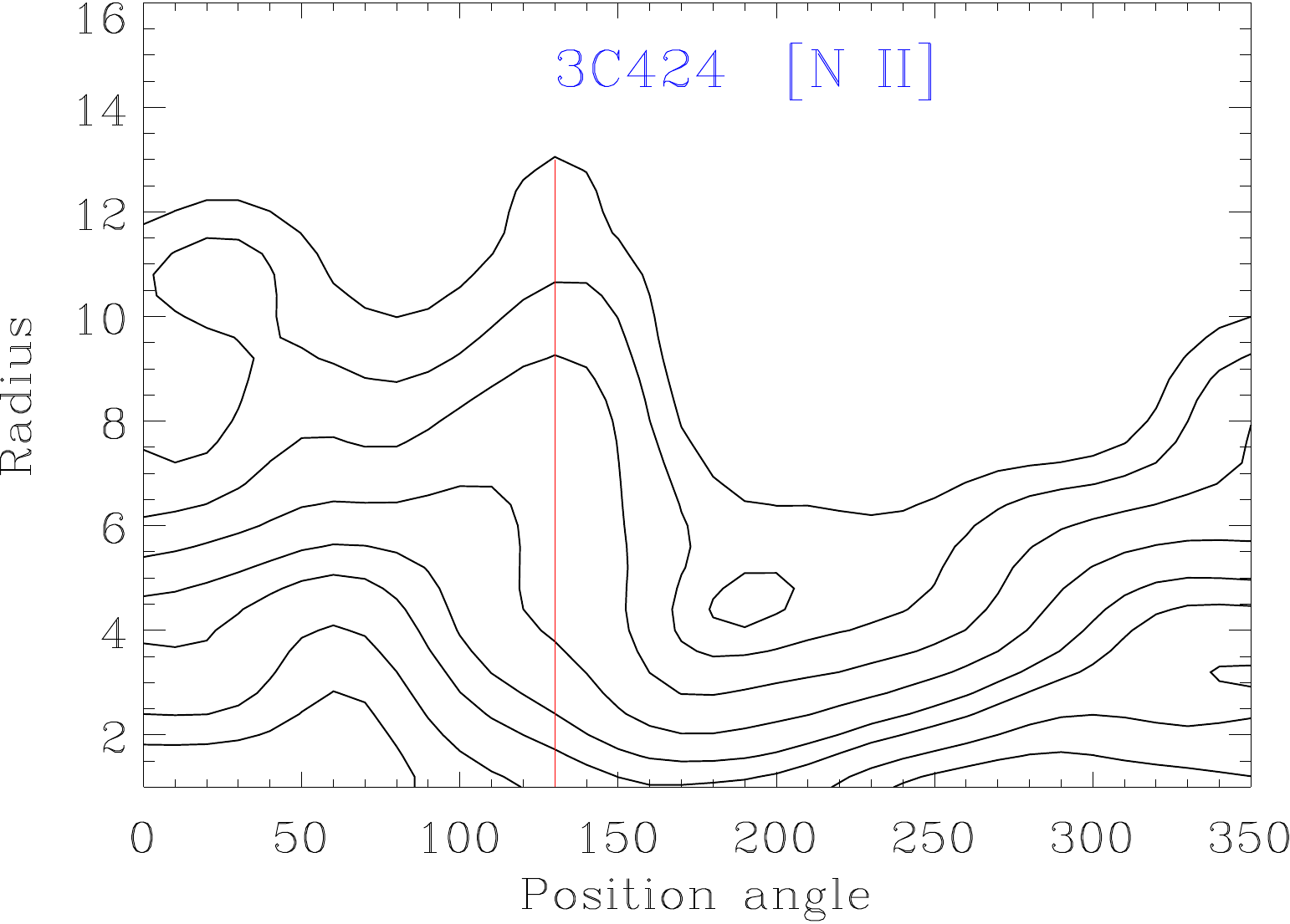}
\includegraphics[width=6.0cm]{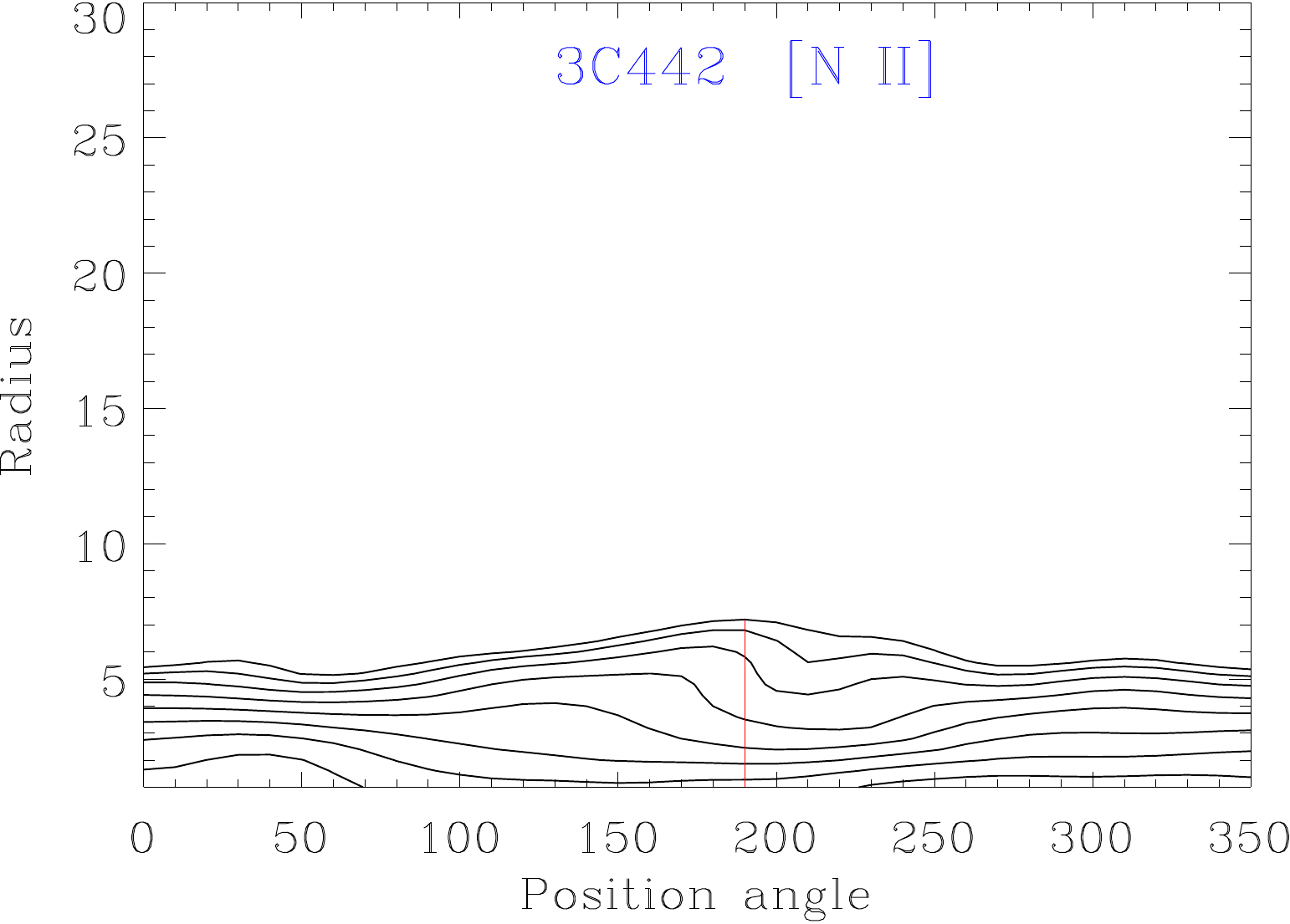}
\includegraphics[width=6.0cm]{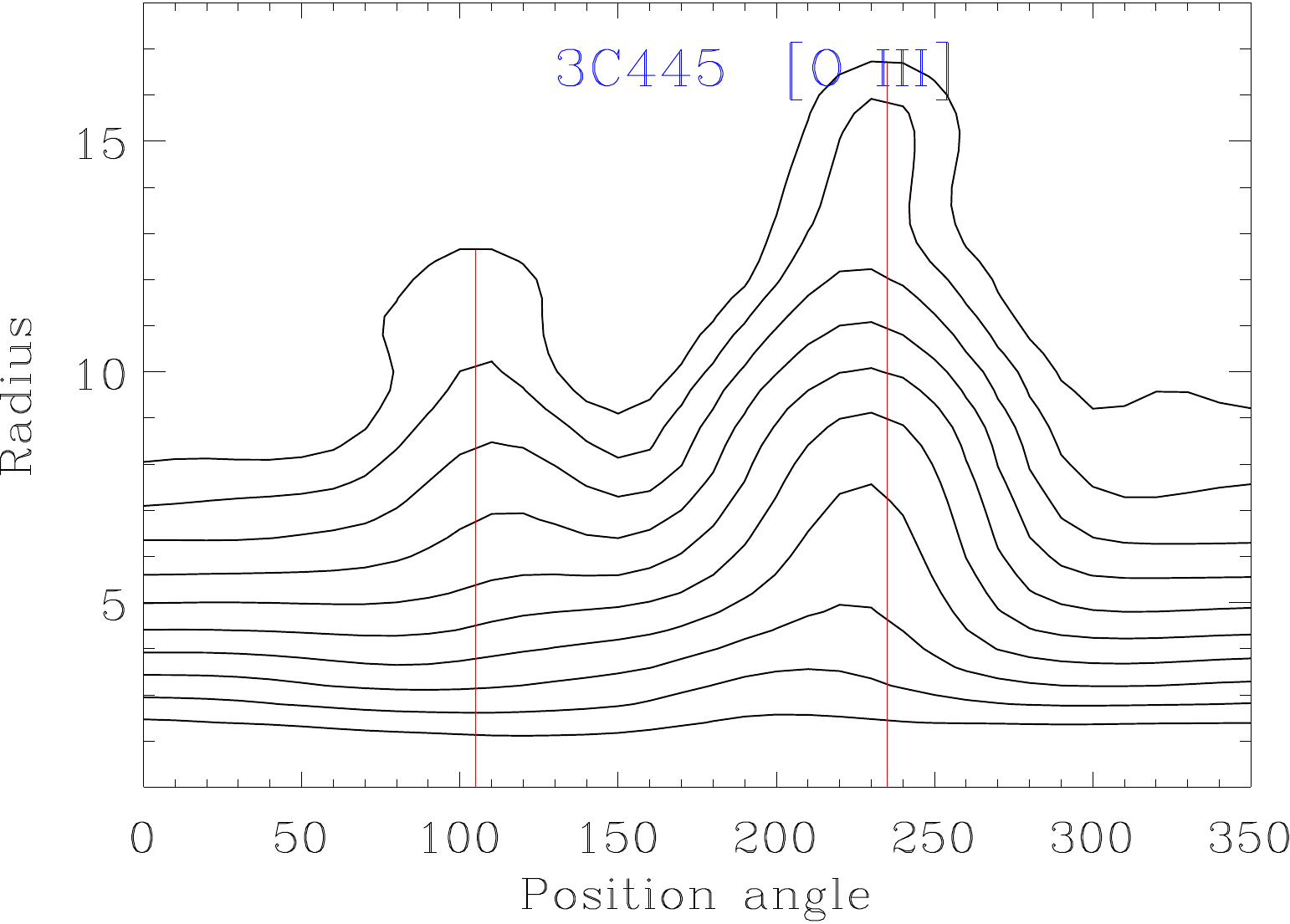}
\includegraphics[width=6.0cm]{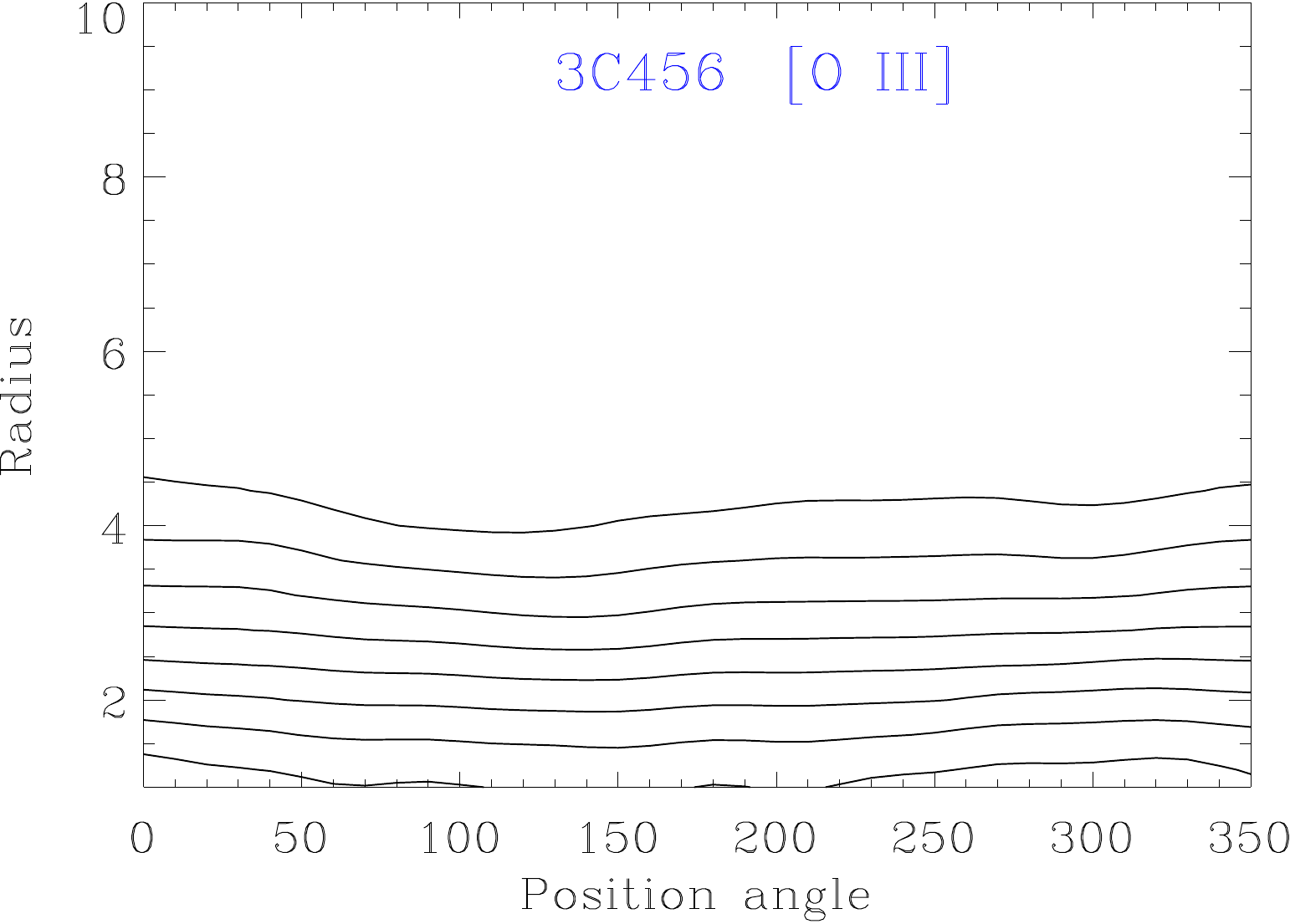}
\includegraphics[width=6.0cm]{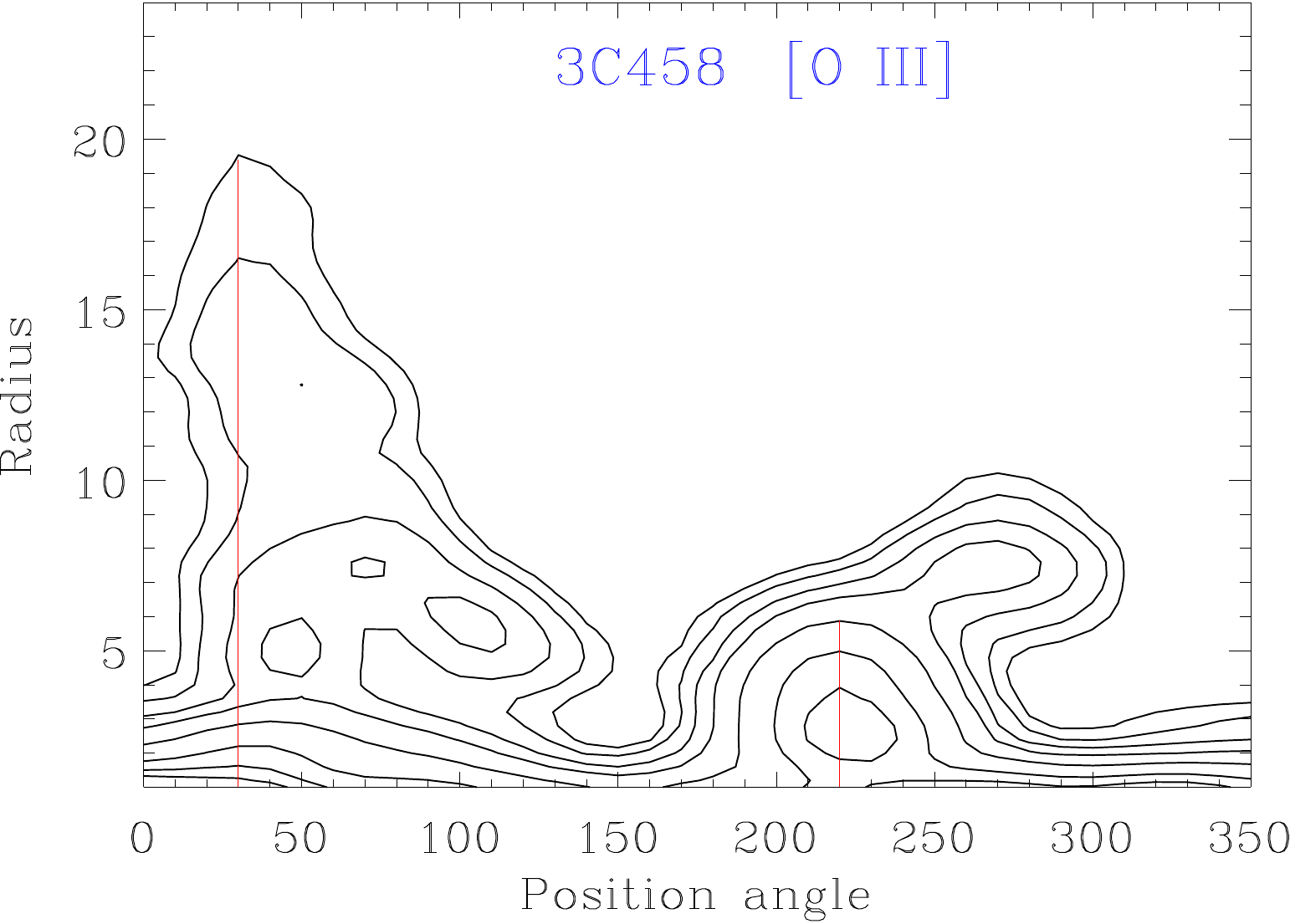}
\includegraphics[width=6.0cm]{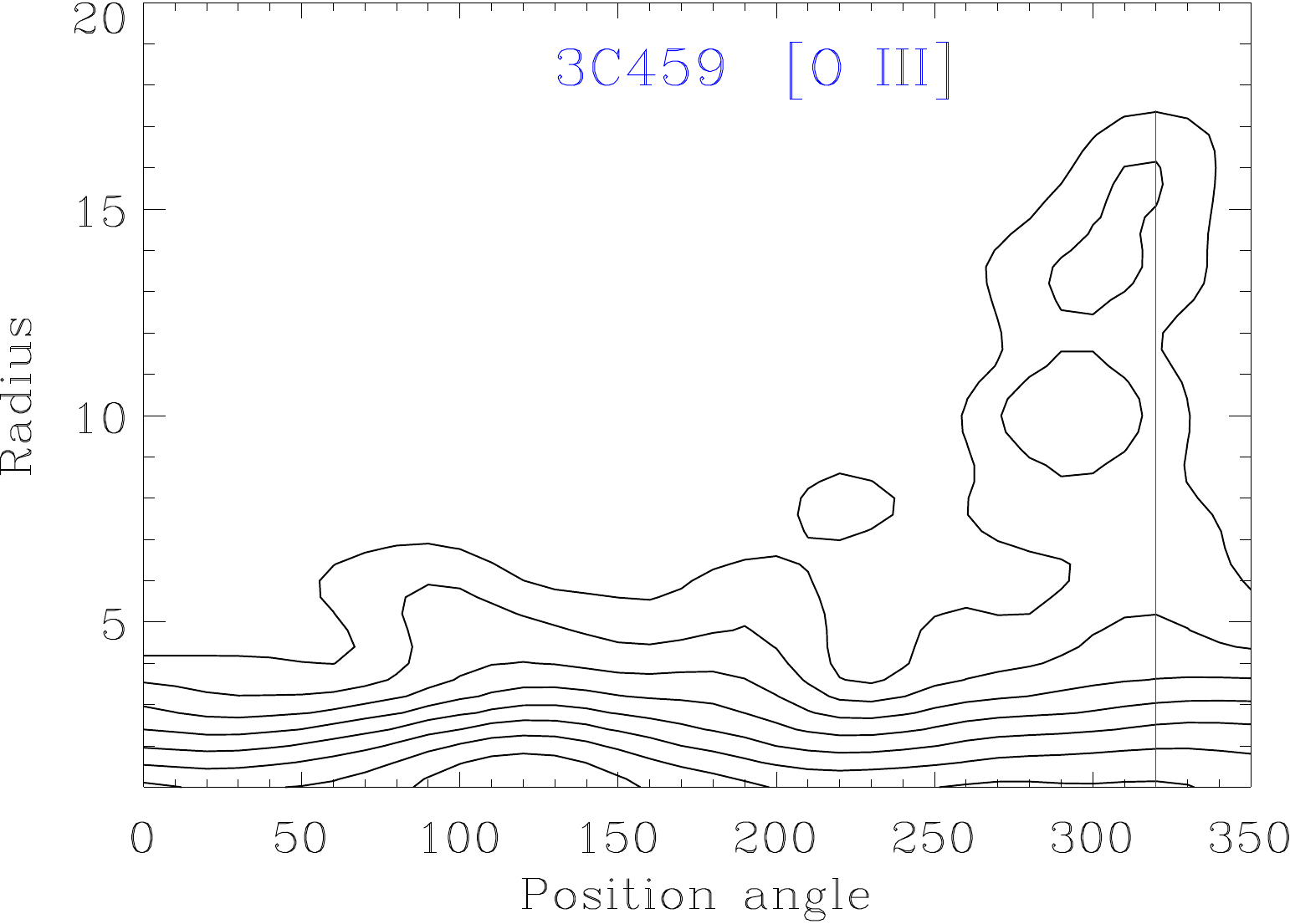}
\caption{- continued.}}
\end{figure*}

\newpage
\clearpage
\section{Kinematics of the nuclear gas.}
\label{appC}  
\begin{figure*}  
\centering{ 
\includegraphics[width=8.5cm]{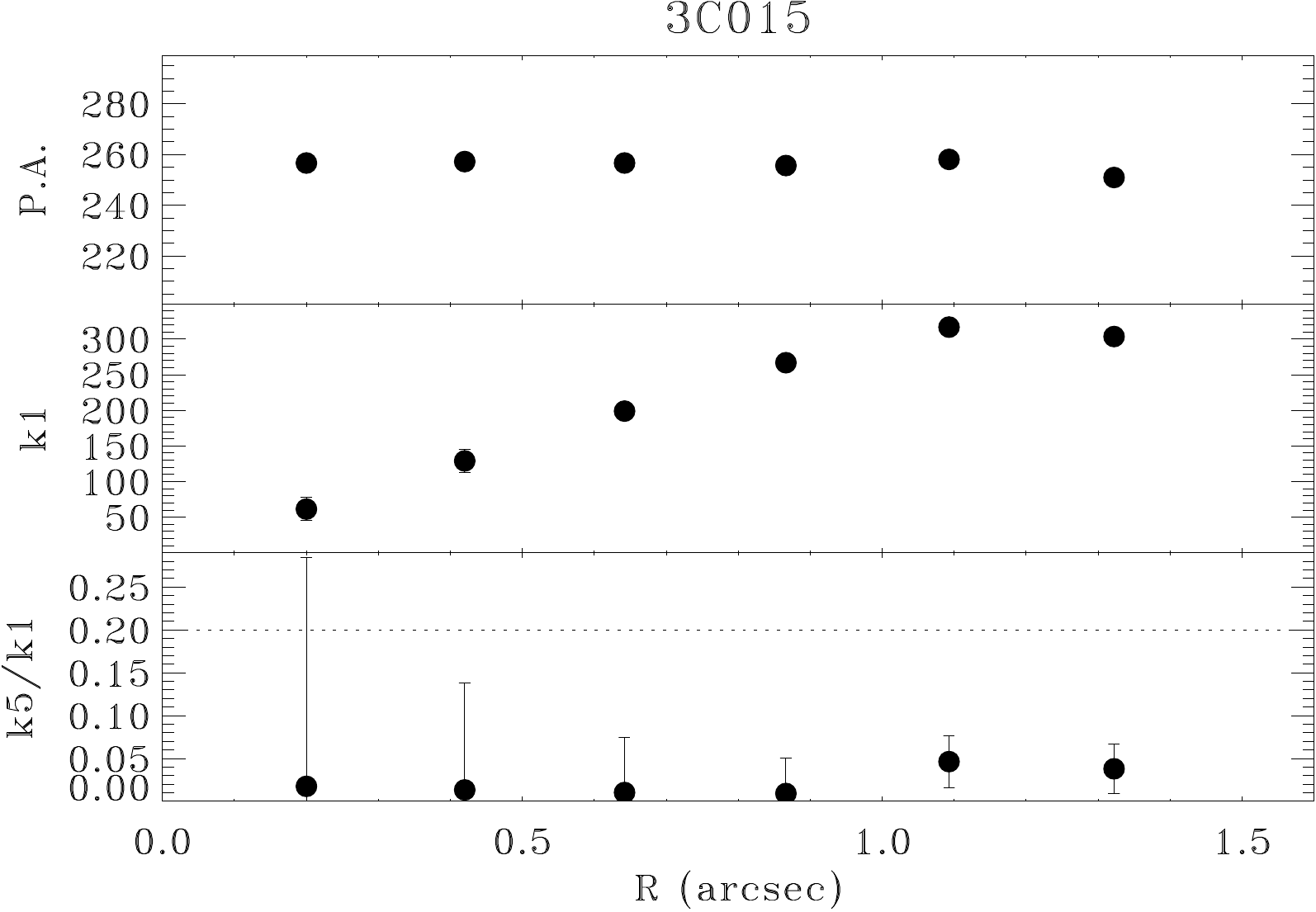}   
\includegraphics[width=8.5cm]{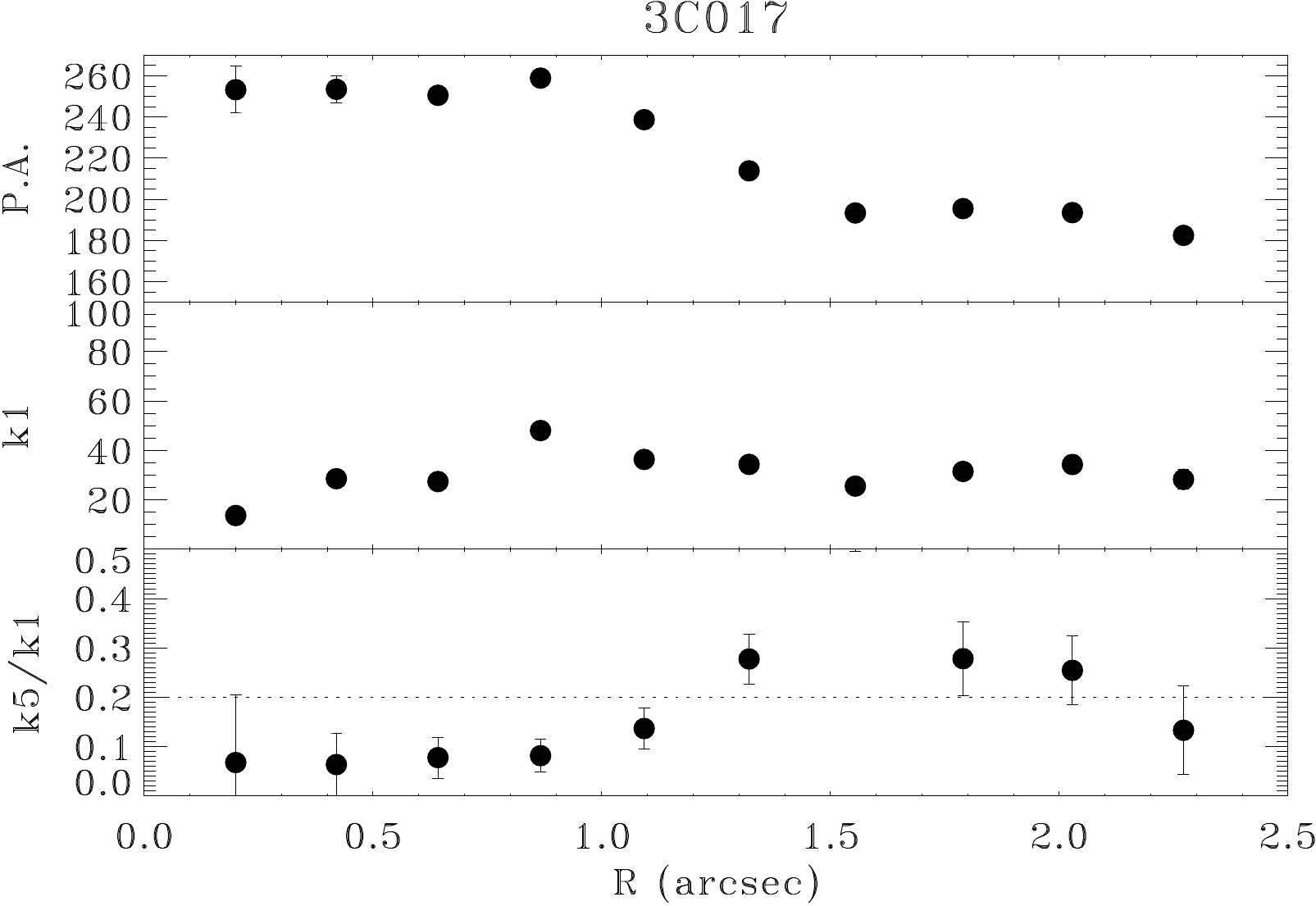}
\includegraphics[width=8.5cm]{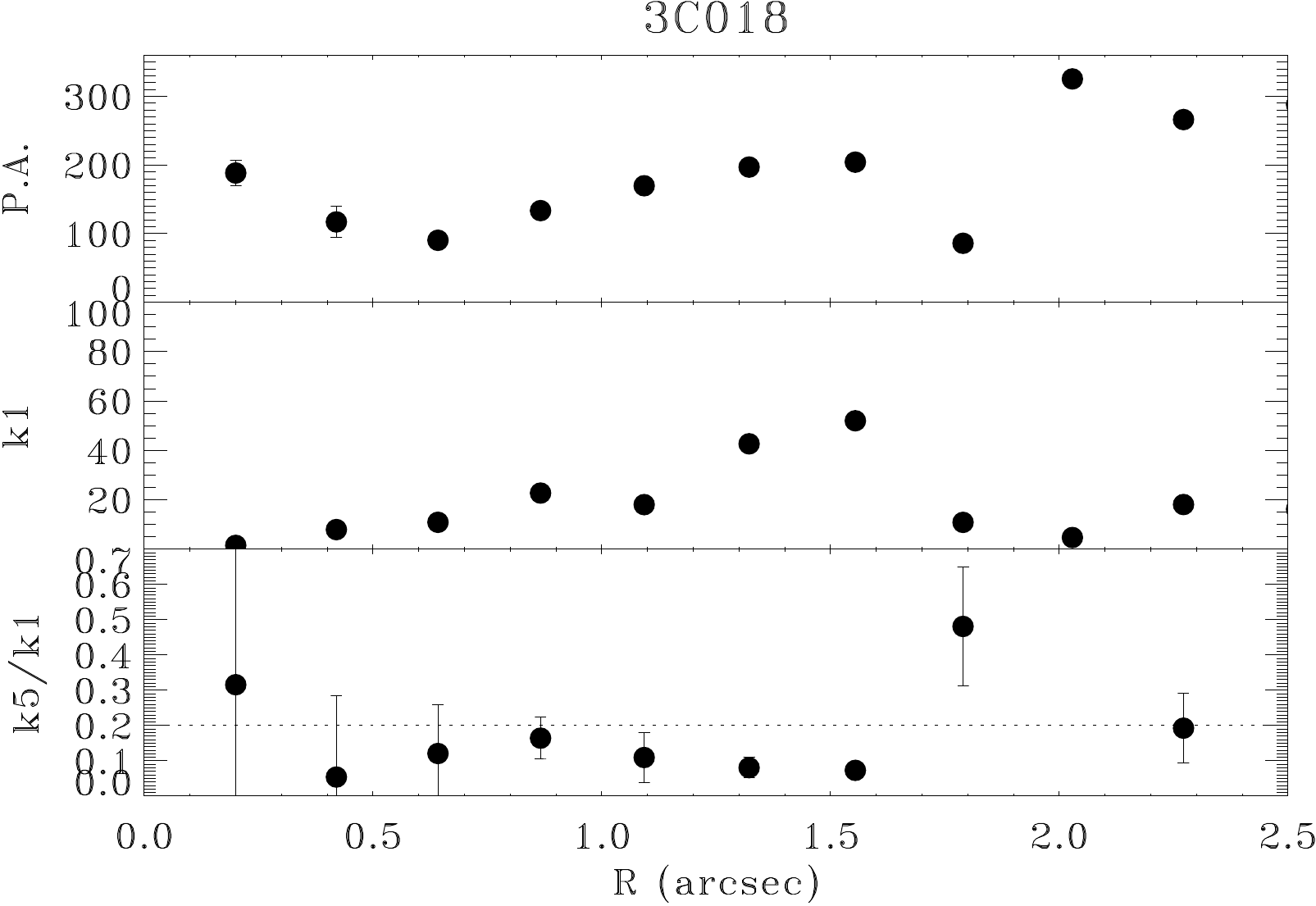}
\includegraphics[width=8.5cm]{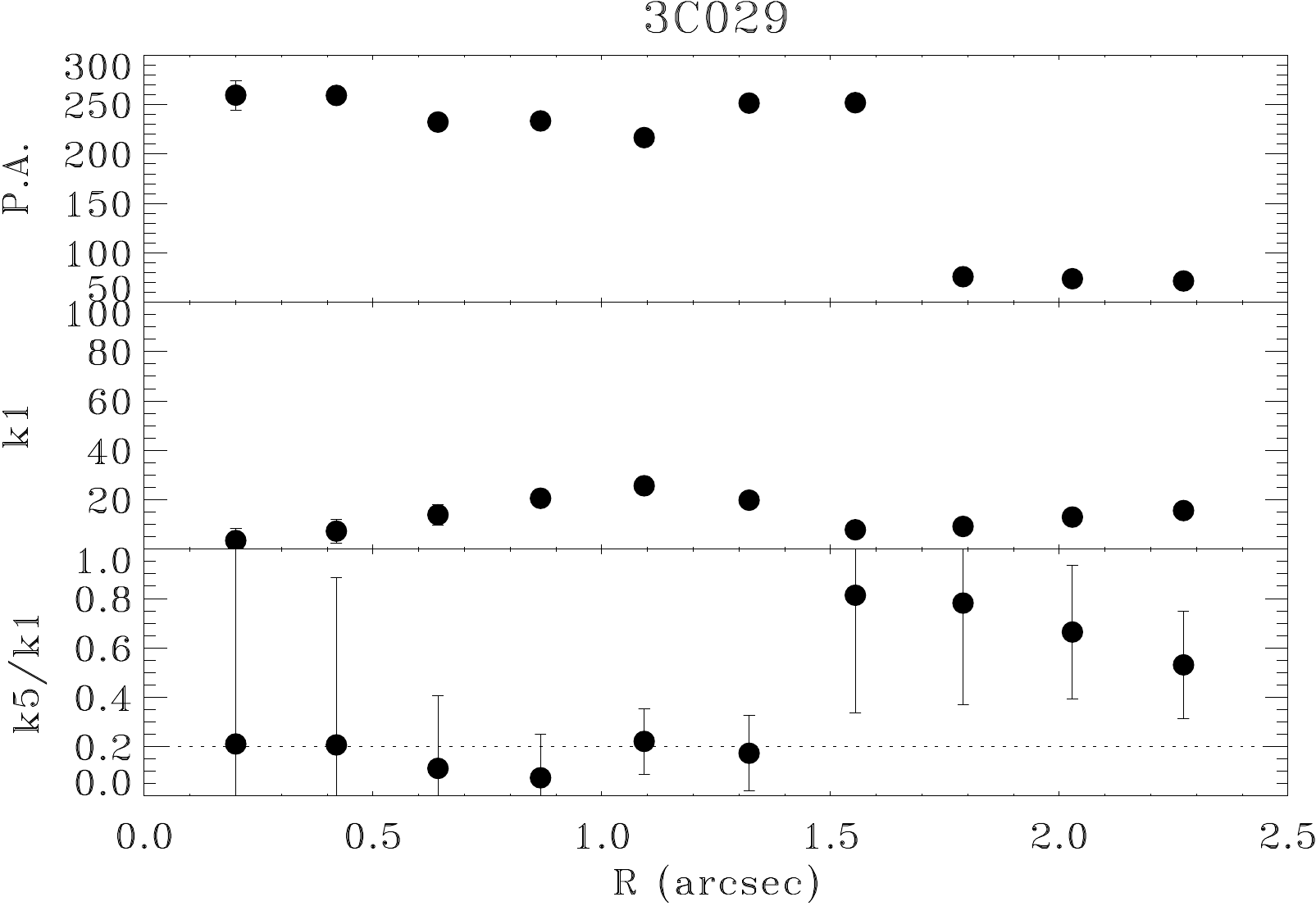}
\includegraphics[width=8.5cm]{3C033kin-crop.pdf}
\includegraphics[width=8.5cm]{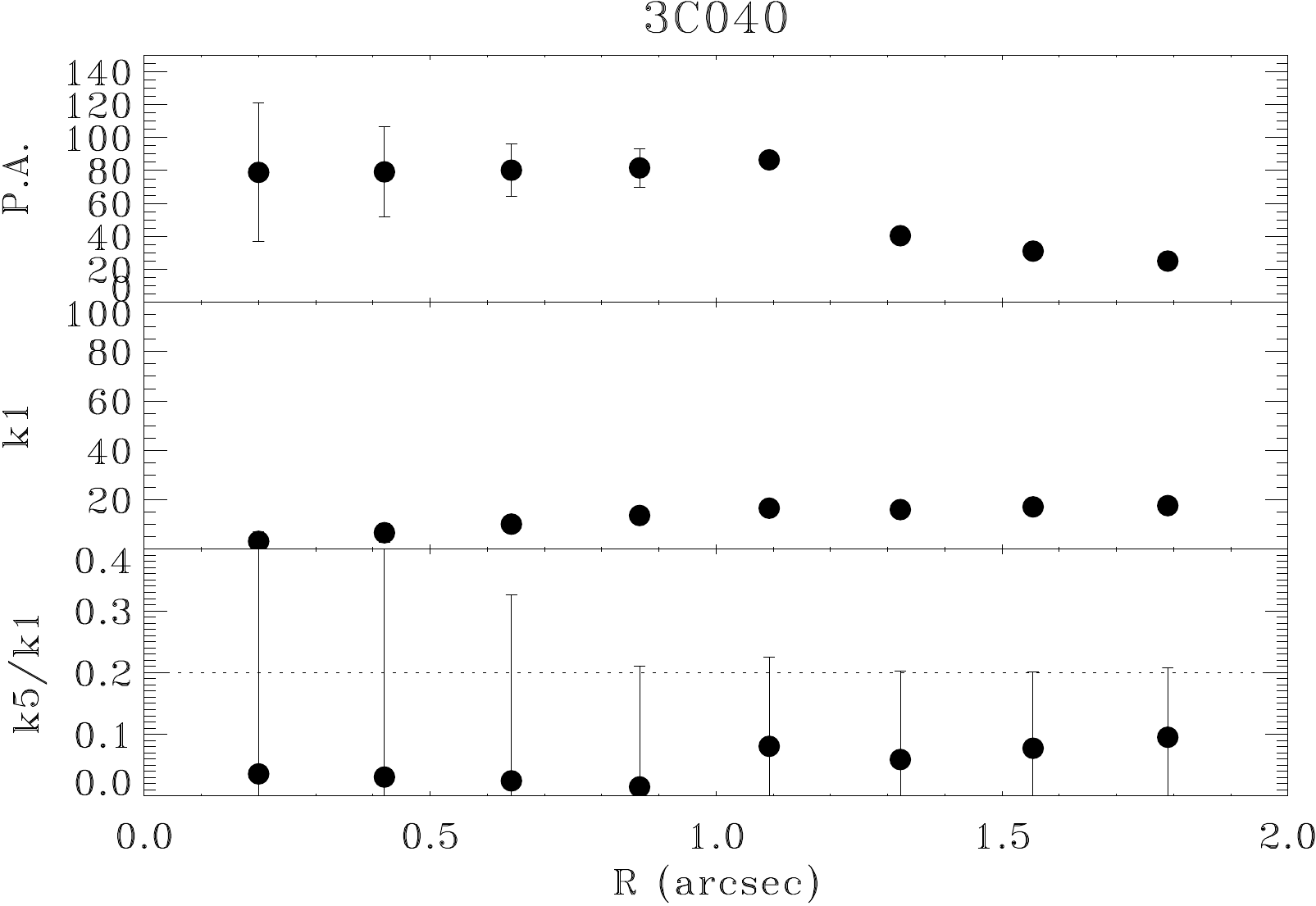}
\includegraphics[width=8.5cm]{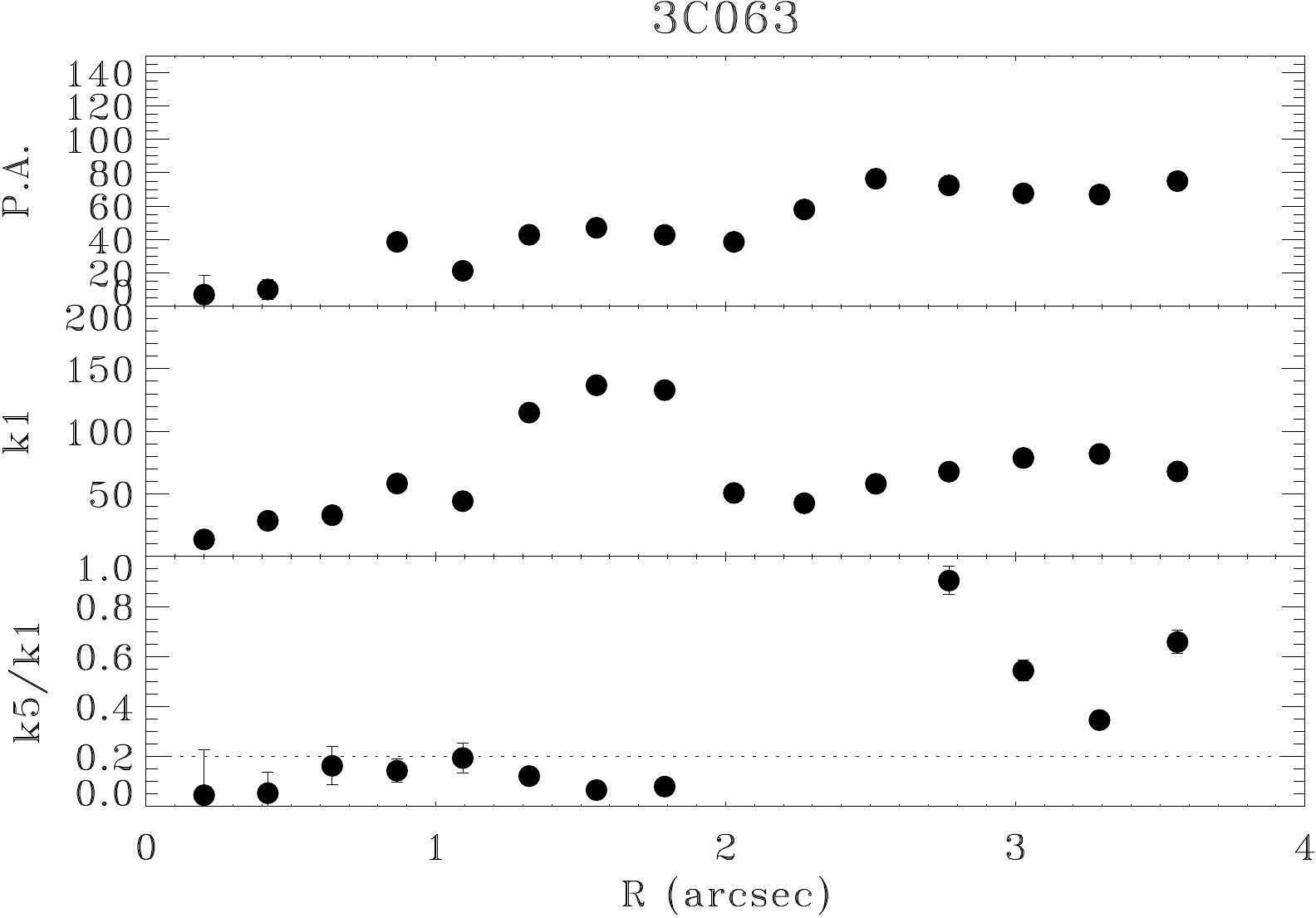}
\includegraphics[width=8.5cm]{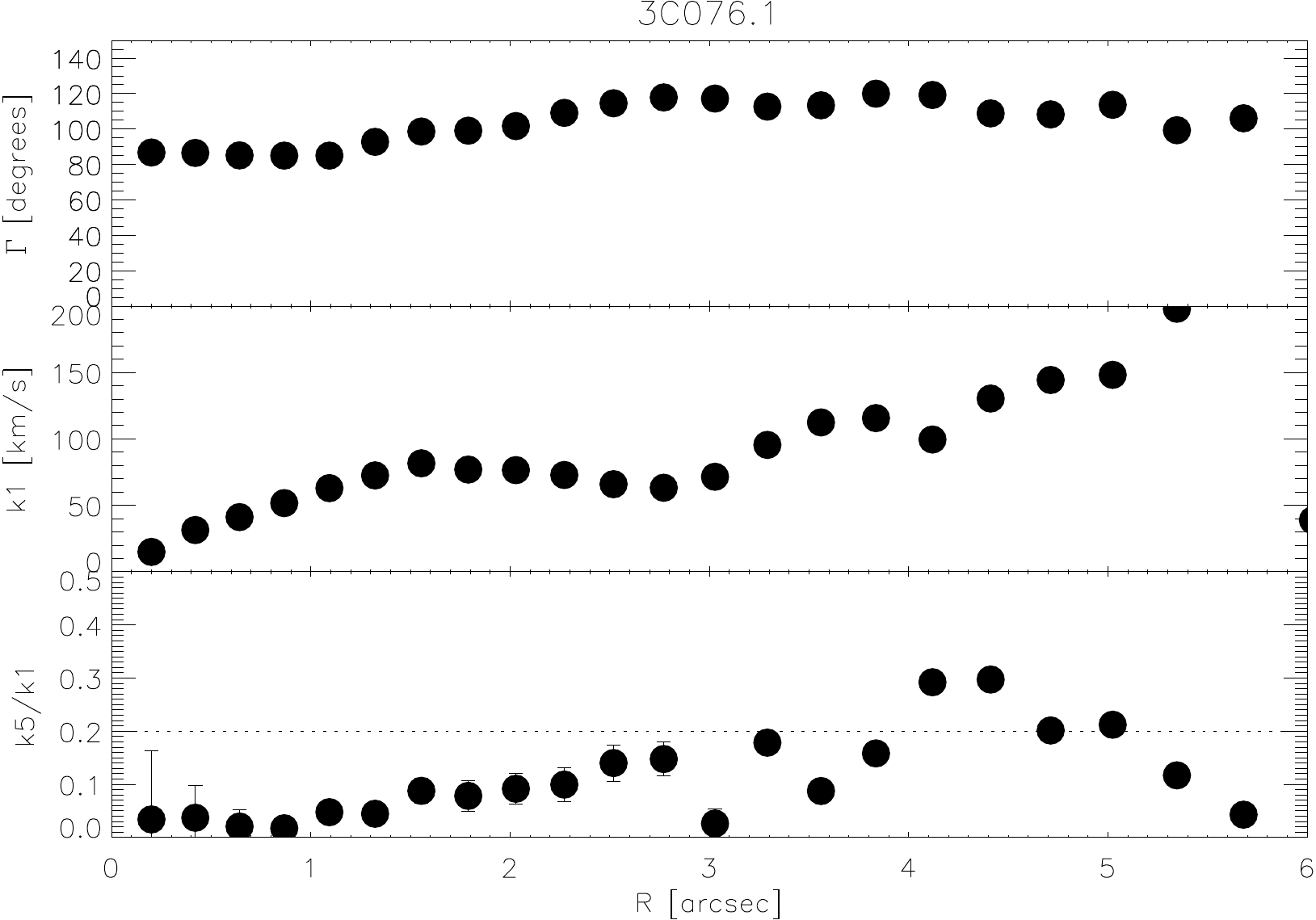}
}
\caption{Results obtained for 36 radio galaxies (all but 3C~089) with
  the {kinemetry} software. From top to bottom in each panel: the
  kinematic position angle $P.A.$ (in degrees), the amplitude of the
  rotation curve (in \kms), and the ratio between the fifth and first
  coefficient $k5/k1$, which quantifies the deviations from simple
  rotation.  }
\end{figure*}

\addtocounter{figure}{-1}
\begin{figure*}  
\centering{ 
\includegraphics[width=8.5cm]{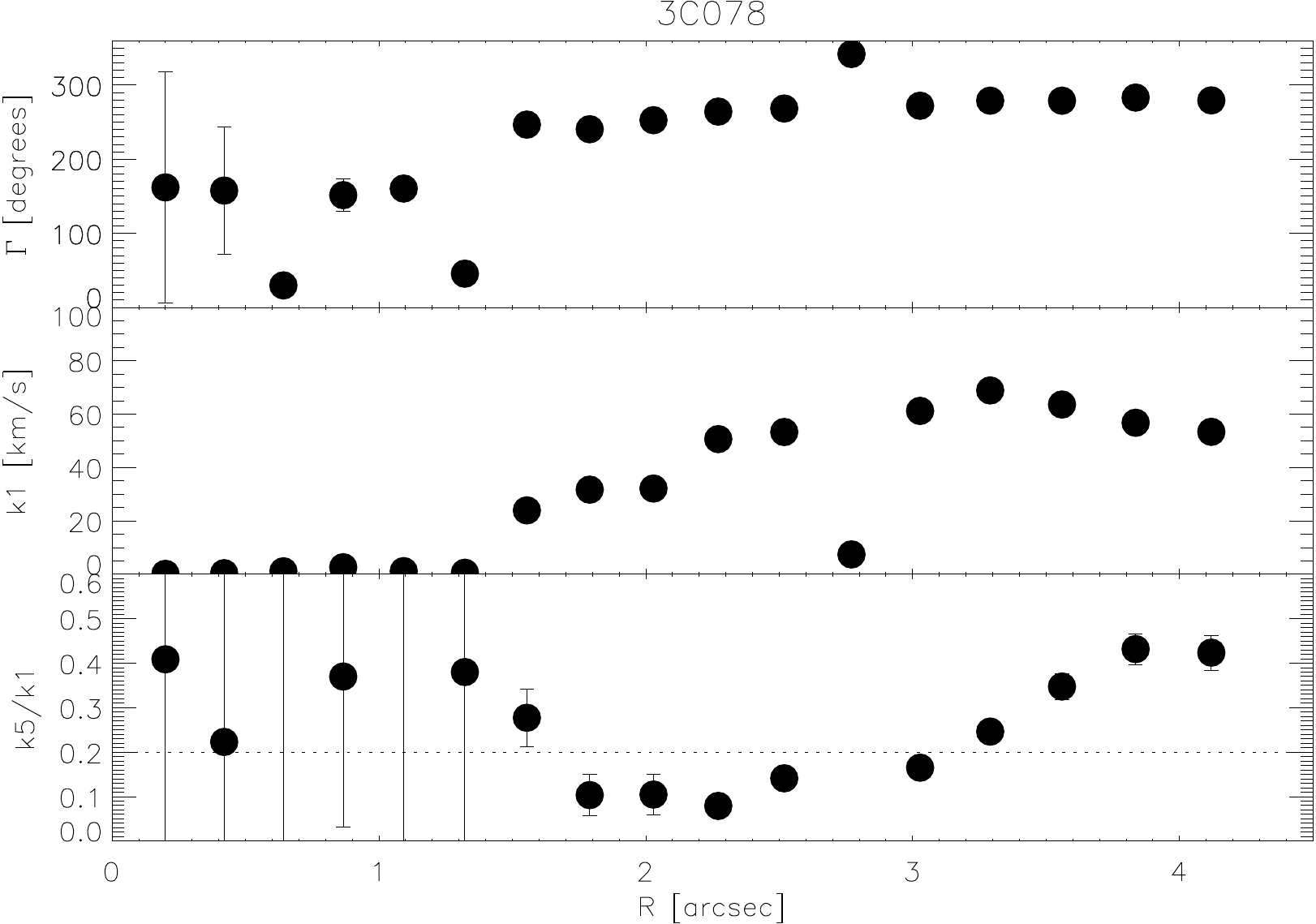}
\includegraphics[width=8.5cm]{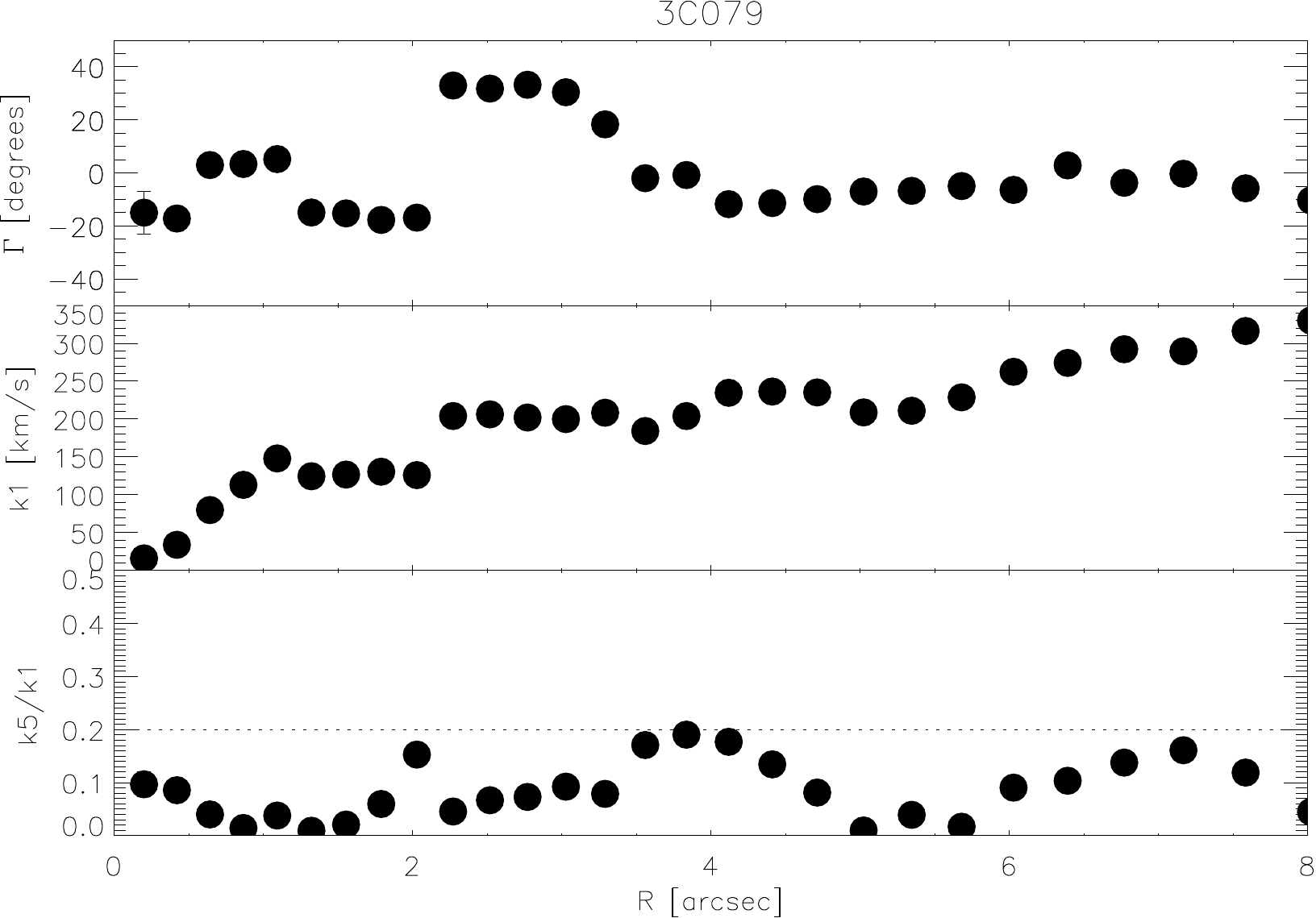}
\includegraphics[width=8.5cm]{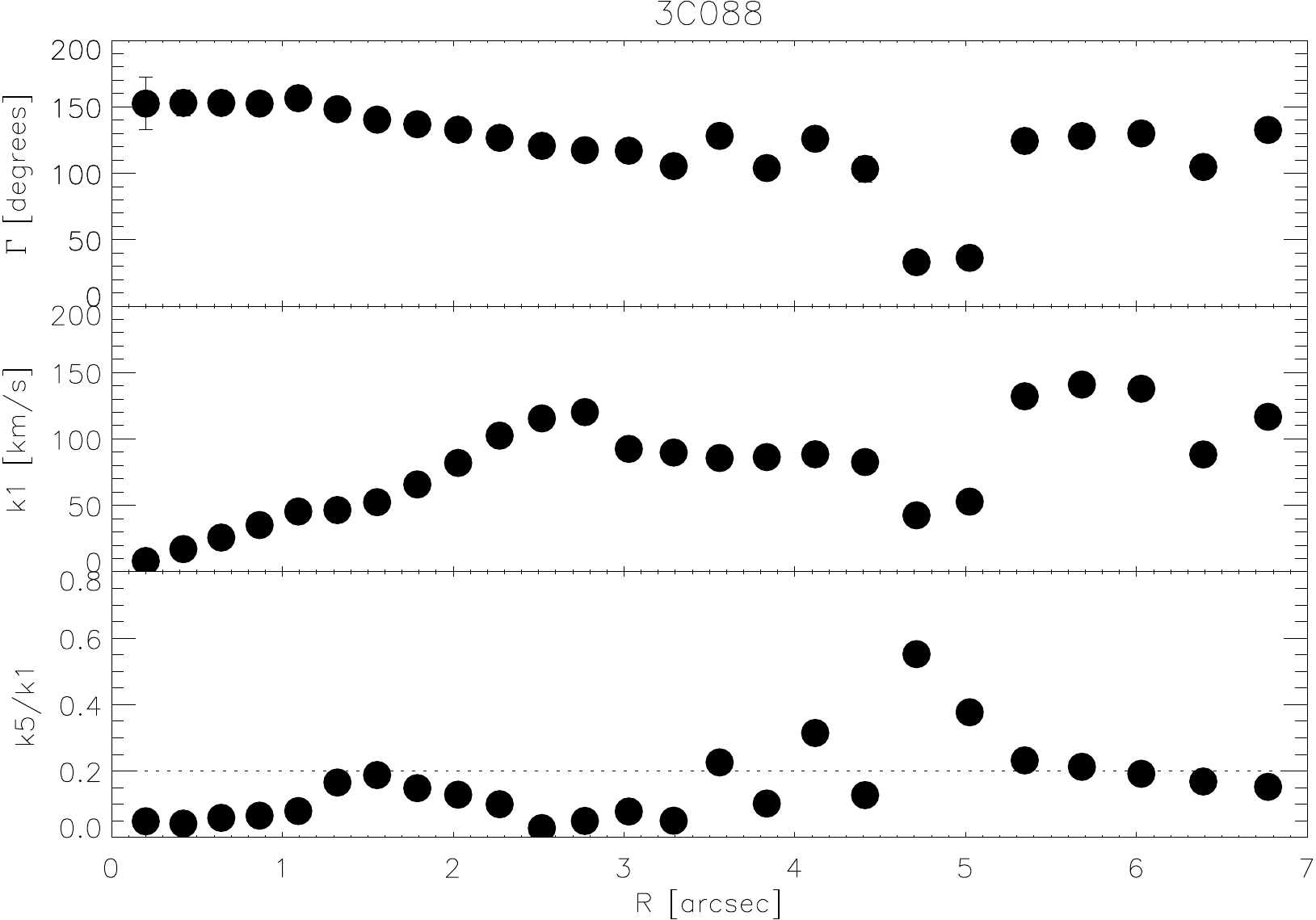}
\includegraphics[width=8.5cm]{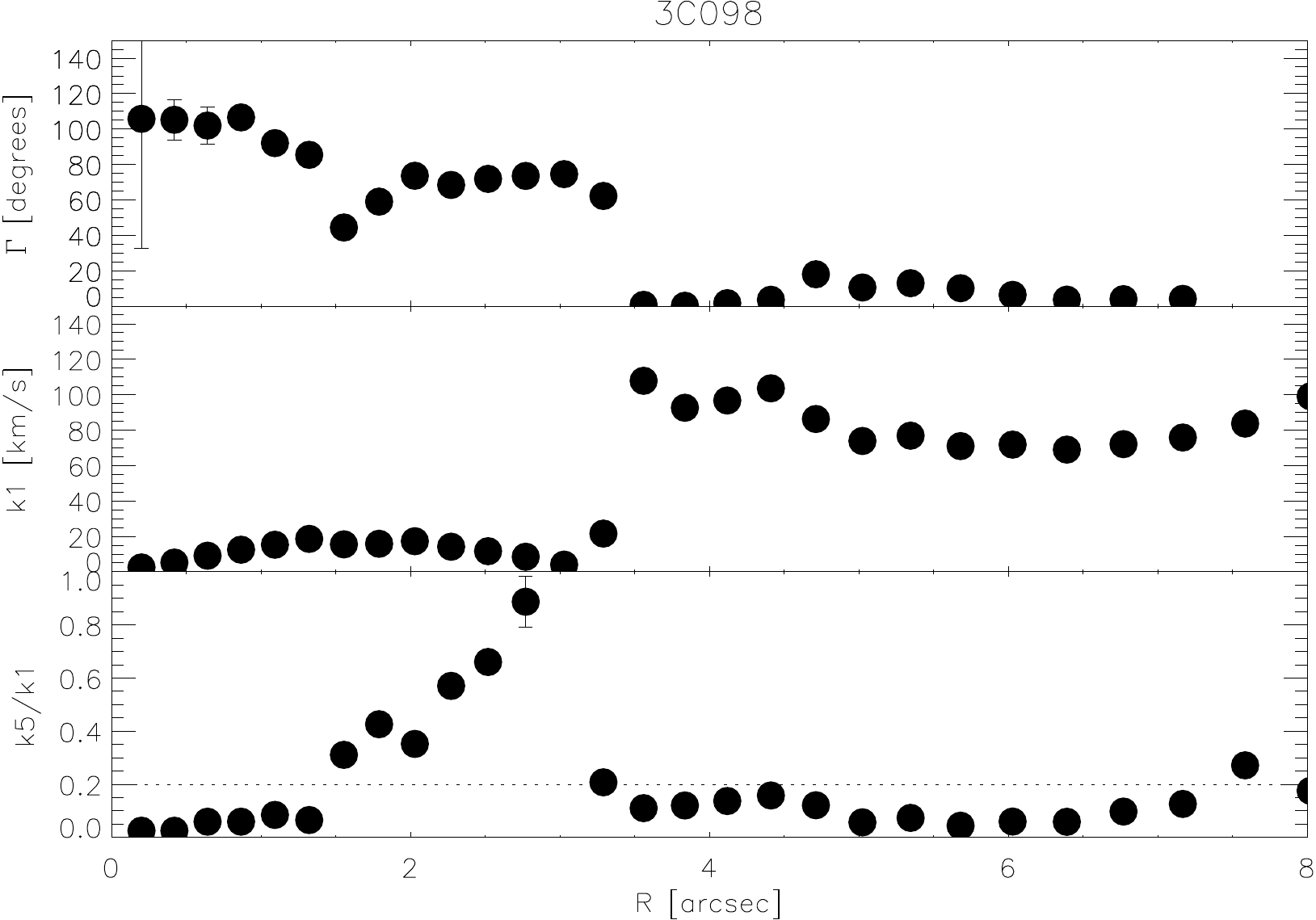}
\includegraphics[width=8.5cm]{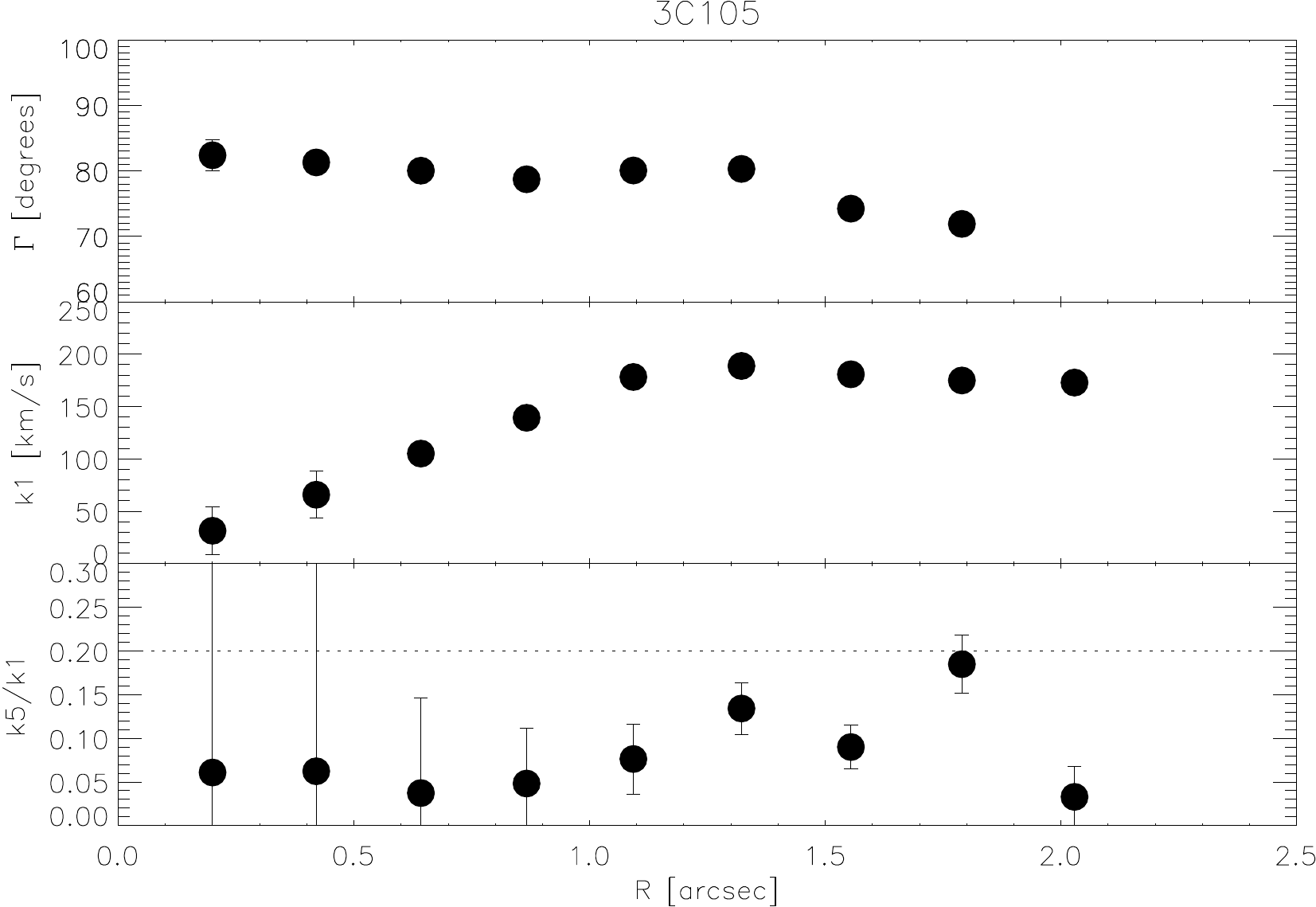}
\includegraphics[width=8.5cm]{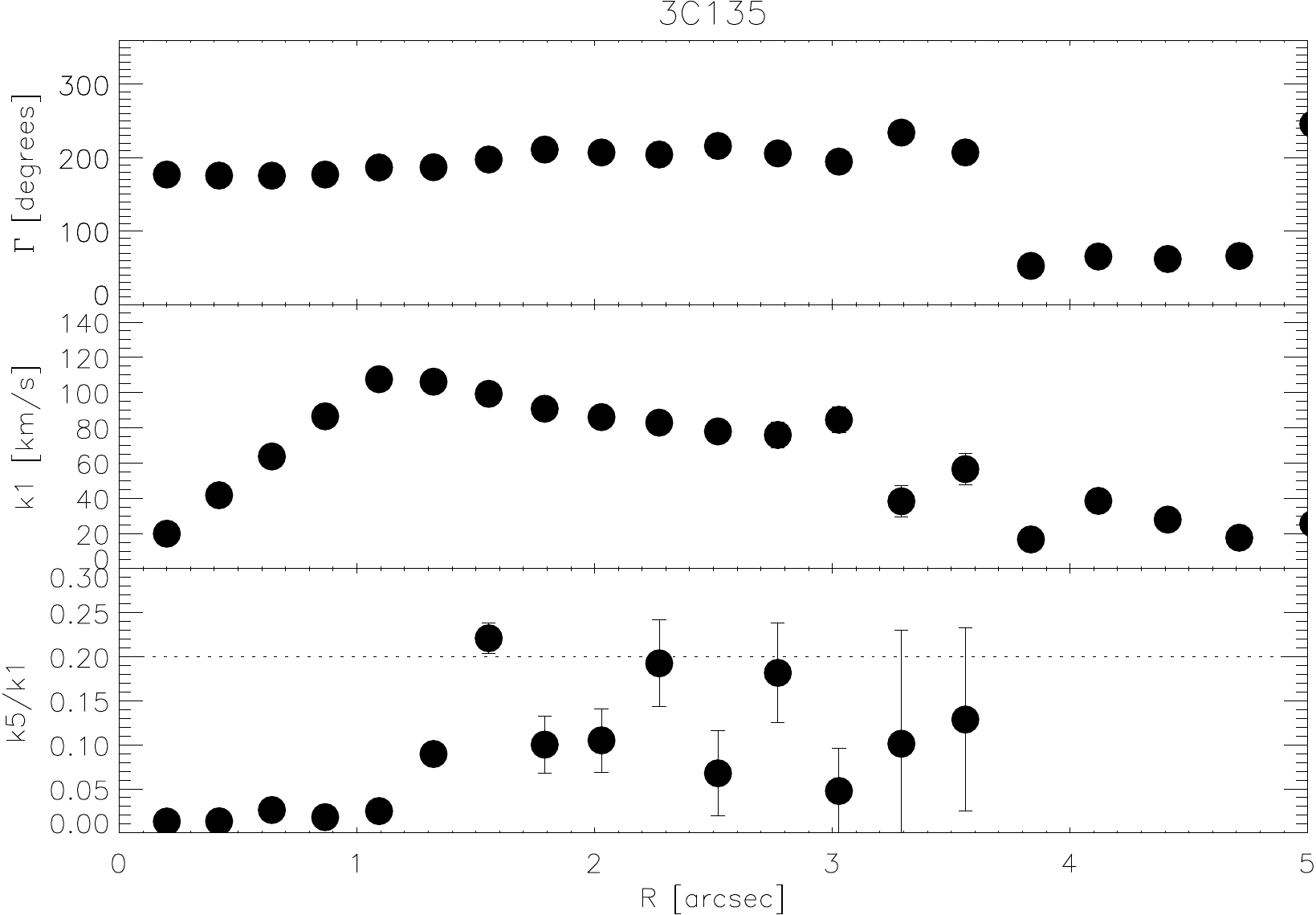}
\includegraphics[width=8.5cm]{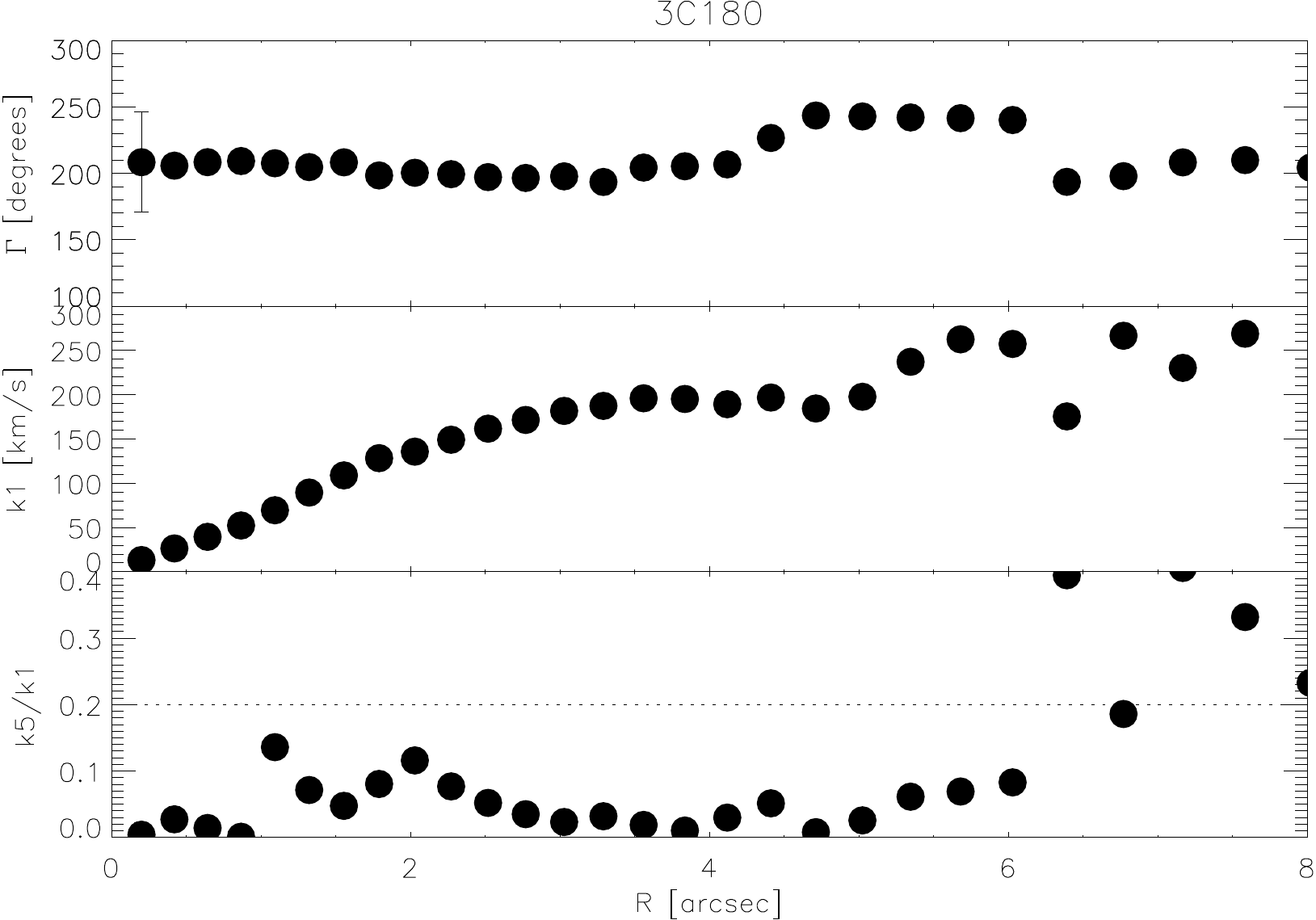}
\includegraphics[width=8.5cm]{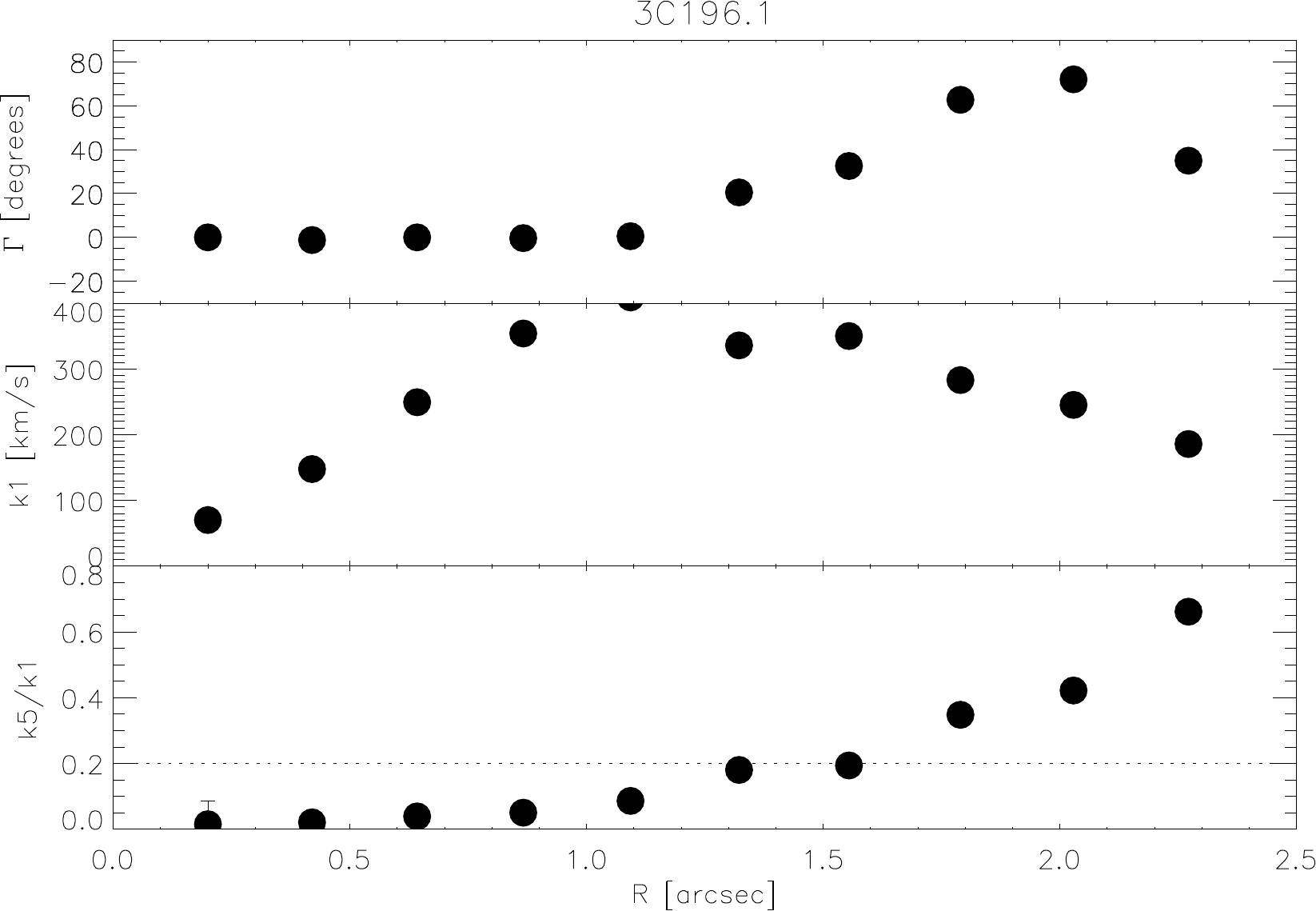}
}
\caption{- continued.}
\end{figure*}   

\addtocounter{figure}{-1}
\begin{figure*}  
\centering{ 
\includegraphics[width=8.5cm]{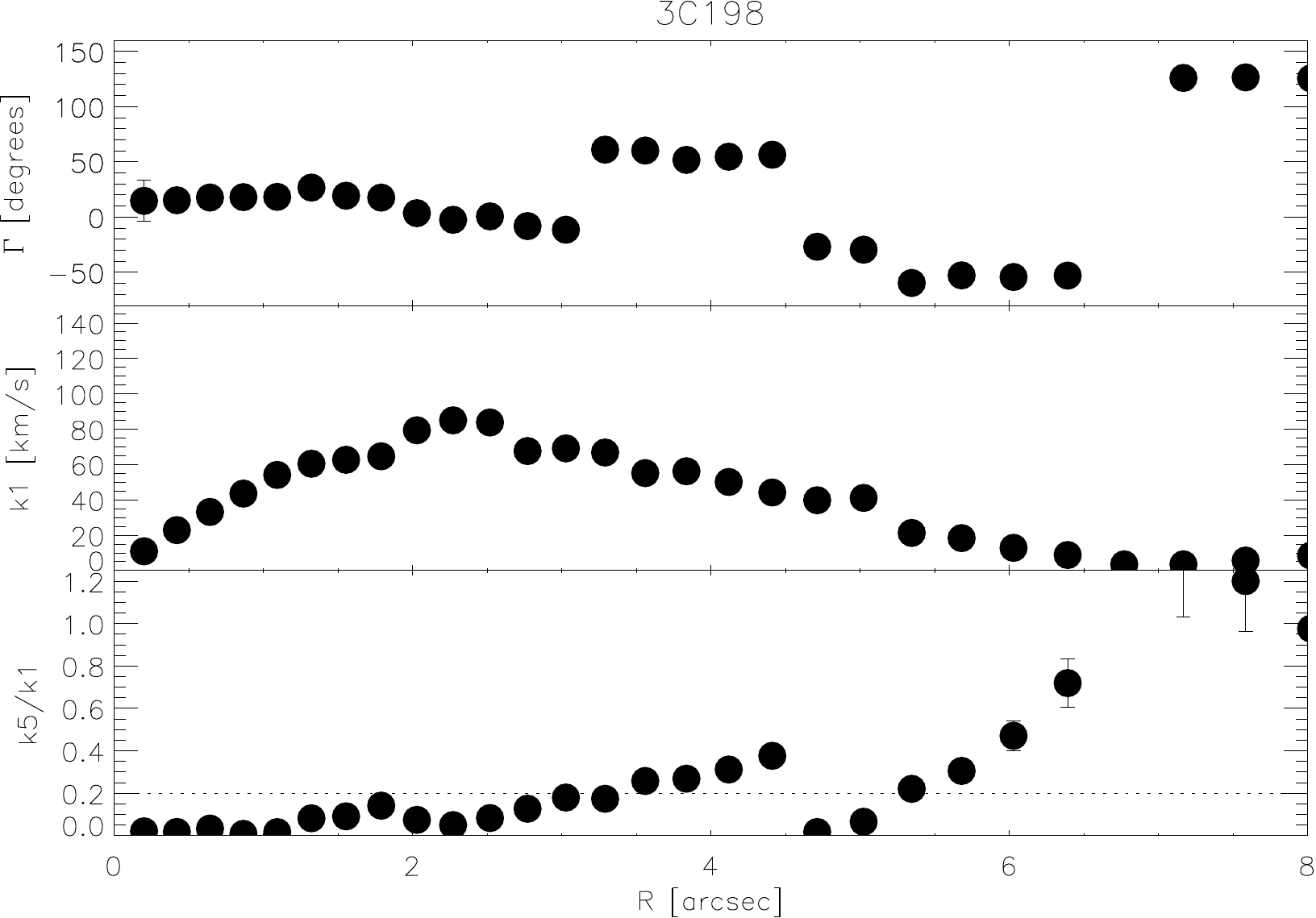}
\includegraphics[width=8.5cm]{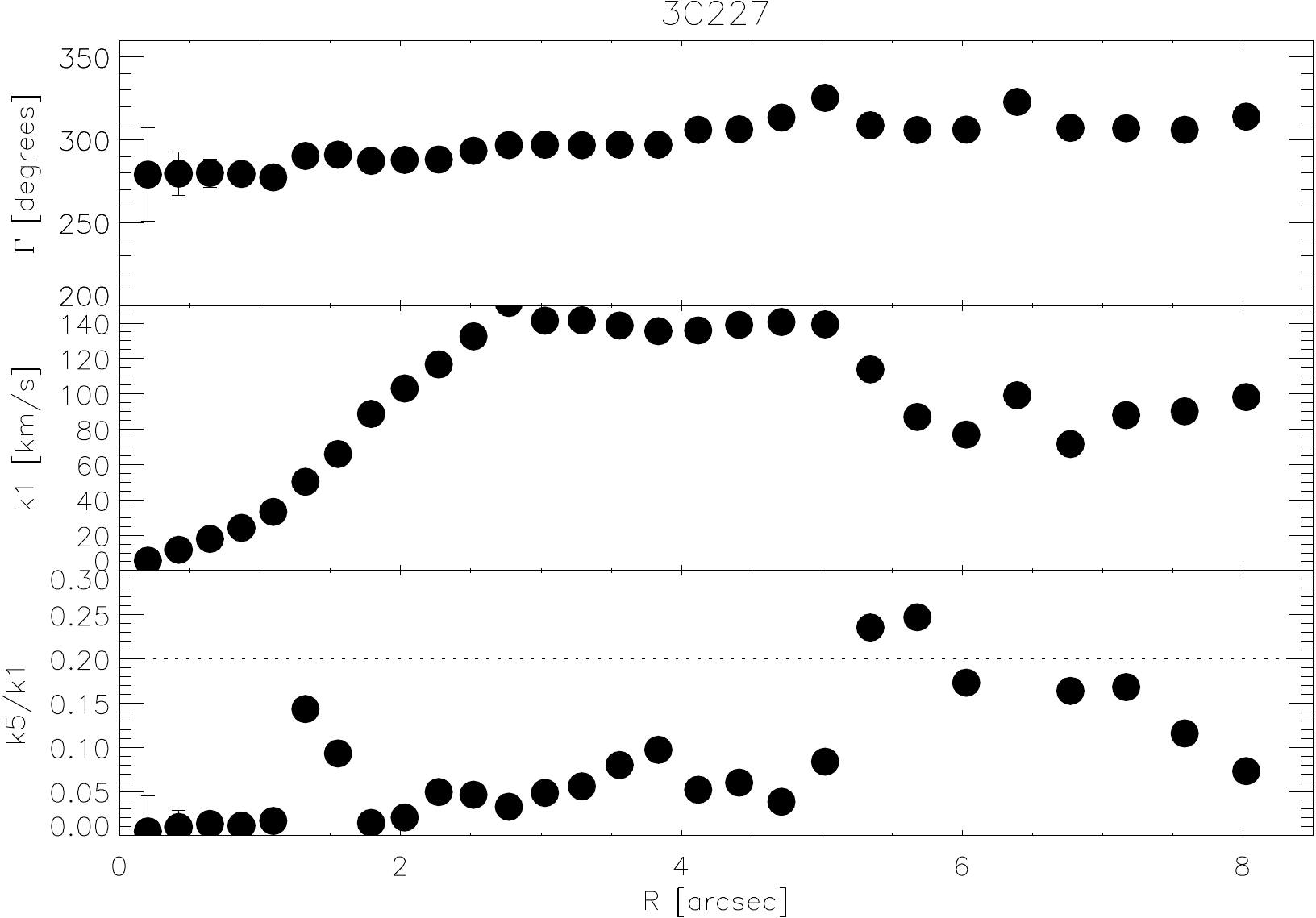}
\includegraphics[width=8.5cm]{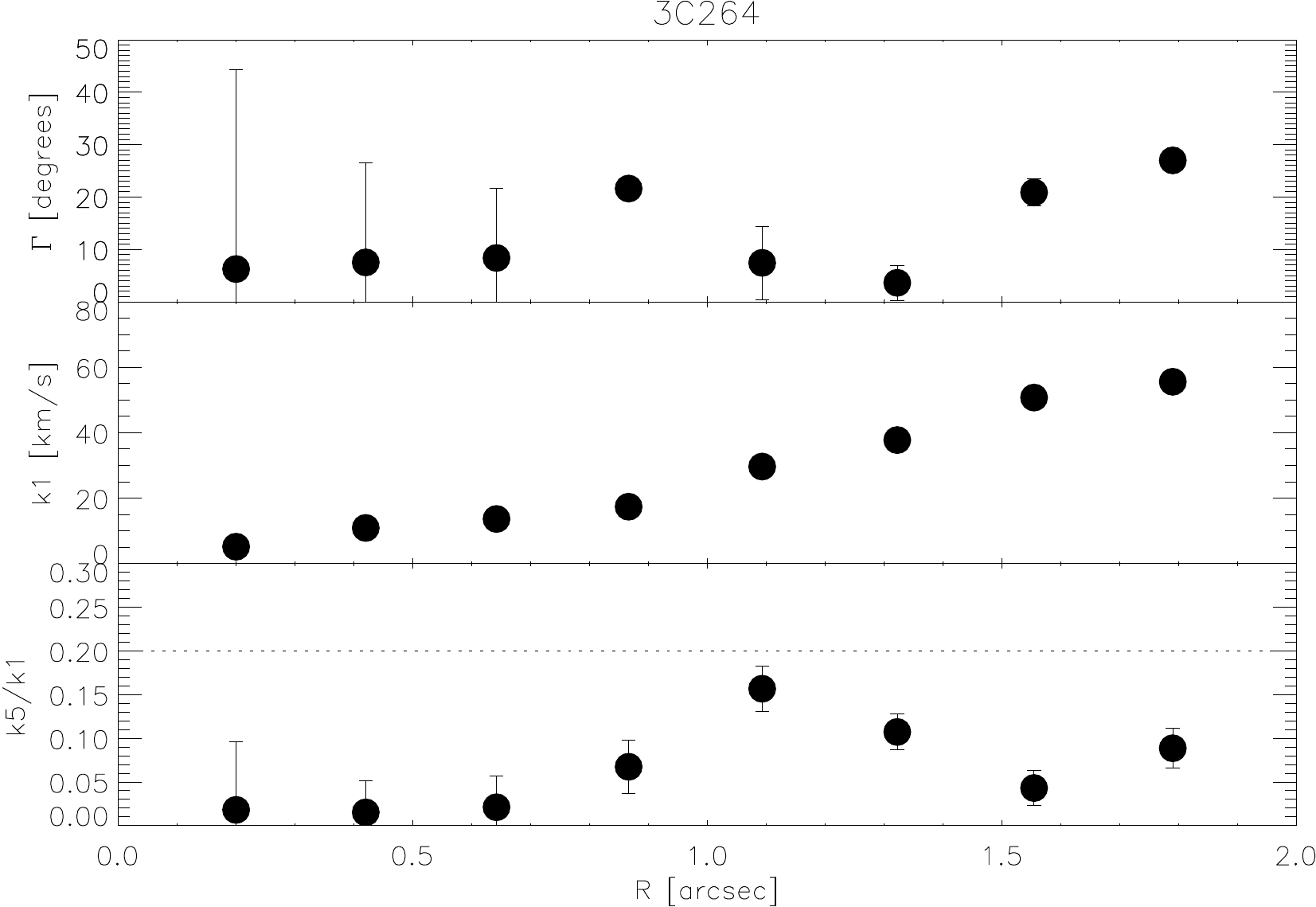}
\includegraphics[width=8.5cm]{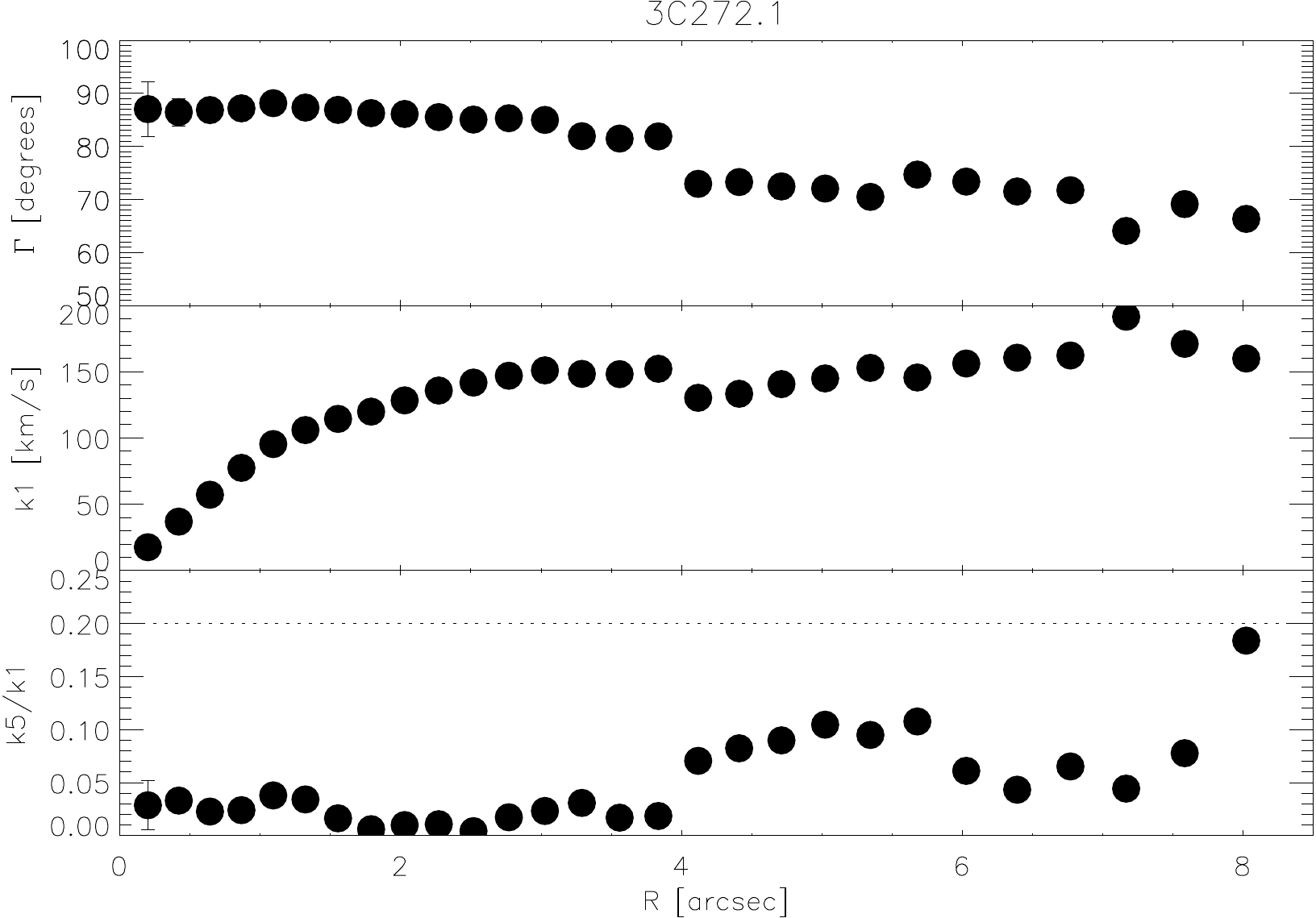}
\includegraphics[width=8.5cm]{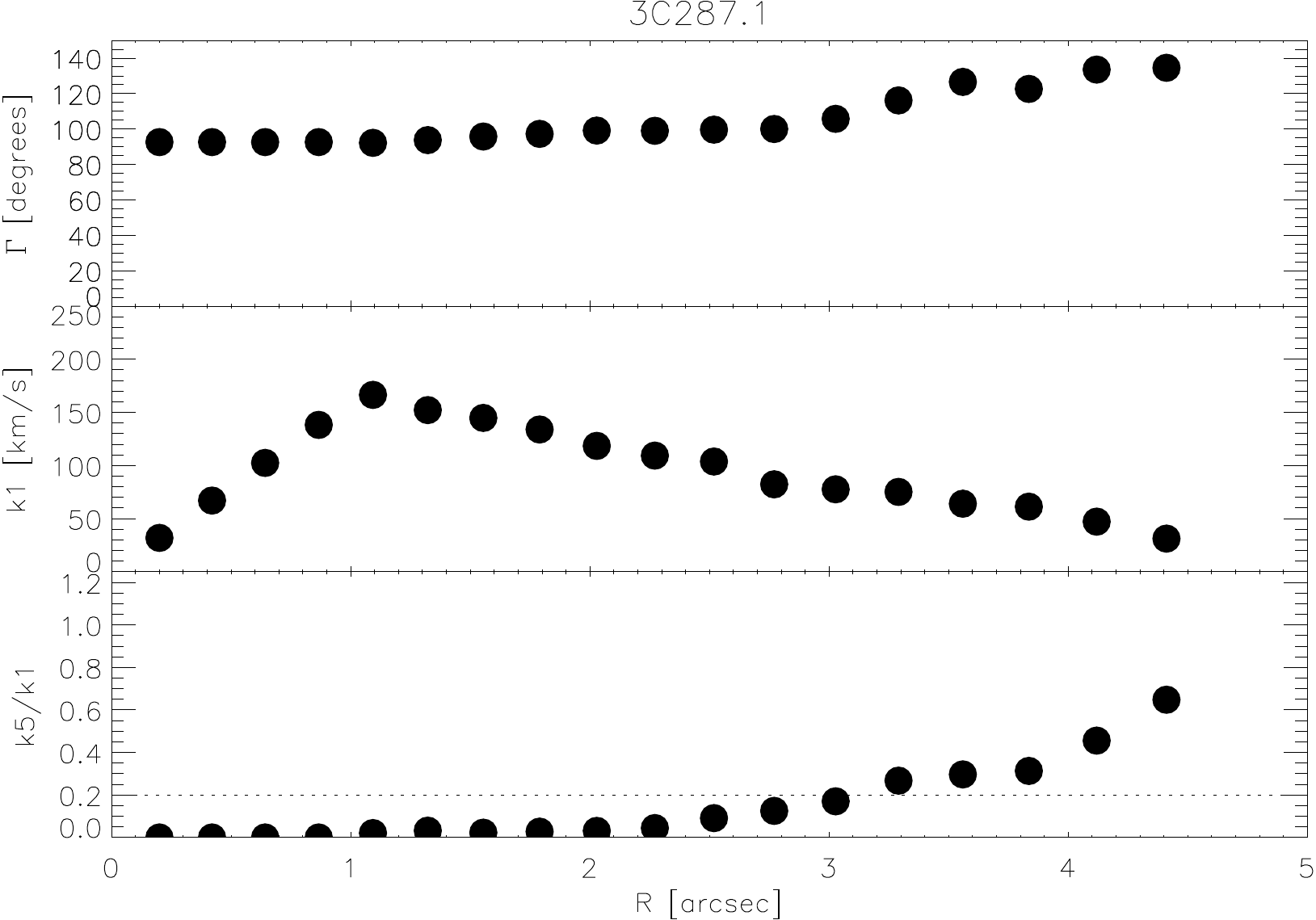}
\includegraphics[width=8.5cm]{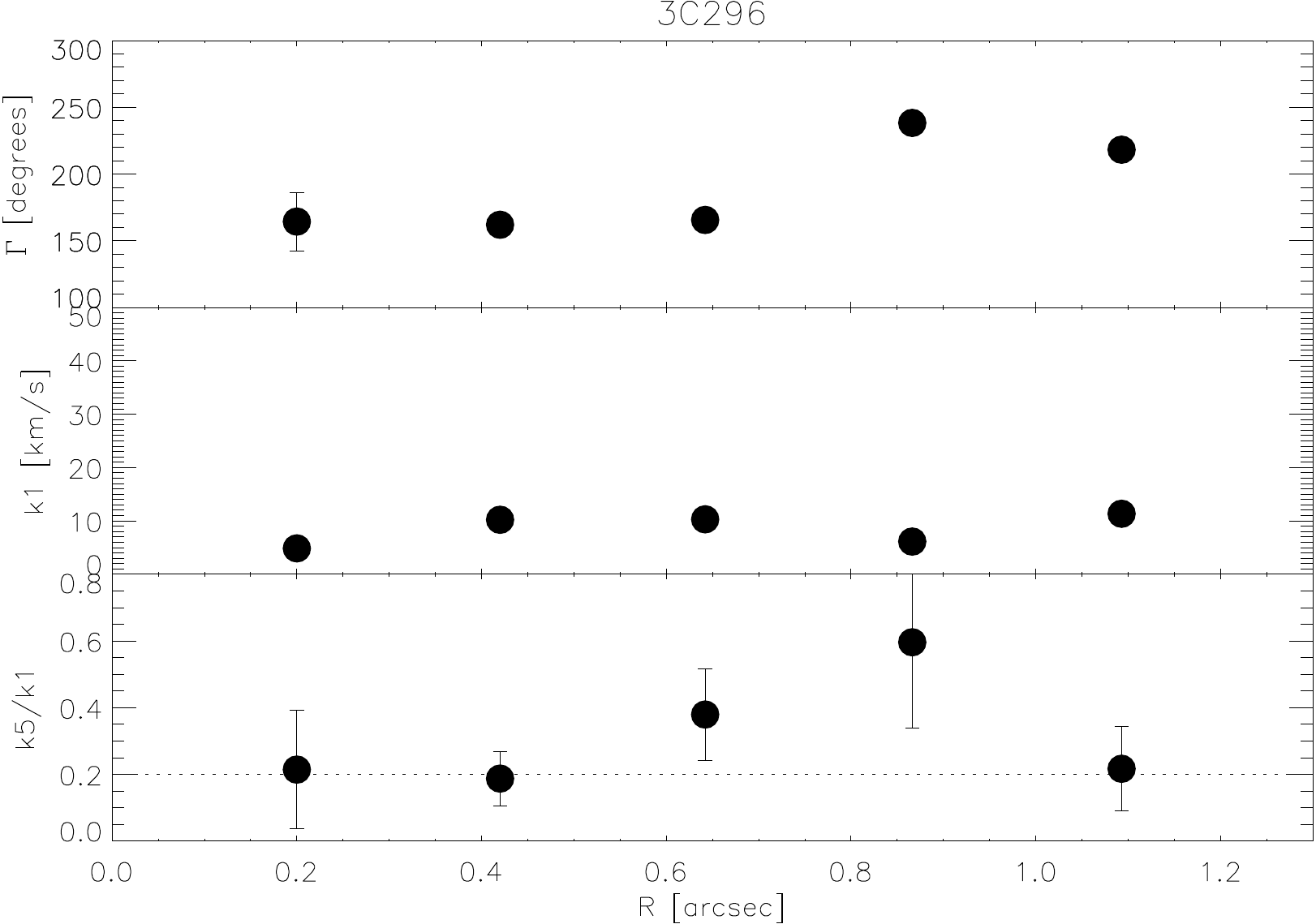}
\includegraphics[width=8.5cm]{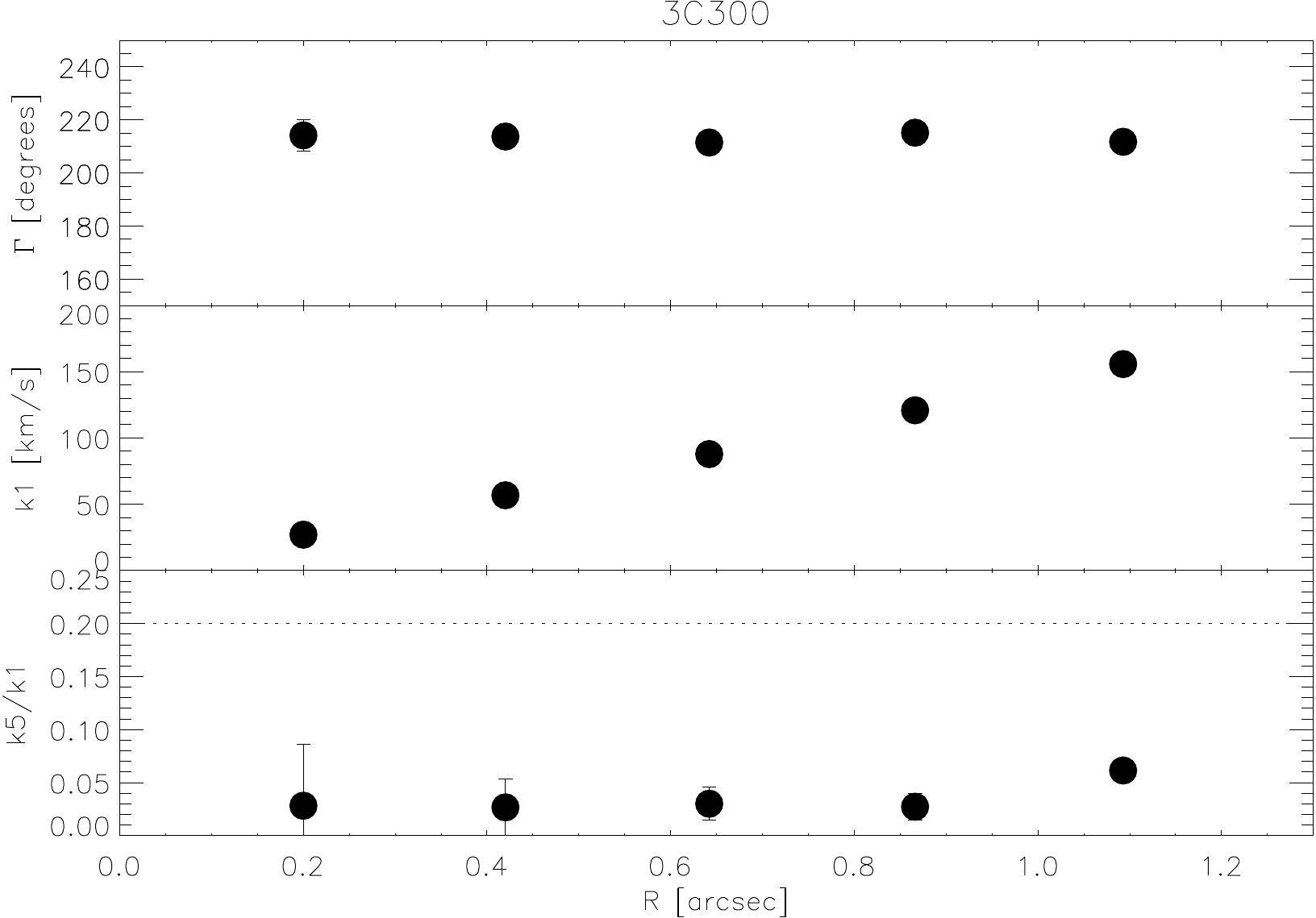}
\includegraphics[width=8.5cm]{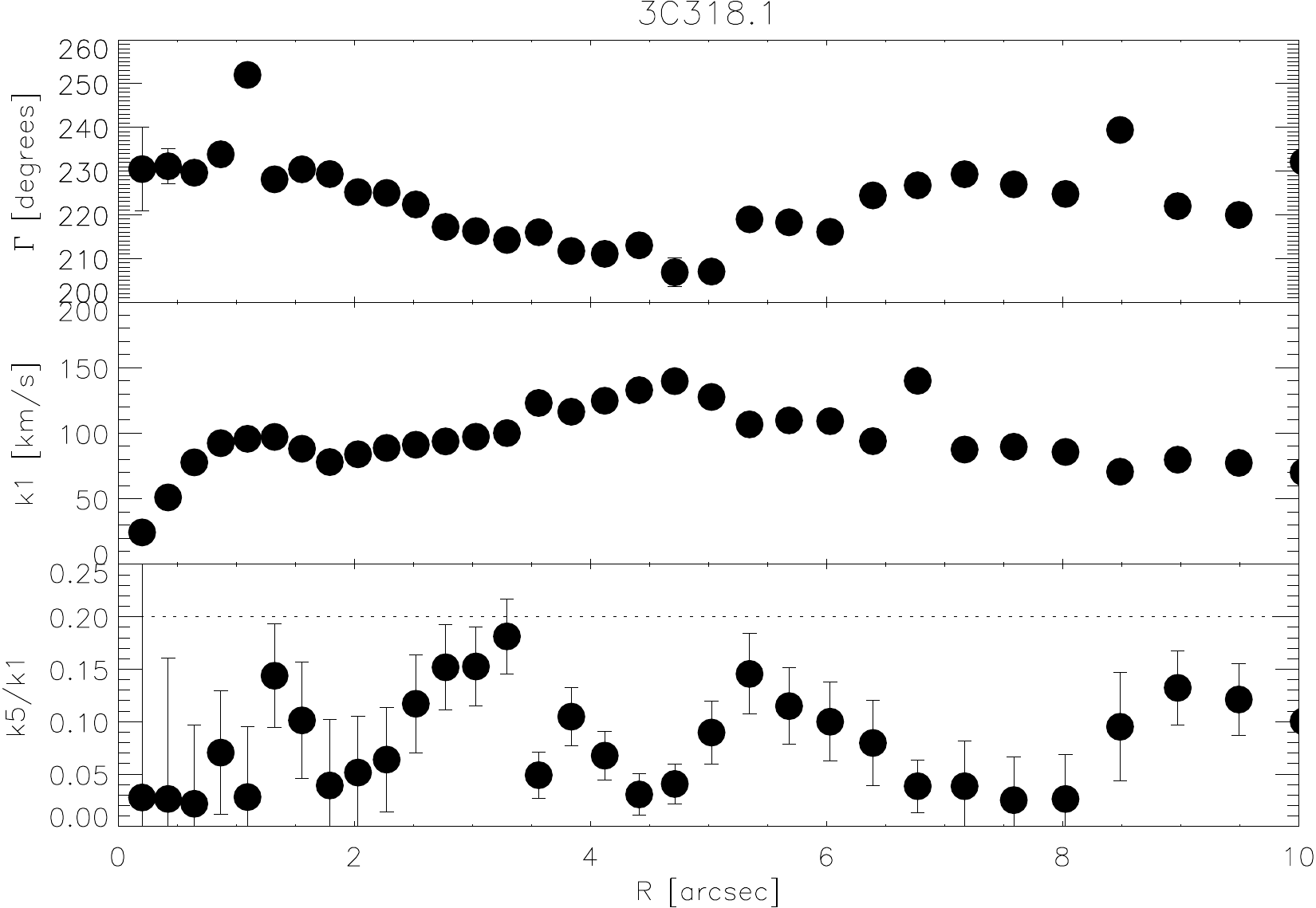}
}
\caption{- continued.}
\end{figure*}

\addtocounter{figure}{-1}
\begin{figure*}  
\centering{ 
\includegraphics[width=8.5cm]{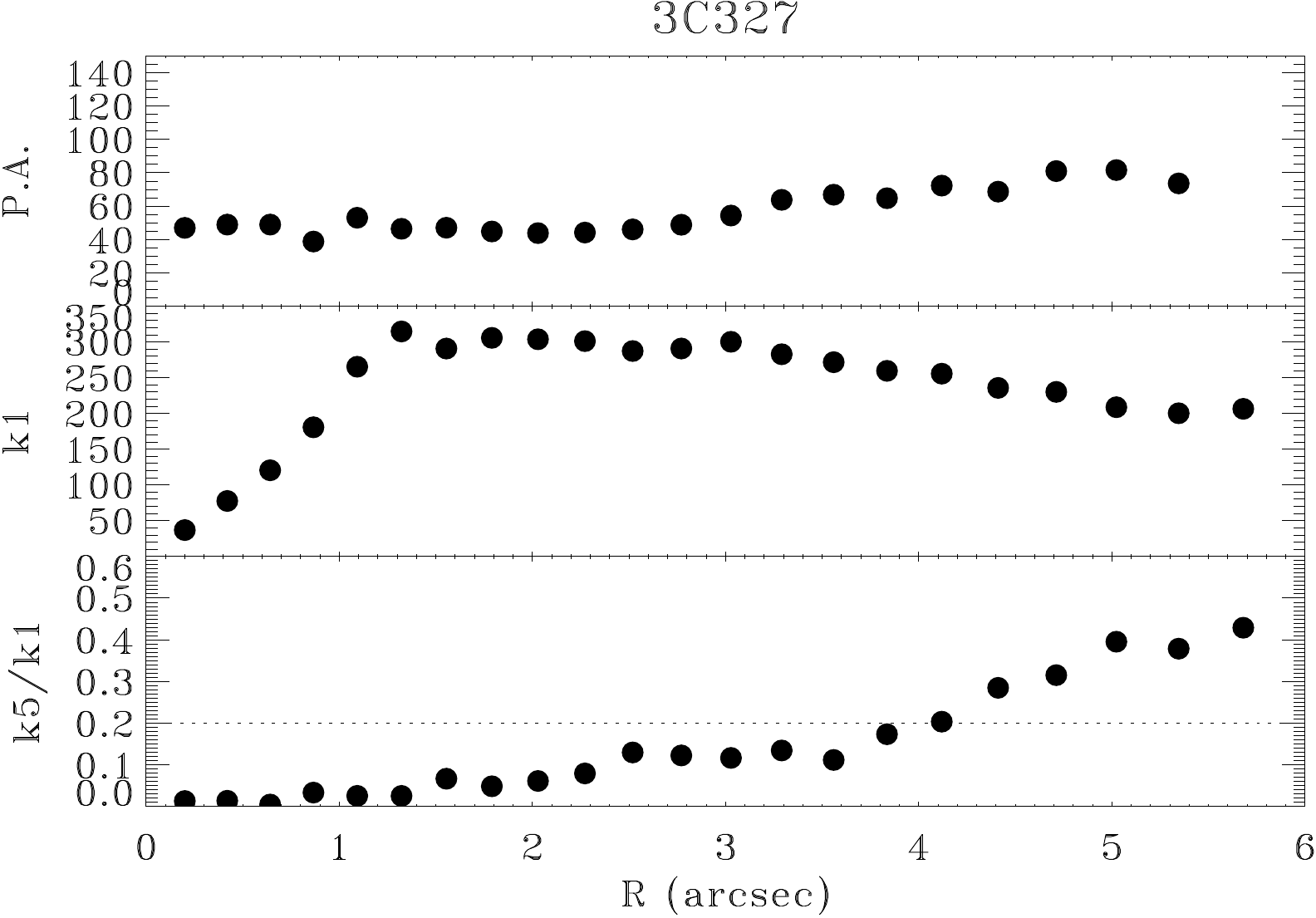}
\includegraphics[width=8.5cm]{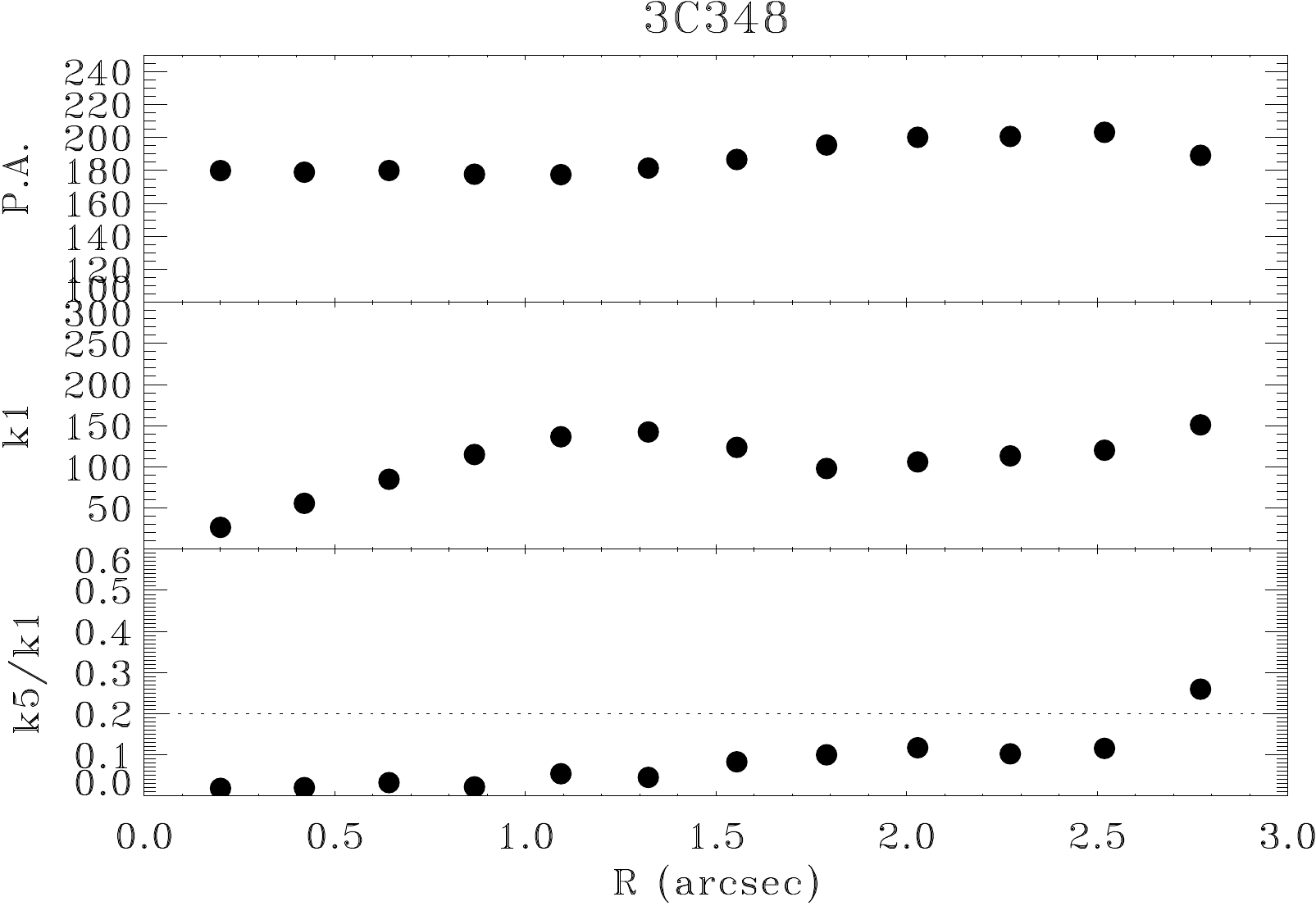}
\includegraphics[width=8.5cm]{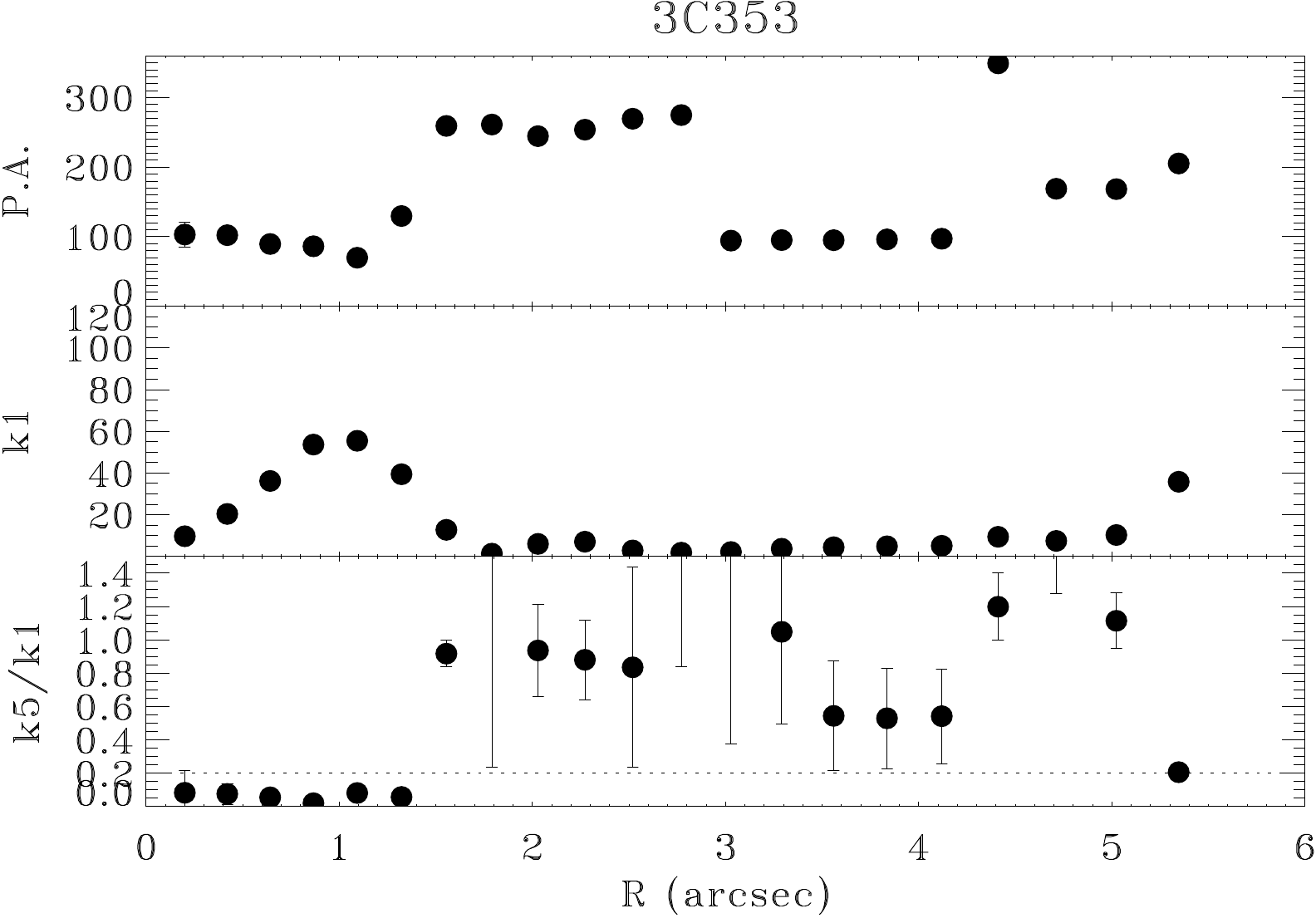}
\includegraphics[width=8.5cm]{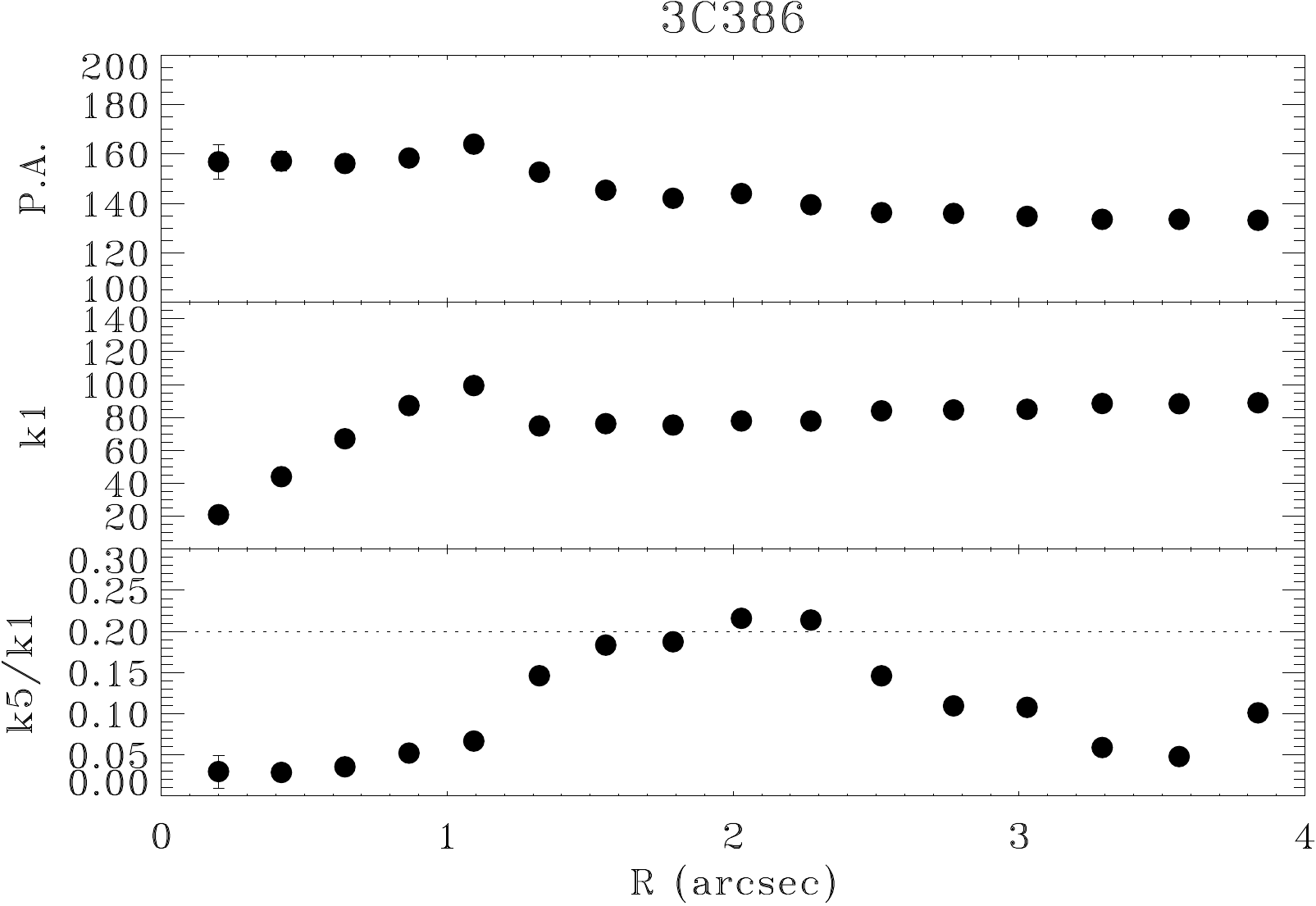}
\includegraphics[width=8.5cm]{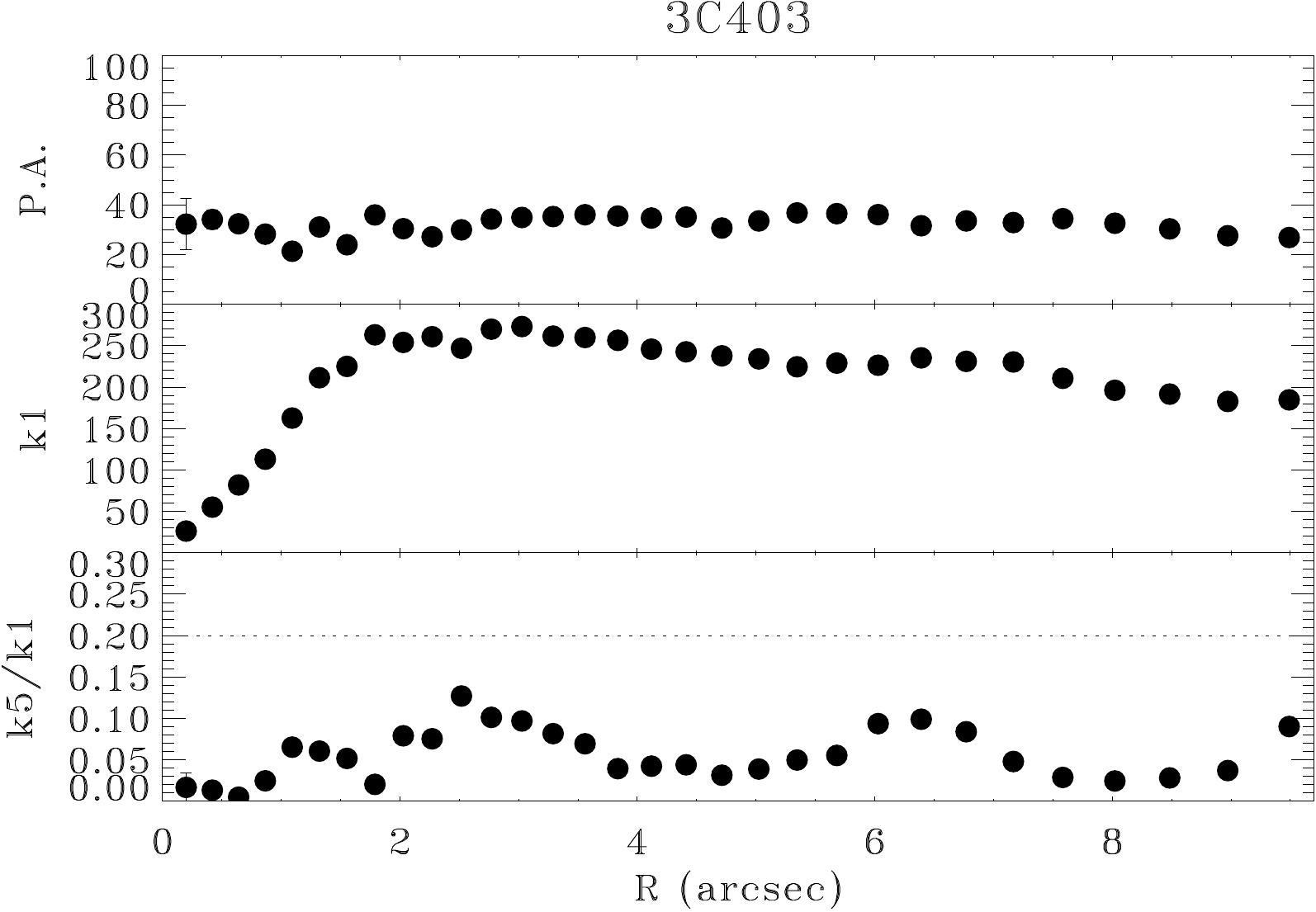}
\includegraphics[width=8.5cm]{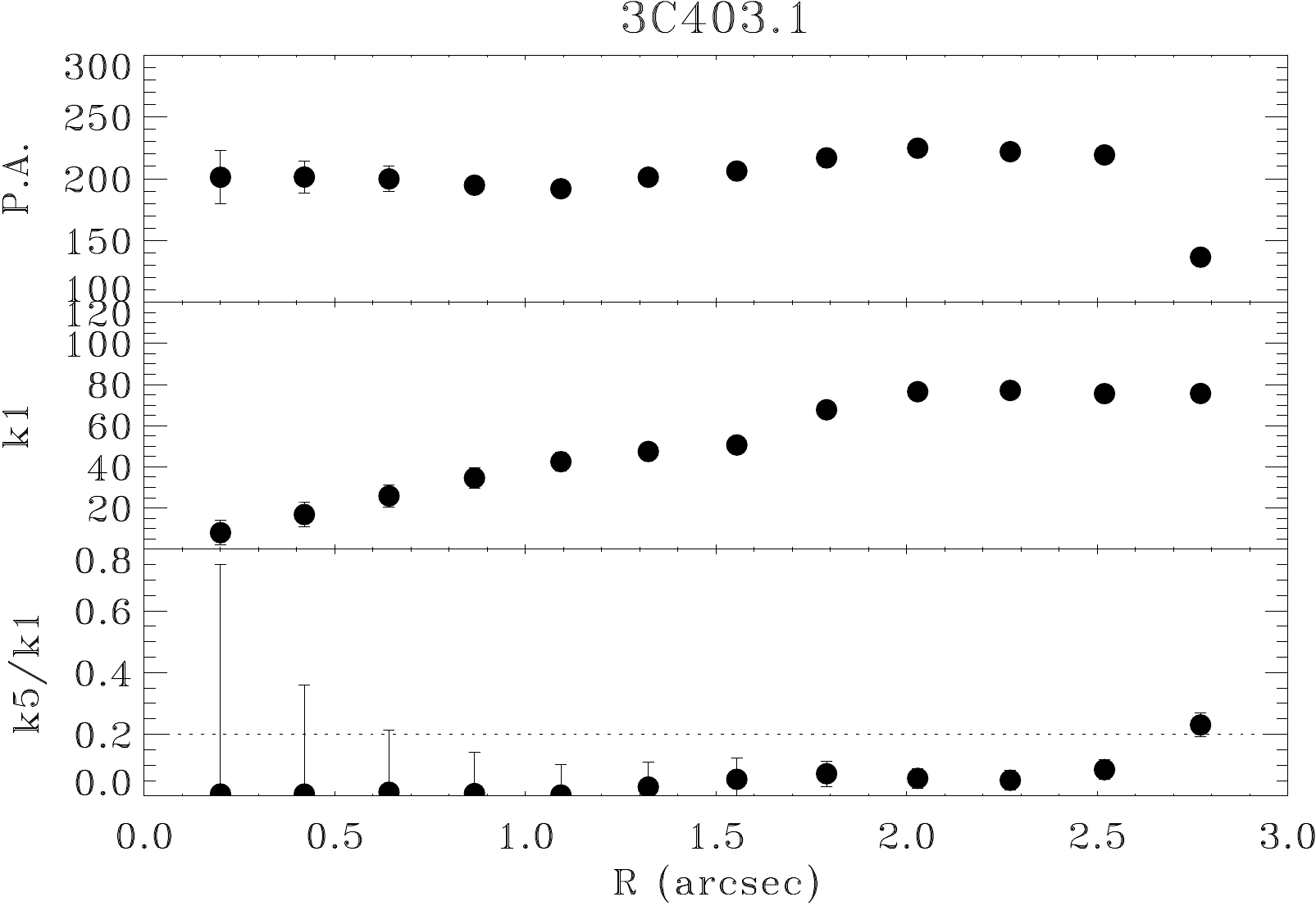}
\includegraphics[width=8.5cm]{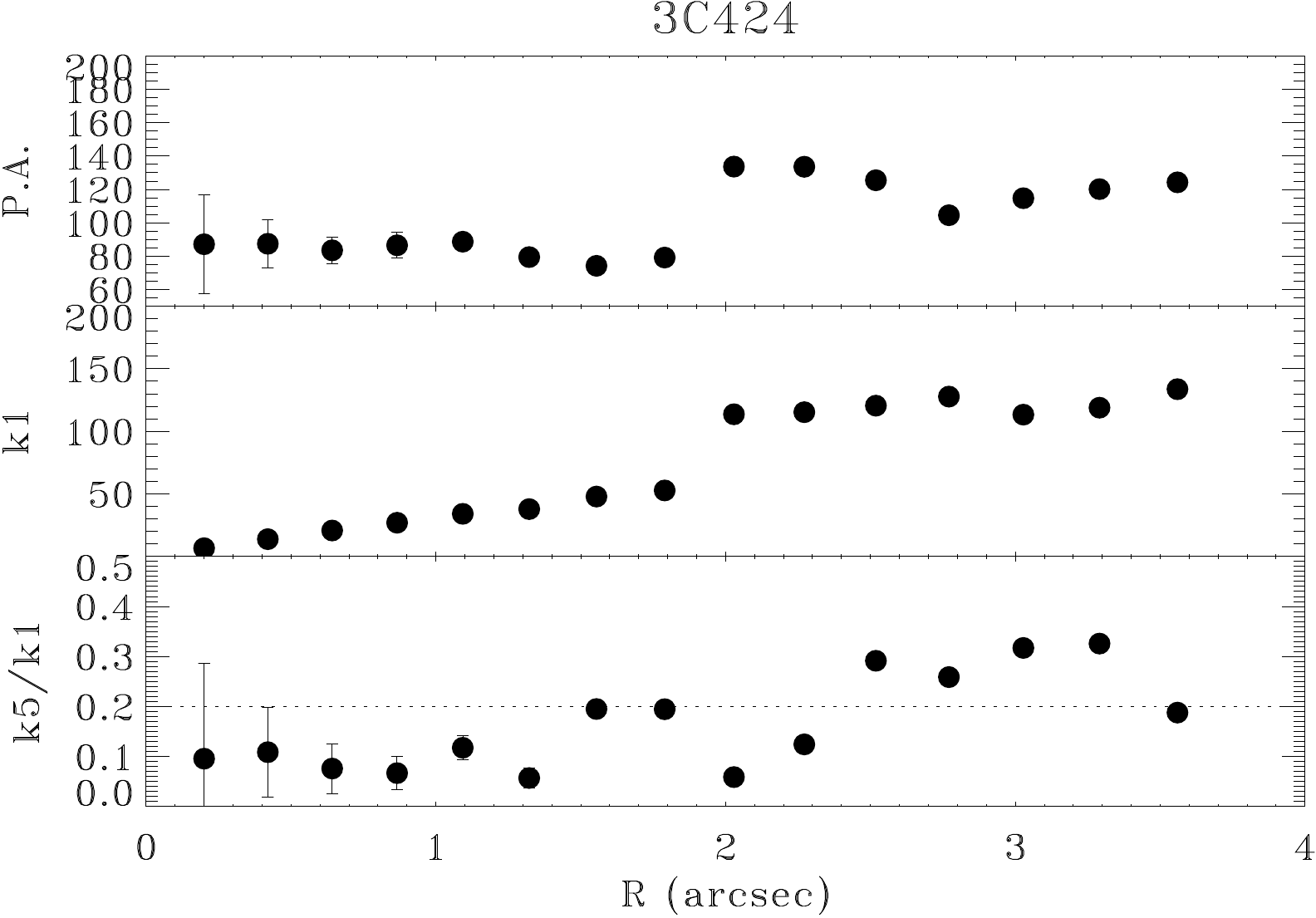}
\includegraphics[width=8.5cm]{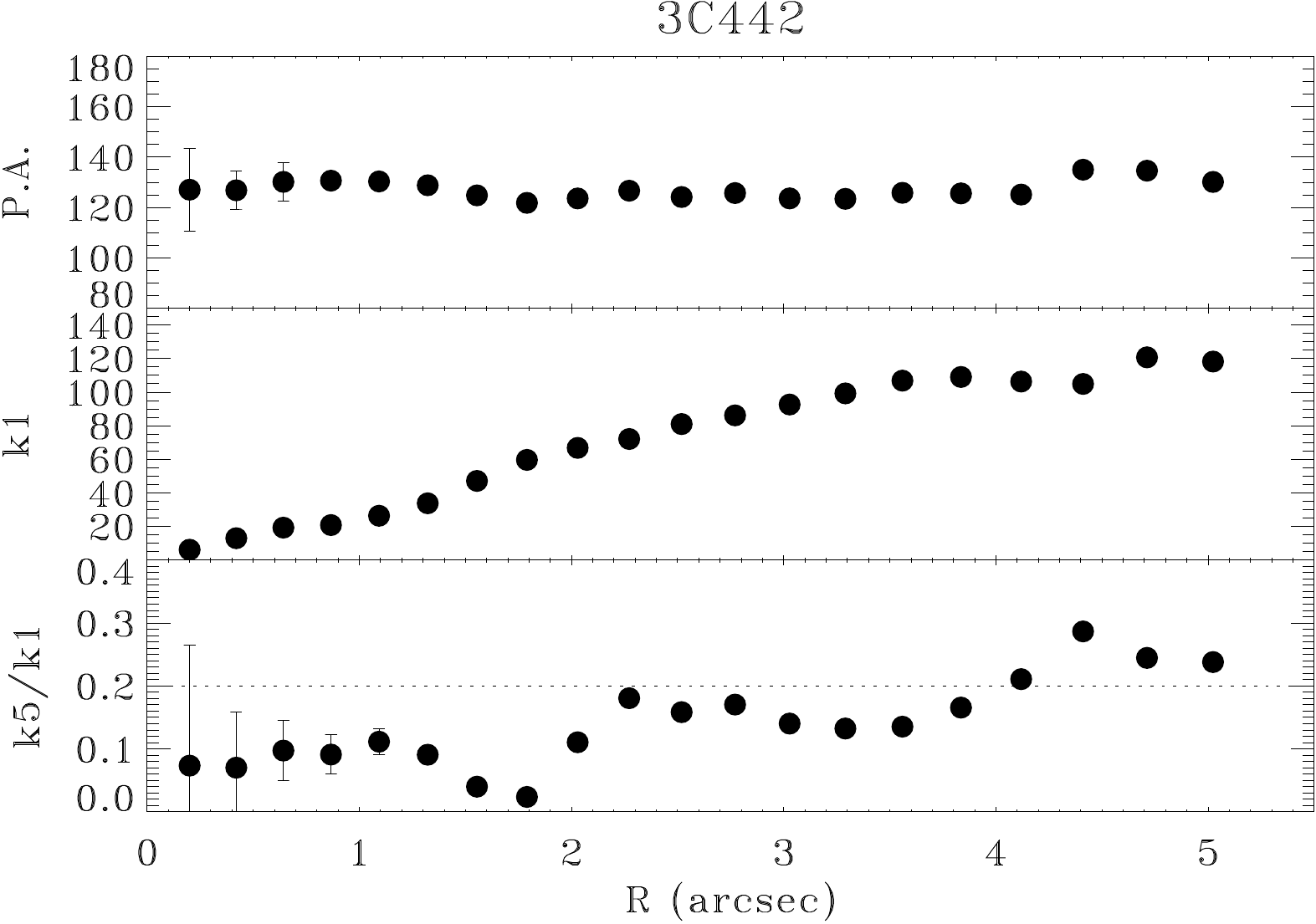}
}
\caption{- continued.}
\end{figure*}   

\addtocounter{figure}{-1}
\begin{figure*}  
\centering{ 
\includegraphics[width=8.5cm]{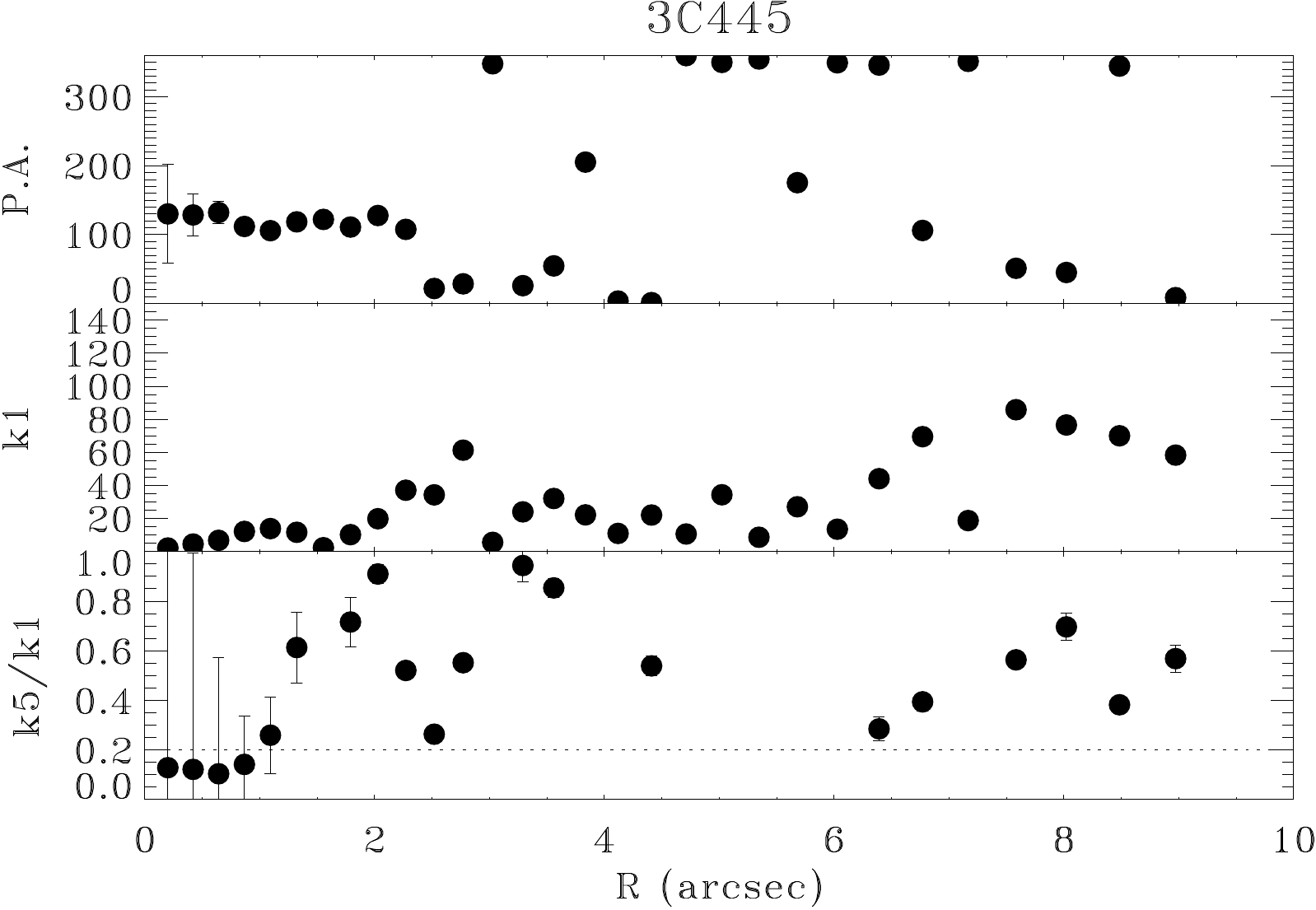}
\includegraphics[width=8.5cm]{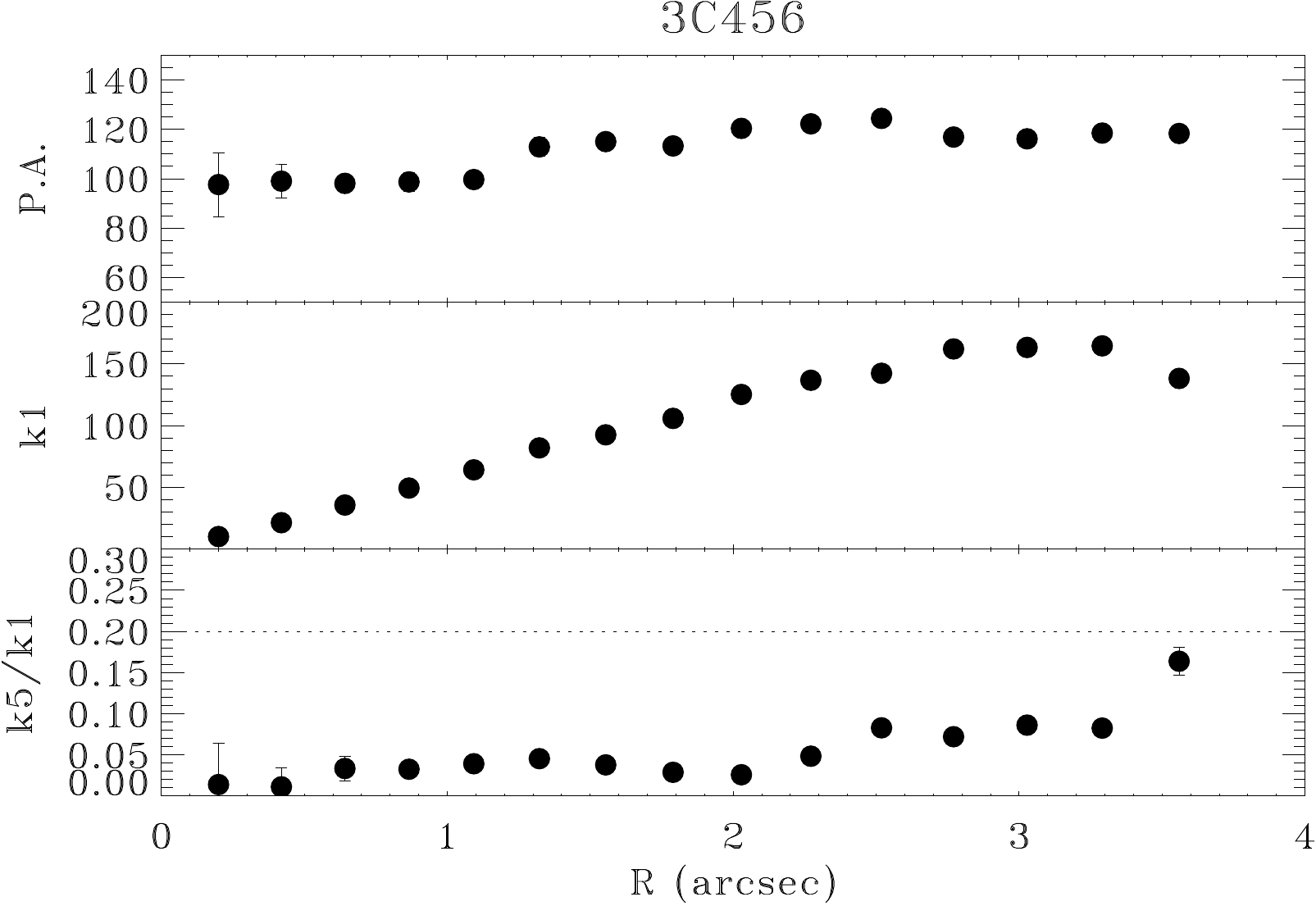}
\includegraphics[width=8.5cm]{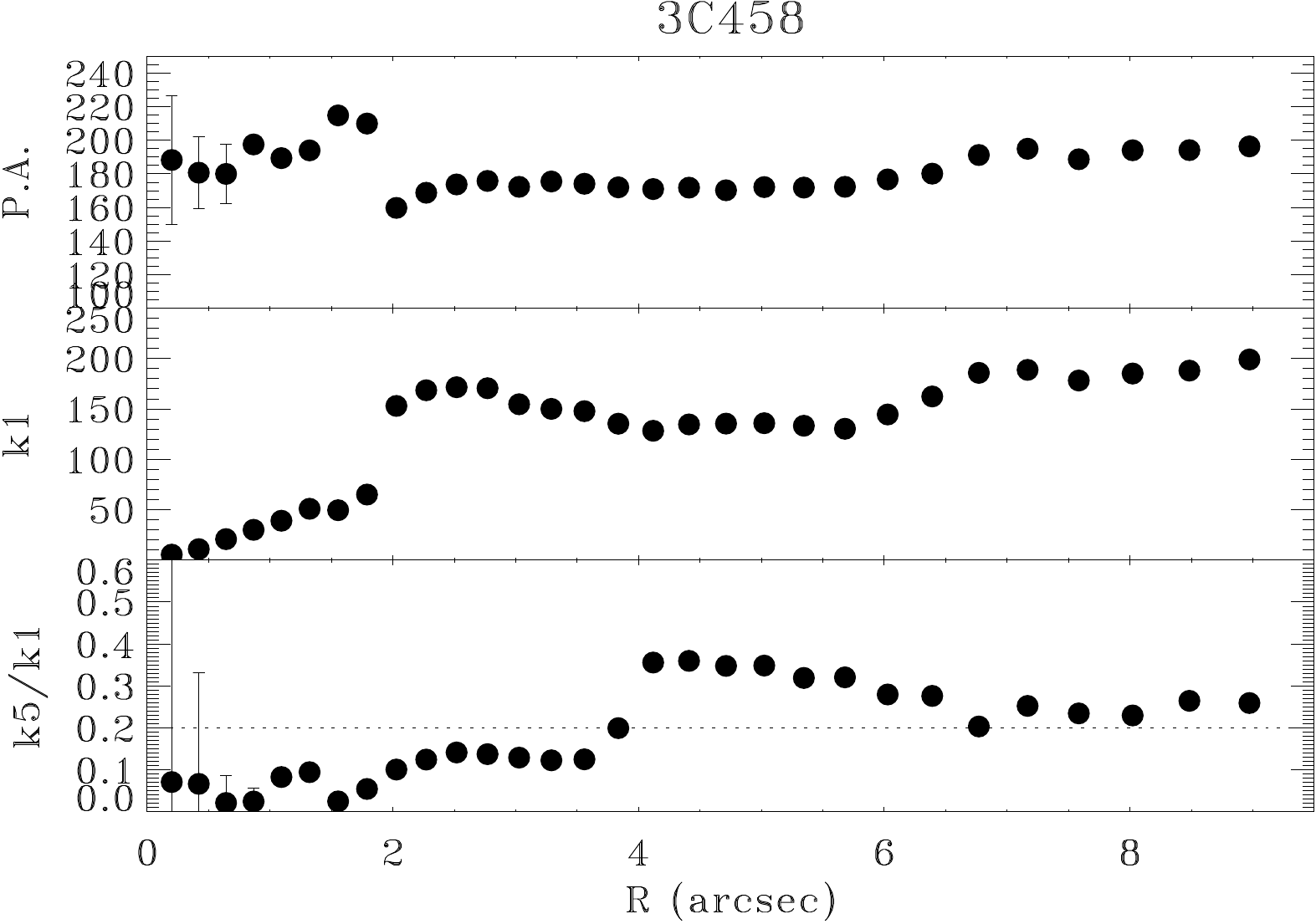}
\includegraphics[width=8.5cm]{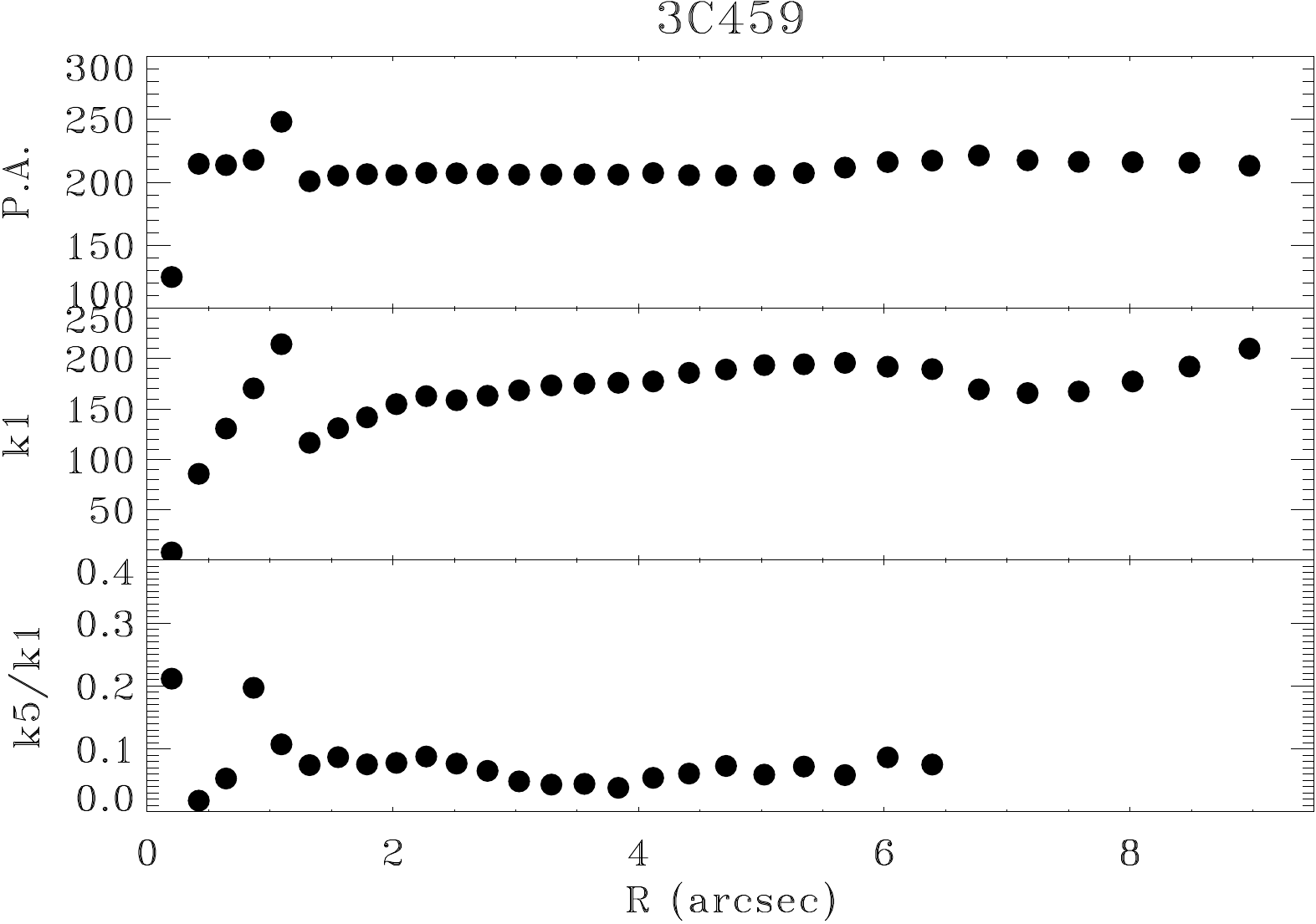}
}
\caption{- continued.}
\end{figure*}   

\newpage
\clearpage
\section{Superposition of radio contours onto the emission line images}
\label{appD}  

\begin{figure*}  
\centering{ 
\includegraphics[width=0.6\columnwidth]{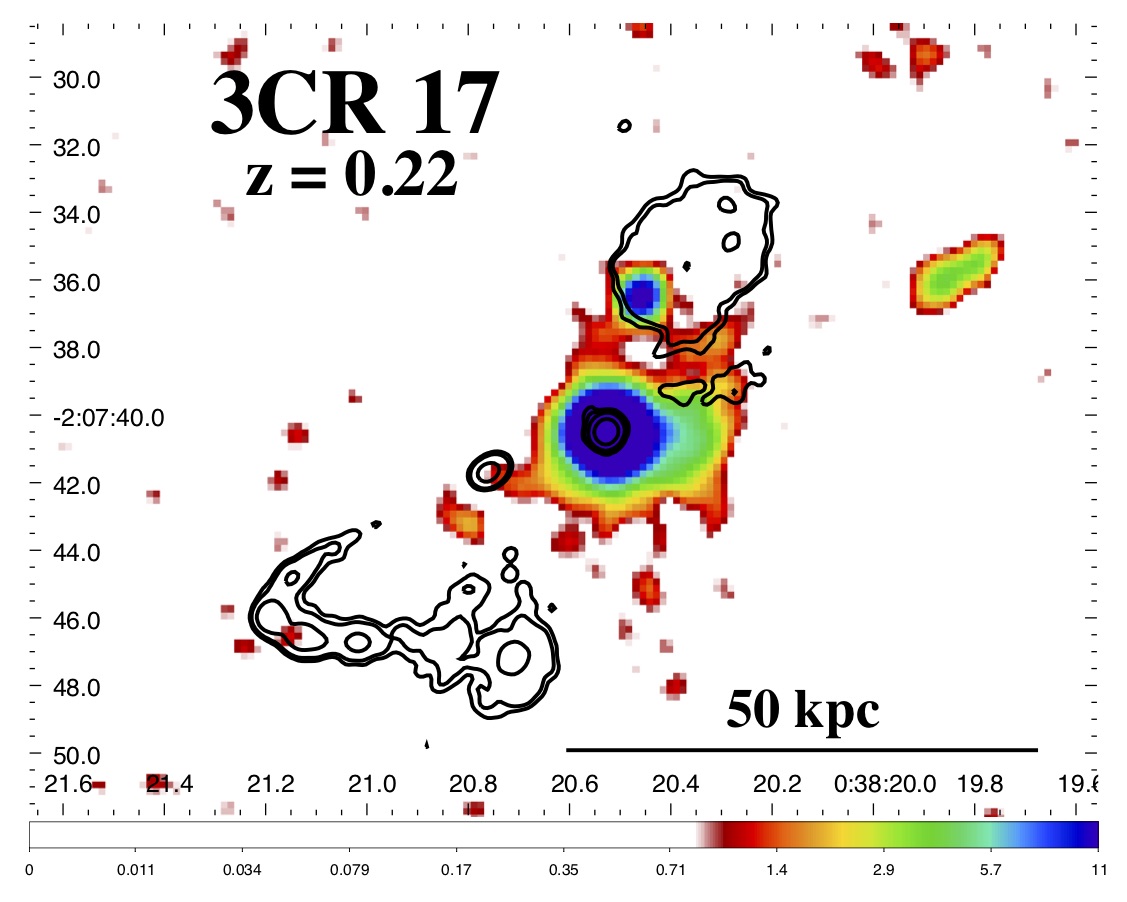}
\includegraphics[width=0.6\columnwidth]{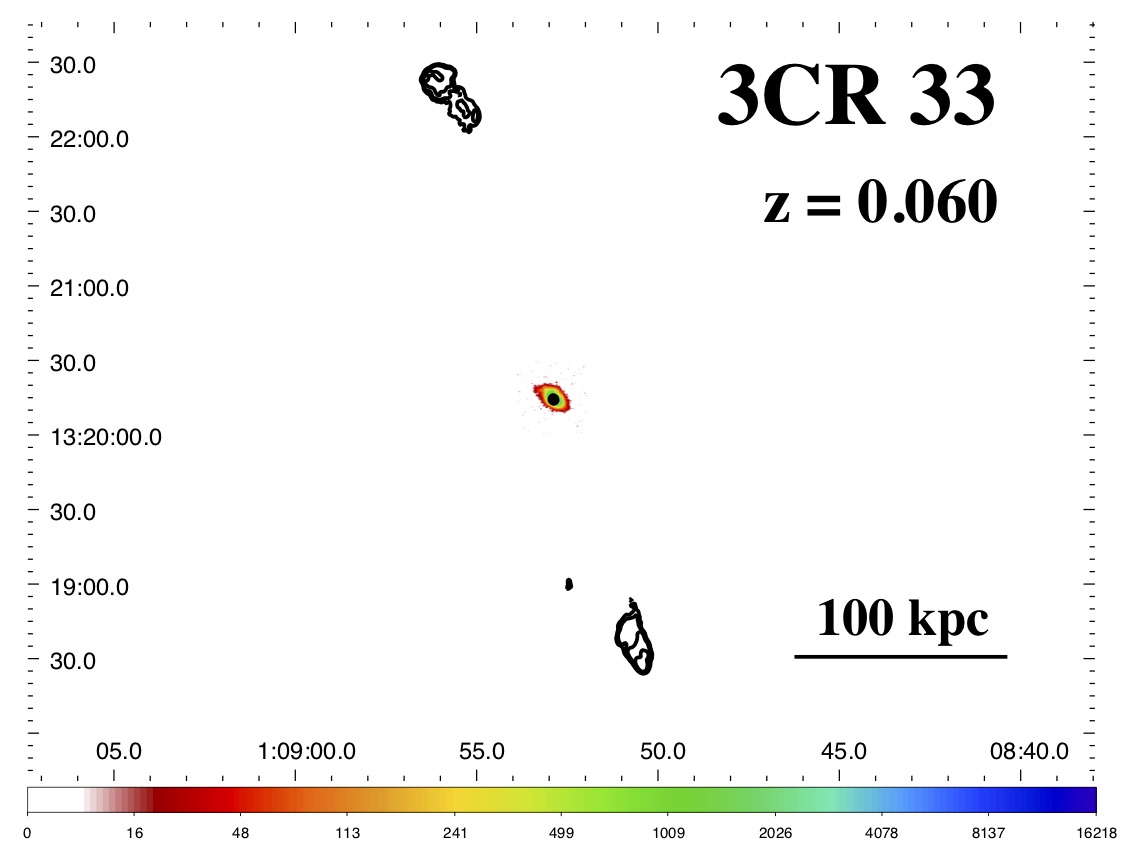}
\includegraphics[width=0.6\columnwidth]{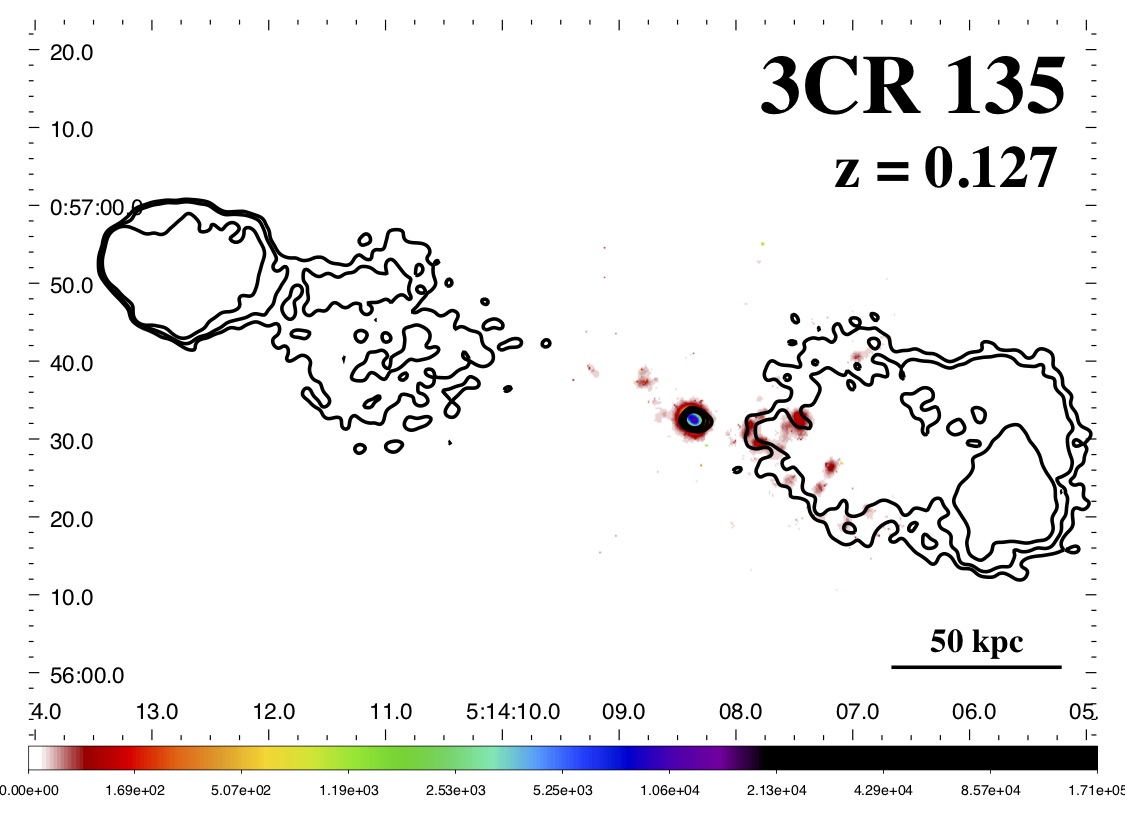}
\includegraphics[width=0.6\columnwidth]{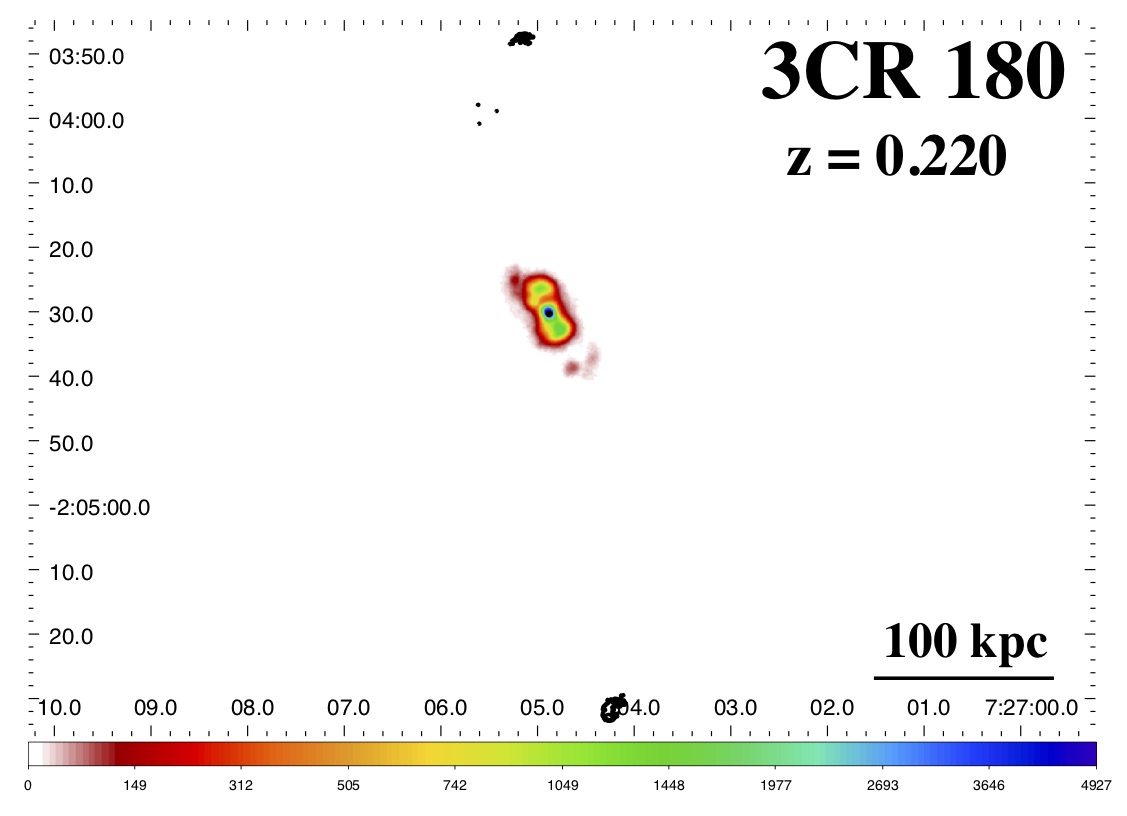}
\includegraphics[width=0.6\columnwidth]{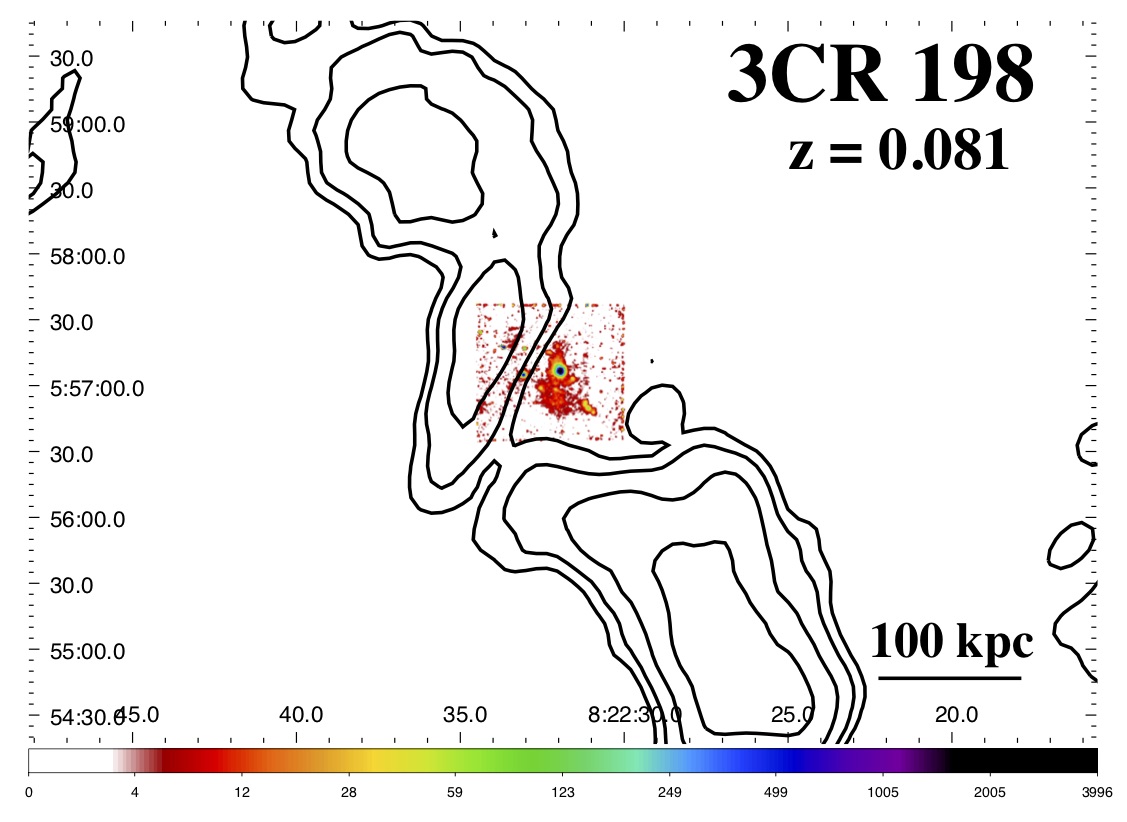}
\includegraphics[width=0.6\columnwidth]{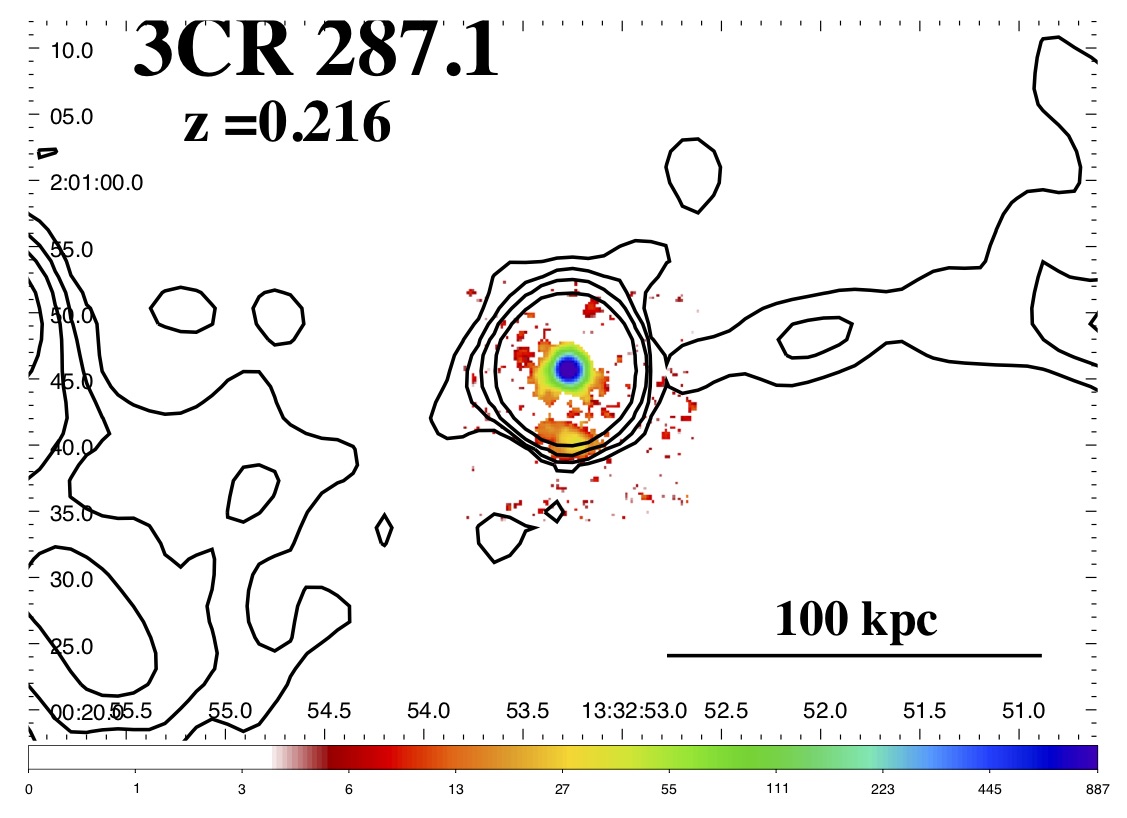}
\includegraphics[width=0.6\columnwidth]{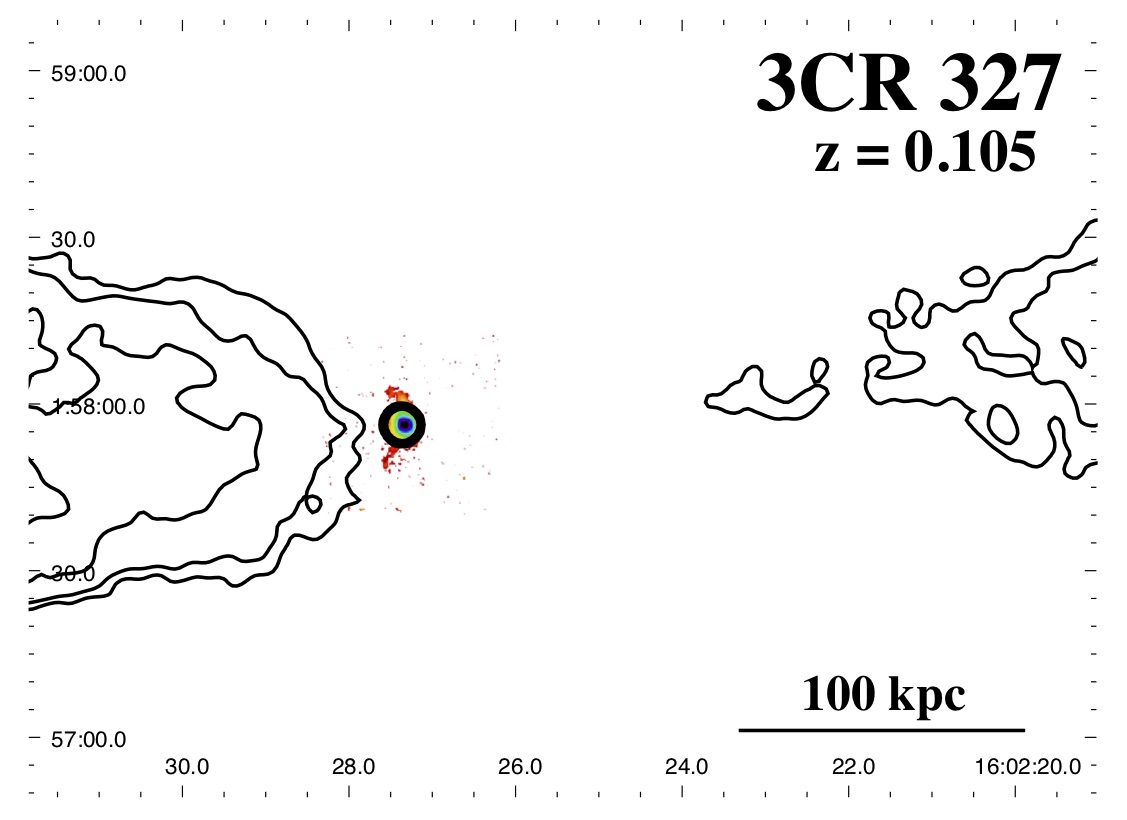}
\includegraphics[width=0.6\columnwidth]{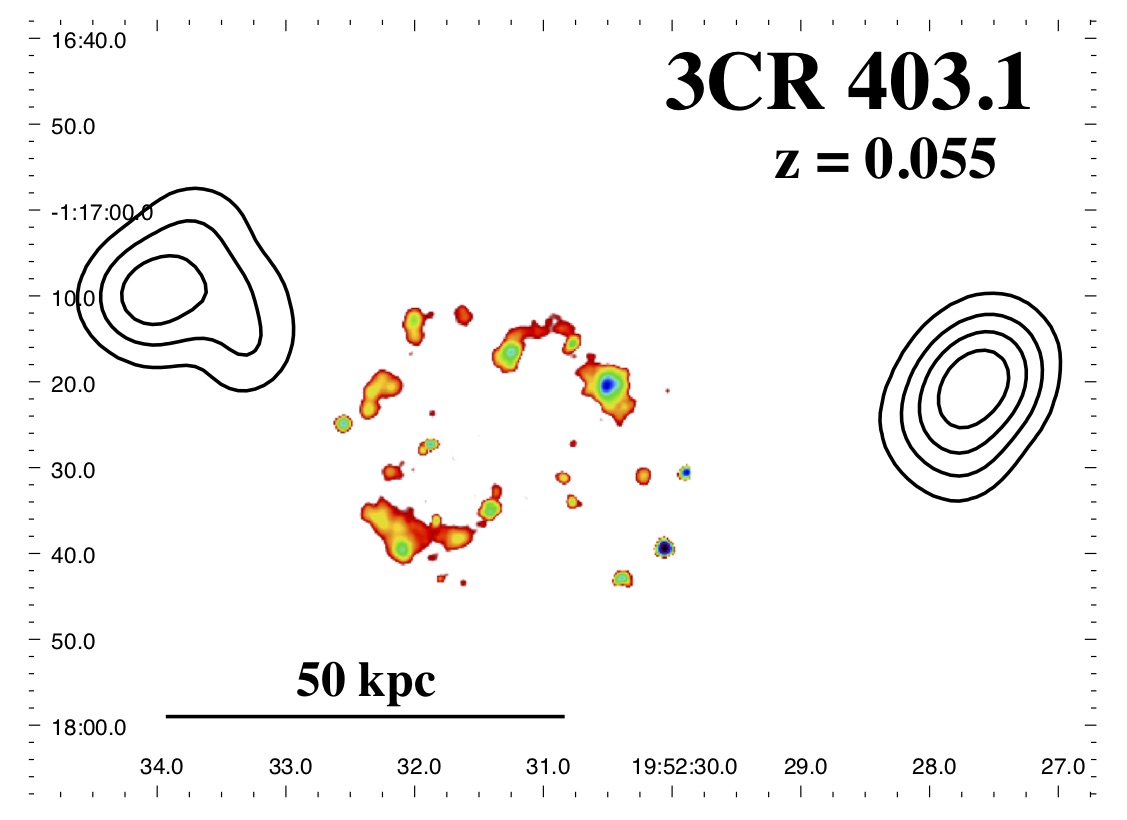}    
\includegraphics[width=0.6\columnwidth]{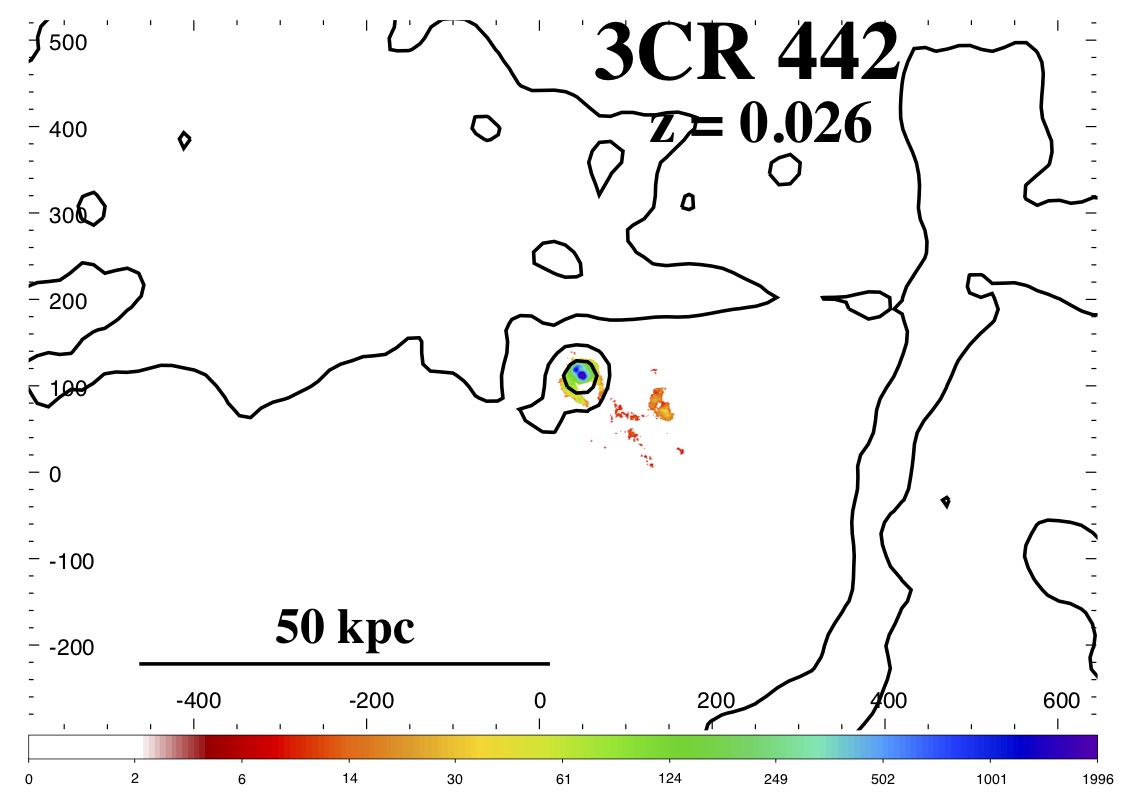}
\includegraphics[width=0.6\columnwidth]{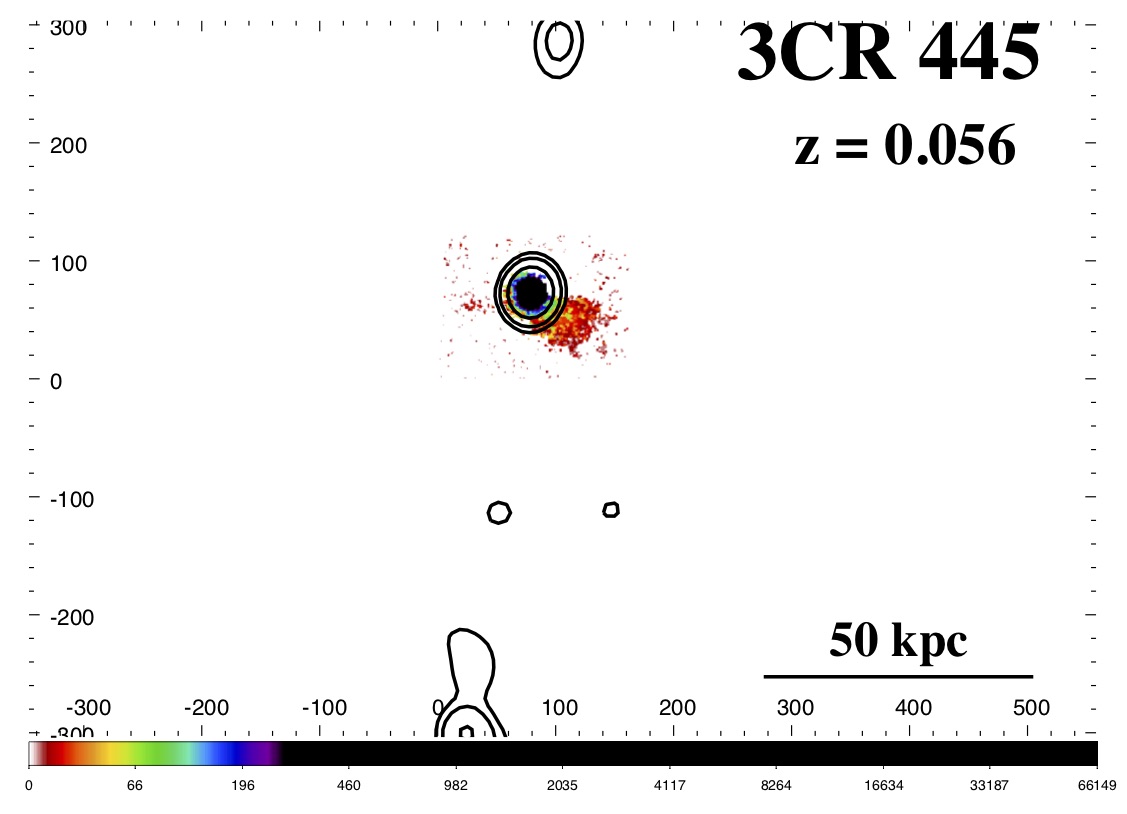}
\includegraphics[width=0.6\columnwidth]{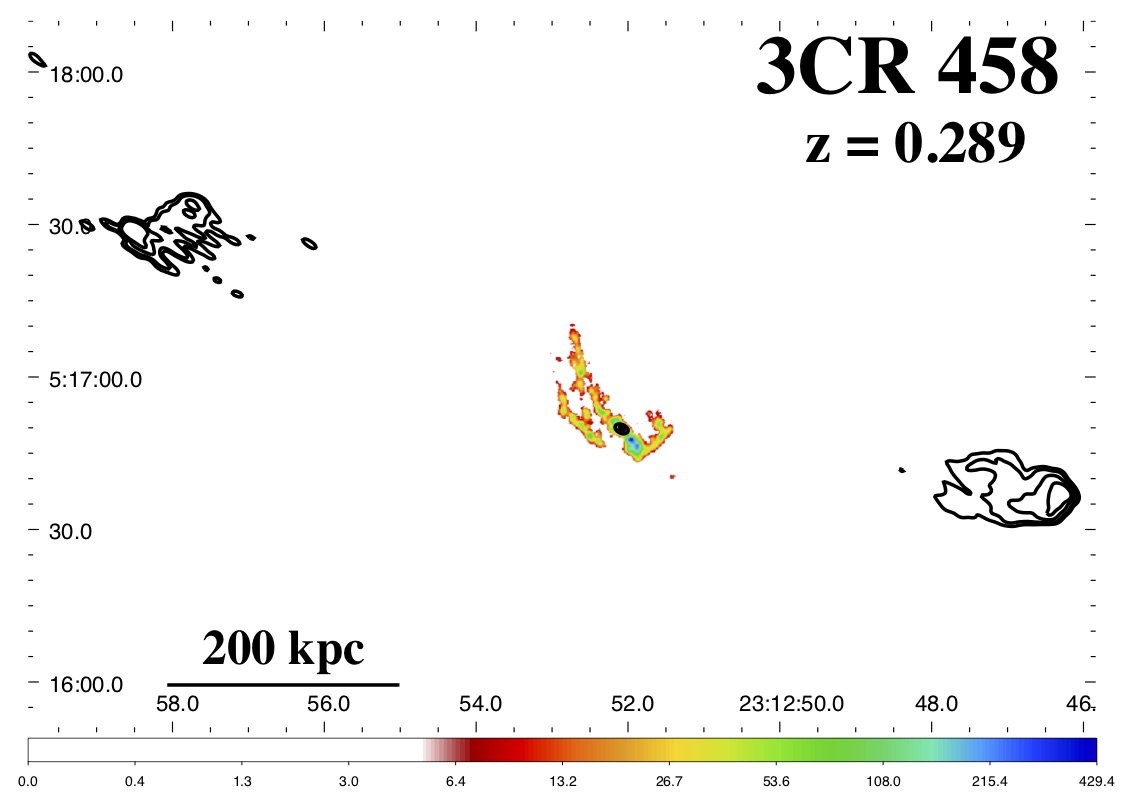}
\includegraphics[width=0.6\columnwidth]{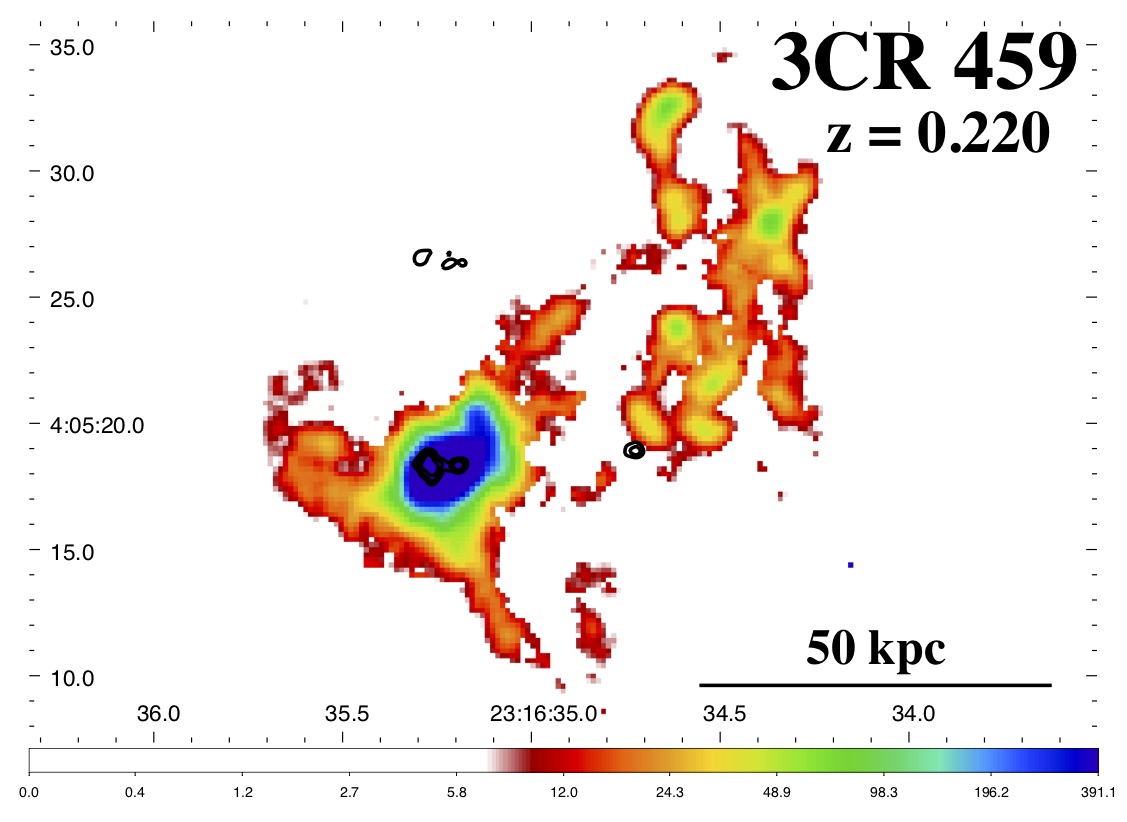}
}
\caption{Superposition of the radio contours onto the gas velocity
  maps for the sources with extended emission line regions not already
  shown in Fig. \ref{best}.}
\label{best2}
\end{figure*}  

\newpage
\clearpage
\section{Offnuclear spectra.}
\label{appE}  
\begin{figure*}  
\centering{ 
  \includegraphics[width=7cm]{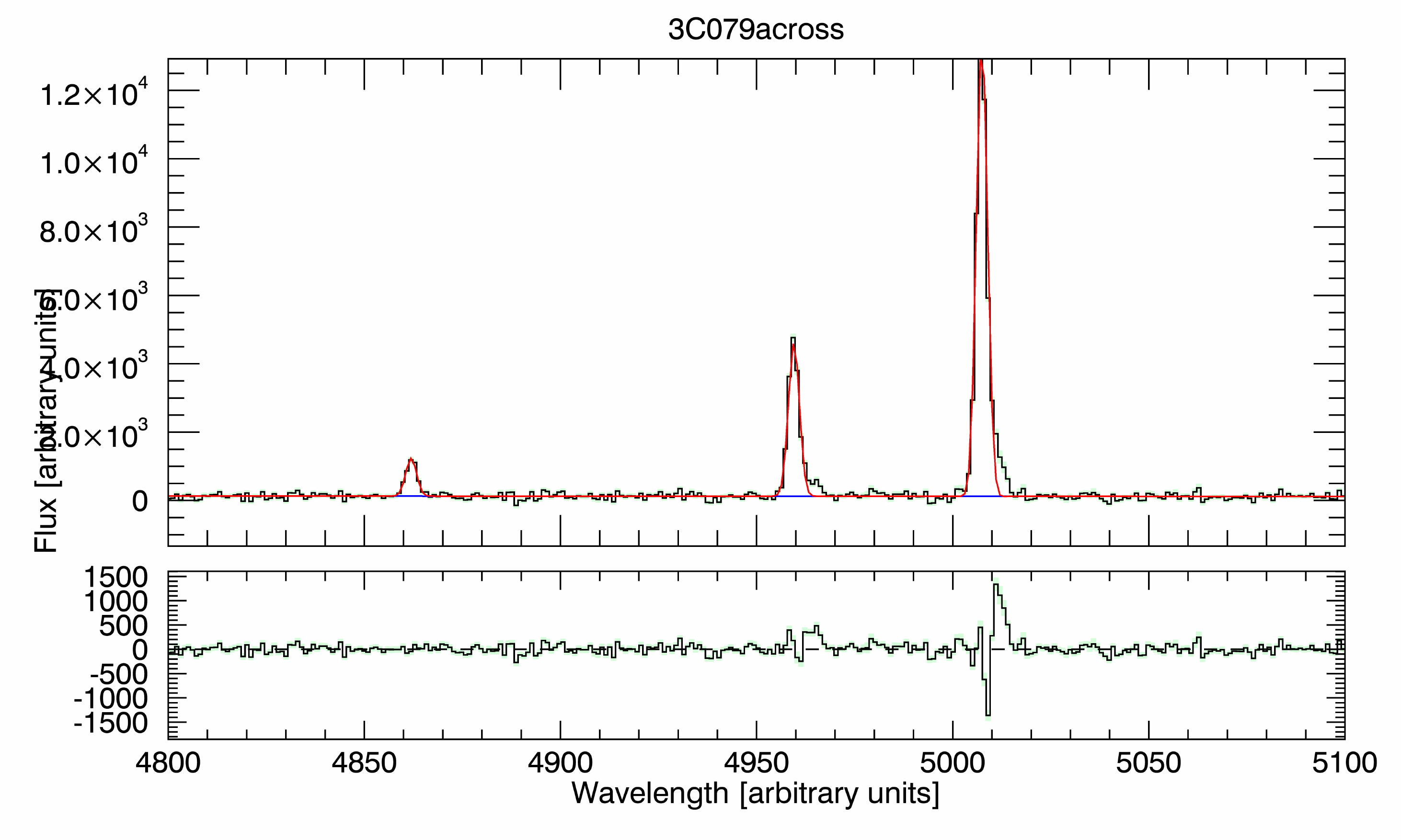}
  \includegraphics[width=7cm]{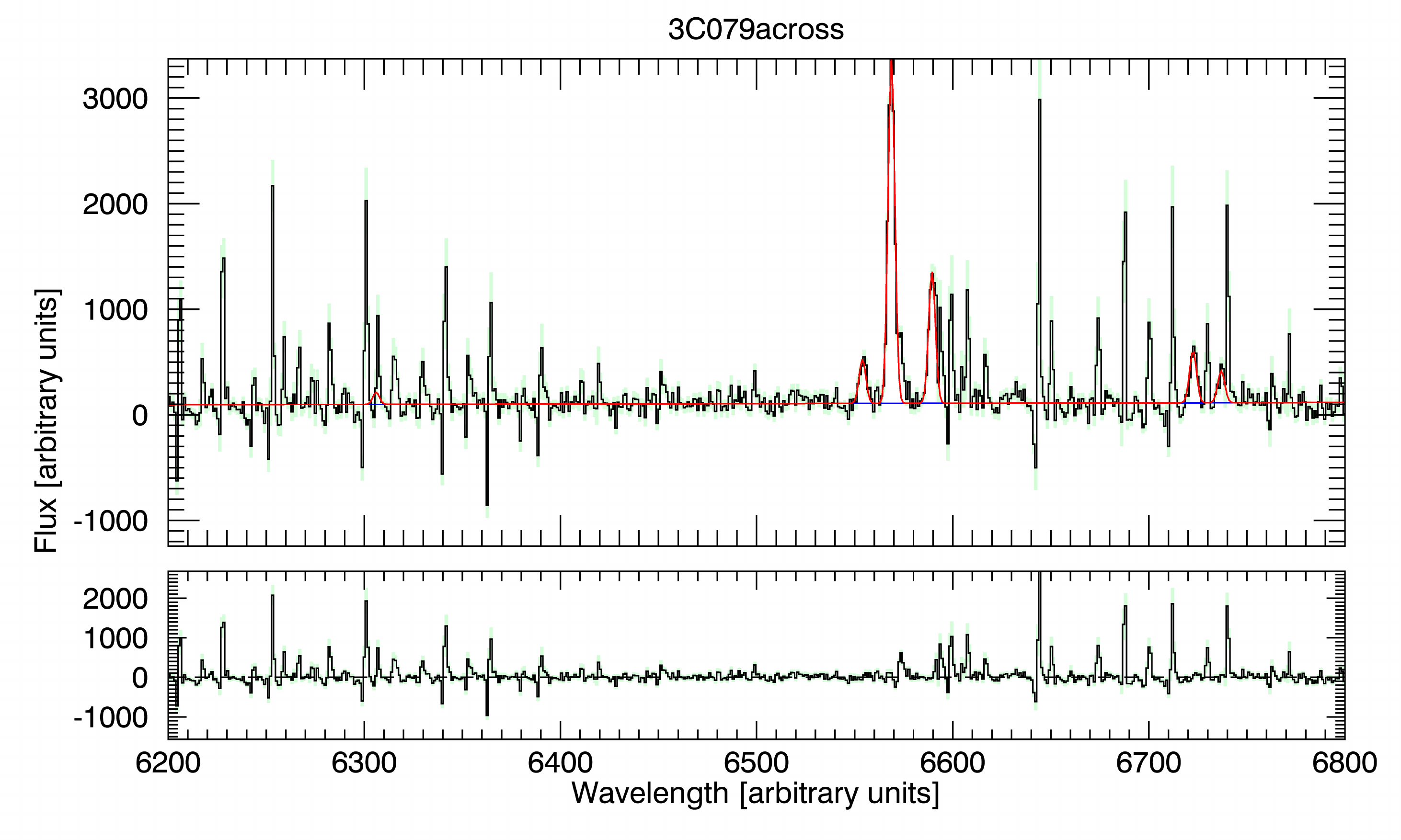}   
  \includegraphics[width=7cm]{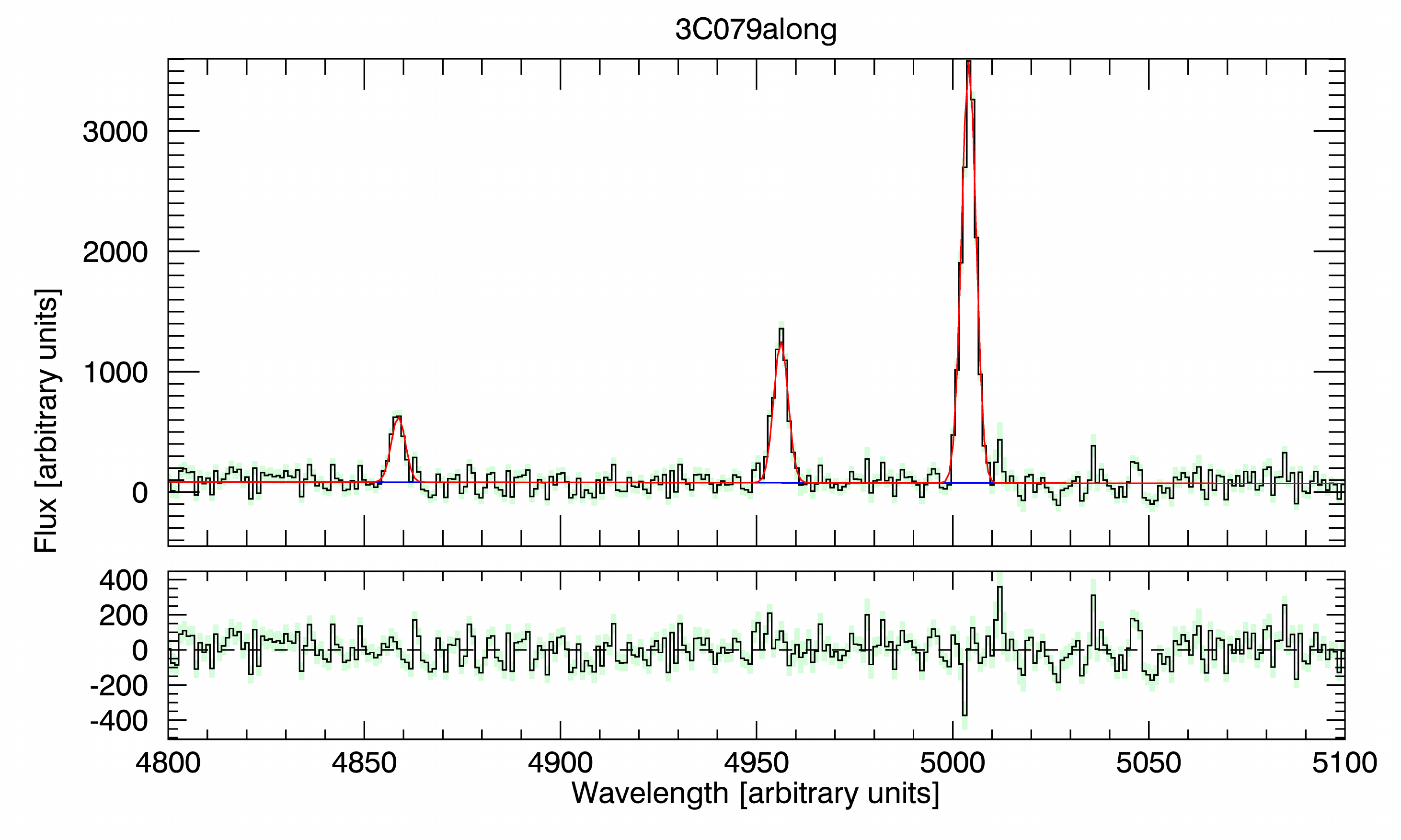}
  \includegraphics[width=7cm]{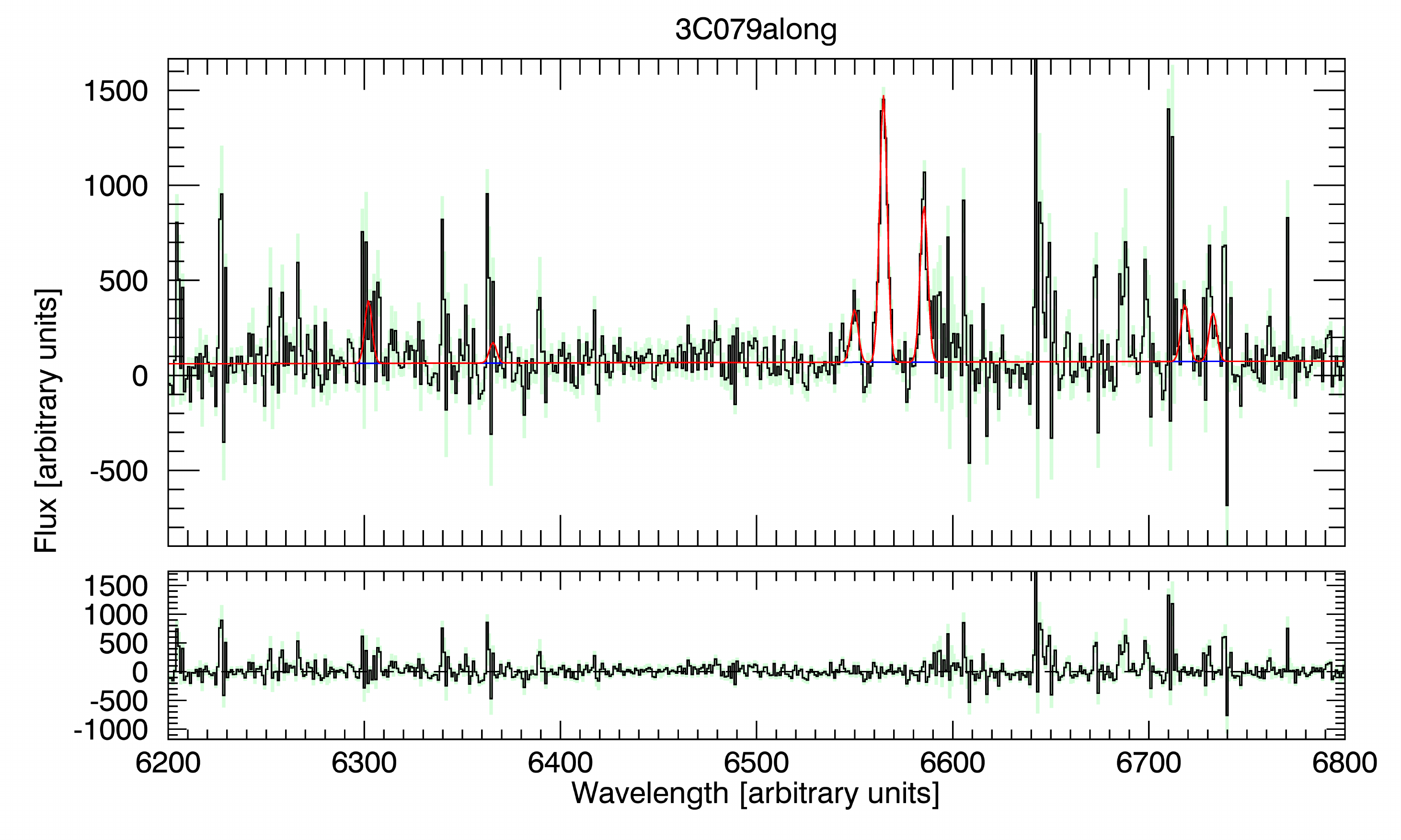}
  \includegraphics[width=7cm]{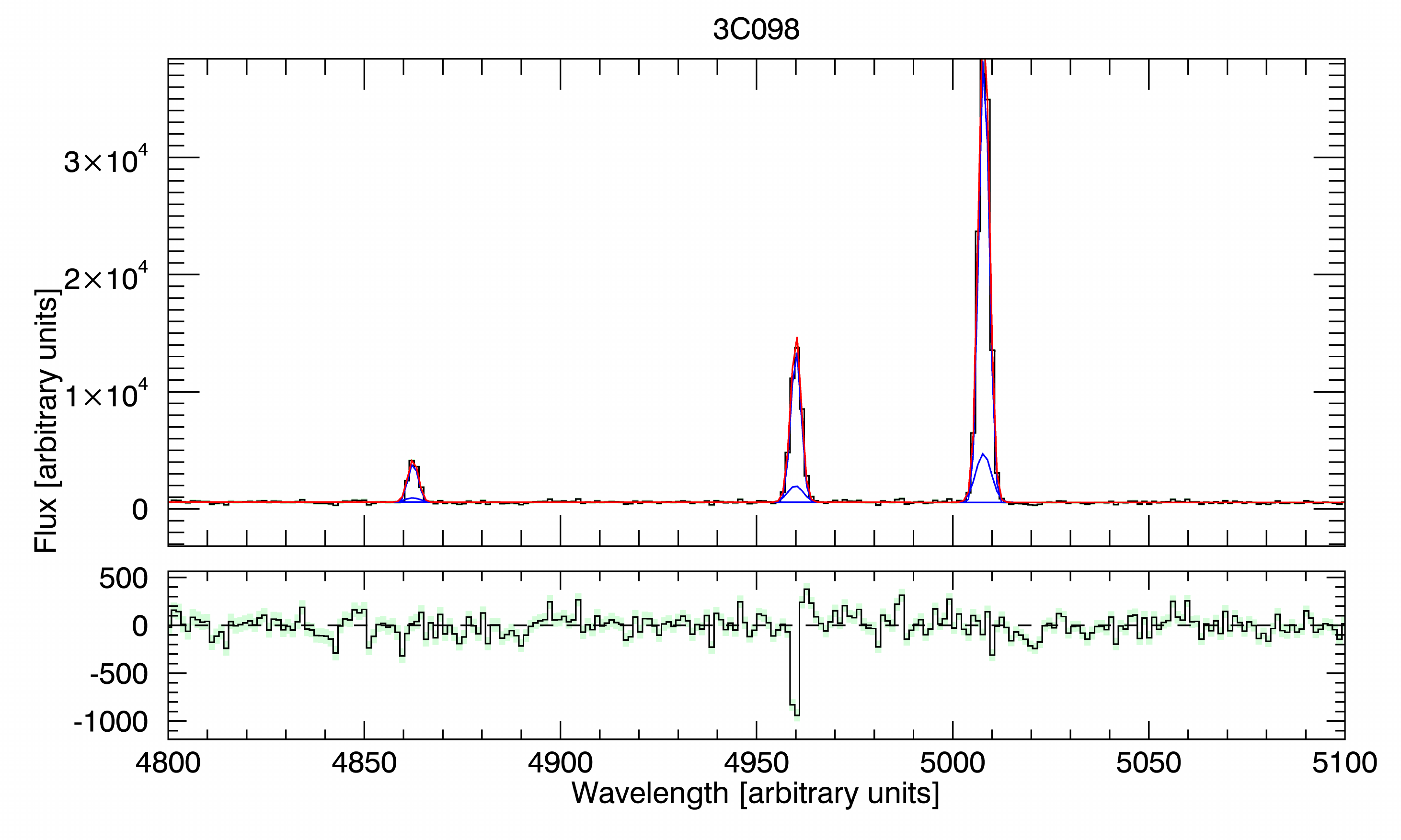}
  \includegraphics[width=7cm]{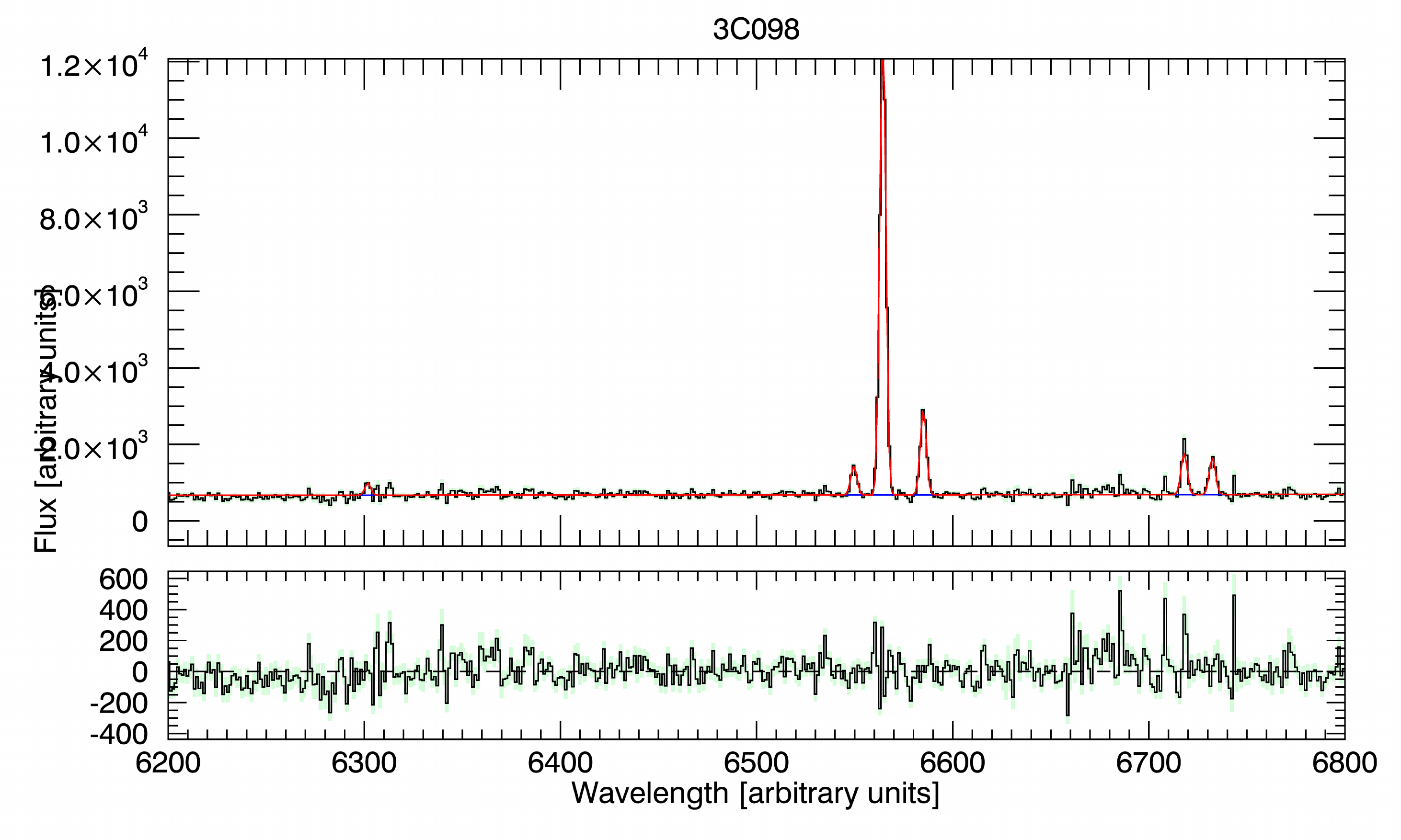}
  \includegraphics[width=7cm]{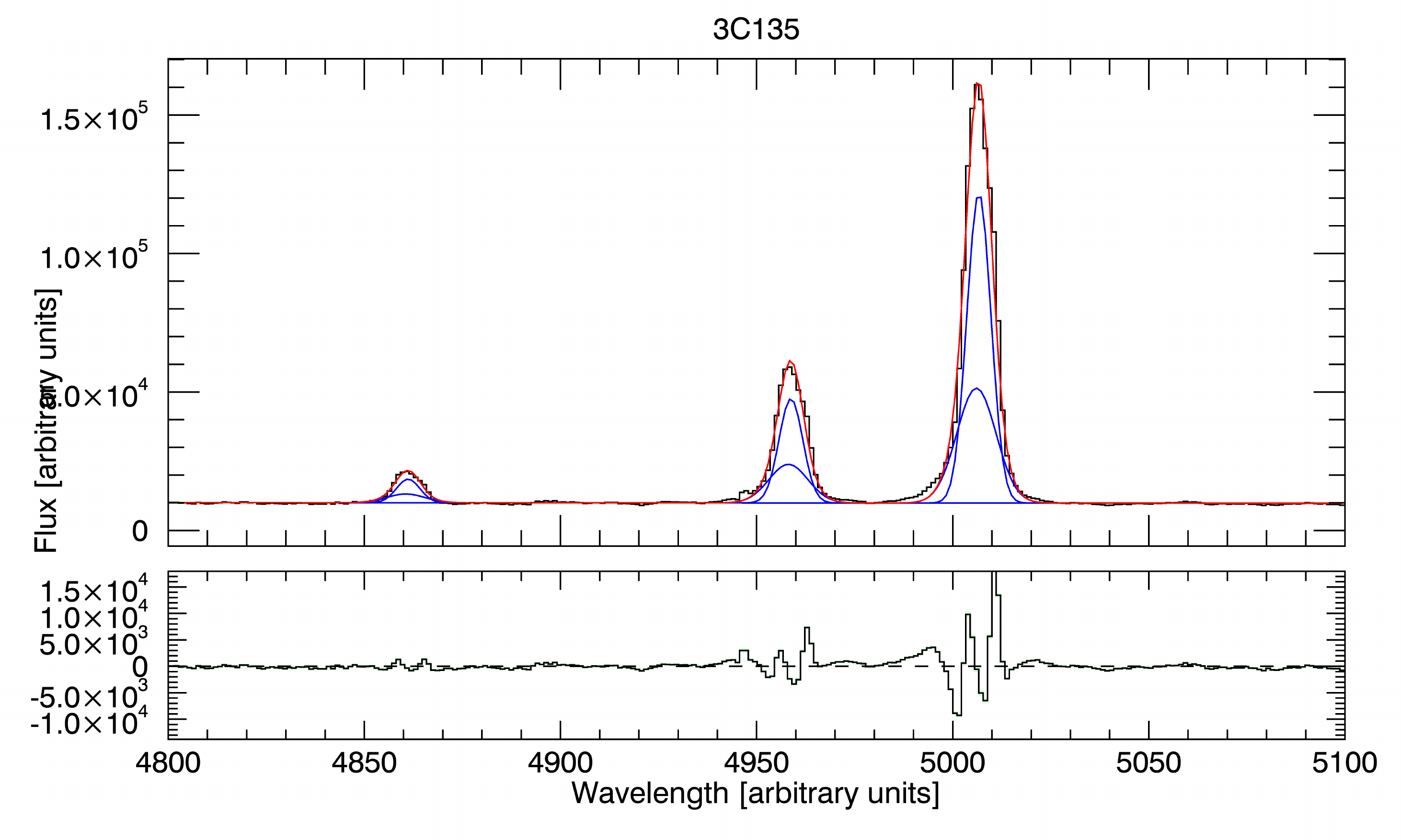}
  \includegraphics[width=7cm]{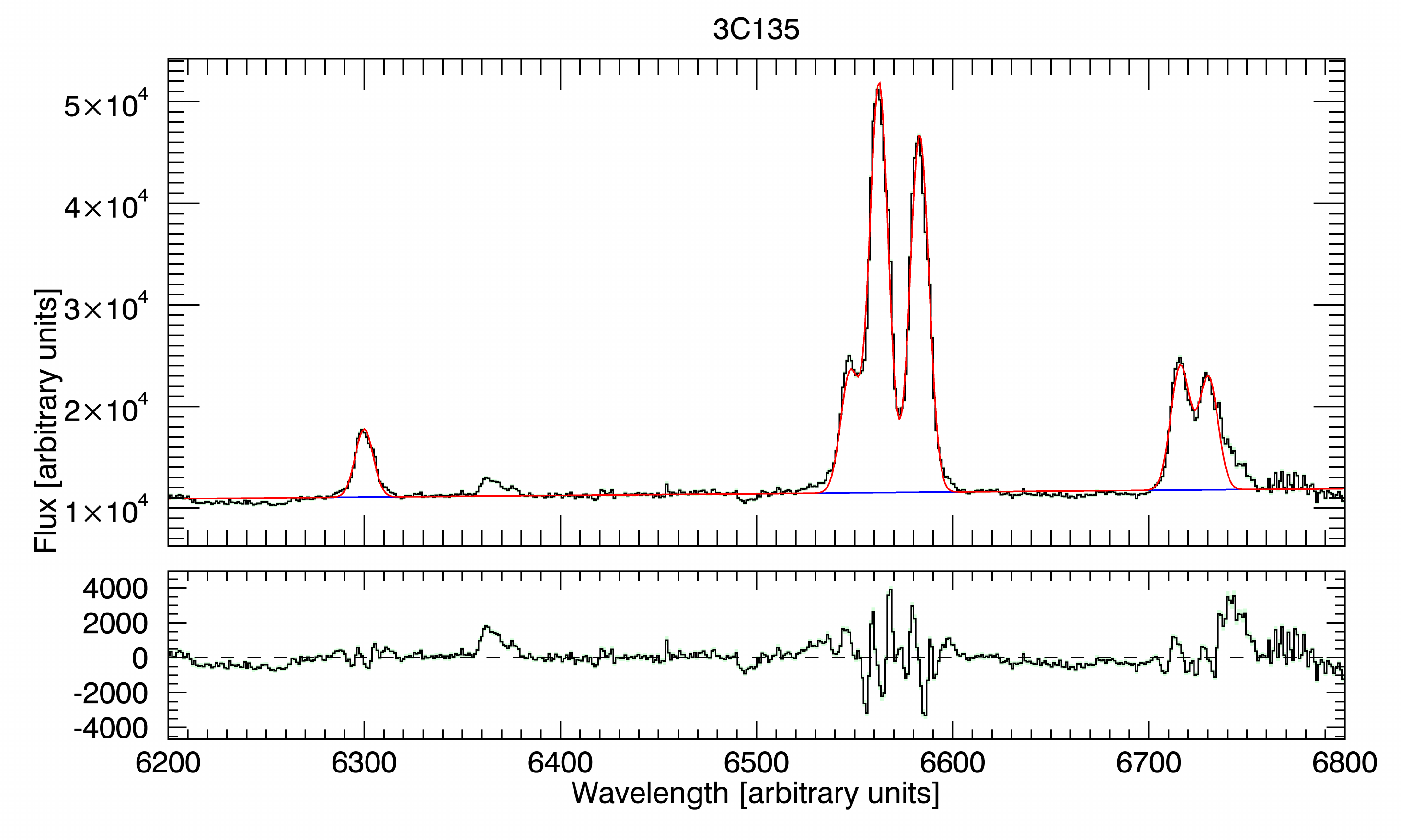}
  \includegraphics[width=7cm]{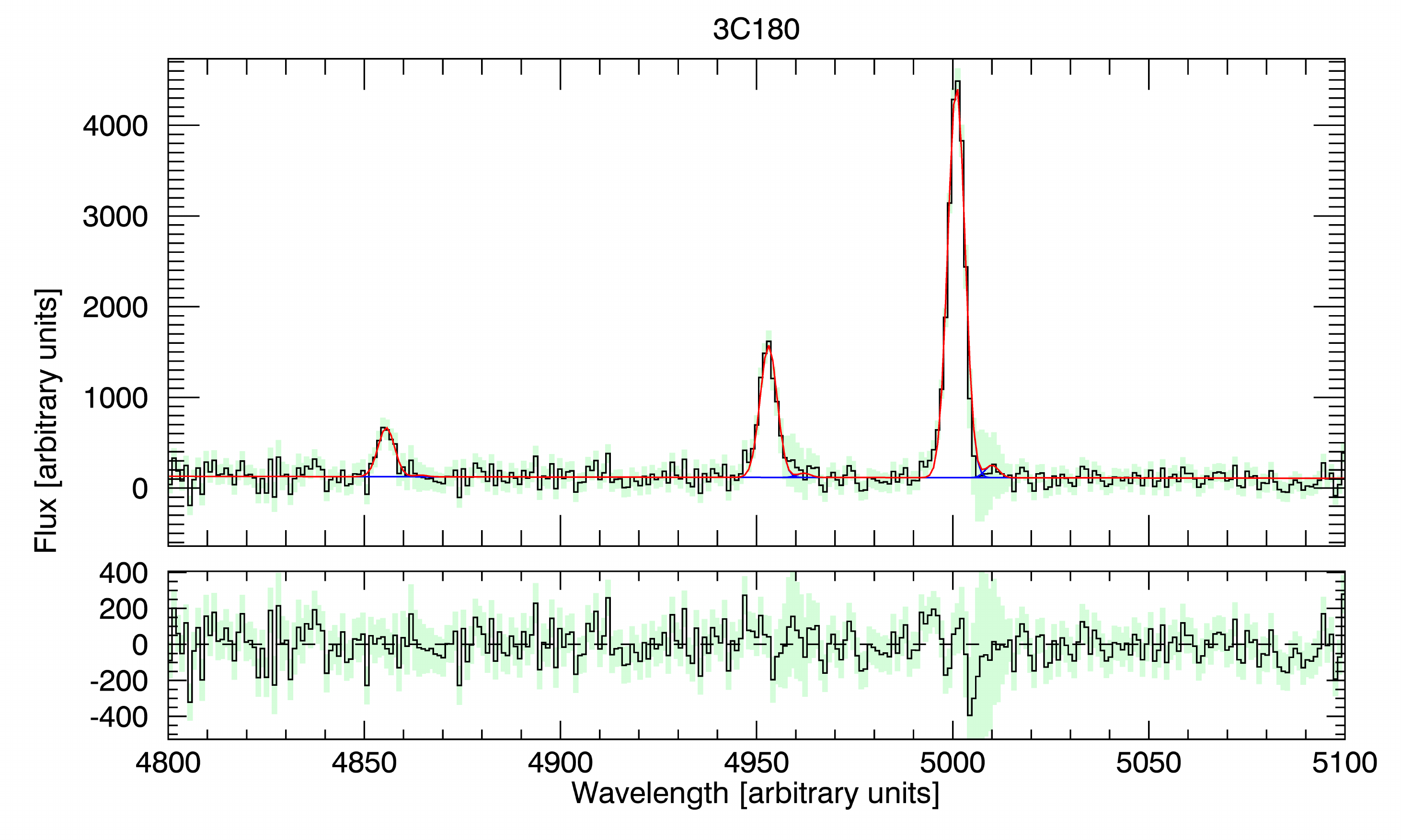}
  \includegraphics[width=7cm]{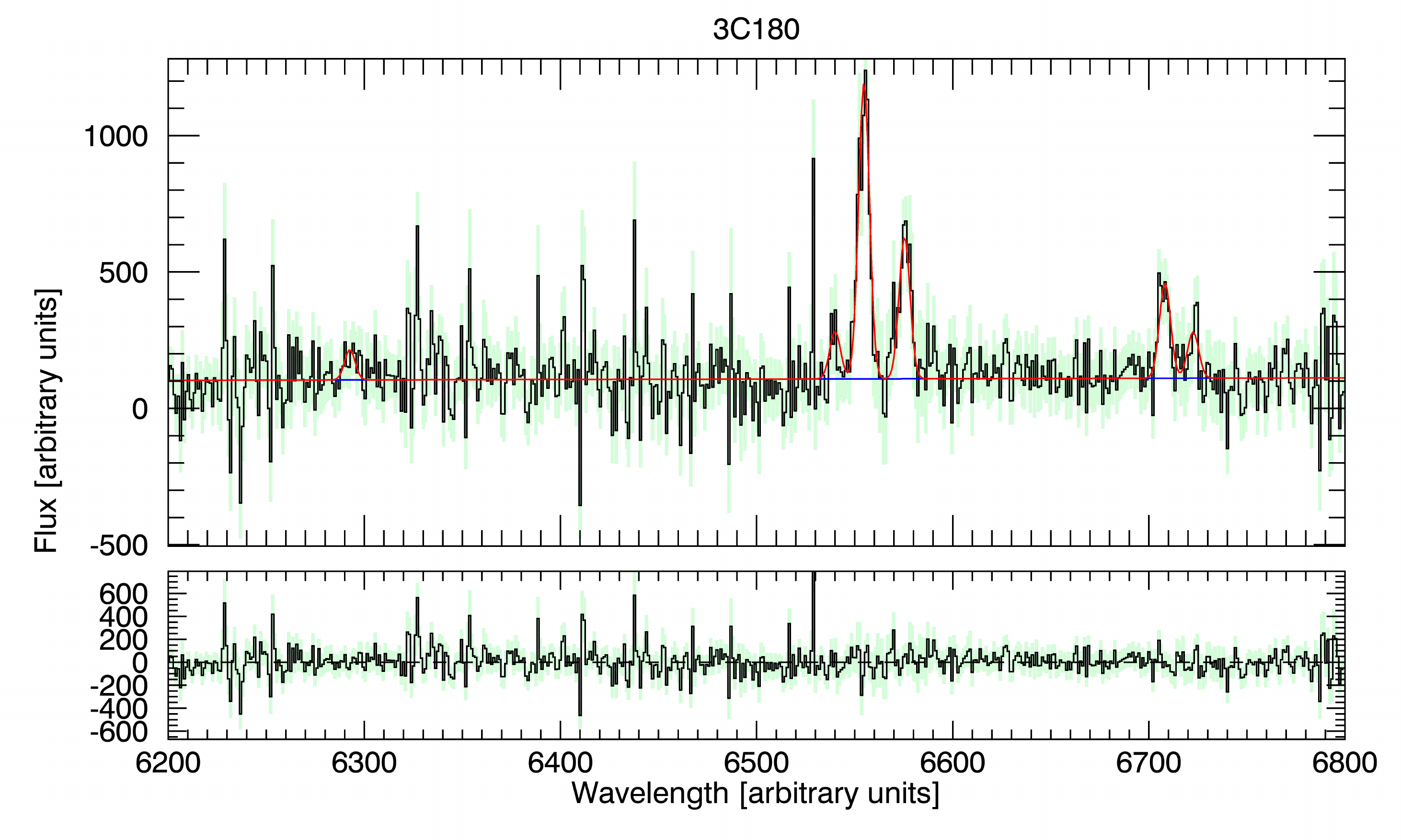}
}
\caption{Off-nuclear spectra of the sources for which it was possible
  to obtain this information. The blue portion of the spectra are on
  the left side, and the red portion of spectra are on the right
  side. The spectra are in black, and the emission line fit is in red
  (when two components are required, they are shown separately in
  blue). The gray areas represent the errors in the spectra. The
  bottom panels show the residuals.}
\end{figure*}

\addtocounter{figure}{-1}

\begin{figure*}  
\centering{ 
  \includegraphics[width=7cm]{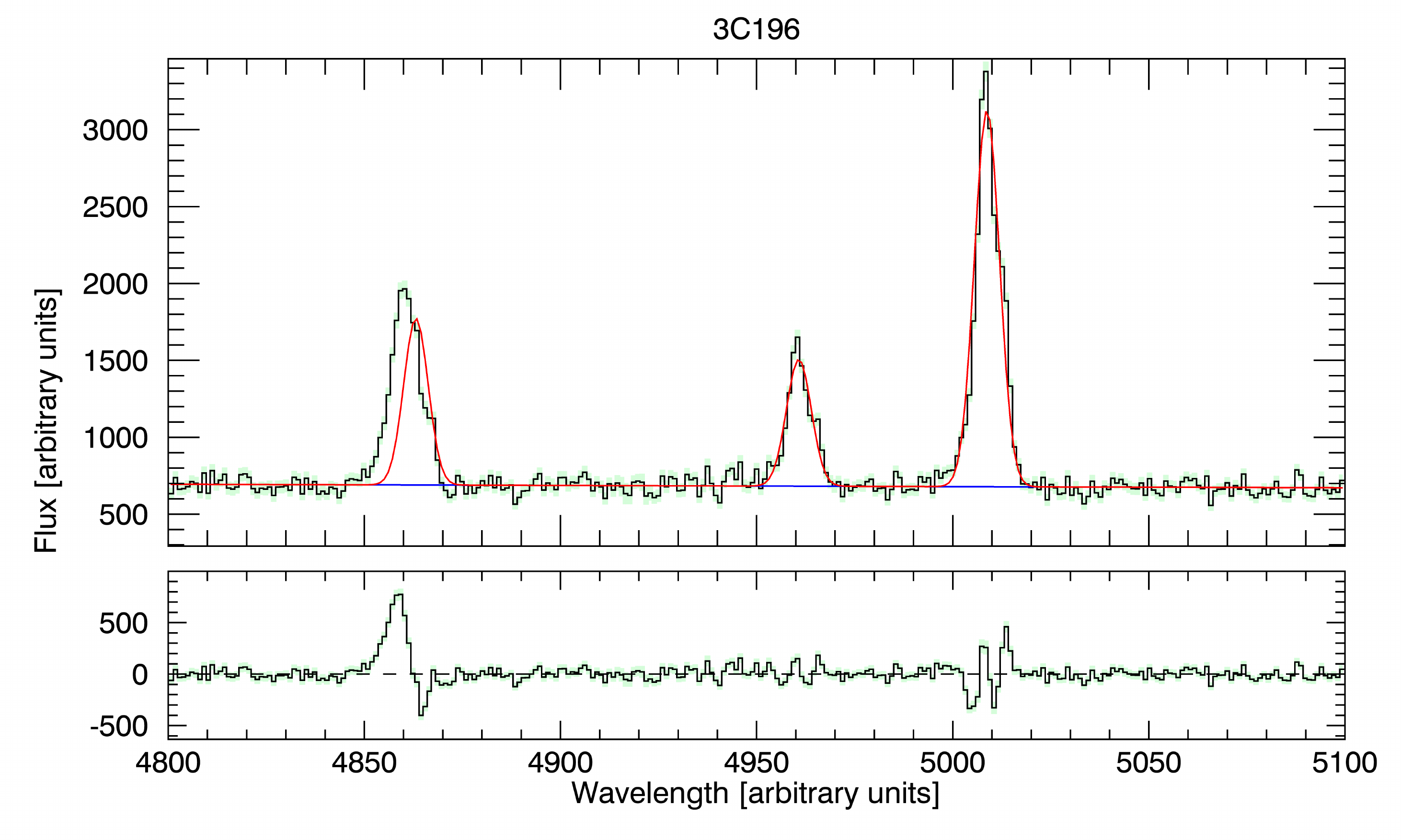}
  \includegraphics[width=7cm]{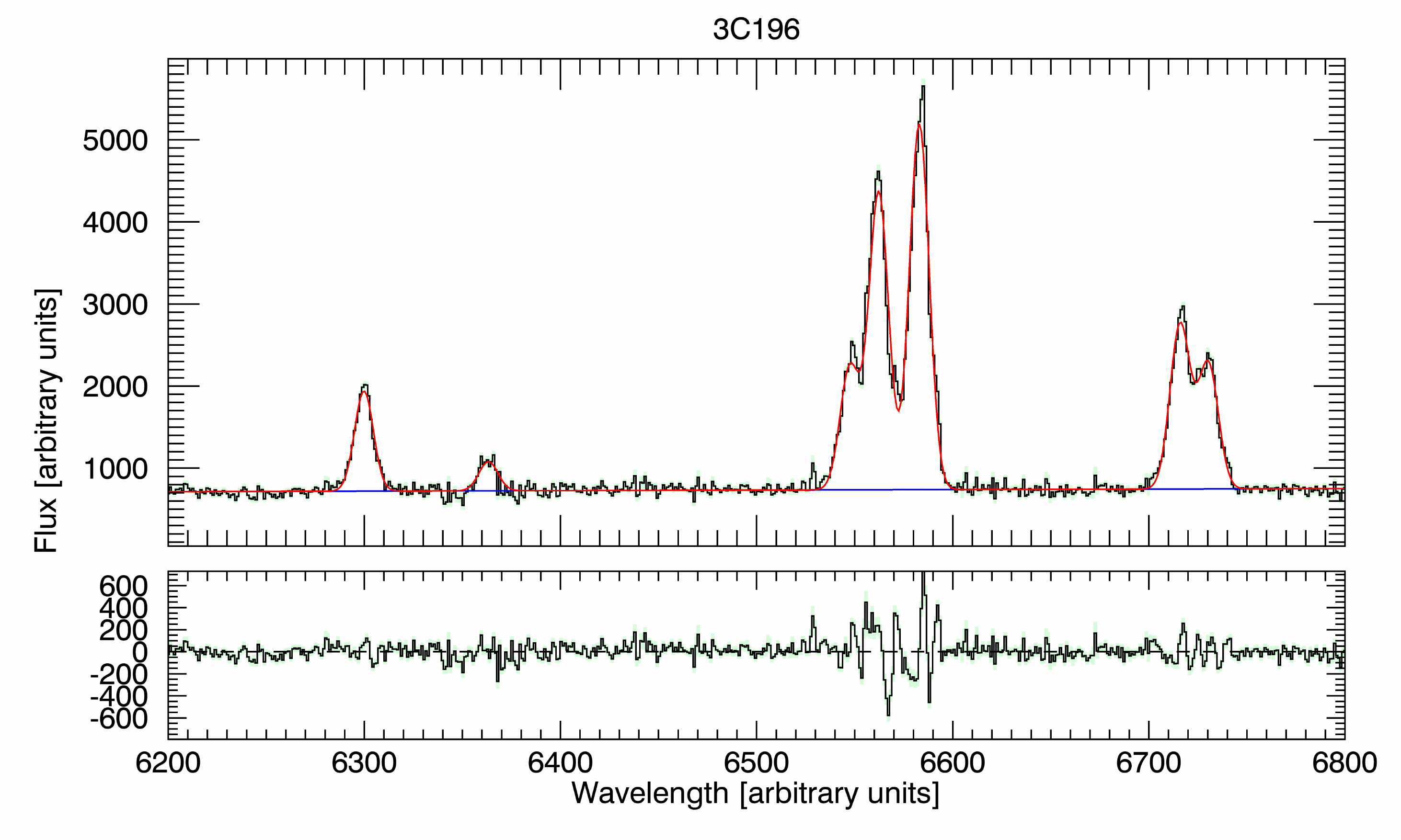}
  \includegraphics[width=7cm]{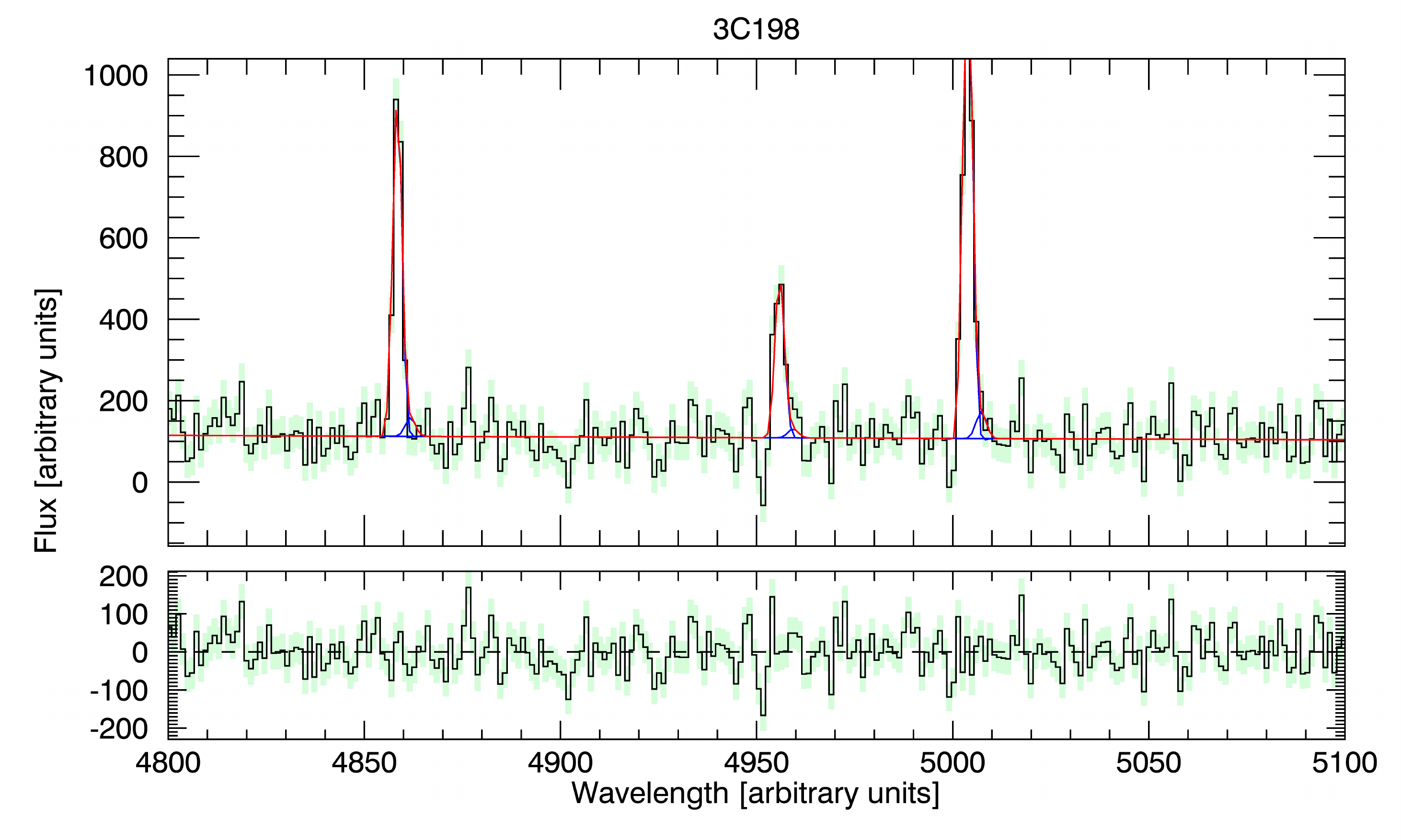}
  \includegraphics[width=7cm]{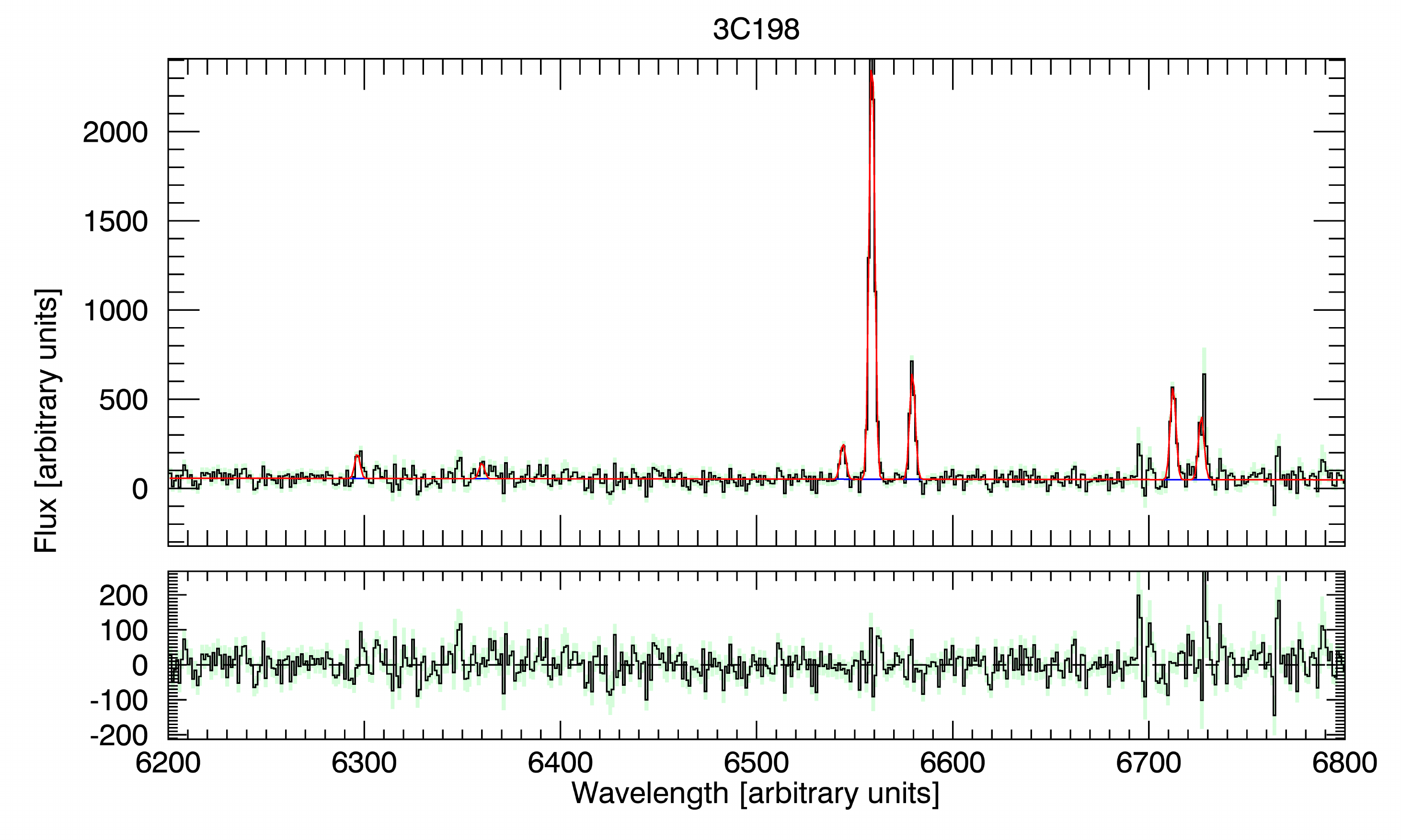}
  \includegraphics[width=7cm]{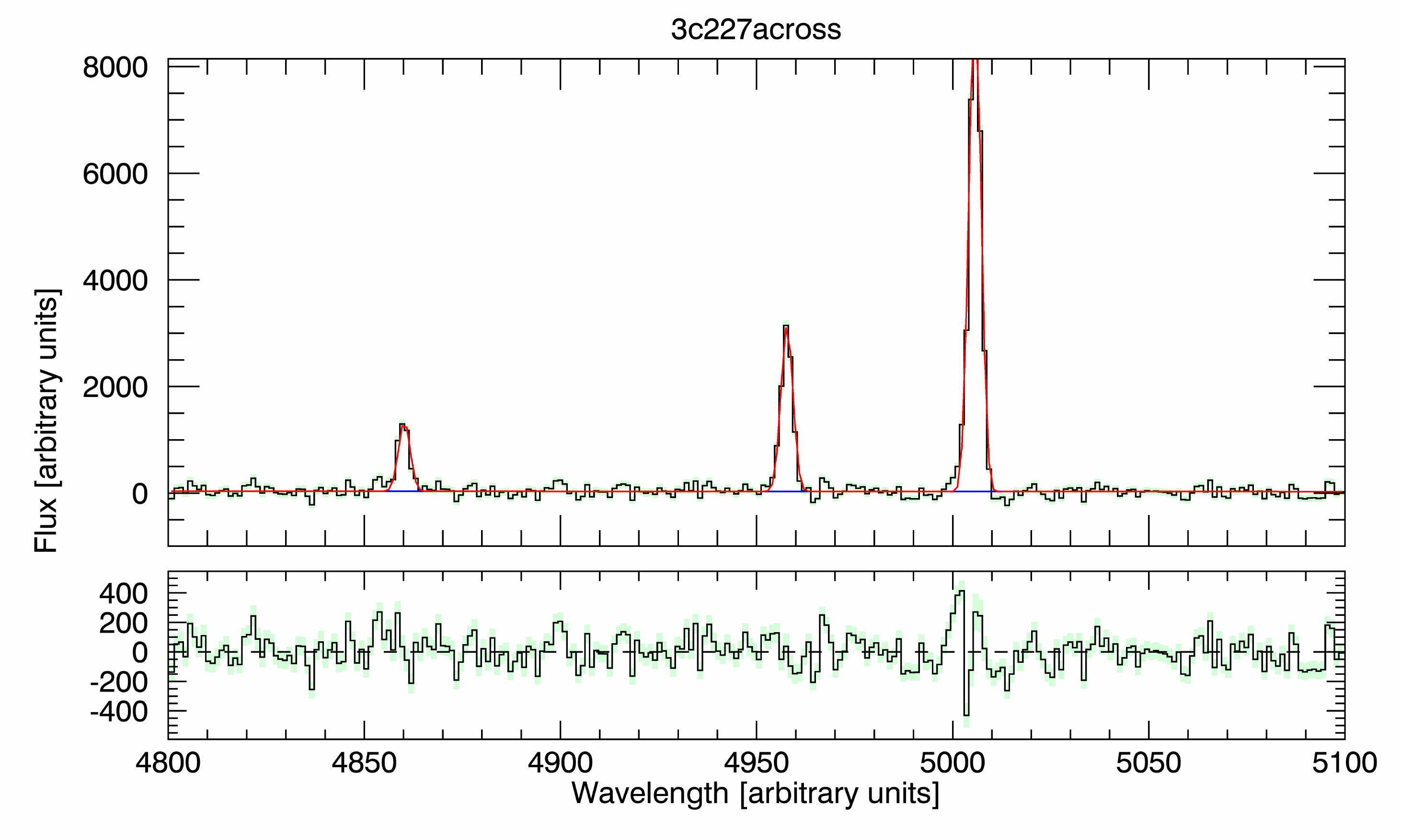}
  \includegraphics[width=7cm]{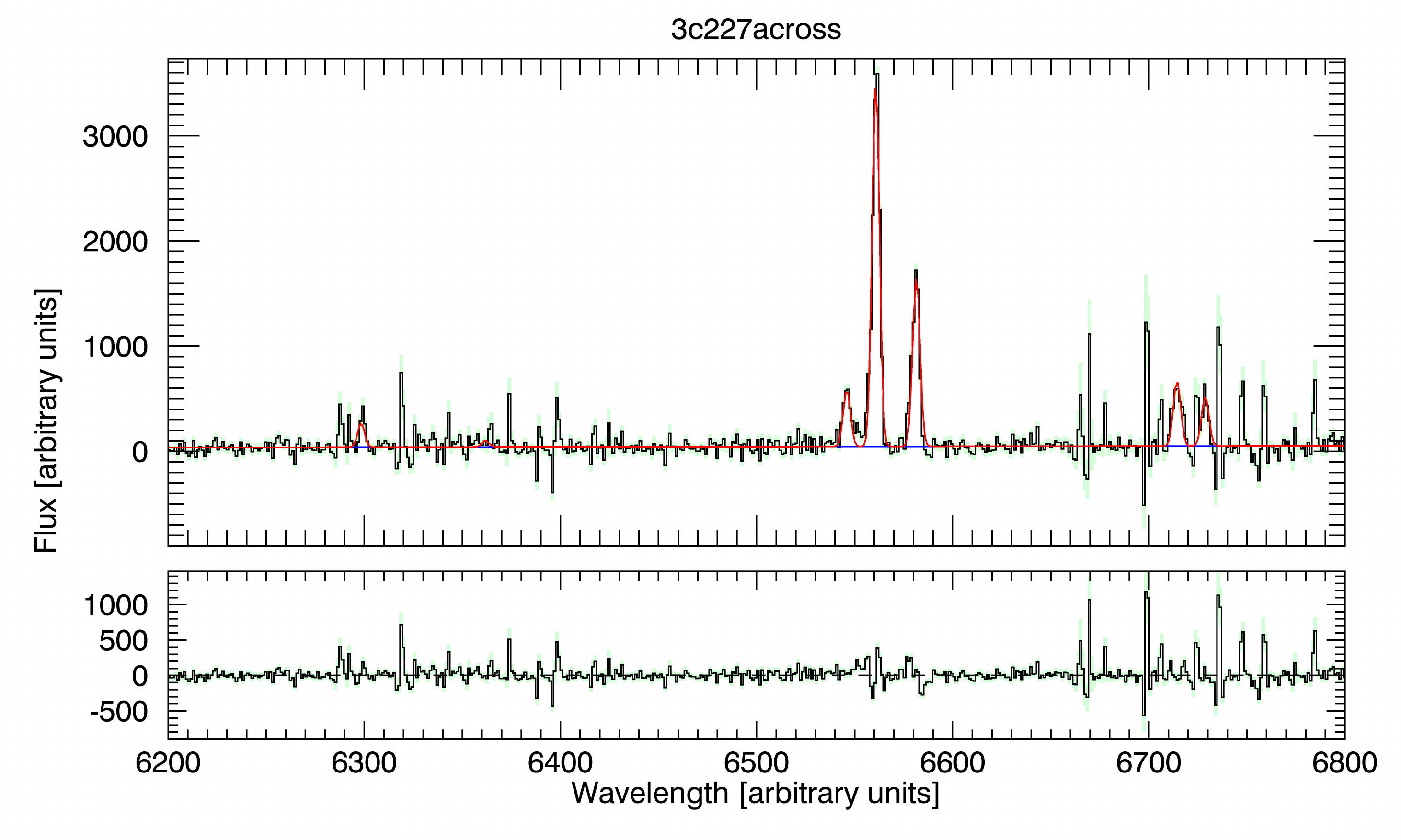}
  \includegraphics[width=7cm]{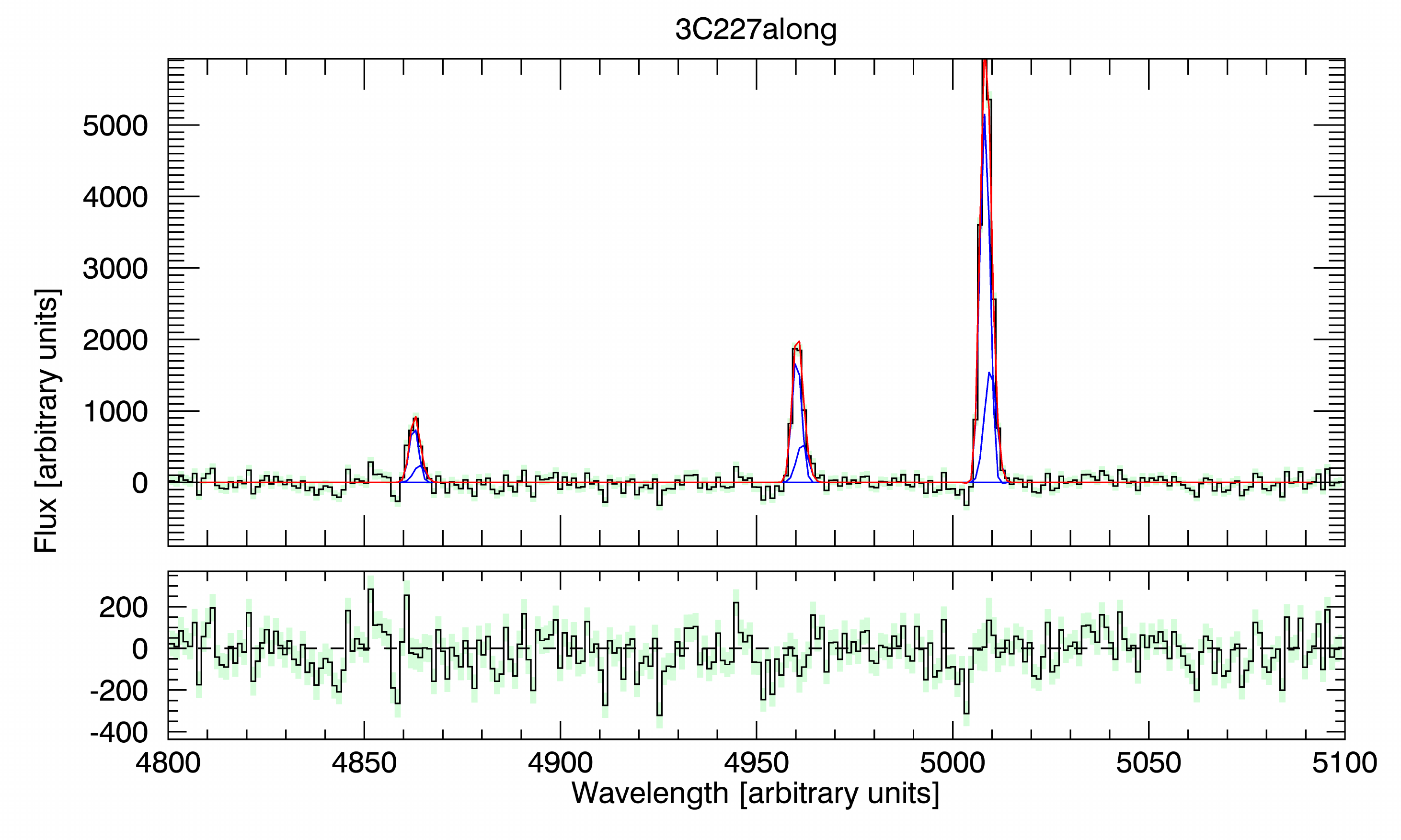}
  \includegraphics[width=7cm]{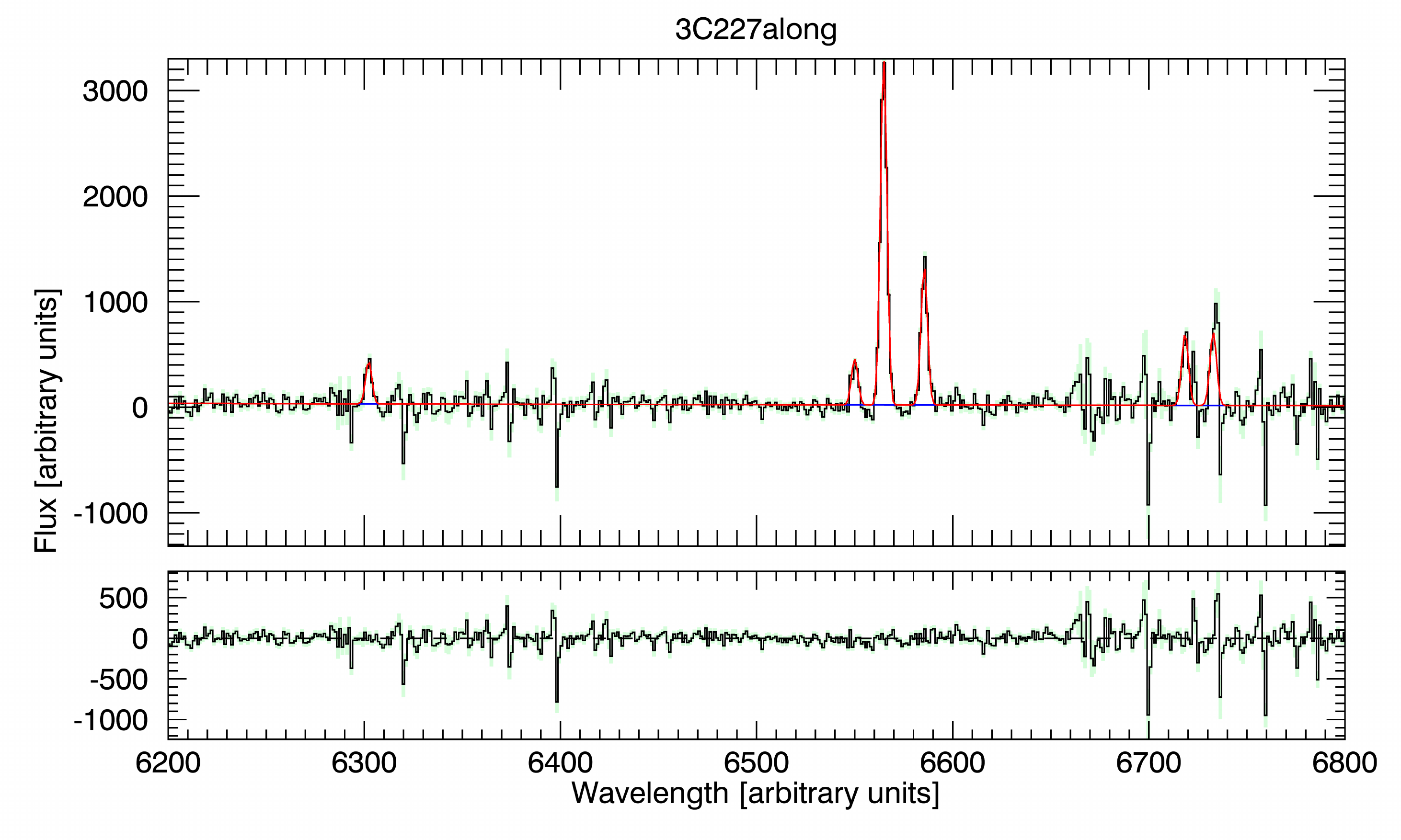}
  \includegraphics[width=7cm]{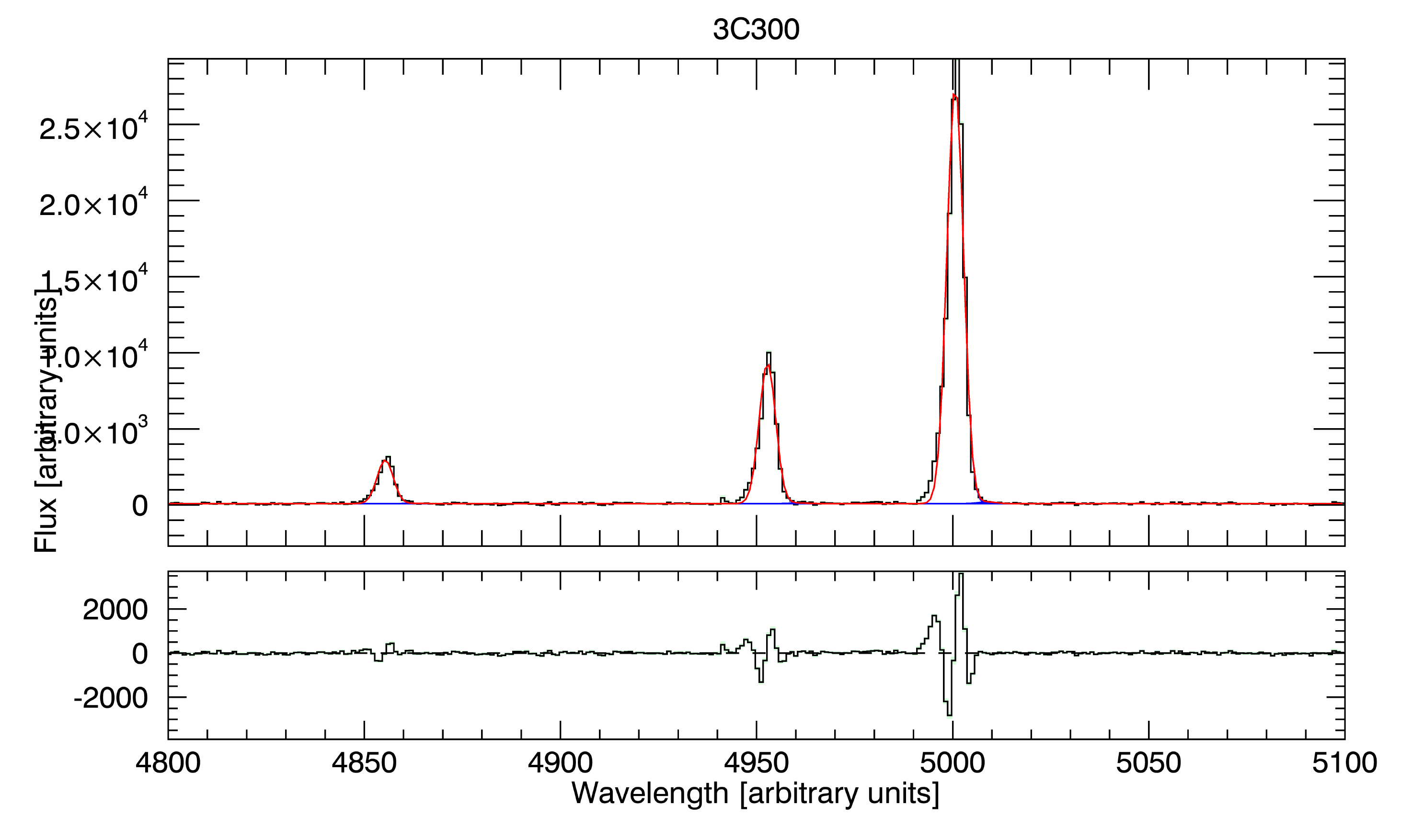}
  \includegraphics[width=7cm]{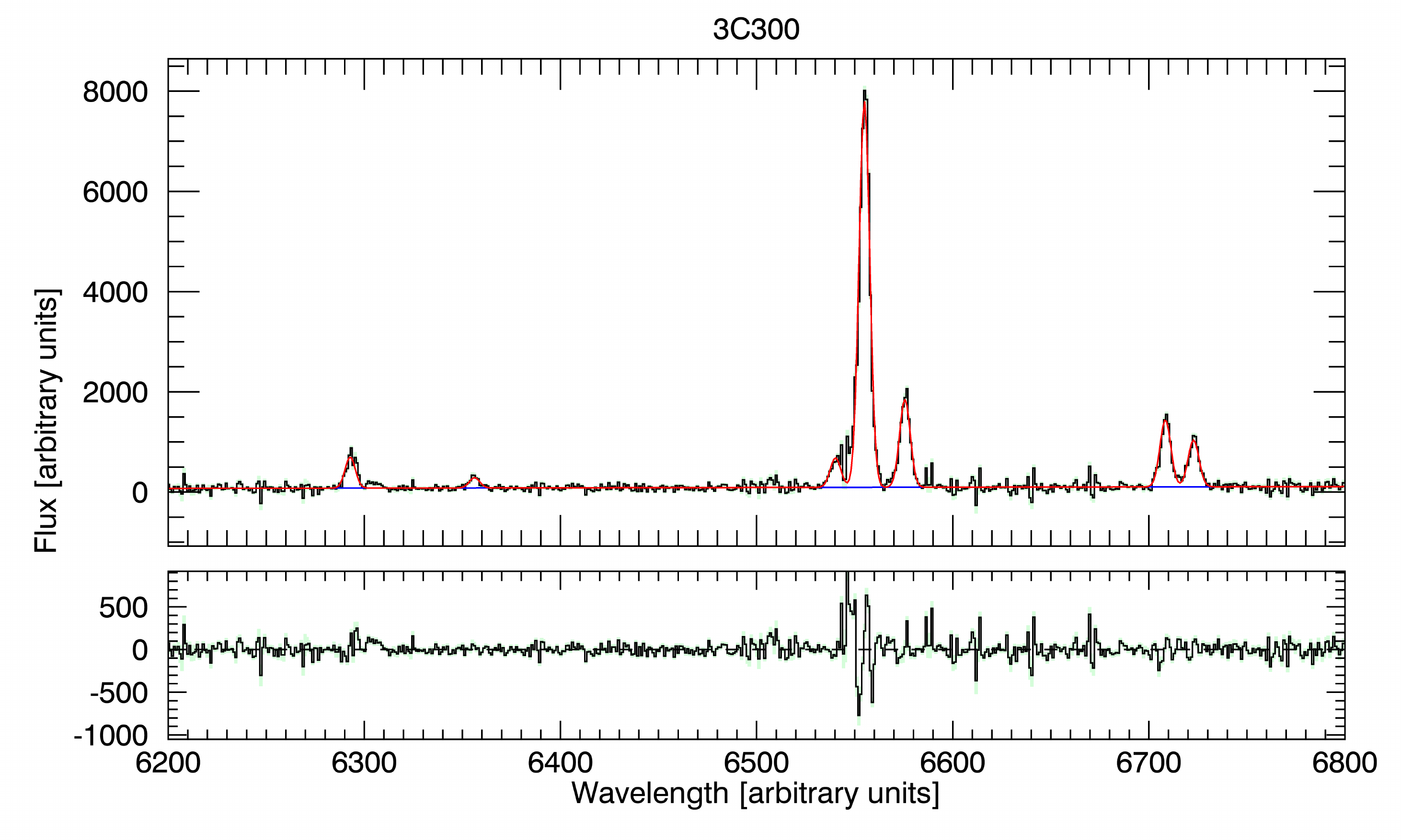}
}
\caption{Spectra
}
\end{figure*}   

\end{appendix}

\end{document}